\documentstyle[aps,eqsecnum]{revtex}

\newcommand{\balpha}{{\mbox{\boldmath$\alpha$}}}
\newcommand{\bnabla}{{\mbox{\boldmath$\nabla$}}}
\newcommand{\be}{\begin{eqnarray}}
\newcommand{\ee}{\end{eqnarray}}
\newcommand{\la}{\langle}
\newcommand{\ra}{\rangle}
\newcommand{\lbr}{\langle}
\newcommand{\rbr}{\rangle}
\newcommand{\bfx}{{\bf x}}
\newcommand{\bfy}{{\bf y}}

\newcommand{\bfz}{{\bf z}}

\newcommand{\bfk}{{\bf k}}
\newcommand{\bfp}{{\bf p}}
\newcommand{\veps}{\varepsilon}
\newcommand{\vare}{\varepsilon}
\newcommand{\eps}{\epsilon}
\newcommand{\beps}{{\mbox{\boldmath$\epsilon$}}}

\newcommand{\bmu}{{\mbox{\boldmath$\mu$}}}
\newcommand{\electronline}{\line(0,1){100}}

\newcommand{\hyperfine}{    
\multiput(0,0)(2,0){10}{\line(1,0){1}}    
\put(20,0){\circle*{3}} 
}    
\newcommand{\hyperfineleft}{    
\multiput(-20,0)(2,0){10}{\line(1,0){1}}    
\put(-20,0){\circle*{3}}
}    

\newcommand{\timeline}{    
\multiput(0,0)(6,0){20}{\line(1,0){2}}       
}

\newcommand{\photonlineright}{    
\multiput(0,0)(8,0){5}{\oval(4,4)[b]}    
\multiput(4,0)(8,0){4}{\oval(4,4)[t]}}

\newcommand{\pphotarctop}{     
  \multiput(10,0)(10,10){2}{\oval(10,10)[tl]}    
  \multiput(0,0)(10,10){3}{\oval(10,10)[br]}    
  \multiput(0,40)(10,-10){3}{\oval(10,10)[tr]}    
  \multiput(10,40)(10,-10){2}{\oval(10,10)[bl]}    
}    
    
\newcommand{\pphotarcbottom}{     
  \multiput(0,0)(-10,10){3}{\oval(10,10)[bl]}    
  \multiput(-10,0)(-10,10){2}{\oval(10,10)[tr]}    
  \multiput(-10,40)(-10,-10){2}{\oval(10,10)[br]}    
  \multiput(0,40)(-10,-10){3}{\oval(10,10)[tl]}    
}

\newcommand{\photonrightup}{     
   \multiput(0,0)(8,8){5}{\oval(8,8)[tl]}    
   \multiput(0,8)(8,8){4}{\oval(8,8)[br]}    
}    

\newcommand{\photonleftup}{     
   \multiput(0,0)(-8,8){5}{\oval(8,8)[tr]}    
   \multiput(0,8)(-8,8){4}{\oval(8,8)[bl]}        
}

\newcommand{\photonvac}{    
\multiput(0,0)(8,0){4}{\oval(4,4)[b]}    
\multiput(4,0)(8,0){3}{\oval(4,4)[t]}}

\begin{document}
\title{Two-time Green function method in quantum electrodynamics
of high-Z few-electron atoms}
\author{V. M. Shabaev \\
       Department of Physics, St. Petersburg State University, \\
        Oulianovskaya Street 1, Petrodvorets,
198504 St.  Petersburg, \\  Russia }
\maketitle
\begin{abstract}
The two-time Green function  method in quantum electrodynamics
of high-Z few-electron atoms is described in detail. This method
provides a simple procedure for deriving  formulas
for the energy shift of a single level and for the energies
and wave functions of degenerate and quasi-degenerate states.
It also allows one to derive formulas for the
transition and scattering amplitudes. Application of the
method to resonance scattering processes yields 
a systematic theory for the spectral line shape.
The practical ability of the method is demonstrated by
deriving formulas for the QED and
interelectronic-interaction  corrections to energy levels
and transition and scattering amplitudes
 in one-, two-, and three-electron atoms.
Numerical calculations of the Lamb shift, the hyperfine
splitting, the bound-electron $g$ factor, and the radiative
recombination cross section in heavy ions  are also reviewed.
\newline
PACS number(s): 12.20.-m, 12.20.Ds, 31.30. Jv
\end{abstract}
\newpage

\tableofcontents

\newpage

\section{Introduction}
A great progress in experimental investigations of high-Z few-electron
systems (see, e.g., \cite{beyer97,beyer99})
stimulated theorists to perform
accurate calculations for these systems in the framework of
 quantum electrodynamics (QED). 
 The calculations of QED and interelectronic-interaction
corrections in high-Z few-electron systems are conveniently
 divided into two stages. The first stage consists in deriving
 formal expressions
for these corrections from the first principles of QED.
The second comprises numerical evaluations
of these expressions.
The present paper is mainly focused on the first stage.
As to the numerical calculations, we give only
a short overview of them in this paper.
For more details we refer to
\cite{shabaev00a,mohr98a,sapir99,beier00,lindgren99}.

Historically, the first method suitable for deriving
the formal expressions for the energy shift of a bound-state
level was formulated by Gell-Mann, Low, and Sucher
\cite{gellmann,sucher}. This method is based on introducing
an adiabatically damped factor, $\exp{(-\lambda|t|)}$,
 in the interaction Hamiltonian and expressing the
energy shift in terms of so-called adiabatic 
 $S_{\lambda}$ matrix elements. Due to its simple formulation,
the Gell-Mann--Low--Sucher formula for the energy
shift gained wide spreading in the literature related
to high-Z few-electron systems
\cite{labz70,braun80,dmit84,braun84,mohr89,lindgren89,labz93,sapir98}.
However, the practical use of this method showed that it
has several serious drawbacks. One of them is
the very complicated derivation of the formal expressions
for so-called 
{\it reducible} diagrams. By "reducible diagrams"
we denote those diagrams where an intermediate-state 
energy of the atom coincides with the reference-state energy.
(This terminology is quite natural since it can be considered
as an extension of  the definitions
introduced  by Dyson \cite{dyson49} 
and by Bethe and Salpeter \cite{bethe51} 
to high-Z few-electron atoms.)
As to {\it irreducuble} diagrams, i.e. those diagrams where
the intermediate-state energies differ from the 
reference-state energy, the derivation of the formal expressions
can easily be reduced to the ``usual ($\lambda=0$)
 $S$-matrix'' elements in each method, including the
Gell-Mann--Low--Sucher method as well (see, e.g., 
\cite{dmit84,labz93}).
Another serious drawback of the Gell-Mann--Low--Sucher
method is the fact that this method requires   
 special investigation of the renormalization procedure since
the adibatic $S_{\lambda}$-matrix suffers from 
ultraviolet divergences. The adiabatically damped factor,
$\exp{(-\lambda |t|)}$, is non-covariant and, therefore,
 the ultraviolet divergences can
not be removed from $S_{\lambda}$ if $\lambda\ne 0$.
However, from the physical point of view one may expect 
these divergenes to cancel each other in the expression
for the energy shift. Therefore, they may be disregarded in the
calculation of the energy shift for a single level.
For the case of degenerate levels, however,
 this problem remains since we can not expect that 
the standard
renormalization procedure makes the secular operator
finite in the ultraviolet limit \cite{braun80,braun84}.
In addition, we should note that at present there is no
 formalism based on the Gell-Mann--Low--Sucher approach 
which would provide a proper treatment of quasidegenerate 
levels. Also no formalism in the framework of this
approach was developed for calculation of the transition 
or scattering amplitudes up to now. 
The same difficulties emerge in the evolution
operator method developed in Refs.
\cite{hubbard57,schweber61,vasil'ev75,zapryagaev85,sapirstein87}.
Attempts to solve some of these problems by 
modifying the Gell-Mann--Low--Sucher method 
were recently undertaken  in \cite{dmi97,dmi98,lin00}.

Another way to formulate a perturbation theory for
 high-Z few-electron systems  consists
in using Green functions. These functions contain the complete
information about the energy levels and the transition
and scattering amplitudes.  In this way,
the renormalization problem
does not appear, since Green functions
can be renormalized from the very beginning (see, e.g.,
\cite{itzykson}). Up to now, various versions of
the Green function formalism were developed which differ 
from each other by the methods of extracting
the physical information  from Green functions, i.e. 
the energy
levels and the transition and scattering amplitudes.
 One of these methods was worked 
out in 
\cite{shab88a,shab88b,shab90a,shab90b,shab91}. It was 
 successfully employed in many practical calculations 
\cite{kar92,sh93,sh94a,sh94b,sh95,yer97,art97,sh98,art99,yer99,art00,sh00,yer00}.
Since one of the key elements of this method consists
in using two-time Green functions, in what follows 
we will call it the two-time Green function (TTGF) method.
This method, which provides a solution of all the problems
appearing in the other formalisms indicated above,
will  be considered in detail in the present work.

As to other versions of the Green function method
\cite{braun84,sapir98,br77,br86,br87,sh85,sh88,zap87,fel87,ful89,yen89},
a detailed discussion of them would be beyond the scope
of the present paper. We  note only that some of these
methods are also based on  employing
 two-time Green functions but yield other forms of 
perturbation theory. 
In Refs. \cite{braun84,br86,br87,sh85,sh88}, the two-time
Green functions were used for constructing quasipotential
equations for high-Z few-electron systems.
This corresponds to the perturbation theory
in the Brillouin-Wigner form while the method of Refs.
 \cite{shab88a,shab88b,shab90a,shab90b,shab91}
yields the perturbation theory in the Rayleigh-Schr\"odinger form.
Various versions of the Bethe-Salpeter equation derived
from the 2$N$-time Green function formalism
  for high-Z few-electron systems
can be found in \cite{braun84,zap87}.
In \cite{fel87,ful89} the perturbation theory
in the Rayleigh-Schr\"odinger form is constructed
for the case of a one-electron system where the problem
of relative time coordinates for the electrons does not occur.

The relativistic unit system ( $ \hbar = c = 1$ ) and the
Heaviside charge unit ($\alpha=\frac{e^{2}}{4\pi}, e<0$)
are used in the paper.

\section{Energy levels of atomic systems}

In this section we formulate the perturbation theory
for the calculation of the energy levels in high-Z few-electron
atoms. In these systems the number of electrons
denoted by $N$  is much smaller than the nuclear charge
number $Z$. It follows that the interaction of the electrons
with each other and with the quantized electromagnetic field
is much smaller (by factors $1/Z$ and $\alpha$, respectively)
 than the interaction of the
electrons with the Coulomb field of the nucleus.
Therefore, it is natural to assume that in zeroth approximation 
the electrons interact only with the Coulomb field of
the nucleus and obey the Dirac equation
   \begin{eqnarray}
 (-i\mbox{\boldmath $\alpha$}\cdot \mbox{\boldmath $\nabla$}
+\beta m+V_{\rm C}(\bfx))\psi_{n}(\bfx)= 
 \varepsilon_{n}\psi_{n}(\bfx)\,.
 \label{dirac}
    \end{eqnarray}  
The interaction of the electrons with each other and with
the quantized electromagnetic field is accounted for by perturbation
theory. In this way we obtain quantum electrodynamics in the 
Furry picture. It should be noted that we could start also with
the Dirac equation with an effective potential $V_{\rm eff}({\bf x})$
 which 
approximately describes the interaction with the other electrons.
In this case the interaction with the potential 
$\delta V(\bfx)=V_{\rm C}(\bfx)-V_{\rm eff}(\bfx)$
must be accounted for perturbatively. 
 Using the effective potential provides an extension of the 
theory to many-electron atoms where, for instance,
 a local version
of the Hartree-Fock potential can be used as $V_{\rm eff}(\bfx)$.
 However, for simplicity,
 in what follows we will assume that in zeroth
 approximation the electrons interact only with the Coulomb
field of the nucleus.

In the present paper we will mainly consider the perturbation theory
 with the standard QED vacuum. The transition to the formalism
in which closed shells are regarded as belonging to the vacuum 
is realized by replacing $i0$ with $-i0$ in the electron 
propagator denominators corresponding to the closed shells. 

Before formulating  the perturbation theory
for calculations of the interelectronic interaction and
radiative corrections to the energy levels,
we consider standard equations of 
the Green function approach in quantum electrodynamics.

\subsection{2$N$-time Green function}

It can be shown that the complete information about the energy levels
of an $N$-electron atom
is contained in the Green function defined as
\begin{eqnarray} \label{green1}
G(x_{1}^{\prime},\dots x_{N}^{\prime};x_{1},\dots x_{N})
=\langle 0|T\psi(x_{1}^{\prime})\cdots\psi(x_{N}^{\prime})\overline{\psi}
(x_{N})\cdots\overline{\psi}(x_{1})|0\rangle\,,
\end{eqnarray}
where $\psi(x)$ is the electron-positron field operator in the Heisenberg
representation, $\overline{\psi}(x)=\psi^{\dag}\gamma^{0}$,
 and $T$ is the time-ordered product operator.
The basic equations of  quantum electrodynamics in the Heisenberg
representation are summarized in Appendix A. 
Equation (\ref{green1}) presents a standard definition of
the $2N$-time
Green function which is a fundamental object of  quantum
electrodynamics. It can be shown (see, e.g., 
\cite{itzykson,bjorken}) that in the interaction 
representation the Green function is given by
\begin{eqnarray} \label{green2}
\lefteqn{G(x_{1}^{\prime},\dots x_{N}^{\prime};x_{1},\dots x_{N})
\;\;\;\;\;\;\;\;\;}\nonumber\\
&=&\frac{
\langle 0|T\psi_{\rm in}(x_{1}^{\prime})\cdots\psi_{\rm in}(x_{N}^{\prime})
\overline{\psi
}_{\rm in}
(x_{N})\cdots\overline{\psi}_{\rm in}(x_{1})
\exp{\{-i\int d^{4}z \; {\cal H}_{I}
(z)\}}
|0\rangle}
{\langle 0|T\exp{\{-i\int d^{4}z \; {\cal H}_{I}
(z)\}}
|0\rangle}\\
&=&\Bigl \{ \sum_{m=0}^{\infty}\frac{(-i)^m}{m!}\int
d^4y_1\cdots d^4y_m \;
\langle 0|T\psi_{\rm in}(x_{1}^{\prime})\cdots\psi_{\rm in}(x_{N}^{\prime})
\overline{\psi
}_{\rm in}
(x_{N})\cdots\overline{\psi}_{\rm in}(x_{1})\nonumber\\
&&\times
{\cal H}_{I}(y_1)\cdots
{\cal H}_{I}(y_m)|0\rangle\Bigr\}
\Bigl \{ \sum_{l=0}^{\infty}\frac{(-i)^l}{l!}\int
d^4z_1\cdots d^4z_l \;
\langle 0|T{\cal H}_{I}(z_1)\cdots
{\cal H}_{I}(z_l)|0\rangle\Bigr\}^{-1}
\label{green3}
\end{eqnarray}
where 
\be \label{intham}
{\cal H}_{I} (x)=\frac{e}{2}\,[\overline{\psi}_{\rm in}(x)
\gamma_{\mu},\psi_{\rm in}(x)]
A_{in}^{\mu}(x)-\frac{\delta m}{2}\,
[\overline{\psi}_{\rm in}(x),\psi_{\rm in}(x)] 
\ee
is the interaction Hamiltonian. The  commutators in equation (\ref{intham})
refer to operators only.
The first term in (\ref{intham})
 describes the interaction of the electron-positron field
with the quantized electromagnetic field
and the second one is the mass renormalization
counterterm.  
 We consider here that
the interaction of the electrons with the Coulomb field of the nucleus
is included in the unperturbed Hamiltonian, i.e. the Furry picture.
However, there is also an alternative method to get
the Furry picture. In that method the interaction with the Coulomb
field of the nucleus is included in the interaction Hamiltonion
and the Furry picture is obtained by summing infinite sequences
of Feynman diagrams describing the interaction of the electrons with
the Coulomb potential.  As a result of this summation,
the free-electron propagators are replaced by bound-electron 
propagators.
 This method is very convenient for studying
processes involving continuum-electron states.   
It  will be used in the section
concerning the radiative recombination process.

The Green function $G$ is constructed by perturbation theory
according to equation (\ref{green3}). This is carried
out with the aid of the Wick theorem (see, e.g., \cite{itzykson}).
According to this theorem the time-ordered product of field
operators is equal to the sum of  normal-ordered products
with all possible contractions between the operators
\be \label{wick}
T(ABCD\cdots)&=&N(ABCD\cdots) +N(A^aB^aCD\cdots)+
N(A^aBC^aD\cdots)\nonumber\\
&&+ {\rm\; all\;\, possible\;\, 
contractions}\,,
\ee
where $N$ is the normal-ordered product operator and
the superscripts denote the contraction between the
corresponding operators. The contraction between 
 neighbouring operators is defined by
\be
A^aB^a=T(AB)-N(AB)=\langle 0|T(AB)|0\rangle\,.
\ee
If the contracted operators are boson operators, they can
be put one next to another. If the contracted operators
are fermion operators, they also can be put one
next to another but in this case the expression must 
be multiplied with the parity of the permutation of the fermion
operators. Since in the Green function the vacuum expectation
value is calculated, only the term with all operators contracted
remains on the right-hand side of equation
(\ref{wick}).
In contrast to the free-electron QED, in the Furry picture
the time-ordered product of two fermion operators
 must be defined also for the equal-time
case to obtain the correct vacuum polarization terms.
As was noticed in \cite{mohr85}, the definition
\be
T[A(t) B(t)]=\frac{1}{2}A(t)B(t)-\frac{1}{2}B(t)A(t)
\ee
provides the following
 simple rule for dealing with the interaction operator.
It can be written as
\be \label{intham1}
{\cal H}_{I} (x)=e\,\overline{\psi}_{\rm in}(x)
\gamma_{\mu}\psi_{\rm in}(x)
A_{in}^{\mu}(x)-\delta m\,
\overline{\psi}_{\rm in}(x)\psi_{\rm in}(x)
\ee
and then the Wick theorem is applied
 with contractions between all operators,
including equal-time  operators. We note that the
problem of the definition of the time-ordered product of 
fermion operators at equal times
 does not appear at all 
if the alternative method for obtaining the Furry picture
discussed above is employed.

The contractions between the electron-positron fields
and between the photon fields
 lead to the following propagators
\be \label{propel}
\langle 0|T\psi_{\rm in}(x)\overline{\psi}_{\rm in}(y)|0\rangle
=\frac{i}{2\pi}\int_{-\infty}^{\infty} d\omega \;
\sum_{n}\frac{\psi_n(\bfx)\overline{\psi}_n(\bfy)}{\omega
-\varepsilon_n(1-i0)}\exp{[-i\omega(x^0-y^0)]}
\ee
and
\be \label{propph}
\langle 0|T A_{\rm in}^{\mu}(x)A_{\rm in}^{\nu}(y)|0\rangle
=-i g^{\mu \nu}\int \frac{d^4 k}{(2\pi)^4} \;
\frac{\exp{[-ik\cdot (x-y)]}}{k^2+i0}\,.
\ee
Here the Feynman gauge is considered. In equation (\ref{propel})
the index $n$ runs over all bound and continuum states.

The denominator in equation (\ref{green2}) describes 
unobservable vacuum-vacuum transitions and,
as can be shown (see, e.g., \cite{itzykson}), it
 cancels disconnected vacuum-vacuum subdiagrams from 
the numerator.
Therefore, we can simply omit all diagrams containing
disconnected
vacuum-vacuum subdiagrams in the numerator and replace 
the denominator by 1.

In practical calculations  of the Green function
it is  convenient to work
with the Fourier transform with respect to time variables,
\begin{eqnarray} \label{green2np}
 \lefteqn{G((p_{1}^{\prime 0},\bfx_1'),
\ldots,(p_{N}^{\prime 0},\bfx_N');(p_{1}^{0},\bfx_1),
\ldots,(p_{N}^{0},\bfx_N)) 
   \;\;\;\;\;\;\;\;\;\;\;\;\;\;\;\;\;\;\;\;\;\;\;}  \nonumber\\  
  & =& (2\pi)^{-2N} \int_{-\infty}^{\infty}
  dx_{1}^{0}\cdots dx_{N}^{0} dx_{1}^{\prime 0}\cdots dx_{N}^{\prime 0} 
      \nonumber  \\ & &
   \times \exp{(ip_{1}^{\prime 0}x_{1}^{\prime 0}+
\cdots+ip_{N}^{\prime 0}x_{N}^{\prime 0}-
  ip_{1}^{0}x_{1}^{0}-\cdots-ip_{N}^{0}x_{N}^{0} )} 
   \nonumber \\ &&
    \times 
G(x_1',...,x_N';x_1,...,x_N)\,.
\end{eqnarray}
For the Green function  
$G((p_{1}^{\prime 0},\bfx_1'),
\ldots,(p_{N}^{\prime 0},\bfx_N');(p_{1}^{0},\bfx_1),
\ldots,(p_{N}^{0},\bfx_N))$, the following Feynman rules
can be derived:\newline
(1) External electron line \newline \ \ 
\newline
\setlength{\unitlength}{0.5mm}
\thicklines
\begin{picture}(60,7)(0,0)
  \put(15,6){\line(1,0){45}}
  \put(15,0){${\bf x}$}
  \put(58,0){${\bf y}$}
  \put(37,6){\vector(-1,0){1}}
  \label{extelline}
\end{picture}
$\;\;\;\;\;\;\; \frac{i}{2\pi}S(\omega ,{\bf x},{\bf y})$\,, \newline \\
\newline   where     
    \begin {eqnarray}    S(\omega ,{\bf x},{\bf y})
                       =\sum_{n} \frac{\psi_{n}({\bf x})
                    \overline{\psi}_{n}
                      ({\bf y})}{\omega -\varepsilon_{n}(1-i0)}\,,
        \label{e2e5}
   \end{eqnarray}
 $ \psi_{n}({\bf x})$  are solutions of the Dirac equation
(\ref{dirac}).\newline 
(2) Internal electron line \newline  \\ 
\newline 
\begin{picture}(60,7)(0,0)
  \put(15,6){\line(1,0){45}}
  \put(15,0){${\bf x}$}
  \put(58,0){${\bf y}$}
  \put(37,6){\vector(-1,0){1}}
  \label{intelline}
\end{picture}
          $\;\;\;\;\;\;\; \frac{i}{2\pi} \int_{-\infty}^{\infty}d\omega \; $
          $   S(\omega ,{\bf x},{\bf y})$\,.
\newline \ \ 
\newline
(3) Disconnected electron line 
 \newline  \\ 
\newline
\begin{picture}(60,7)(0,0)
  \put(15,6){\line(1,0){45}}
  \put(15,0){${\bf x}$}
  \put(58,0){${\bf y}$}
  \put(37,6){\vector(-1,0){1}}
  \label{sepelline}
\end{picture}
$\;\;\;\;\;\;\;
 \frac{i}{2\pi}S(\omega ,{\bf x},{\bf y})\delta (\omega-\omega^{'})$\,.
\newline \ \ 
\newline
(4) Internal photon line \newline \\
\newline
\begin{picture}(60,7)(0,0)
\put(15,6){\photonlineright}
  \put(10,-2){${\bf x}$}
  \put(49,-2){${\bf y}$}
  \label{intphotline}
\end{picture}
          $\;\; \frac{i}{2\pi} \int_{-\infty}^{\infty}d\omega \; D_{\rho\sigma}
            (\omega,{\bf x}-{\bf y})$\,, \newline \ \ 
\newline
where, for zero photon mass,  
        $  D_{\rho\sigma} (\omega,{\bf x}-{\bf y})$
is given by
     \begin{eqnarray}
            D_{\rho\sigma} (\omega,{\bf x}-{\bf y})=
              -g_{\rho\sigma}\int \frac{d{\bf k}}
                   {(2\pi)^{3}}\;
\frac{\exp{(i{\bf k}\cdot({\bf x}-{\bf y}))}}
                           {\omega^{2}-{\bf k}^{2}+i0}             
         \label{e2e7}
     \end{eqnarray} 
in the Feynman gauge 
 and by  \newline
\be 
 D_{00} (\omega,{\bf x}-{\bf y})&=&\frac{1}
                                       {4\pi|{\bf x}-{\bf y}|} \; ,
  \;\;  D_{i0}=D_{0i}=0 \;\;\;\;\;\;\;  (i=1,2,3)\,,\\
     D_{il}(\omega,{\bf x}-{\bf y})&=& 
               \int \frac{d{\bf k}}
                   {(2\pi)^{3}} \;
\frac{\exp{(i{\bf k}\cdot({\bf x}-{\bf y}))}}
                        {\omega^{2}-{\bf k}^{2}+i0}
              \Bigl(\delta_{il}-\frac{k_{i}k_{l}}
                                {{\bf k}^{2}}\Bigr)            
   \;\;\;\;\;  (i,l=1,2,3)\,,    
           \label{e2e8}
\end{eqnarray}    
 in the Coulomb gauge. (In this work we assume that the Coulomb gauge
is used only for diagrams which do not involve a renormalization
procedure. The renormalization in the Coulomb gauge is discussed
in Refs. \cite{adkins1,adkins2}.)
\newline
 (5) Vertex \newline
\begin{picture}(60,70)(0,0)
  \put(50,30){\line(-1,2){15}}
  \put(50,30){\line(-1,-2){15}}
\put(52,30){\photonlineright}
  \put(42,28){$\bfx$}
  \put(70,35){$\omega_{2}$}
  \put(28,15){$\omega_{1}$}
  \put(28,45){$\omega_{3}$}
  \put(66,32){\vector(1,0){1}}
  \put(44,42){\vector(-1,2){1}}
  \put(44,18){\vector(1,2){1}}
\label{vertex}
\end{picture}  
       $\;\;\;\;\;
 -2\pi ie\gamma^{\rho}\delta(\omega_{1}-\omega_{2}-\omega_{3})
            \int d{\bf x} $\,. \newline \ \ 
\newline
(6) The mass counterterm
\newline \ \ 
\newline
\setlength{\unitlength}{0.5mm}
\thicklines
\begin{picture}(60,60)(0,0)
  \put(50,30){\line(0,1){20}}
  \put(50,30){\line(0,-1){20}}
  \put(40,15){$\omega$}
  \put(40,45){$\omega'$}
 \put(40,28){$\bfx$}
  \put(50,42){\vector(0,1){1}}
  \put(50,21){\vector(0,1){1}}
  \put(46,26){\line(1,1){8}}
  \put(46,34){\line(1,-1){8}}
  \label{counterterm}
\end{picture}
$\;\;\;2\pi i\delta(\omega-\omega')\delta m\int d{\bfx}$\,.
\newline
(7) Symmetry factor $ (-1)^{P}$, where $ P $ is the parity of the 
   permutation
of the final electron coordinates with respect to the initial
ones. \newline
(8) Factor ($-1$) for every closed electron loop.
\newline
(9) If, in addition, an external potential $\delta V(\bfx)$
is present, an additional vertex appears,
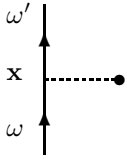
\newline
\begin{picture}(60,60)(0,0)
  \put(50,30){\line(0,1){20}}
  \put(50,30){\line(0,-1){20}}
  \multiput(50,30)(2,0){10}{\line(1,0){1}}
  \put(40,15){$\omega$}
  \put(40,45){$\omega'$}
 \put(40,30){$\bfx$}
 \put(70,30){\circle*{3}}
  \put(50,42){\vector(0,1){1}}
  \put(50,21){\vector(0,1){1}}
\label{vertex2}
\end{picture}  
       $ -2\pi i\gamma^{0}\delta(\omega-\omega')
            \int d{\bf x}\,\delta V(\bfx) $\,. \newline \ \ 
\newline

In principle, the Green function $G$ 
contains the complete information
about the energy levels of the atomic system. 
This can be shown by deriving the spectral representation
for $G$. However, 
it is a hard task to extract this information directly from
$G$ since it depends on $2(N-1)$ relative times (energies)
 in the time (energy) representation. As we will see in the
next section, the two-time Green function defined as
\be
\widetilde{G}(t',t)\equiv G(t_{1}^{\prime}=t_{2}^{\prime}=\cdots
t_{N}^{\prime}\equiv t';t_{1}=t_{2}=\cdots t_{N}\equiv t)
\ee
also contains the complete information about the energy levels, and
it is a much simpler task to extract the energy levels from
 $\widetilde{G}$. 

\subsection{Two-time Green function (TTGF) and its analytical
properties}

Let us introduce the Fourier transform of the two-time Green function
by
\begin{eqnarray} \label{gfirst}
{\cal G}(E;{\bf x}_1^{\prime},...{\bf x}_N^{\prime};
 {\bf x}_1,...{\bf x}_N) \delta (E-E^{\prime}) = 
\frac{1}{2\pi i}\frac{1}{N!}\int_{-\infty}^
{\infty}dx^0dx'^0 \; \exp{(iE^{\prime}x'^0-iEx^0) }\nonumber\\
\times \langle 0|T\psi (x'^0,{\bf x}_1
^{\prime})\cdots\psi (x'^0,{\bf x}_N^{\prime})\overline{\psi}
(x^0,{\bf x}_N)\cdots\overline{\psi}
(x^0,{\bf x}_1)|0\rangle\, ,
\end{eqnarray}
where, as in (\ref{green1}),
 the Heisenberg representation for the electron-positron
field operators is used. Defined by equation (\ref{gfirst}) for real $E$,
the Green function ${\cal G}$ can be continued analytically to the 
complex $E$ plane. 
 Analytical properties of this type of Green
functions in the complex $E$ plane 
were studied in various fields of physics (see, e.g.,
\cite{bonch,thouless,migdal}).
In quantum field theory they were considered  in detail
by Logunov and Tavkhelidze in \cite{logunov}
(see also \cite{faustov}), where
the two-time Green function was employed
 for constructing a quasipotential equation.
To study the analytical properties of the two-time Green function
we derive the spectral representation  for $\cal G$.
 Using the time-shift transformation rule for the Heisenberg operators
(see Appendix A)
\begin{eqnarray}
\psi(x^{0},{\bf x})=\exp{(iHx^{0})}\psi(0,{\bf x})\exp{(-iHx^{0})}\, 
\ee
and the equations
\be
H|n\rangle=E_{n}|n\rangle \,,\qquad\qquad\sum_{n}|n\rangle \langle n|=I \,,
\end{eqnarray}
where $H$ is the Hamiltonian of the system in the Heisenberg
representation, we find
\begin{eqnarray} \label{e221}
\lefteqn{{\cal G}(E;\bfx'_1,...,\bfx'_N;\bfx_1,...,\bfx_N)
\delta(E-E')\;\;\;\;\;\;}\nonumber\\
&=&\frac{1}{2\pi i}\frac{1}{N!}\int_{-\infty}
^{\infty}dx^{0}dx^{\prime 0}\,\exp{(iE'x^{\prime 0}-iEx^{0})}
\nonumber\\
&&\times \Bigl 
\{\theta(x^{\prime 0}-x^{0})\sum_{n}\exp{[i(E_{0}-E_{n})(x^{\prime 0}
-x^{0})]}
\langle 0|\psi(0,{\bf x}_{1}^{\prime})\cdots\psi(0,{\bf x}_{N}^{\prime})
|n\rangle\nonumber\\
&&\times \langle 
n|\overline{\psi}(0,{\bf x}_{N})\cdots\overline{\psi}(0,{\bf x}_{1})
|0\rangle+(-1)^{N^{2}}\theta(x^{0}-x^{\prime 0})\sum_{n}\exp{[i(E_{0}-E_{n})
(x^{0}-x^{\prime 0})]}\nonumber\\
&&\times \langle 0|\overline{\psi}(0,{\bf x}_{N})\cdots\overline{\psi}
(0,{\bf x}_{1})|n\rangle \langle n|\psi(0,{\bf x}_{1}^{\prime})\cdots\psi(0,
{\bf x}_{N}^{\prime})|0\rangle \Bigr \}\, .
\end{eqnarray}
Assuming, for simplicity, $E_{0}=0$
(it corresponds to choosing the vacuum energy as the origin of reference)
 and taking into account that
\begin{eqnarray}
\int_{-\infty}^{\infty}dx^{0}dx^{\prime 0}\,\theta(x^{\prime 0}-
x^{0})\exp{[-iE_{n}(x^{\prime 0}-x^{0})]}\exp{[i(E'x^{\prime 0}-Ex^{0})]}
\nonumber\\
=2\pi \delta(E'-E)\frac{i}{E-E_{n}+i0} \,,\\
\int_{-\infty}^{\infty}dx^{0}dx^{\prime 0}\,\theta(x^{0}-
x^{\prime 0})\exp{[-iE_{n}(x^{ 0}-x^{\prime 0})]}
\exp{[i(E'x^{\prime 0}-Ex^{0})]}\nonumber\\
=-2\pi \delta(E'-E)\frac{i}{E+E_{n}-i0}\, ,
\end{eqnarray}
we obtain 
\begin{eqnarray} \label{spect1}
{\cal G }(E) = \sum_{n}\frac{\Phi_n\overline {\Phi}_n}{E-E_n+i0}-
(-1)^N\sum_{n}\frac{\Xi_{n}\overline {\Xi}_{n}}{E+E_n-i0}\, ,
\end{eqnarray} 
where the variables $\bfx'_1,...,\bfx'_N,\bfx_1,...,\bfx_N$
are implicit and
\begin{eqnarray} \label{phi}
\Phi_n({\bf x}_1,...{\bf x}_N) = \frac{1}{\sqrt{N!}}
\langle 0|\psi(0,{\bf x}_1)\cdots\psi(0,{\bf x}_N)|n\rangle \, ,\\
\Xi_n({\bf x}_1,...{\bf x}_N) = \frac{1}{\sqrt{N!}}
\langle n|\psi(0,{\bf x}_1)\cdots\psi(0,{\bf x}_N)|0\rangle\, . 
\label{xi}
\end{eqnarray}
In equation (\ref{spect1}) the summation runs over all
bound and continuum  states of the system of the interacting
fields.
Let us introduce the functions
\be \label{aaa}
A(E;\bfx'_1,...,\bfx'_N;\bfx_1,...,\bfx_N)&=&
\sum_{n}\delta(E-E_n)\Phi_n(\bfx'_1,...,\bfx'_N)
\overline{\Phi}_n(\bfx_1,...,\bfx_N)\,,\\
B(E;\bfx'_1,...,\bfx'_N;\bfx_1,...,\bfx_N)&=&
\sum_{n}\delta(E-E_n)\Xi_n(\bfx'_1,...,\bfx'_N)
\overline{\Xi}_n(\bfx_1,...,\bfx_N)\,.
\label{bbb}
\ee
These functions satisfy the conditions
\be \label{intaaa}
\int_{-\infty}^{\infty} dE \;
A(E;\bfx'_1,...,\bfx'_N;\bfx_1,...,\bfx_N)&=&
\frac{1}{N!} \langle 0|\psi(0,{\bf x}'_{1})\cdots{\psi}
(0,{\bf x}'_{N})\nonumber\\
&&\times\overline{\psi}(0,{\bf x}_{N})\cdots
\overline{\psi}(0,
{\bf x}_{1})|0\rangle
\,,\\
\int_{-\infty}^{\infty} dE \;
B(E;\bfx'_1,...,\bfx'_N;\bfx_1,...,\bfx_N)&=&
\frac{1}{N!} \langle 0|\overline{\psi}(0,{\bf x}_{N})\cdots\overline{\psi}
(0,{\bf x}_{1})\nonumber\\
&&\times\psi(0,{\bf x}_{1}^{\prime})\cdots\psi(0,
{\bf x}_{N}^{\prime})|0\rangle\,.
\label{intbbb}
\ee
In terms of these functions, the equation (\ref{spect1})
is
\be \label{spect2}
{\cal G}(E)=\int_{0}^{\infty} dE' \; \frac{A(E')}{E-E'+i0}
-(-1)^N\int_{0}^{\infty}dE' \;\frac{B(E')}{E+E'-i0}\,,
\ee
where we have omitted the variables 
$\bfx_1,...,\bfx_N,\bfx'_1,...,\bfx'_N$
and have taken into account that $A(E')=B(E')=0$ for $E'<0$
since $E_n\ge 0$.
In fact, due to  charge conservation,  only states
with an electric charge of $eN$ contribute to $A$ in the sum over
$n$ in the right-hand side of equation (\ref{aaa})
and only states with an electric charge of $-eN$ contribute
to $B$ in the sum over $n$ in the right-hand side of equation
(\ref{bbb}). This can easily be shown by using the
following commutation relations 
\be
[Q,\psi(x)]=-e\psi(x)\,,\;\;\;\;\;\;\;
 [Q,\overline{\psi}(x)]=e\overline{\psi}(x)\,,
\ee
where $Q$ is the charge operator in the Heisenberg representation.
Therefore, the equation (\ref{spect2}) can be written as
\be \label{spect3}
{\cal G}(E)=\int_{E_{\rm min}^{(+)}}^{\infty} dE' \;\frac{A(E')}{E-E'+i0}
-(-1)^N\int_{E_{\rm min}^{(-)}}^{\infty}dE' \;\frac{B(E')}{E+E'-i0}\,,
\ee
where $E_{\rm min}^{(+)}$ is the minimal energy of states with
electric charge $eN$ and $E_{\rm min}^{(-)}$ is the minimal energy
of states with electric charge $-eN$. So far we considered 
${\cal G}(E)$ for real $E$.
 The equation (\ref{spect3})
shows that the Green function ${\cal G}(E)$ is the sum of
 Cauchy-type integrals.  Using   the fact that 
 the integrals
${\displaystyle\int_{E_{\rm min}^{(+)}}^{\infty} dE \; A(E)}$ and
${\displaystyle\int_{E_{\rm min}^{(-)}}^{\infty} dE \; B(E)}$ converge
(see equations (\ref{intaaa}), (\ref{intbbb})),
one can show with the help
of standard mathematical methods  
that the equation 
\be \label{spect4}
{\cal G}(E)=\int_{E_{\rm min}^{(+)}}^{\infty} dE' \;\frac{A(E')}{E-E'}
-(-1)^N\int_{E_{\rm min}^{(-)}}^{\infty}dE' \;\frac{B(E')}{E+E'}\,
\ee
defines an analytical function of $E$
in the complex $E$ plane with the cuts $(-\infty,E_{\rm min}^{(-)}]$
and $[E_{\rm min}^{(+)},\infty)$ (see Fig. 1).
This equation provides the analytical continuation of the 
Green function to the complex $E$ plane.
According to (\ref{spect3}), to get the Green function for real $E$
we have to approach the right-hand cut from the upper half-plane and
the left-hand cut from the lower half-plane.

In what follows we will be interested in bound states
of  the system. According to equations (\ref{spect1})-(\ref{spect4}),
the bound states correspond to the poles of the function ${\cal G}(E)$
on the right-hand real semiaxis. If
the interaction between the electron-positron field
and the electromagnetic field is switched off, the poles 
corresponding to bound states are isolated (see Fig. 2).
Switching on the interaction between the fields  transforms
the isolated poles into  branch points. This is caused by
the fact that due to zero photon mass the bound states
are no longer isolated points of the spectrum.
Disregarding the instability of excited states,
the singularities of the Green function ${\cal G}(E)$
 are shown in Fig. 3. The poles corresponding to the bound states
lie on the upper boundary of the cut starting from the pole
corresponding to the ground state. It is natural to assume
that ${\cal G}(E)$ can be continued analytically under the cut, 
to the second sheet of the Riemann surface. As a result
of this continuation the singularities of ${\cal G}(E)$ can be 
turned down as displayed in Fig. 4.

In fact due to instability of excited states the energies
of these states have small imaginary components and, 
therefore,
the related poles lie slightly  below the right-hand real semiaxis
(Fig. 5). However, in calculations
of the energy levels and the transition and scattering amplitudes
of non-resonance processes we will neglect the instability
of the excited states and, therefore, will assume that the
poles lie on the real axis. The  imaginary parts of the energies
will be taken into account when we will consider the
resonance scattering processes. 

To formulate the perturbation theory for calculations of the energy
levels and the transition and scattering amplitudes
 we will need to isolate  the poles corresponding to the
bound states from the related cuts. It can be done by
introducing a non-zero photon mass $\mu$ which is generally
 assumed to be 
larger than the energy shift (or the energy splitting) of the
level (levels) under consideration and much smaller than the
distance to other levels. The singularities of ${\cal G}(E)$ 
with non-zero photon mass, including one- and
two-photon spectra, are shown in Fig. 6. As one can see 
from this figure, introducing the photon mass makes the 
poles corresponding to the bound states to be isolated.

In every finite order of perturbation theory 
the singularities of the Green function ${\cal G} (E)$
in the complex E-plane are defined by the unperturbed Hamiltonian.
In quantum mechanics this fact easily follows from
the expansion of the Green function $(E-H)^{-1}=(E-H_0-\delta V)^{-1}$
in powers of the perturbation potential $\delta V$
\be \label{resolv}
(E-H)^{-1}=\sum_{n=0}^{\infty}(E-H_0)^{-1}[\delta V (E-H_0)^{-1}]^n\,.
\ee
As one can see from this equation, to $n$-th order of 
perturbation theory the Green function has poles of all orders
till $n+1$ at the unperturbed positions of the bound state energies.
This fact remains also valid in quantum electrodynamics
for ${\cal G}(E)$ defined above.  It can easily be checked for 
every specific diagram in  first and second order in $\alpha$.
A general proof for an arbitrary diagram is given in Appendix B.

\subsection{Energy shift of a single level}

In this section we are interested in the energy shift
 $ \Delta E_{a}=E_{a}-E_{a}^{(0)}$ 
of a single isolated level $a$ of an $N$-electron atom
 due to the perturbative interaction.
 The unperturbed energy $E_a^{(0)}$
is equal to the sum of the one-electron Dirac-Coulomb energies
\be
E_a^{(0)}=\veps_{a_1}+\cdots +\veps_{a_N}\,,
\ee
which are defined by the Dirac equation (\ref{dirac}).
In the simplest case the unperturbed wave function 
$u_a(\bfx_1,...,\bfx_N)$ is a one-determinant function 
\be \label{onedet}
u_a(\bfx_1,...,\bfx_N)=\frac{1}{\sqrt{N!}}\sum_{P}(-1)^P
\psi_{Pa_1}(\bfx_1)\cdots\psi_{Pa_N}(\bfx_N)\,,
\ee
where $\psi_n$ are the one-electron Dirac wave functions
defined by equation (\ref{dirac}) and
 $P$ is the permutation operator.
 In the general case
the unperturbed wave function is a linear combination
of the one-determinant functions
\be \label{manydet}
 u_a(\bfx_1,...,\bfx_N)=\sum_b C_a^b 
\frac{1}{\sqrt{N!}}\sum_{P}(-1)^P
\psi_{Pb_1}(\bfx_1)\cdots\psi_{Pb_N}(\bfx_N)\,.
\ee
We introduce the Green function $g_{aa}(E)$ by 
\begin{eqnarray} \label{gsmall}
  g_{aa}(E)&=&\langle u_{a}|{\cal G}(E)\gamma_1^0\cdots
\gamma_N^0|u_{a} \rangle\nonumber\\
&\equiv&\int d\bfx_1\cdots d\bfx_N d\bfx'_1\cdots d \bfx'_N \;
u^{\dag}_a(\bfx'_1,...,\bfx'_N)\nonumber\\
&&\times {\cal G}(E,\bfx'_1,...,
\bfx'_N;\bfx_1,...,\bfx_N)\gamma_1^0\cdots\gamma_N^0
u_a(\bfx_1,...,\bfx_N)\,.
\end{eqnarray}
>From the spectral representation for ${\cal G}(E)$
(see equations (\ref{spect1})-(\ref{spect4})) we have 
\begin{eqnarray}
 g_{aa}(E)=  \frac{A_{a}}{E-E_{a}} 
 + \mbox{terms that are regular at } E \sim E_{a}\,,
\end{eqnarray}
where
\begin{eqnarray}
 A_{a}&=&\frac{1}{N!} \int d{\bf x}_{1} \cdots d{\bf x}_{N}
		d{\bf x}_{1}^{\prime} \cdots d{\bf x}_{N}^{\prime}\,
	u^{\dag}_{a}({\bf x}_{1}^{\prime},\ldots,{\bf x}_{N}^{\prime}) 
\langle 0|\psi (0,{\bf x}_{1}^{\prime})
\cdots \psi(0,{\bf x}_{N}^{\prime})|a\rangle 
                      \nonumber \\ 
       &&\times \langle a|\psi^{\dag} (0,{\bf x}_{N})
			\cdots \psi^{\dag}(0,{\bf x}_{1})
                         |0 \rangle 
	u_{a}({\bf x}_{1},\ldots,{\bf x}_{N})\,.
  \label{AAA}
\end{eqnarray}
We assume here that a non-zero photon mass $\mu$ is introduced
to isolate the pole corresponding to the bound state $a$ from
the related cut. We  consider that the photon mass is larger
than the energy shift under consideration and much smaller than 
the distance to other levels.
 To generate the perturbation series for
$E_a$ it is convenient to use a contour integral formalism
developed first in operator theory by Sz\"okefalvi-Nagy and Kato
\cite{nagy,kato1,kato2,messia,reed,lepage}.
  Choosing a contour $\Gamma$ in the complex $E$ plane
 in a way that it surrounds the pole corresponding to the level $a$
and keeps outside all other singularities (see Fig. 7), we have
\begin{equation}
   \frac{1}{2\pi i}\oint_{\Gamma} dE \; E g_{aa}(E)=E_{a}A_{a}\,,
  \label{e2e5r}
\end{equation}
\begin{equation}
   \frac{1}{2\pi i}\oint_{\Gamma} dE \; g_{aa}(E)=A_{a}\,.
  \label{e2e6r}
\end{equation}
Here we have assumed that the contour $\Gamma$ is oriented
anticlockwise.
Dividing the equation (\ref{e2e5r}) by (\ref{e2e6r}), we obtain
\begin{equation}
  E_{a} = \frac{\displaystyle
\frac{1}{2\pi i}\oint_{\Gamma} dE \; E g_{aa}(E)}
          {\displaystyle
\frac{1}{2\pi i}\oint_{\Gamma} dE \; g_{aa}(E)}
  \label{e2e7r}
\end{equation}
It is convenient to transform the equation (\ref{e2e7r}) to a
form that directly yields the energy shift $ \Delta
E_{a}=E_{a}-E_{a}^{(0)}$. 
In zeroth order, substituting  the operators 
\be
\psi_{\rm in}(0,\bfx)=\sum_{\veps_n>0}b_n\psi_n(\bfx)
+\sum_{\veps_n<0}d_n^{\dag}\psi_n(\bfx)\,,\\
\overline{\psi}_{\rm in}(0,\bfx)=\sum_{\veps_n>0}b_n^{\dag}
\overline{\psi}_n(\bfx)
+\sum_{\veps_n<0}d_n\overline{\psi}_n(\bfx)
\ee
into equations (\ref{phi}) and
(\ref{xi}) 
instead of $\psi (0,\bfx)$ and $\overline{\psi}(0,\bfx)$,
respectively, 
and considering the states $|n\rangle$ in (\ref{phi}) and
(\ref{xi}) as unperturbed states in the Fock space,
from equations (\ref{spect1})-(\ref{xi}) and  (\ref{gsmall})
we find  
\begin{equation} \label{green00}
  g_{aa}^{(0)}=\frac{1}{E-E_{a}^{(0)}}\,.
\end{equation}
This equation can also be derived using the Feynman rules
for $G$ (see subsection II(E1)).
Denoting $ \Delta g_{aa}=g_{aa}-g_{aa}^{(0)}$, from
(\ref{e2e7r}) we obtain the desired formula \cite{shab88a}
\begin{equation}
  \Delta E_{a} = \frac{\displaystyle
\frac{1}{2\pi i}
        \oint_{\Gamma} dE \; (E-E_{a}^{(0)}) \Delta g_{aa}(E)}
      {\displaystyle
1+\frac{1}{2\pi i}\oint_{\Gamma} dE \; \Delta g_{aa}(E)}\,.
  \label{e2e9r}
\end{equation}
The Green function $\Delta g_{aa}(E)$ is constructed by
perturbation theory 
\be
\Delta g_{aa}(E)= \Delta g_{aa}^{(1)}(E)+\Delta g_{aa}^{(2)}(E)+\cdots\,,
\ee
where the superscript denotes the order in $\alpha$.
If we represent the energy shift  as a series in $\alpha$
\be
\Delta E_{a}= \Delta E_{a}^{(1)}+\Delta E_{a}^{(2)}+\cdots\,,
\ee
the formula (\ref{e2e9r}) yields
     \be
       \Delta E_{a}^{(1)}&=& 
        \frac{1}{2\pi i}\oint_{\Gamma} dE \;\Delta E
                    \,\Delta g_{aa}^{(1)}(E)\,,
           \label{e3e11}
\\
       \Delta E_{a}^{(2)}&=&  
          \frac{1}{2\pi i}\oint_{\Gamma} dE \;\Delta E
                    \,\Delta g_{aa}^{(2)}(E)-
                     \biggl(\frac{1}{2\pi i}\oint_{\Gamma} dE \;\Delta E
                   \, \Delta g_{aa}^{(1)}(E)\biggr)
                    \;\biggl(\frac{1}{2\pi i}\oint_{\Gamma} dE 
                    \; \Delta g_{aa}^{(1)}(E)\biggr)\,,
           \label{e3e12}
\ee
where       $\Delta E\equiv E-E_{a}^{(0)}$. 

  Deriving equations (\ref{e2e7r}) and (\ref{e2e9r}) we have assumed
that a non-zero photon mass $\mu$ is introduced.
This allows  taking all the cuts outside the contour $\Gamma$
as well as regularizing the infrared singularities 
of individual diagrams. In the Feynman gauge,
the photon propagator with non-zero photon mass $\mu$ is 
     \begin{equation} 
D_{\rho\sigma} (\omega,{\bf x}-{\bf y})=
              -g_{\rho\sigma}\int \;\frac{d{\bf k}}
                   {(2\pi)^{3}}\;
\frac{\exp{(i{\bf k}\cdot({\bf x}-{\bf y}))}}
                           {\omega^{2}-{\bf k}^{2}-\mu^{2}+i0}   
         \label{e2e9} 
     \end{equation}
or, after integration,
  \begin{equation}
 D_{\rho\sigma} (\omega,{\bf x}-{\bf y})= 
         g_{\rho\sigma}\frac{\exp{(i\sqrt{\omega^{2}-\mu^{2}+i0}\;|
        {\bf x}-{\bf y}|)}}
         {4\pi|{\bf x}-{\bf y}|}\,,          
                 \label{e2e10}
\end{equation}
   where ${\rm Im}\sqrt{\omega^{2}-\mu^{2}+i0}>0 $. 
$ D_{\rho\sigma} (\omega,{\bf x}-{\bf y})$ is an analytical function of
$\omega$ in the complex $\omega $ plane with cuts 
beginning at the points $\omega=-\mu+i0$ and $\omega=\mu-i0$
(Fig. 8). The related expressions for the photon propagator
with non-zero photon mass in other covariant gauges are
presented, e.g., in \cite{itzykson}.

As was noted in the previous subsection, the singularities of
the two-time Green function in the complex $E$ plane
are defined by the unperturbed Hamiltonian
if it is constructed by perturbation theory. In particular, it
means that in $n$-th order of perturbation theory
 $g_{aa}(E)$ has poles of all orders till $n+1$ at the 
position of the unperturbed energy level under consideration. Therefore,
in calculations by  perturbation theory it is sufficient
to consider the photon mass as a very small parameter
which provides a separation of the pole from the related cut.
At the end of the
calculations after taking into account a whole gauge invariant set
of Feynman diagrams  we can put  $\mu \rightarrow 0$. 
The possibility of taking the limit $\mu \rightarrow 0$
follows, in particular, from the fact that the contour $\Gamma$ can be
shrunk continuosly to the point $E=E_a^{(0)}$ (see Fig. 7).

Generally speaking, the energy shift of an excited level
derived by formula 
(\ref{e2e9r}) contains an imaginary component which is caused
by its instability. This component defines the width 
of the spectral line in the Lorentz approximation
(see sections III(F,G) for details).

For practical calculations it is convenient to express the Green
function $g_{aa}(E)$ in terms of the Fourier transform of the $2N-$time
Green function defined by equation (\ref{green2np}).
 By using the identity 
\begin{equation}
  \frac{1}{2\pi} \int_{-\infty}^{\infty} dx \,\exp{(i \omega x)}=\delta(\omega)
  \label{e2e10r}
\end{equation} 
one easily finds (see Appendix C)
\begin{eqnarray}       
 g_{aa}(E) \delta(E-E')  &=&
  \frac{2\pi}{i}\frac{1}{N!}
           \int_{-\infty}^{\infty}dp_{1}^{0}\cdots dp_{N}^{0}
    dp_{1}^{\prime 0}\cdots dp_{N}^{\prime 0}\nonumber\\
&& \times \delta(E-p_{1}^{0} -\cdots - p_{N}^{0})
            \delta(E'-p_{1}^{\prime 0} -\cdots - p_{N}^{\prime 0}) 
\nonumber\\  
 && \times  \langle u_{a}|
  G(p_{1}^{\prime 0} ,\ldots ,p_{N}^{\prime 0};p_{1}^{0},
\ldots ,p_{N}^{0})
	     \gamma_{1}^{0} \ldots \gamma_{N}^{0}
	     |u_{a}\rangle\,,
       \label{e2e3}
\end{eqnarray}
where
\be
\lefteqn{\langle u_{a}|
  G(p_{1}^{\prime 0} ,\ldots ,p_{N}^{\prime 0};p_{1}^{0},\ldots ,p_{N}^{0})
   \gamma_{1}^{0} \ldots \gamma_{N}^{0}
    |u_{a}\rangle}\;\;\;\;\;\;\;\;\;\;\;\;\;\;\;\;\;\nonumber\\
& \equiv &  \int d\bfx_1\cdots d\bfx_N d\bfx'_1\cdots d\bfx'_N \;
 u_{a}(\bfx'_1,...,\bfx'_N)\nonumber\\
&&\times   G((p_{1}^{\prime 0},\bfx'_1),
\ldots ,(p_{N}^{\prime 0},\bfx'_N);(p_{1}^{0},\bfx_1),
\ldots ,(p_{N}^{0},\bfx_N))\nonumber\\
&&\times   \gamma_{1}^{0} \ldots \gamma_{N}^{0}
     u_{a}(\bfx_1,...,\bfx_N)\,.
\ee

According to equation (\ref{manydet}) the calculation of the
matrix elements in (\ref{e2e3}) is reduced to the calculation 
of the matrix elements between the one-determinant wave functions
\begin{equation}            
u_{i}=\frac{1}{\sqrt{N!}}\sum_{P} (-1)^{P}
        \psi_{P i_{1}}({\bf x}_{1}) \cdots \psi_{P i_{N}}
({\bf x}_{N})\,,
      \label{e2e11i} 
\end{equation}
\begin{equation}            
u_{k}=\frac{1}{\sqrt{N!}}\sum_{P} (-1)^{P}
        \psi_{P k_{1}}({\bf x}_{1}) \cdots \psi_{P k_{N}}
({\bf x}_{N})\,.
      \label{e2e11k} 
\end{equation}
To simplify the summation
procedure over the permutations in (\ref{e2e3}) which arise from 
the wave functions as well as from the Green function
 $ G(p_{1}^{\prime 0} ,\ldots ,p_{N}^{\prime 0};
p_{1}^{0},\ldots ,p_{N}^{0}) $,
 it is convenient
to transform equation (\ref{e2e3}) in the following way. 
Denoting $\overline{G}=G\gamma_{1}^{0}\ldots\gamma_{N}^{0}$,
 we can write
\begin{eqnarray}
 \overline{G}((p_{1}^{\prime 0},\xi_{1}^{\prime})  ,
     \ldots ,(p_{N}^{\prime 0},\xi_{N}^{\prime});
          (p_{1}^{0},\xi_{1})
           \ldots ,(p_{N}^{0},\xi_{N}))\;\;\;\;\;\;\;\;\;\;\;\;\;\;\;
\;\;\;\;\;\;\;\;\;\;\;\;\;\;\;\;\;\;\;\;\;\;\;\;\;\nonumber \\ 
         = \sum_{P}(-1)^{P}
       \widehat{G}((p_{P1}^{\prime 0},\xi_{P1}^{\prime})  ,
          \ldots ,(p_{PN}^{\prime 0},\xi_{PN}^{\prime});
          (p_{1}^{0},\xi_{1}),
           \ldots ,(p_{N}^{0},\xi_{N}))\,,
     \label{e2e12}    
\end{eqnarray}
 where $\xi\equiv ({\bf x},\alpha)$ and $\alpha $ is the bispinor index
 $(\alpha=1,2,3,4)$. Substituting
(\ref{e2e12}) in (\ref{e2e3}) and using the symmetry
of $\widehat{G}$ with respect to the electron coordinates,
 for 
$g_{ik}(E)=\langle u_{i}|{\cal G}(E)\gamma_1^0\cdots \gamma_N^0
|u_{k}\rangle$ one can obtain 
(see Appendix D)
\begin{eqnarray}
 g_{ik}(E)\delta(E-E')&=&\frac{2\pi}{i}\sum_{P}(-1)^{P}
    \psi_{P i_{1}}^{*}(\xi_{1}^{\prime})\ldots\psi_{P i_{N}}^{*}
     (\xi_{N}^{\prime}) 
      \int_{-\infty}^{\infty}dp_{1}^{0}\ldots dp_{N}^{0}
	     dp_{1}^{\prime 0}\ldots dp_{N}^{\prime 0} \nonumber \\ &
  & \times    \delta(E-p_{1}^{0} -\cdots - p_{N}^{0})
\delta(E'-p_{1}^{\prime 0} -\cdots - p_{N}^{\prime 0})              
    \nonumber \\ &
   & \times  \widehat{G}((p_{1}^{\prime 0},\xi_{1}^{\prime})  ,
          \ldots ,(p_{N}^{\prime 0},\xi_{N}^{\prime});
          (p_{1}^{0},\xi_{1}),
           \ldots ,(p_{N}^{0},\xi_{N})) \nonumber \\ &
&\times     \psi_{k_{1}}(\xi_{1})\ldots\psi_{k_{N}}(\xi_{N})\,,
        \label{e2e13}
 \end{eqnarray}
where repeated variables $\{\xi\}$ imply integration (the integration
over ${\bf x}$  and  the summation over $\alpha $ ).   
In practical calculations by perturbation theory, the formula
(\ref{e2e13}) must be only employed for symmetric sets of
Feynman diagrams since the symmetry property was used in its 
derivation.

\subsection{Perturbation theory for degenerate and
 quasidegenerate levels}

In this section we are interested in the atomic levels with energies
$ E_1,...,E_s$ 
arising from unperturbed degenerate or quasidegenerate  levels
with energies $E_1^{(0)},...,E_s^{(0)}$.
As usual, we assume that the energy shifts of the levels under 
consideration or their splitting caused by the 
interaction are much smaller 
than the distance to other levels. The unperturbed eigenstates
form an $s$-dimensional subspace $\Omega$. We denote the projector
 on $\Omega$ by 
\be
P^{(0)}=\sum_{k=1}^{s} P_k^{(0)}=\sum_{k=1}^{s} u_{k}u_{k}^{\dag}\,,
\ee
 where $\{u_k\}_{k=1}^{s}$ are the unperturbed
wave functions which, in a general case, are linear combinations
of one-determinant functions (see equation (\ref{manydet}) ).
 We project the Green function ${\cal G}(E)$ on the
subspace $\Omega$
\begin{eqnarray} \label{gsmall2}
g(E)=P^{(0)} {\cal G }(E) \gamma_{1}^{0}...\gamma_{N}^{0}P^{(0)}\, ,
\end{eqnarray}
where, as in (\ref{gsmall}), the integration over the
electron coordinates is implicit.
As in the case of a single level, to isolate the poles
of $g(E)$ corresponding to the bound states under consideration,
 we introduce
a non-zero photon mass $\mu$. We assume that
the photon mass $\mu$ is larger than the energy distance
between the levels under consideration
and much smaller than the distance to other levels.
In this case we can choose a contour $\Gamma$ in the complex $E$ plane
in a way that it surrounds all the poles corresponding 
to the considered states $(E_1,...E_s)$
and keeps outside all other singularities, including
the cuts starting from the lower-lying bound states (see Fig. 9).
In addition, if we neglect the instability of the
states under consideration, the spectral representation 
(see equations (\ref{spect1})-(\ref{spect4})) gives
\widetext
\begin{eqnarray} \label{spect5}
g(E)=\sum_{k=1}^{s}\frac{\varphi_{k}\varphi_{k}^{\dag}}{E-E_k}+ 
\mbox{ terms that are regular inside of }  \Gamma\, ,
\end{eqnarray}
where
\begin{eqnarray} \label{varphi}
\varphi_{k}=P^{(0)}\Phi_{k}\, , \qquad
 \varphi_{k}^{\dag}=\Phi_{k}^{\dag}P^{(0)}\, .
\end{eqnarray}
As in the case of a single level, in  zeroth approximation 
one easily finds
\begin{eqnarray}
g^{(0)}(E)=\sum_{k=1}^s \frac{P_k^{(0)}}{E-E_k^{(0)}}\, .
\end{eqnarray}
We introduce the operators $K$ and $P$ by
\begin{eqnarray}\label{kkk}
K&\equiv&\frac{1}{2\pi i}\oint_{\Gamma}dE\;Eg(E)\, ,\\ 
P&\equiv&\frac{1}{2\pi i}\oint_{\Gamma}dE\;g(E) \, .
\label{ppp}
\end{eqnarray}
Using equation (\ref{spect5}), we obtain
\begin{eqnarray}\label{kkk2}
K&=&\sum_{i=1}^{s}E_i\varphi_i\varphi_{i}^{\dag}\, ,\\
P&=&\sum_{i=1}^{s}\varphi_i\varphi_{i}^{\dag}\, .
\label{ppp2}
\end{eqnarray}
We note here that, generally speaking, the operator $P$ is not
 a projector (in particular, $P^2\ne P$). If the perturbation
goes to zero, the vectors $\{\varphi_{i}\}_{i=1}^{s}$ approach
 the correct
linearly independent combinations of the vectors $\{u_{i}\}_{i=1}^{s}$.
Therefore, it is natural to assume that the vectors 
$\{\varphi_{i}\}_{i=1}^{s}$ are also linearly independent.
It follows that one can find such vectors $\{v_{i}\}_{i=1}^{s}$
that
\begin{eqnarray} \label{biort}
\varphi_{i}^{\dag}v_{k}=\delta_{ik}\, .
\end{eqnarray}
Indeed, let
\begin{eqnarray}
\varphi_{i}=\sum_{j=1}^{s}a_{ij}u_{j} \,,
 \qquad v_{k}=\sum_{l=1}^{s}x_{kl}u_{l}\,.
\end{eqnarray}
The biorthogonality condition (\ref{biort}) gives
\begin{eqnarray} \label{axxx}
\sum_{j=1}^{s}a_{ij}x_{kj}=\delta_{ik}\,.
\end{eqnarray}
Since the determinant of the matrix $\{a_{ij}\}$ is nonvanishing
due to the linear independence of $\{\varphi_{i}\}_{i=1}^{s}$, 
the system (\ref{axxx}) has a unique solution
 for any fixed $k=1,...,s$.
>From (\ref{kkk2})-(\ref{biort}) we have
\begin{eqnarray}
Pv_{k}&=&\sum_{i=1}^{s}\varphi_{i}\varphi_{i}^{\dag}v_{k}=\varphi_{k}\,,\\
Kv_{k}&=&\sum_{i=1}^{s}E_i\varphi_{i}\varphi_{i}^{\dag}v_{k}=
E_k\varphi_{k}\, .
\end{eqnarray}
Hence we obtain the equation for $v_{k}$,  $E_k$
\cite{shab88a}
\begin{eqnarray} \label{kkpp}
Kv_{k}=E_k Pv_{k}\, .
\end{eqnarray}
According to (\ref{biort})
 the vectors $v_{k}$ are normalized by the condition
\begin{eqnarray} \label{normgen}
v_{k'}^{\dag}Pv_{k}=\delta_{k'k}\, .
\end{eqnarray}
The solvability  of equation (\ref{kkpp}) yields an equation for 
 the atomic energy levels
\begin{eqnarray}
{\rm det}\,(K-EP)=0 \,.
\end{eqnarray}
The generalized eigenvalue problem (\ref{kkpp}) with the normalization
condition (\ref{normgen})
 can be transformed by the substitution $ \psi_{k}=P^{\frac{1}{2}}v_{k}$
 to the ordinary eigenvalue problem ("Schr\"odinger-like equation")
\cite{shab91}
\begin{eqnarray} \label{schr}
H\psi_{k}=E_k\psi_{k}
\end{eqnarray}
with the ordinary normalization condition
\begin{eqnarray}
\psi_{k}^{\dag}\psi_{k'}=\delta_{kk'}\, ,
\end{eqnarray}
where $H\equiv P^{-\frac{1}{2}}(K)P^{-\frac{1}{2}}$.

The energy levels are determined from the equation
\be \label{detschr}
{\rm det} (H-E)=0\,.
\ee
Generally speaking, the energies determined by this equation
contain imaginary components which are due to the instability
of excited states. In the case when the imaginary components
are much smaller than the energy distance between the levels (or
the levels have different quantum numbers), they define
the widths of the spectral lines in the Lorentz approximation.
In the opposite case, when the imaginary components are 
comparable with the energy distance between the levels which have
the same quantum numbers, the spectral line shape depends on
the process of the formation of the states under consideration
even in the resonance approximation  (see sections
III(F,G) for details). In what follows, calculating the energy
levels we neglect the instability of excited states and
 assume $H\equiv (H+H^{\dag})/2$
in equations (\ref{schr}), (\ref{detschr}).

The operators $K$ and $P$ are constructed by the formulas
(\ref{kkk}) and (\ref{ppp}) using perturbation theory
\be
K&=&K^{(0)}+K^{(1)}+K^{(2)}+\cdots\,,\\
P&=&P^{(0)}+P^{(1)}+P^{(2)}+\cdots\,,
\ee
where the superscript denotes the order in $\alpha$.
The operator $H$ is
\be
H=H^{(0)}+H^{(1)}+H^{(2)}+\cdots\,,
\ee
where
\be
\label{H0}
H^{(0)}&=&K^{(0)}\,,\\
\label{H1}
H^{(1)}&=&K^{(1)}-\frac{1}{2}P^{(1)}K^{(0)}-\frac{1}{2}K^{(0)}P^{(1)}\,,\\
\label{H2}
H^{(2)}&=&K^{(2)}-\frac{1}{2}P^{(2)}K^{(0)}-\frac{1}{2}K^{(0)}P^{(2)}
\nonumber\\
&&-\frac{1}{2}P^{(1)}K^{(1)}-\frac{1}{2}K^{(1)}P^{(1)}\nonumber\\
&&+\frac{3}{8}P^{(1)}P^{(1)}K^{(0)}+\frac{3}{8}K^{(0)}P^{(1)}P^{(1)}
\nonumber\\
&&+\frac{1}{4}P^{(1)}K^{(0)}P^{(1)}\,.
\ee
It is evident that in  zeroth order
\be
K_{ik}^{(0)}&=&E_i^{(0)}\delta_{ik}\,,\\
P_{ik}^{(0)}&=&\delta_{ik}\,,\\
H_{ik}^{(0)}&=&E_i^{(0)}\delta_{ik}\,.
\ee

To derive the equations (\ref{kkpp})-(\ref{schr})
we have introduced a non-zero photon mass $\mu$ which
was assumed to be larger than the energy distance
between the levels under consideration and much smaller
than the distance to other levels. At the end of the calculations
after taking into account a whole gauge invariant set of
Feynman diagrams, we can put $\mu \rightarrow 0$.
The possibility of taking this limit
in the case of quasidegenerate states follows
from the fact that the cuts can be drawn to the related poles
by a deformation of the contour $\Gamma$ as shown in Fig. 10.

As in the case of a single level,
for practical calculations we express the Green
function $g(E)$ in terms of the Fourier transform of the $2N$-time
Green function
\begin{eqnarray}
g(E)\delta (E-E')&=& \frac{2 \pi}{i} \frac{1}{N!}\int_{-\infty}^
{\infty} dp_{1}^{0}...dp_{N}^{0}dp_{1}^{\prime 0}...dp_{N}^
{\prime 0}\;\delta(E-p_{1}^{0}\cdots -p_{N}^{0})\delta(E'-p_{1}^
{\prime 0}\cdots -p_{N}^{\prime 0})\nonumber\\
&&\times P^{(0)}G(p_{1}^{\prime 0},...p_{N}^{\prime 0};p_{1}^{0},...
p_{N}^{0})\gamma_{1}^{0}...\gamma_{N}^{0}P^{(0)}\,,
\end{eqnarray}
where $G(p_{1}^{\prime 0},...,p_{N}^{\prime 0};
p_{1}^{0},...,p_{N}^{0})$
is defined by equation (\ref{green2np}).

\subsection{Practical calculations}

In this section we consider some practical applications
of the method in the lowest orders of perturbation
theory. In what follows we will use the notation
\be \label{inter2}
I(\omega)=e^2\alpha^{\rho}\alpha^{\sigma}D_{\rho \sigma}(\omega)\,,
\ee 
where $\alpha^{\rho}\equiv\gamma^0 \gamma^{\rho}=(1,\balpha)$.
In addition
 we will employ the following symmetry properties of the photon
propagator 
\be \label{symI}
I(\omega)=I(-\omega)\,,\;\;\;\;\;\;\;\;\;I'(\omega)=-I'(-\omega)\,,
\ee
which, in particular, are valid in the Feynman and Coulomb gauges.
Here $I'(\omega)\equiv dI(\omega)/d\omega\,$.

\subsubsection{Zeroth order approximation}

Let us derive first 
the formula (\ref{green00}) using the Feynman rules for $G$.
According to the equations (\ref{e2e3})-(\ref{e2e13}) we have
\be 
 g_{aa}(E) \delta(E-E')&=&\frac{2\pi}{i}\sum_{P}(-1)^{P}
\int d\bfx_1\cdots d\bfx_N d\bfx'_1\cdots d\bfx'_N
\nonumber\\
&&\times
    \psi_{P a_{1}}^{\dag}(\bfx_{1}^{\prime})\cdots\psi_{P a_{N}}^{\dag}
     (\bfx_{N}^{\prime}) 
      \int_{-\infty}^{\infty}dp_{1}^{0}\cdots dp_{N}^{0}
	     dp_{1}^{\prime 0}\cdots dp_{N}^{\prime 0} \nonumber \\ &
  & \times   \delta(E-p_{1}^{0} -\cdots - p_{N}^{0})
 \delta(E'-p_{1}^{\prime 0} -\cdots - p_{N}^{\prime 0})       
    \nonumber \\ &
   & \times  \frac{i}{2\pi}S_1(p_1^0,\bfx_{1}^{\prime},\bfx_1) 
\delta(p_1'^0-p_1^0)
          \cdots \frac{i}{2\pi} S_N(p_{N}^{\prime 0},\bfx_{N}^{\prime},\bfx_N)
\delta(p_N'^0-p_N^0)\nonumber\\
&&\times   \gamma_1^0\cdots \gamma_N^0
  \psi_{a_{1}}(\bfx_{1})\cdots\psi_{a_{N}}(\bfx_{N})\\
  &=&
  \frac{2\pi}{i}\int_{-\infty}^{\infty}dp_{1}^{0}\cdots dp_{N}^{0}
    dp_{1}^{\prime 0}\cdots dp_{N}^{\prime 0}\nonumber\\
&& \times \delta(E-p_{1}^{0} -\cdots - p_{N}^{0})
            \delta(E'-p_{1}^{\prime 0} -\cdots - p_{N}^{\prime 0}) 
\nonumber\\  
 && \times  \frac{i}{2\pi}\frac{1}{p_1^0-\veps_{a_{1}}+i0}\delta(p_1'^0-p_1^0)
\cdots
\frac{i}{2\pi}\frac{1}{p_N^0-\veps_{a_{N}}+i0}\delta(p_N'^0-p_N^0)\,.
\end{eqnarray}
Integrating over the energies one easily obtains equation
 (\ref{green00}).

\subsubsection{One-electron atom}

Formal expressions for the energy shift in the case of a
one-electron atom (or in the case of an atom with one electron 
over closed shells) can be derived by various methods.
In particular, the Dyson-Schwinger equation can be employed
for such a derivation.
Therefore, the one-electron system is not the best
example to demonstrate the advantages of the method
under consideration. However, we start with 
a detailed description of this simple case since it
may serve as the simplest introduction to the technique.

Let us consider first a diagram describing the interaction of 
a one-electron atom with an external potential $\delta V(\bfx)$
to first order in $\delta V(\bfx)$ (Fig. 11).
According to equation (\ref{e2e3}) we have
\be
\Delta g_{aa}^{(1)}(E)\delta (E-E')&=&\frac{2\pi}{i}
\int d\bfx' d\bfz d\bfx \;
\psi_a^{\dag}(\bfx')\frac{i}{2\pi}
\sum_{n_1}\frac{\psi_{n_1}(\bfx')\overline{\psi}_{n_1}(\bfz)}
{E'-\veps_{n_1}(1-i0)}\nonumber\\
&& \times \frac{2\pi}{i} \gamma^0  \delta V(\bfz)\delta(E'-E)
\frac{i}{2\pi}\sum_{n_2}
\frac{\psi_{n_2}(\bfz)\overline{\psi}_{n_2}(\bfx)}
{E-\veps_{n_2}(1-i0)}\gamma^0 \psi_a(\bfx)\nonumber\\
&=& \la a|\sum_{n_1} \frac{|n_1\ra \la n_1|}{E-\veps_{n_1}
(1-i0)}
\delta V\sum_{n_2} \frac{|n_2\ra \la n_2|}{E-\veps_{n_2}
(1-i0)}|a\ra \delta(E'-E)\nonumber\\
&=& \frac{1}{(E-\veps_a)^2}\la a|\delta V|a \ra \delta (E'-E).
\label{gvvv1}
\ee
Substituting (\ref{gvvv1}) into (\ref{e3e11}), we obtain
\be
\Delta E_a^{(1)}&=&\frac{1}{2\pi i} \oint_{\Gamma} dE \; (E-\veps_a) 
\Delta g_{aa}^{(1)} (E)\nonumber\\
&=&\frac{1}{2\pi i} \oint_{\Gamma} dE \;\frac{\la a|\delta V |a\ra}
{E-\veps_a} 
=\la a|\delta V|a\ra\,.
\ee

To first order in $\alpha$, the QED corrections 
are defined by  the self energy (SE) and
 vacuum polarization (VP)  diagrams which are shown in Figs. 12 and
13, respectively. Consider
first the SE diagram. We find
\be
\Delta g_{aa}^{(1)}(E)\delta (E-E')&=&\frac{2\pi}{i}
\int d\bfx'd\bfy d\bfz d\bfx \; \psi_a^{\dag}(\bfx')\frac{i}{2\pi}
\sum_{n_1}\frac{\psi_{n_1}(\bfx')\overline{\psi}_{n_1}(\bfy)}
{E'-\veps_{n_1}(1-i0)}
 \frac{i}{2\pi}\int_{-\infty}^{\infty}dp^0 \;
\frac{i}{2\pi}\int_{-\infty}^{\infty}d\omega
\nonumber\\
&& \times e\gamma^{\rho}\frac{2\pi}{i} \delta(E'-p^0-\omega)
\sum_{n}
\frac{\psi_{n}(\bfy)\overline{\psi}_{n}(\bfz)}
{p^0-\veps_{n}(1-i0)} D_{\rho \sigma} (\omega,\bfy-\bfz)
\nonumber\\
&& \times e\gamma^{\sigma} \frac{2\pi}{i}
\delta(p^0+\omega-E)
\frac{i}{2\pi}\sum_{n_2}
\frac{\psi_{n_2}(\bfz)\overline{\psi}_{n_2}(\bfx)}
{E-\veps_{n_2}(1-i0)}\gamma^0 \psi_a(\bfx)\nonumber\\
&=&\frac{1}{(E-\veps_a)^2}e^2 \frac{i}{2\pi}\int_{-\infty}^{\infty}
d\omega \;
\int d\bfy d\bfz  \; \psi_a^{\dag}(\bfy)
\alpha^{\rho}\sum_{n}
\frac{\psi_{n}(\bfy)\overline{\psi}_{n}(\bfz)}
{E-\omega-\veps_{n}(1-i0)}\nonumber\\
&&\times D_{\rho \sigma} (\omega,\bfy-\bfz)
\alpha^{\sigma} \psi_a(\bfz)\delta(E'-E)\,,
\label{gse1}
\ee
where $\alpha^{\rho}\equiv \gamma^{0}\gamma^{\rho}
=(1,\balpha)$. Denoting
\be
\la a|\Sigma(E)|b\ra=\frac{i}{2\pi} \int_{-\infty}^{\infty}
d\omega \;
\sum_{n}\frac{\la an|e^2\alpha^{\rho}\alpha^{\sigma}
D_{\rho \sigma}(\omega)|n b\ra}{E-\omega-\veps_n(1-i0)} \,,
\ee
we get
\be \label{gse2}
\Delta g_{aa}^{(1)}(E)=\frac{\la a |\Sigma (E)|a\ra}
{(E-\veps_a)^2}\,.
\ee
Substituting (\ref{gse2}) into (\ref{e3e11}), we obtain
\be
\Delta E_a^{(1)}&=&\frac{1}{2\pi i} \oint_{\Gamma} dE \; (E-\veps_a) 
\Delta g_{aa}^{(1)} (E)\nonumber\\
&=&\frac{1}{2\pi i} \oint_{\Gamma} dE \; \frac{\la a| \Sigma(E) |a\ra}
{E-\veps_a} 
=\la a|\Sigma(\veps_a)|a\ra\,.
\label{se1}
\ee
Here we have taken into account the fact that, for a non-zero
photon mass $\mu$, $\Delta g(E)$ has isolated poles at $E=\veps_a$
in every order of perturbation theory. In the final expression
one can put $\mu\rightarrow 0$. 

The expression (\ref{se1}) suffers from an ultraviolet divergence
and has to be considered together with a counterterm diagram
(Fig. 14). Taking into account the counterterm results in a replacement
\be \label{eq102}
\la a|\Sigma(\veps_a)|a\ra \rightarrow \la a|\Sigma_R(\veps_a)|a\ra
=\la a|(\Sigma(\veps_a)-\gamma^0 \delta m)|a \ra\,.
\ee

The corresponding calculation for the VP diagram (Fig. 13) yields
\be
\Delta g_{aa}^{(1)}(E)\delta (E-E')&=&-\frac{2\pi}{i}
\int d\bfx  d\bfy \;  \overline{\psi}_a(\bfx)\frac{i}{2\pi}
\frac{1}{E'-\veps_{a}} \frac{i}{2\pi}
\int_{-\infty}^{\infty}d\omega'
\nonumber\\
&& \times e\gamma^{\rho}\frac{2\pi}{i} \delta(E'+\omega'-E)
 D_{\rho \sigma} (\omega',\bfx-\bfy)
e \frac{2\pi}{i}
\delta(\omega')
\nonumber\\
&&\times
\frac{i}{2\pi}\int_{-\infty}^{\infty} d\omega \;
{\rm Tr}\Bigl[\sum_n
\frac{\psi_{n}(\bfy)\overline{\psi}_{n}(\bfy)}
{\omega-\veps_{n}(1-i0)}\gamma^{\sigma}\Bigr]
\frac{i}{2\pi}\frac{1}{E-\veps_a}\psi_a(\bfx)\,.
\label{gvp1}
\ee
Introducing the VP potential by
\be
U_{\rm VP}(\bfx)=\frac{e^2}{2\pi i}
\int d \bfy \; \alpha^{\rho} D_{\rho \sigma}(0,\bfx-\bfy)
\int_{-\infty}^{\infty} d\omega \; {\rm Tr}\Bigl[
\sum_n\frac{\psi_{n}(\bfy)\psi_{n}^{\dag}(\bfy)}
{\omega-\veps_{n}(1-i0)}\alpha^{\sigma}\Bigr]\,,
\ee
we find
\be
\Delta g_{aa}^{(1)}(E)=\frac{\la a|U_{\rm VP}| a\ra}{(E-\veps_a)^2}
\ee
and, therefore,
\be \label{vp1}
\Delta E_a^{(1)}=\la a|U_{\rm VP}|a\ra \,.
\ee
In reality, due to a spherical symmetry of the Coulomb potential
of the nucleus, only zeroth components of the $\alpha$ matrices
contribute to $U_{\rm VP}$,
\be \label{vppot}
U_{\rm VP}(\bfx)=\frac{\alpha}{2\pi i}
\int d \bfy \; \frac{1}{|\bfx-\bfy|}
\int_{-\infty}^{\infty} d\omega \; {\rm Tr}[
G_{\rm C}(\omega,\bfy,\bfy)]\,,
\ee
where 
\be
G_{\rm C}(\omega,\bfx,\bfy)=\sum_n\frac{\psi_{n}(\bfx)\psi_{n}^{\dag}(\bfy)}
{\omega-\veps_{n}(1-i0)}
\ee
is the Dirac-Coulomb Green function.
The expression (\ref{vppot}) is ultraviolet divergent.
The charge renormalization makes it finite.

Let us now consider the combined $\delta V$-SE corrections
described by the Feynman diagrams presented in Fig. 15.
For the diagrams "a" and "b" one easily finds
\be
\Delta g_{aa}^{(2,a+b)}(E)&=&\frac{1}{(E-\veps_a)^2}
\sum_{n}\la a|\delta V|n\ra \frac{1}{E-\veps_n}
\la n|\Sigma(E)|a\ra\nonumber\\
&&+\frac{1}{(E-\veps_a)^2}
\sum_{n}\la a|\Sigma(E)|n\ra \frac{1}{E-\veps_n}
\la n|\delta V|a\ra\,.
\ee
This contribution is conveniently
divided into two parts: {\it irreducible} ($\veps_n \not = \veps_a$)
and {\it reducible} ($\veps_n=\veps_a$).
For the irreducible part one obtains
\be
\frac{1}{2\pi i}\oint_{\Gamma} 
dE \; (E-\veps_a) \Delta g_{aa}^{(2,a+b,{\rm irr})}(E)
&=&\frac{1}{2\pi i}\oint_{\Gamma} dE \; \frac{1}{E-\veps_a}
\sum_{n}^{(n\not = a)} \Bigl[\frac{\la a|\delta V|n\ra 
\la n|\Sigma(E)|a\ra}{E-\veps_n}\nonumber\\
&&+\frac{\la a|\Sigma(E)|n\ra 
\la n|\delta V|a\ra}{E-\veps_n}\Bigr] \nonumber\\
&=&\sum_{n}^{(n\not = a)} \Bigl[\frac{\la a|\delta V|n\ra 
\la n|\Sigma(\veps_a)|a\ra}{\veps_a-\veps_n}\nonumber\\
&&+\frac{\la a|\Sigma(\veps_a)|n\ra 
\la n|\delta V|a\ra}{\veps_a-\veps_n}\Bigr]
\ee
Here we have taken into account that, due to the spherical symmetry
of the Coulomb potential, 
matrix elements  $\la a|\Sigma(\veps_a)|b\ra$ are equal to
zero if $\veps_a=\veps_b$ and $a\not =b$.
The reducible part is
\be
\frac{1}{2\pi i}\oint_{\Gamma}
 dE \; (E-\veps_a) \Delta g_{aa}^{(2,a+b,{\rm red})}(E)
&=&\frac{1}{2\pi i}\oint_{\Gamma} dE \; \frac{1}{(E-\veps_a)^2}
\Bigl[\la a|\delta V|a\ra 
\la a|\Sigma(E)|a\ra\nonumber\\
&&+\la a|\Sigma(E)|a\ra 
\la a|\delta V|a\ra\Bigr] \nonumber\\
&=&2\la a|\delta V|a\ra \la a|\Sigma'(\veps_a)|a\ra\,,
\ee
where
$\Sigma'(\veps_a)\equiv (d\Sigma (E)/dE)_{E=\veps_a}$\,.
The reducible contribution should be considered
together with the related contribution from
the second term in equation
(\ref{e3e12}). Taking into account that
\be 
\frac{1}{2\pi i}\oint_{\Gamma}
 dE \; (E-\veps_a) \Delta g_{aa}^{(1,\delta V)}(E)
=\la a|\delta V|a\ra\,
\ee
and
\be 
\frac{1}{2\pi i}\oint_{\Gamma} dE \; \Delta g_{aa}^{(1,{\rm SE})}(E) 
=\frac{1}{2\pi i}\oint_{\Gamma} dE \;
\frac{ \la a|\Sigma(E)|a\ra}{(E-\veps_a)^2}
=\la a|\Sigma'(\veps_a)|a\ra\,
\ee
we obtain
\be
-\Bigl(\frac{1}{2\pi i}\oint_{\Gamma} dE \; (E-\veps_a)
 \Delta g_{aa}^{(1,\delta V)}(E)\Bigr)\Bigl(\frac{1}
{2\pi i}\oint_{\Gamma} dE \; \Delta g_{aa}^{(1,{\rm SE})}(E)\Bigr) 
=-\la a|\delta V|a\ra \la a|\Sigma'(\veps_a)|a\ra\,.
\ee
For the total contribution of the diagrams "a" and "b" we
find
\be 
\Delta E_a^{(2,a+b)} \label{abxx1}
&=&\sum_{n}^{(n\not = a)} \Bigl[\frac{\la a|\delta V|n\ra 
\la n|\Sigma(\veps_a)|a\ra}{\veps_a-\veps_n}
+\frac{\la a|\Sigma(\veps_a)|n\ra 
\la n|\delta V|a\ra}{\veps_a-\veps_n}\Bigr]\nonumber\\
&&+\la a|\delta V|a\ra \la a|\Sigma'(\veps_a)|a\ra\,.
\ee

For the vertex contribution (the diagram "c")  we obtain
\be
\Delta g_{aa}^{(2,c)}(E)&=& \frac{1}{(E-\veps_a)^2} e^2
\frac{i}{2\pi}\int_{-\infty}^{\infty} d\omega \;
\int d\bfx\, d\bfy\, d\bfz \, \psi_a^{\dag}(\bfy)\alpha^{\rho}
\sum_{n_1}\frac{\psi_{n_1}(\bfy)\psi_{n_1}^{\dag}(\bfx)}
{E-\omega-\veps_{n_1}(1-i0)} \delta V(\bfx)\nonumber\\
&&\times
\sum_{n_2}\frac{\psi_{n_2}(\bfx)\psi_{n_2}^{\dag}(\bfz)}
{E-\omega-\veps_{n_2}(1-i0)}D_{\rho \sigma}(\omega,\bfy-\bfz)
\alpha^{\sigma} \psi_{a}(\bfz)\,.
\ee
This diagram is irreducible.
 A simple evaluation yields
\be \label{abxx2}
\Delta E_{a}^{(2,c)}&=& e^2
\frac{i}{2\pi}\int_{-\infty}^{\infty} d\omega \;
\int d\bfx\, d\bfy\, d\bfz \;\psi_a^{\dag}(\bfy)\alpha^{\rho}
G_{\rm C}(\veps_a-\omega,\bfy,\bfx) \delta V(\bfx)\nonumber\\
&&\times
G_{\rm C}(\veps_a-\omega,\bfx,\bfz)
D_{\rho \sigma}(\omega,\bfy-\bfz)
\alpha^{\sigma} \psi_{a}(\bfz)\,.
\ee

The related mass-counterterm diagrams (Fig. 16) are accounted
for by the replacement
 $\Sigma \rightarrow \Sigma_R=\Sigma-\gamma^0\delta m$
in equation (\ref{abxx1}). This replacement makes the
irreducible contribution in equation (\ref{abxx1}) to be
finite. As to the reducible contribution, its ultraviolet 
and infrared divergences are cancelled by the corresponding
divergences of the vertex contribution given by
equation (\ref{abxx2}).

\subsubsection{Atom with one electron over closed shells}

The consideration given above can easily be adopted to
the case of an atom with one electron over closed shells 
by regarding the closed shells as belonging to a redefined vacuum. 
The redefinition of the vacuum results in replacing $i0$
 by $-i0$ in the electron
propagator denominators corresponding to the closed shells.
In other words, it means replacing the standard Feynman contour
of integration over the electron energy $C$ by a new contour $C'$
(Fig. 17). In this formalism the one-electron radiative
 corrections are
incorporated together
with the interelectronic-interaction corrections
and the total energy of the closed shells is considered as the origin of
reference.
The difference of the integrals along $C'$ and $C$
is an integral along the contour $C_{\rm int}$. It describes
the interaction of the valence electron with the closed-shell
electrons.
Therefore, to find the interelectronic-interaction corrections
we have to replace the contour $C$ in the expressions
for the one-electron radiative corrections by the contour $C_{\rm int}$.
For example, in the case of one electron over the $(1s)^2$ shell 
in a lithiumlike ion, the first order interelectronic-interaction
corrections are obtained from the formulas for the SE and VP
corrections derived above
by the replacement
\be
\sum_n\frac{|n\ra \la n|}
{\omega-\veps_{n}(1-i0)} \rightarrow 
-\frac{2\pi}{i}\delta(\omega-\veps_{1s})
\sum_{c}^{(\veps_c=\veps_{1s})} |c\ra \la c|\,.
\ee  
As a result of this replacement, one obtains for the
interelectronic-interaction correction
\be
\Delta E_a^{(1,{\rm int})}=\sum_{c}^{(\veps_c=\veps_{1s})}
[\la a c|I(0)|ac\ra-\la a c|I(\veps_a-\veps_{1s})|c a\ra]\,,
\ee
where $I(\omega)$ is defined by equation (\ref{inter2}).
In Ref. \cite{sh95} this formalism was employed to derive formal
expressions for the interelectronic-interaction corrections
to the hyperfine splitting in lithiumlike ions (see section IV(C2)).

\subsubsection{Two-electron atom}

Let us consider now the energy shift of a single level $(n)$ in
 a two-electron atom. To first order
in $\alpha$, in addition to the
 one-electron SE and VP contributions
(Figs. 18,19)
we have to consider the one-photon exchange diagram (Fig. 20).
The derivation of the energy shift due to the SE and VP diagrams
is easily reduced to the case of a one-electron atom
by a simple integration over the energy variable
 of a disconnected electron
propagator. Therefore,
below we discuss only the one-photon exchange diagram.

For simplicity, we assume that the unperturbed wave 
function of the state under consideration
is the one-determinant function
  \begin{equation}
u_{n}(\bfx_1,\bfx_2)
=\frac{1}{\sqrt{2}}\sum_{P}(-1)^{P}\psi_{P a}({\bf x_{1}})
        \psi_{P b}({\bf x_{2}})\,.
    \label{e3e1}
\end{equation}
The transition to the general case of a many-determinant function
(\ref{manydet}) causes no problem and can be done in the final
expression for the energy shift.

According to equation (\ref{e2e3}), for the one-photon
 exchange diagram we have
   \begin{eqnarray}
\Delta g_{nn}^{(1)}&=&
      \Bigl(\frac{i}{2\pi}\Bigr)^{2}
        \int_{-\infty}^{\infty}dp_{1}^{0}dp_{1}^{\prime 0}\;
                           \sum_{P}(-1)^{P}
             \frac{1}{p_{1}^{\prime 0}-\veps_{P a}+i0}\,  
            \frac{1}{E-p_{1}^{\prime 0}-\veps_{P b}+i0}
\nonumber\\
&&\times       \frac{1}{p_{1}^{0}-\veps_{a}+i0}\,
               \frac{1}{E-p_{1}^{0}-\veps_{b}+i0} 
            \langle P a P b|I(p_{1}^{\prime 0}-p_{1}^{0})|ab\rangle\,.
   \label{ddgg}
\end{eqnarray}
Formula (\ref{e3e11}) gives
   \begin{eqnarray}
\Delta E_{n}^{(1)}&=&
    \frac{1}{2\pi i}\oint_{\Gamma} dE \;\Delta E  
\Bigl(\frac{i}{2\pi}\Bigr)^{2}
        \int_{-\infty}^{\infty}dp_{1}^{0}dp_{1}^{\prime 0}\;
                           \sum_{P}(-1)^{P}
             \frac{1}{p_{1}^{\prime 0}-\veps_{P a}+i0}  
 \nonumber \\ 
&& \times
            \frac{1}{E-p_{1}^{\prime 0}-\veps_{P b}+i0}\,
              \frac{1}{p_{1}^{0}-\veps_{a}+i0}\,
               \frac{1}{E-p_{1}^{0}-\veps_{b}+i0}\,\nonumber\\ 
&&\times  \langle P a P b|I(p_{1}^{\prime 0}-p_{1}^{0})|ab\rangle\,,
   \label{e3e13}
\end{eqnarray}
where, as in (\ref{e3e11}), $\Delta E\equiv E-E_n^{(0)}$.
        Transforming
\begin{eqnarray}  \label{iden1}
  \frac{1}{p_{1}^{\prime 0}-\veps_{P a}+i0}
       \frac{1}{E-p_{1}^{\prime 0}-\veps_{P b}+i0} 
    & =& \frac{1}{\Delta E}
\Bigl(\frac{1}{p_{1}^{\prime 0}-\veps_{P a}+i0}
  +    \frac{1}{E-p_{1}^{\prime 0}-\veps_{P b}+i0}\Bigr)\,, \\
 \frac{1}{p_{1}^{0}-\veps_{a}+i0}
       \frac{1}{E-p_{1}^{0}-\veps_{b}+i0}
    & =& \frac{1}{\Delta E}
\Bigl( \frac{1}{p_{1}^{0}-\veps_{a}+i0}+
       \frac{1}{E-p_{1}^{0}-\veps_{b}+i0}\Bigr) \,, 
     \label{iden2}
\end{eqnarray}
      we obtain     
     \begin{eqnarray}
\Delta  E_{n}^{(1)}&=&
   \frac{1}{2\pi i}\oint_{\Gamma} dE \; \frac{1}{\Delta E}
   \Bigl\{\Bigl(\frac{i}{2\pi}\Bigr)^{2}
 \int_{-\infty}^{\infty}dp_{1}^{0}dp_{1}^{\prime 0}\;
             \sum_{P}(-1)^{P} 
 \Bigl( \frac{1}{p_{1}^{\prime 0}-\veps_{P a}+i0}
              +   \frac{1}{E-p_{1}^{\prime 0}-\veps_{P b}+i0}\Bigr)
\nonumber\\
&&\times   \Bigl( \frac{1}{p_{1}^{0}-\veps_{a}+i0}+
           \frac{1}{E-p_{1}^{0}-\veps_{b}+i0}\Bigr) 
 \langle P a P b|I(p_{1}^{\prime 0}-p_{1}^{0})|ab\rangle 
\Bigr\}\,.
   \label{e3e15}
\end{eqnarray}
    The expression in the curly braces of (\ref{e3e15})  is a
regular
function of $E$ inside the contour $\Gamma$, if the photon mass 
$\mu$ is chosen as indicated above. 
A direct way to check this fact
consists in integrating over $p_1^0$ and $p_1'^0$ by
using the apparent expression for
 the photon propagator given by equation (\ref{e2e9}).
It can also be understood 
by observing that the integrand in this expression
is the sum of terms which contain singularities 
 from the electron propagators in $p_1^0$
($p_1'^0$) 
only above or only
below the real axis (for real $E$). Therefore, in each term we can vary $E$
in the complex $E$ plane within the contour $\Gamma$, keeping
the same order of bypassing the singularities in the
$p_1^0$ ($p_1'^0$) integration by moving slightly the contour
of the $p_1^0$ ($p_1'^0$) integration into the complex plane.
(In contrast to that, the contour of the $p_1^0$ ($p_1'^0$)
 integration in equation (\ref{e3e13}) is sandwiched between
two poles and, therefore, can not be moved to the complex 
plane.)
The branch points of the photon propagators are moved outside
the contour $\Gamma$ due to the non-zero photon mass.
However, instead of investigating the analytical properties
for every specific diagram, it is more convenient to
 use  the general
analytical properties of the Green function $g_{nn}(E)$  
in the complex $E$ plane in every order of perturbation theory
(see the related discussion above and Appendix B).
 According to these properties, the function $\Delta g_{nn}^{(1)}(E)$
can not have poles at the point $E=E_n^{(0)}$ of order higher than 2
and, therefore,
the expression in the curly braces of (\ref{e3e15})  is
a regular function of $E$ at this point.
So, we have to calculate
the first order residue at the point $E=E_n^{(0)}$.
We also stress that we do not need any apparent form
for the analytical continuation  of the expression in the curly
 braces to the complex $E$ plane
since we calculate it only for real $E$,
at the point $E=E_n^{(0)}$, where the present expression is
valid.
Calculating the $ E $ residue we obtain
\begin{eqnarray}
\Delta E_{n}^{(1)}&=& \Bigl(\frac{i}{2\pi}\Bigr)^{2}
             \int_{-\infty}^{\infty}dp_{1}^{0}dp_{1}^{\prime 0}\;
             \sum_{P}(-1)^{P}  
            \Bigl( \frac{1}{p_{1}^{\prime 0}-\veps_{P a}+i0}
            + \frac{1}{-(p_{1}^{\prime 0}-\veps_{P a})+i0}\Bigr)
       \nonumber \\ 
&& \times      \Bigl( \frac{1}{p_{1}^{0}-\veps_{a}+i0}+
                \frac{1}{-(p_{1}^{0}-\veps_{a})+i0}\Bigr) 
               \langle P a P b|I(p_{1}^{\prime 0}-p_{1}^{0})|ab\rangle\,.
   \label{e3e16}
\end{eqnarray}
      Taking into account the identity
\begin{equation}
         \frac{i}{2 \pi}\Bigl(\frac{1}{x+i0}+
                      \frac{1}{-x+i0}\Bigr)=\delta (x)
    \label{e3e17}
\end{equation}
     we find
 \begin{equation} 
           \Delta E_{n}^{(1)}=\sum_{P}(-1)^{P}
            \langle P a P b|I(\veps_{Pa}-\veps_{a})|ab\rangle\,.
   \label{e3e18}
\end{equation}

\subsubsection{General rules for practical calculations}

Let us consider now some general remarks to the derivation
given above. In order to perform first the integration over
$E$ we have separated the singularity in $\Delta E$ by employing the
identities (\ref{iden1})  and (\ref{iden2}). 
Another way could consist in transforming
 the left-hand sides of equations (\ref{iden1}) and (\ref{iden2})
by the identity
\be \label{iden3}
 \frac{1}{p^{0}-\veps_{a}+i0}
       \frac{1}{E-p^{0}-\veps_{b}+i0}
    & =& \frac{2\pi}{i}\delta(p^0-\veps_a)\frac{1}{\Delta E}
\nonumber\\
&&+ \frac{1}{p^{0}-\veps_{a}-i0}
       \frac{1}{E-p^{0}-\veps_{b}+i0}\,,
\ee
where we have used equation (\ref{e3e17}).
Using this identity allows one to separate
contributions  singular
 in $1/\Delta E$ from non-singular ones.
 The singular contributions result only
 from the first term in the right-hand
side of equation (\ref{iden3}). 
It can easily be understood 
by observing that the second term has both singularities in $p^0$
 above the real axis (for real $E$).
 It follows that the contour
of the  $p^0$ integration in the expression for the energy shift
can be moved slightly into the complex $E$
plane keeping the same order of bypassing the singularities.
It means that we can vary $E$ in the complex $E$ plane within
the contour $\Gamma$ and, therefore, the integrand is a regular
function of $E$ within this contour.
Identities like (\ref{iden3}) are very useful in calculations
of three- and more electron atoms (see section IV(B3)).

We also want to note that in all cases
 the order of the singularity
in $1/\Delta E $ is quite evident from the type of the diagram
under consideration. If the diagram is irreducible, the
factors $1/\Delta E$ may come only from the initial and final
propagators. In this case the second term in the right-hand
side of equation (\ref{iden3}) does not contribute to the
energy shift and, therefore, the derivation of the formal
expression for the energy shift becomes trivial. 
For reducible diagrams the factors $1/\Delta E$ arise
also from internal electron propagators.

We can formulate the following simple rule for deriving
the energy shift from a certain diagram. Using identities
like (\ref{iden1}), (\ref{iden2}) or (\ref{iden3}), we separate
all singularities in $1/\Delta E$ and then  integrate
over $E$ assuming that the rest is a regular function of $E$
within the contour $\Gamma$. As is discussed above, 
the order of the singularity is quite evident for every
specific diagram and it is a simple task to separate
 the factor $1/\Delta E$ to the right power. However, even if
we separate this factor to a power which is larger or smaller
than the real order of the singularity, it is impossible to miss
the correct result. In the first case (the power is larger than
the real order of the singularity) the result of the calculation
remains the same as in the case when we separate the factor
$1/\Delta E$ to the right power. In the second case (the power
is smaller than the real order of the singularity) we obtain
an infinite result ($\sim 1/0$). It means that we should increase
the power of the separated singularity and repeat the calculation
until we get a finite result.

\subsubsection{Two-photon exchange diagrams for the ground
state of a heliumlike atom}

The two-photon exchange diagrams are presented in Fig. 21.
Here we derive the energy shift from these diagrams for the
case of the ground state of a heliumlike atom. The case
of a single excited  state of a two-electron atom
 is considered in detail in \cite{sh94a}.
The wave function of the ground state is given by
\be
u_1(\bfx_1,\bfx_2)
=\frac{1}{\sqrt{2}}\sum_P (-1)^P \psi_{Pa}(\bfx_1)
\psi_{Pb}(\bfx_2)\,.
\ee 
The unperturbed energy is $E_1^{(0)}=\veps_a+\veps_b$,
where $\veps_a=\veps_b$.

Consider first the two-photon ladder diagram (Fig. 21a).
For the first term in (\ref{e3e12}) we have

\begin{eqnarray}
 \Delta E_{\rm lad}^{(2)} &=& 
           \frac{1}{2\pi i}\oint_{\Gamma} dE\; \Delta E 
            \sum_{P}(-1)^{P}   \Bigl(\frac{i}{2\pi}\Bigr)^{3}
       \int_{-\infty}^{\infty}dp_{1}^{0}dp_{1}^{\prime 0}d\omega 
                   \nonumber  \\ 
  &&   \times \sum_{n_{1}n_{2}} 
      \langle P a P b|I(p_{1}^{\prime 0}-\omega)|n_{1}n_{2}\rangle
   \langle n_{1}n_{2}|I(\omega-p_{1}^{0})|ab\rangle \nonumber\\ 
&&\times  \frac{1}{p_{1}^{\prime 0}-\veps_{P a}+i0}
              \frac{1}{E-p_{1}^{\prime 0}-\veps_{P b}+i0} 
                    \frac{1}{\omega-\veps_{n_{1}}(1-i0)}
\nonumber\\
&&\times  \frac{1}{E-\omega-\veps_{n_{2}}(1-i0)} 
  \frac{1}{p_{1}^{0}-\veps_{a}+i0}
               \frac{1}{E-p_{1}^{0}-\veps_{b}+i0}\,.
        \label{e3e21}
\end{eqnarray}
   Let us divide this contribution into irreducible 
($\veps_{n_1}+\veps_{n_2}\neq \veps_a+\veps_b $ )
and reducible ( $\veps_{n_1}
+\veps_{n_{2}} = \veps_a+\veps_b $)  parts
 \begin{equation}
     \Delta E_{\rm lad}^{(2)}=
   \Delta E_{\rm lad}^{(2,{\rm irred})}+
\Delta E_{\rm lad}^{(2,{\rm red})}\,.
       \label{e3e22}
\end{equation}
 Using the identities (\ref{iden1}) and (\ref{iden2})
  we obtain for the irreducible part
\begin{eqnarray}
\frac{1}{2\pi i}\oint_{\Gamma} dE \;
 \Delta E \Delta g_{11}^{(2,{\rm irred})}(E) &=& 
   \frac{1}{2\pi i}\oint_{\Gamma} dE \; \frac{1}{\Delta E} 
       \Bigl\{   \sum_{P}(-1)^{P}   \Bigl(\frac{i}{2\pi}\Bigr)^{3}
       \int_{-\infty}^{\infty}dp_{1}^{0}dp_{1}^{\prime 0}d\omega 
                       \nonumber  \\ 
 && \times  
\sum_{n_1,n_2}^{\veps_{n_{1}}+\veps_{n_{2}}\neq \veps_a+\veps_b} 
   \langle P a P b|I(p_{1}^{\prime 0}-\omega)|n_{1}n_{2}\rangle 
 \langle n_{1}n_{2}|I(\omega-p_{1}^{0})|ab\rangle\nonumber\\
&& \times \Bigl( \frac{1}{p_{1}^{\prime 0}-\veps_{P a}+i0}
              +   \frac{1}{E-p_{1}^{\prime 0}-\veps_{P b}+i0}\Bigr) 
\nonumber\\
&&\times \frac{1}{\omega-\veps_{n_{1}}(1-i0)}
\frac{1}{E-\omega-\veps_{n_{2}}(1-i0)} \nonumber\\
&&\times \Bigl( \frac{1}{p_{1}^{0}-\veps_{a}+i0}+
        \frac{1}{E-p_{1}^{0}-\veps_{b}+i0}\Bigr) \Bigl\}\,.
          \label{e3e23}
\end{eqnarray}
The expression in the curly braces of  (\ref{e3e23}) is a 
regular  function of $ E $ inside the contour $ \Gamma $ if 
the photon mass $\mu$   is chosen as indicated above.
(If it were not so, we would get an infinite result; see the
related discussion in the previous subsection.)
 Calculating the $ E $  residue we find
\begin{eqnarray}
\Delta E_{\rm lad}^{(2,{\rm irred})} & = &
             \sum_{P}(-1)^{P} \frac{i}{2\pi}
       \int_{-\infty}^{\infty}d\omega \; 
   \sum_{n_1,n_2}^{\veps_{n_{1}}+\veps_{n_{2}}\neq \veps_a+\veps_b}   
      \langle P a P b|I(\veps_{P a}-\omega)|n_{1}n_{2}\rangle
               \nonumber\\ 
 & &\times \langle n_{1}n_{2}|I(\omega-\veps_{a})|ab\rangle
       \frac{1}{\omega-\veps_{n_{1}}(1-i0)}\,
                \frac{1}{E_1^{(0)}-\omega-\epsilon_{n_{2}}(1-i0)}\,.
  \label{e3e24}
\end{eqnarray}
This derivation shows that the energy shift from an 
irreducible diagram is obtained by evaluation
of the  "usual S-matrix" element.
 For the numerical evaluation of
(\ref{e3e24}) it is convenient to rotate the contour of the
integration in the complex $\omega$ plane \cite{blund93}.

For the reducible contribution we have
\begin{eqnarray}
\frac{1}{2\pi i}\oint_{\Gamma} dE \;
 \Delta E \Delta g_{11}^{(2,{\rm red})}(E)
&=&\frac{1}{2\pi i}\oint_{\Gamma} dE \; \frac{1}{(\Delta E)^{2}} 
           \Bigl\{ \sum_{P}(-1)^{P} \Bigl(\frac{i}{2\pi}\Bigr)^{3}
       \int_{-\infty}^{\infty}dp_{1}^{0}dp_{1}^{\prime 0}d\omega 
                \nonumber  \\
   &&  \times 
\sum_{n_1,n_2}^{\veps_{n_{1}}+\veps_{n_{2}} = \veps_a+\veps_b} 
     \langle P a P b|I(p_{1}^{\prime 0}-\omega)|n_{1}n_{2}\rangle 
           \langle n_{1}n_{2}|I(\omega-p_{1}^{0})|ab\rangle
       \nonumber\\
   & &  \times   \Bigl( \frac{1}{p_{1}^{\prime 0}-\veps_{P a}+i0}+
                   \frac{1}{E-p_{1}^{\prime 0}-\veps_{P b}+i0}\Bigr)
       \nonumber \\
    &&  \times \Bigl(\frac{1}{\omega-\veps_{n_{1}}+i0}+
                \frac{1}{E-\omega-\veps_{n_{2}}+i0)}\Bigr) 
            \nonumber \\
    && \times        \Bigr( \frac{1}{p_{1}^{0}-\veps_{a}+i0}+
                \frac{1}{E-p_{1}^{0}-\veps_{b}+i0}\Bigl) \Bigl\}\,.
          \label{e3e25}
\end{eqnarray}        
      The expression in the curly braces of  (\ref{e3e25}) 
is a regular function within the contour $\Gamma$. 
Calculating the $E$ residue and taking into account
that $\veps_a=\veps_b$, we get
\begin{eqnarray}
\frac{1}{2\pi i}\oint_{\Gamma} dE  \;
 \Delta E \Delta g_{11}^{(2,{\rm red})}(E) & =& 
           -\frac{i}{2\pi} 
            \sum_{P}(-1)^{P}
 \sum_{n_1,n_2}^{\veps_{n_{1}}+\veps_{n_{2}} = 2\veps_a} 
\Bigl\{  \int_{-\infty}^{\infty}dp_{1}^{\prime 0}\;                    
      \langle P a P b|I(p_{1}^{\prime 0}-\veps_a)|n_{1}n_{2}\rangle
               \nonumber\\          
 &&  \times     \langle n_{1}n_{2}|I(0)|
      ab\rangle
       \frac{1}{(\veps_{a}-p_{1}^{\prime 0}+i0)^{2}}+
        \int_{-\infty}^{\infty}dp_{1}^{0}\;                    
   \langle P a P b|I(0)|n_{1}n_{2}
   \rangle
               \nonumber\\
           &&
       \times    \langle n_{1}n_{2}|I(\veps_{a}-p_{1}^{0})|ab
        \rangle
       \frac{1}{(\veps_{a}-p_{1}^{0}+i0)^{2}}\nonumber\\
&&+   \int_{-\infty}^{\infty}d\omega\;            
      \langle P a P b|I(\veps_{a}-\omega)|n_{1}n_{2}\rangle
       \nonumber\\
          && \times \langle n_{1}n_{2}|I(\omega-\veps_{a})|ab\rangle
       \frac{1}{(\veps_{a}-\omega+i0)^{2}} \Bigl\}\,.
     \label{e3e26}   
\end{eqnarray}
This contribution should be considered together with the second
term in equation (\ref{e3e12}). As was obtained above,
the first factor in this term is
\be
\frac{1}{2\pi i}\oint_{\Gamma} dE \; \Delta E \Delta g_{11}^{(1)}(E)
=\sum_{P}(-1)^P\la Pa Pb|I(0)|ab\ra\,.
\ee
A simple calculation of the second factor yields
\be
\frac{1}{2\pi i}\oint_{\Gamma} dE \; \Delta g_{11}^{(1)}(E)
&=&\frac{1}{2\pi i}\oint_{\Gamma} dE \; \frac{1}{(\Delta E)^2}
\Bigl\{\Bigl(\frac{i}{2\pi}\Bigr)^2\int_{-\infty}^{\infty}
dp_1^0 dp_1'^0
\nonumber\\
&&\times 
\sum_{P}(-1)^P \Bigr( \frac{1}{p_{1}'^{0}-\veps_{a}+i0}+
                \frac{1}{E-p_{1}'^{0}-\veps_{a}+i0}\Bigl)
\nonumber\\
&&\times \Bigr( \frac{1}{p_{1}^{0}-\veps_{a}+i0}+
                \frac{1}{E-p_{1}^{0}-\veps_{a}+i0}\Bigl)
\la Pa Pb|I(p_1'^0-p_1^0)|ab\ra\nonumber\\
&=&
-\frac{i}{2\pi}\sum_P(-1)^P\Bigl\{\int_{-\infty}^{\infty}
dp_1'^0 \;\frac{1}{(p_1'^0-\veps_a-i0)^2}
\la Pa Pb|I(p_1'^0-\veps_a)|ab\ra\nonumber\\
&&+\int_{-\infty}^{\infty}
dp_1^0 \; \frac{1}{(p_1^0-\veps_a-i0)^2}
\la Pa Pb|I(p_1^0-\veps_a)|ab\ra\Bigr\}\,.
\ee
For the total reducible contribution we obtain
\be \label{eq139}
\Delta E_{\rm lad}^{(2,{\rm red})}&=&  
          \frac{1}{2\pi i}\oint_{\Gamma} dE \;\Delta E
                    \,\Delta g_{11}^{(2,{\rm red})}(E)-
                     \biggl(\frac{1}{2\pi i}\oint_{\Gamma} dE \;\Delta E
                   \, \Delta g_{11}^{(1)}(E)\biggr)
                    \;\biggl(\frac{1}{2\pi i}\oint_{\Gamma} dE 
                    \;\Delta g_{11}^{(1)}(E)\biggr)\nonumber\\
&=&-\sum_P(-1)^P\sum_{n_1,n_2}^{\veps_{n_1}+\veps_{n_2}=2\veps_a}
\frac{i}{2\pi} \int_{-\infty}^{\infty} d\omega \;
 \la PaPb| I(\omega-\veps_a)
|n_1n_2\ra\nonumber\\
&&\times \la n_1 n_2 |I(\omega-\veps_a)|ab\ra 
\frac{1}{(\omega-\veps_a -i0)^2}\,.
\ee

A similar calculation of
the two-photon crossed-ladder diagram (Fig. 21b) gives
\be \label{eq140}
\Delta E_{\rm cross}^{(2)}  
&=&\sum_P(-1)^P\sum_{n_1,n_2}
\frac{i}{2\pi} \int_{-\infty}^{\infty} d\omega \;
 \la Pan_2| I(\omega-\veps_a)
|n_1b\ra \la n_1 Pb |I(\omega-\veps_a)|an_2\ra \nonumber\\
&& \times
\frac{1}{\omega-\veps_{n_1}(1-i0)}
\frac{1}{\omega-\veps_{n_2}(1-i0)}\,.
\ee

The contribution $\Delta E_{\rm lad}^{(2,{\rm red})}$ contains an
infrared divergent term which is cancelled by a related term
($\veps_{n_1}=\veps_{n_2}=\veps_a$)
from the contribution $\Delta E_{\rm cross}^{(2)}$.
In the individual contributions
the infrared singularities are regularized by a non-zero photon
mass $\mu$. If the ladder and crossed-ladder 
contributions  are merged by the common $\omega$-integration,
the integral is convergent and 
we can put $\mu\rightarrow 0$ before the integration over $\omega$
(see \cite{sh94a} for details). 
However, to show how the calculation for a non-zero photon
mass can be performed, let us calculate
 the reducible contribution for a finite $\mu$.
We have to calculate the integral
\be
I_1=\frac{i}{2\pi}\int_{-\infty}^{\infty} d\omega \;
\frac{\exp{[i\sqrt{\omega^2-\mu^2+i0}\,(r_{12}+r_{34})]}}{r_{12}r_{34}}
\frac{1}{(\omega-i0)^2}\,.
\ee
Using the identity
\be 
\exp{[i\sqrt{\omega^2-\mu^2+i0}\,\;r]}
=-\frac{2}{\pi}\int_{0}^{\infty}dk \; k \frac{\sin{(kr)}}{(\omega^2-
k^2-\mu^2+i0)}\,,
\ee
we obtain
\be
I_1=-\frac{2}{\pi}\frac{i}{2\pi}\frac{1}{r_{12}r_{34}}
\int_{-\infty}^{\infty}d\omega \;
\int_{0}^{\infty}dk \; k \frac{\sin{(k(r_{12}+r_{34}))}}{(\omega^2-
k^2-\mu^2+i0)}\frac{1}{(\omega-i0)^2}\,.
\ee
Decomposing the denominator
\be
\omega^2-k^2-\mu^2+i0=(\omega- \sqrt{k^2+\mu^2}+i0)
(\omega+\sqrt{k^2+\mu^2}-i0)
\ee
and integrating over $\omega$, we find
\be
I_1=-\frac{1}{\pi}\frac{1}{r_{12}r_{34}}
\int_{0}^{\infty}dk \,k
\frac{\sin{(k(r_{12}+r_{34}))}}{(k^2+\mu^2)^{3/2}}\,.
\ee
According to \cite{grad}, the last integral is
\be
I_1=-\frac{1}{\pi}\frac{r_{12}+r_{34}}{r_{12}r_{34}}
K_0[\mu (r_{12}+r_{34})]\,,
\ee
where
\be
K_0(z)=-\log{(z/2)}\sum_{k=0}^{\infty}\frac{(z/2)^{2k}}
{(k!)^2}+\sum_{k=0}^{\infty}\frac{z^{2k}}
{2^{2k}(k!)^2}\psi(k+1)\,.
\ee
Considering $\mu\rightarrow 0$ we find
\be
\Delta E_{\rm lad}^{(2,{\rm red})}&=&-\frac{\alpha^2}{\pi}
\sum_{P}(-1)^P\int d\bfx_1\cdots d\bfx_4 \;\overline{\psi}_{Pa}(\bfx_3)
\overline{\psi}_{Pb}(\bfx_4)\gamma_3^{\rho}\gamma_4^{\sigma}\nonumber\\
&&\times\sum_{n_1,n_2}^{\veps_{n_1}=\veps_{n_2}=\veps_a}\psi_{n_1}(\bfx_3)
\overline{\psi}_{n_1}(\bfx_1)\psi_{n_2}(\bfx_4)
\overline{\psi}_{n_2}(\bfx_2)\gamma_1^{\lambda}\gamma_2^{\nu}
\Bigl(\frac{1}{r_{12}}+\frac{1}{r_{34}}\Bigr)
\nonumber\\
&&\times
[\log{(r_{12}+r_{34})}+\log{\mu}-\log{2}-\psi(1)]
g_{\rho \sigma}g_{\lambda \nu}\psi_a(\bfx_1)\psi_b(\bfx_2)\,.
\ee
The corresponding contribution ($\veps_{n_1}=\veps_{n_2}=\veps_a$)
 from the crossed-ladder diagram
is calculated in the same way. The sum of the reducible contribution
from the ladder diagram and the related contribution from the
crossed-ladder diagram is \cite{shab88a}
\be
\Delta E^{(2,{\rm infr})}&=&-\frac{\alpha^2}{\pi}
\sum_{P}(-1)^P \sum_{n_1,n_2}^{\veps_{n_1}=\veps_{n_2}=\veps_a}
\int d
\bfx_1\cdots d\bfx_4 \; 
\overline{\psi}_{Pa}(\bfx_3)
\overline{\psi}_{Pb}(\bfx_4)
\gamma_3^{\rho}\gamma_4^{\sigma}\nonumber\\
&&\times\psi_{n_1}(\bfx_3)
\overline{\psi}_{n_1}(\bfx_1)\psi_{n_2}(\bfx_4)
\overline{\psi}_{n_2}(\bfx_2)\gamma_1^{\lambda}\gamma_2^{\nu}
\nonumber\\
&&\times\Bigl[g_{\rho \sigma}g_{\lambda \nu} 
 \Bigl(\frac{1}{r_{12}}+\frac{1}{r_{34}}\Bigr)
\log{(r_{12}+r_{34})}\nonumber\\
&&-g_{\rho \nu}g_{\sigma \lambda}
 \Bigl(\frac{1}{r_{14}}+\frac{1}{r_{23}}\Bigr)
\log{(r_{14}+r_{23})}\Bigr]\psi_a(\bfx_1)\psi_b(\bfx_2)\,.
\ee
 The terms containing
the  factor $\log{\mu}-\log{2}-\psi(1)$ have cancelled each other.
For the numerical evaluation of this contribution, 
one can use the following representation 
\cite{sh94a}
\begin{eqnarray}
\Delta E^{(2,{\rm infr})}
&=&   \frac{1}{\pi} \int_{0}^{\infty}\frac{dy}{y} \;
        \langle b a| S^{\prime}(y)|a b\rangle 
    [2 \langle  a b | S(y)|a b\rangle
          -\langle  a a | S(y)|a a\rangle- 
           \langle  b b | S(y)|b b\rangle ]
  \nonumber \\ 
      &&+ \frac{1}{\pi} \int_{0}^{\infty}\frac{dy}{y} \; 
        \langle b a | S(y)|a b\rangle 
        [ 2 \langle  a b | S^{\prime}(y)|a b\rangle 
           -\langle  a a |S^{\prime}(y)|a a\rangle- 
            \langle  b b |S^{\prime}(y)|b b\rangle ]
  \nonumber \\ 
     && - \frac{2}{\pi} \int_{0}^{\infty}\frac{dy}{y}\; 
        \langle b a | S(y)|a b\rangle
        \langle b a | S^{\prime}(y)|a b\rangle \,,
   \label{e3e46}
\end{eqnarray}
    where   
\begin{eqnarray}
  S(y)&=&  \frac{\alpha
     (1-\mbox{\boldmath $\alpha$} _{1}\cdot
\mbox{\boldmath $\alpha$}_{2})}{r_{1 2 }}
              e^{-yr_{1 2}}\,,  \\ 
          S^{\prime}(y)&=& 
   -\alpha(1-\mbox{\boldmath $\alpha$}_{1}\cdot
\mbox{\boldmath $\alpha$}_{2}) e^{-yr_{1 2}}\,.
     \label{e3e47}
\end{eqnarray}
Other representations of this term can be found in
Refs. \cite{blund93,lind95}.

\subsubsection{Self-energy and vacuum-polarization
screening diagrams}

The self-energy and vacuum-polarization screening diagrams
are presented in Figs. 22 and 23, respectively.
The derivation of the calculation formulas for the energy
shift of a single level due to the SE screening diagrams
using the TTGF method
is described in detail in Ref. \cite{yer99}.
If the unperturbed wave function is  given by equation
(\ref{e3e1}), the contribution of the SE screening diagrams
is written as
\be \label{sescr}
\Delta E&=&\Delta E^{\rm irred} +\Delta E^{\rm red}
+\Delta E^{\rm ver}\,,\\
\Delta E^{\rm irred}&=&\sum_{P}(-1)^{P} \Bigl\{
\sum_n^{\varepsilon_{n} \ne \varepsilon_{a}}
\langle PaPb|
I(\Delta )|nb\rangle\frac{1}{\varepsilon_{a}-\varepsilon_{n}}
\langle n|\Sigma(\varepsilon_{a})|a\rangle\nonumber\\
&&
+\sum_n^{\varepsilon_{n} \ne \varepsilon_{b}}
\langle PaPb|I(\Delta )|an\rangle \frac{1}{\varepsilon_{b}-
\varepsilon_{n}}
\langle n|\Sigma(\varepsilon_{b})|b\rangle\nonumber\\
&&
+\sum_n^{\varepsilon_{n} \ne \varepsilon_{Pa}}
\langle Pa|\Sigma(\varepsilon_{Pa})|n\rangle
\frac{1}{\varepsilon_{Pa}-\varepsilon_{n}}
\langle nPb|I(\Delta )|ab\rangle\nonumber\\
&&
+\sum_n^{\varepsilon_{n} \ne \varepsilon_{Pb}}
\langle Pb|\Sigma(\varepsilon_{Pb})|n\rangle
\frac{1}{\varepsilon_{Pb}-\varepsilon_{n}}
\langle Pan|I(\Delta )|ab\rangle\Bigr\}\,,
\label{sescr1}
\\
\Delta E^{\rm red}&=& \la ba|I^{\prime}(\veps_b-\veps_a)|ab\ra
   \Bigl[ \la a|\Sigma(\vare_a)|a\ra - \la b|\Sigma(\vare_b)|b\ra
   \Bigr] \nonumber\\
&&+\sum_P (-1)^P \la PaPb|I(\Delta )|ab\ra
 \Bigl[\la a|\Sigma^{\prime}(\vare_a)|a\ra +
       \la b|\Sigma^{\prime}(\vare_b)|b\ra  \Bigr] \,, 
\label{sescr2}
\\
\Delta E^{\rm ver}&=& \sum_P (-1)^P \sum_{n_1 n_2} \frac{i}{2\pi}
        \int_{-\infty}^{\infty} d\omega \; \Biggl[
\frac{\la n_1 Pb|I(\Delta )|n_2 b\ra
                \la Pa n_2|I(\omega )|n_1 a\ra}
        {[\vare_{Pa} -\omega -\vare_{n_1}(1-i0)]
         [\vare_a -\omega -\vare_{n_2}(1-i0)]    }
         \nonumber \\
&&+\frac{ \la Pa n_1|I(\Delta )|an_2\ra
             \la Pbn_2| I(\omega )|n_1 b\ra }
       {[\vare_{Pb} -\omega -\vare_{n_1}(1-i0)]
        [\vare_b -\omega -\vare_{n_2}(1-i0)]}
\Biggr] \,,
\label{sescr3}
\ee
where $\Delta = \vare_{Pa}-\vare_a$.
The corresponding contribution of the VP screening diagrams
is \cite{art97,art00}
\be \label{vpscr}
\Delta E&=&\Delta E_a^{\rm irred} +\Delta E_a^{\rm red}
+\Delta E_b\,,\\
\Delta E_a^{\rm irred}&=&\sum_{P}(-1)^{P} \Bigl\{
\sum_n^{\varepsilon_{n} \ne \varepsilon_{a}}
\langle PaPb|
I(\Delta )|nb\rangle\frac{1}{\varepsilon_{a}-\varepsilon_{n}}
\langle n|U_{\rm VP}|a\rangle\nonumber\\
&&
+\sum_n^{\varepsilon_{n} \ne \varepsilon_{b}}
\langle PaPb|I(\Delta )|an\rangle \frac{1}{\varepsilon_{b}-
\varepsilon_{n}}
\langle n|U_{\rm VP}|b\rangle\nonumber\\
&&
+\sum_n^{\varepsilon_{n} \ne \varepsilon_{Pa}}
\langle Pa|U_{\rm VP}|n\rangle
\frac{1}{\varepsilon_{Pa}-\varepsilon_{n}}
\langle nPb|I(\Delta )|ab\rangle\nonumber\\
&&
+\sum_n^{\varepsilon_{n} \ne \varepsilon_{Pb}}
\langle Pb|U_{\rm VP}|n\rangle
\frac{1}{\varepsilon_{Pb}-\varepsilon_{n}}
\langle Pan|I(\Delta )|ab\rangle\Bigr\}\,,
\label{vpscr1}
\\
\Delta E_a^{\rm red}&=& \la ba|I^{\prime}(\veps_b-\veps_a)|ab\ra
   \Bigl[ \la a|U_{\rm VP}|a\ra - \la b|U_{\rm VP}|b\ra
   \Bigr] \,,
\label{vpscr2}
\\
\Delta E_b&=&\sum_{P}(-1)^P
\la Pa Pb| I_{\rm VP}(\Delta) | ab\ra 
\,,
\label{vpscr3}
\ee
where
\be \label{uvpb} I_{\rm VP}(\varepsilon,\bfx, \bfy)
&=&\frac{\alpha^2}{2\pi i}\int\limits_{-\infty
}^\infty d\omega \; \int d \bfz_1\int d \bfz_2 \; \frac{\alpha_{1\mu}
\exp{(i|\varepsilon||\bfx-\bfz_1|})}{|\bfx
-\bfz_1|} \frac{\alpha_{2\nu}\exp{(i|\varepsilon||\bfy-\bfz_2|})}
{| \bfy-\bfz_2|}\nonumber \\
&\times&{\rm Tr}[\alpha^{\mu}G(\omega-\frac \varepsilon 2 ,\bfz_1{\bf
,z}_2)\alpha^{\nu}G(\omega+\frac \varepsilon 2,\bfz_2{\bf ,z}_1)]\,.
\ee
The expressions (\ref{sescr})-(\ref{vpscr3}) are
ultraviolet divergent. The renormalization of these expressions
can be performed in the same way as for the first-order
SE and VP contributions (see Refs. \cite{yer97,art97,art99,yer99}
for details).

\subsubsection{Quasidegenerate states}

Let us now consider some applications of the method to
the case of quasidegenerate states of a heliumlike ion.
This case arises, for instance, if one
is interested in the energies of the $(1s2p_{1/2})_1$ and
$(1s2p_{3/2})_1$ states. These  states
are strongly mixed for low and middle $Z$ and, therefore,
must be treated as quasidegenerate. It means that
 the off-diagonal
matrix elements of the energy operator $H$
between these states have to be taken into account.
The unperturbed wave functions are written as
\be \label{u1}
u_i(\bfx_1,\bfx_2)=
A_N \sum_{m_{i_1} m_{i_2}}\langle j_{i_1} m_{i_1} j_{i_2} m_{i_2}|JM\rangle
\sum_{P}(-1)^P \psi_{Pi_1}(\bfx_1)\psi_{Pi_2}(\bfx_2)\,,
\ee
where $A_N$ is the normalization factor 
equal to $1/\sqrt{2}$ for non-equivalent electrons and
to $1/2$ for equivalent electrons,
 $J$ is the total angular momentum, and $M$ is its projection.
 However, in what follows,
to compactify the formulas we will construct the matrix
 elements of  $H$  between the one-determinant wave functions
\be  \label{onedeti}
u_i(\bfx_1,\bfx_2)=
\frac{1}{\sqrt{2}}\sum_{P}(-1)^P \psi_{Pi_1}(\bfx_1)\psi_{Pi_2}(\bfx_2)\,.
\ee
The transition to the wave functions defined by equation
(\ref{u1}) can easily be accomplished in the final formulas.

First we consider the contribution from the one-photon exchange
diagram.
To derive the formulas for $H_{ik}^{(1)}$ we will assume
that $E_{i}^{(0)}\not =E_k^{(0)}$. However,
all the final formulas
 remain to be valid also for
the case  $E_{i}^{(0)}=E_k^{(0)}$ which was considered in detail
above.
According to the Feynman rules and the definition of $g(E)$,
the contribution of the one-photon exchange diagram 
 to $g^{(1)}(E)$ is
   \begin{eqnarray}
             g_{ik}^{(1)}(E)&=&
     \Bigl(\frac{i}{2\pi}\Bigr)^{2}
        \int_{-\infty}^{\infty}dp_{1}^{0}dp_{1}^{\prime 0}\;
                           \sum_{P}(-1)^{P}
             \frac{1}{p_{1}^{\prime 0}-\varepsilon_{P i_1}+i0}
            \frac{1}{E-p_{1}^{\prime 0}-\varepsilon_{P i_2}+i0}
 \nonumber \\ && \times
              \frac{1}{p_{1}^{0}-\varepsilon_{k_1}+i0}
               \frac{1}{E-p_{1}^{0}-\varepsilon_{k_2}+i0}
            \langle P i_1 P i_2|I(p_{1}^{\prime 0}-p_{1}^{0})|k_1 k_2\rangle\,.
   \label{g1}
\end{eqnarray}
Using the identities (\ref{iden1}) and (\ref{iden2}),
 we obtain
     \begin{eqnarray}
     K_{ik}^{(1)}&=&
   \frac{1}{2\pi i}\oint_{\Gamma} dE \;
\frac{E}{(E-E_i^{(0)})(E-E_k^{(0)})}
  \Bigl \{\Bigl(\frac{i}{2\pi}\Bigr)^{2}
 \int_{-\infty}^{\infty}dp_{1}^{0}dp_{1}^{\prime 0} \;
             \sum_{P}(-1)^{P}  \nonumber \\
 &&          \times \Bigl( \frac{1}{p_{1}^{\prime 0}-\varepsilon_{P i_1}+i0}
              +   \frac{1}{E-p_{1}^{\prime 0}-\varepsilon_{P i_2}+i0}\Bigr)
              \Bigl( \frac{1}{p_{1}^{0}-\varepsilon_{k_1}+i0}+
           \frac{1}{E-p_{1}^{0}-\varepsilon_{k_2}+i0}\Bigr) \nonumber \\[5pt]
 &&       \times \langle P i_1 P i_2|
I(p_{1}^{\prime 0}-p_{1}^{0})|k_1 k_2\rangle \Bigr\}\,.
   \label{K1}
\end{eqnarray}
The expression in the curly braces of (\ref{K1}) is a regular
function of $E$ inside the contour $\Gamma$, if the photon mass
$\mu$ is chosen as indicated above.
Calculating the $E$ residues and taking into account the identity
(\ref{e3e17}), we obtain
     \begin{eqnarray} \label{K1a}
K_{ik}^{(1)}&=&  \frac{i}{2\pi}
 \int_{-\infty}^{\infty}dp_{1}^{0} \;
             \sum_{P}(-1)^{P}
\frac{E_i^{(0)}\langle P i_1 P i_2|
I(\varepsilon_{Pi_1}-p_{1}^{0})|k_1 k_2\rangle
}{E_i^{(0)}-E_k^{(0)}}
\nonumber\\
& & \times
\Bigl( \frac{1}{p_{1}^{0}-\varepsilon_{k_1}+i0}+
   \frac{1}{E_i^{(0)}-p_{1}^{0}-\varepsilon_{k_2}+i0}\Bigr) \nonumber \\[5pt]
 &&+
 \frac{i}{2\pi}
 \int_{-\infty}^{\infty}dp_{1}^{\prime 0} \;
             \sum_{P}(-1)^{P}
\frac{E_k^{(0)}\langle P i_1 P i_2|
I(p_{1}^{\prime 0}-\varepsilon_{k_1})|k_1 k_2\rangle
}{E_k^{(0)}-E_i^{(0)}}
\nonumber\\
& & \times
\Bigl( \frac{1}{p_{1}^{\prime 0}-\varepsilon_{P i_1}+i0}
              +   \frac{1}{E_{k}^{(0)}
-p_{1}^{\prime 0}-\varepsilon_{P i_2}+i0}\Bigr)\,.
\end{eqnarray}
In the same way we find
     \begin{eqnarray} \label{P1}
P_{ik}^{(1)}&=&
 \frac{i}{2\pi}
 \int_{-\infty}^{\infty}dp_{1}^{0} \;
             \sum_{P}(-1)^{P}
\frac{\langle P i_1 P i_2|
I(\varepsilon_{Pi_1}-p_{1}^{0})|k_1 k_2\rangle
}{E_i^{(0)}-E_k^{(0)}}
\nonumber\\
& &  \times
  \Bigl( \frac{1}{p_{1}^{0}-\varepsilon_{k_1}+i0}+
   \frac{1}{E_i^{(0)}-p_{1}^{0}-\varepsilon_{k_2}+i0}\Bigr) \nonumber \\[5pt]
 &&+
 \frac{i}{2\pi}
 \int_{-\infty}^{\infty}dp_{1}^{\prime 0} \;
             \sum_{P}(-1)^{P}
\frac{\langle P i_1 P i_2|
I(p_{1}^{\prime 0}-\varepsilon_{k_1})|k_1 k_2\rangle
}{E_k^{(0)}-E_i^{(0)}}
\nonumber\\
& & \times
 \Bigl( \frac{1}{p_{1}^{\prime 0}-\varepsilon_{P i_1}+i0}
              +   \frac{1}{E_{k}^{(0)}
-p_{1}^{\prime 0}-\varepsilon_{P i_2}+i0}\Bigr)\,.
\end{eqnarray}
Symmetrizing equations (\ref{K1a}) and (\ref{P1})
with respect to both electrons we
transform them to the form
\begin{eqnarray} \label{K1aa}
K_{ik}^{(1)}&=&
\sum_{P}(-1)^{P} 
\left\{ 
\begin{array}{c} \\[10pt]  \end{array}
\frac{1}{2}
[\langle P i_1 P i_2|I(\Delta_1) |k_1 k_2\rangle
+\langle P i_1 P i_2|I(\Delta_2) |k_1 k_2\rangle]
\right.
\nonumber\\
& & 
-\frac{(E_i^{(0)}+E_k^{(0)})}{2}
 \frac{i}{2\pi}
 \int_{-\infty}^{\infty}d\omega\,
\langle P i_1 P i_2|I(\omega) |k_1 k_2\rangle
\nonumber\\
&& 
\left.
\times \Bigl[
\frac{1}{(\omega+\Delta_1-i0)
(\omega-\Delta_2-i0)}
+\frac{1}{(\omega+\Delta_2-i0)
(\omega-\Delta_1-i0)}\Bigr]\
\begin{array}{c} \\ \end{array}
\right\}\,,
\nonumber\\
\\
P_{ik}^{(1)}&=&
-\sum_{P}(-1)^{P}
 \frac{i}{2\pi}
 \int_{-\infty}^{\infty}d\omega\,
\langle P i_1 P i_2|I(\omega) |k_1 k_2\rangle
\nonumber\\
&&\times \Bigl[
\frac{1}{(\omega+\Delta_1-i0)
(\omega-\Delta_2-i0)}
+\frac{1}{(\omega+\Delta_2-i0)
(\omega-\Delta_1-i0)}\Bigr]\,,
 \label{P1a}
\ee
where $\Delta_1=\varepsilon_{Pi_1}-\varepsilon_{k_1}$
and $\Delta_2=\varepsilon_{Pi_2}-\varepsilon_{k_2}$.
Substituting (\ref{K1aa}), (\ref{P1a}) into (\ref{H1}),
we get \cite{sh93,mittleman}
\be \label{ope}
H_{ik}^{(1)}=
 \frac{1}{2}\sum_{P}(-1)^P
[\langle P i_1 P i_2|I(\Delta_1) |k_1 k_2\rangle
+\langle P i_1 P i_2|I(\Delta_2) |k_1 k_2\rangle]\,.
\ee

Let us now consider the contribution to $H$ from the
combined $\delta V$ - interelectronic interaction diagrams
presented in Fig. 24. For simplicity, we will assume that
$\delta V$ is a spherically-symmetric potential.
 In the case under consideration,
 the simplest way to derive the formulas for $H_{ik}^{(2)}$
 consists
in using the fact that these diagrams  can be obtained
as the first-order (in $\delta V$) correction
to the one-photon exchange contribution derived above.
So, the contribution from these diagrams 
can be obtained by the following replacements in 
equation (\ref{ope})
\be
|k_1\ra &\rightarrow& |k_1\ra +\delta |k_1 \ra\,, \\
|k_2\ra &\rightarrow& |k_2\ra +\delta |k_2 \ra\,, \\
|Pi_1\ra &\rightarrow& |Pi_1\ra +\delta |Pi_1 \ra\,, \\
|Pi_2\ra &\rightarrow& |Pi_2\ra +\delta |Pi_2 \ra\,, \\
I(\varepsilon_a-\varepsilon_b) &\rightarrow&
I(\varepsilon_a+\delta \varepsilon_a-\varepsilon_b-\delta \varepsilon_b)\,,
\ee
where, to first  order in $\delta V$,
\be
\delta \varepsilon_a&=&\la a|\delta V|a\ra\,,\\
\delta |a \ra&=&\sum_{n}^{\varepsilon_n \ne \varepsilon_a}
\frac{|n\ra \la n|
\delta V|a\ra} {\varepsilon_a-\varepsilon_n}\,.
\ee
Here we have taken into account that, due to the spherical 
symmetry of $\delta V$, $\la n|\delta V|a\ra=0$ if 
$\veps_n =\veps_a$ and $|n\ra \ne |a\ra\,$.
 Decomposing the modified expression
for the one-photon exchange diagram to the first order in 
$\delta V$, we find that the total correction
 is the sum of the irreducible and
reducible parts,
\be \label{hik}
H_{ik}^{(2)}&=&H_{ik}^{\rm (2,irred)}+H_{ik}^{\rm (2, red)}\,,
\ee
where
\be \label{irr}
H_{ik}^{\rm(2,irred)}&=& \frac{1}{2}\sum_{P}(-1)^P
[\langle \delta P i_1 P i_2|I(\Delta_1)+I(\Delta_2) |k_1 k_2\rangle
\nonumber\\
&&+\langle P i_1 \delta P i_2|I(\Delta_1)+I(\Delta_2) |k_1 k_2\rangle
\nonumber\\
&&+\langle P i_1 P i_2|I(\Delta_1)+I(\Delta_2) |\delta k_1 k_2\rangle
\nonumber\\
&&+\langle P i_1 P i_2|I(\Delta_1) +I(\Delta_2) |k_1 \delta k_2\rangle]\,
\ee
and
\be \label{red}
H_{ik}^{\rm(2,red)}&=& \frac{1}{2}\sum_{P}(-1)^P
\{[\la Pi_1|\delta V|Pi_1\ra-\la k_1|\delta V|k_1\ra]
\langle P i_1 P i_2|I'(\Delta_1)|k_1 k_2\rangle
\nonumber\\
&&+[\la Pi_2|\delta V|Pi_2\ra-\la k_2|\delta V|k_2\ra]
\langle P i_1 P i_2|I'(\Delta_2)|k_1 k_2\rangle\}\,.
\ee

Equations (\ref{irr}) and (\ref{red})  provide the matrix
elements between the one-determinant wave functions defined by equation
(\ref{onedeti}). To get the matrix elements between the wave functions
defined by equation (\ref{u1}), we have to multiply these equations
by the Clebsch-Gordan coefficients and to sum over the projections of
the one-electron angular momenta.

The expression for $H_{ik}^{(2)}$ can also be derived
by direct application of the TTGF method. It 
can easily be performed in the same way as for the one-photon
exchange diagram. We note that this derivation
yields a formula for $H_{ik}^{(2)}$ which is slightly different
from the expression given above. In particular, for the
irreducible contribution one finds
\be \label{irr2}
H_{ik}^{\rm(2,irred)}&=& \frac{1}{2}\sum_{P}(-1)^P
[\langle \delta P i_1 P i_2|I(\Delta_1)+I(\Delta_2) |k_1 k_2\rangle
\nonumber\\
&&+\langle P i_1 \delta P i_2|I(\Delta_1)+I(\Delta_2) |k_1 k_2\rangle
\nonumber\\
&&+\langle P i_1 P i_2|I(\Delta_1)+I(\Delta_2) |\delta k_1 k_2\rangle
\nonumber\\
&&+\langle P i_1 P i_2|I(\Delta_1) +I(\Delta_2) |k_1 \delta k_2\rangle]
\nonumber\\
&& +\Delta H_{ik}^{\rm(2,irred)}\,,
\ee
where
\be \label{delh}
\Delta H_{ik}^{\rm(2,irred)}&=&\frac{1}{2}(E_i^{(0)}-E_k^{(0)})
\sum_P(-1)^P\frac{i}{2\pi}\int_{-\infty}^{\infty}d\omega \;
\Bigl\{\frac{\la \delta Pi_1 Pi_2|I(\omega-\veps_{Pi_1})
|k_1 k_2\ra}{(\omega-\veps_{k_1}+i0)(E_i^{(0)}-\omega-\veps_{k_2}-i0)}
\nonumber\\
&&+\frac{\la Pi_1 \delta Pi_2|I(\omega-\veps_{Pi_1})
|k_1 k_2\ra}{(\omega-\veps_{k_1}-i0)(E_i^{(0)}-\omega-\veps_{k_2}+i0)}
\nonumber\\
&&-\frac{\la Pi_1 Pi_2|I(\omega-\veps_{k_1})
|\delta k_1 k_2\ra}{(\omega-\veps_{Pi_1}+i0)
(E_k^{(0)}-\omega-\veps_{Pi_2}-i0)}
\nonumber\\
&&-\frac{\la Pi_1 Pi_2|I(\omega-\veps_{k_1})
|k_1 \delta k_2\ra}
{(\omega-\veps_{Pi_1}-i0)(E_k^{(0)}-\omega-\veps_{Pi_2}+i0)}
\nonumber\\
&&+\sum_{n}^{n\ne k_1}
\frac{\la Pi_1 Pi_2|I(\omega-\veps_{Pi_1})
|n k_2\ra \la n|\delta V|k_1\ra}
{(\omega-\veps_{k_1}-i0)(E_i^{(0)}-\omega-\veps_{k_2}+i0)
(\omega-\veps_n(1-i0))}\nonumber\\
&&+\sum_{n}^{n\ne k_2}
\frac{\la Pi_1 Pi_2|I(\omega-\veps_{Pi_1})
|k_1 n\ra \la n|\delta V|k_2\ra}
{(\omega-\veps_{k_1}+i0)(E_i^{(0)}-\omega-\veps_{k_2}-i0)
(E_i^{(0)}-\omega-\veps_n(1-i0))}\nonumber\\
&&-\sum_{n}^{n\ne Pi_1}
\frac{\la Pi_1|\delta V|n\ra
\la n Pi_2|I(\omega-\veps_{k_1})
|k_1 k_2\ra}
{(\omega-\veps_{Pi_1}-i0)(E_k^{(0)}-\omega-\veps_{Pi_2}+i0)
(\omega-\veps_n(1-i0))}\nonumber\\
&&-\sum_{n}^{n\ne Pi_2}
\frac{\la Pi_2|\delta V|n\ra
\la Pi_1 n|I(\omega-\veps_{k_1})
|k_1 k_2\ra}
{(\omega-\veps_{Pi_1}+i0)(E_k^{(0)}-\omega-\veps_{Pi_2}-i0)
(E_k^{(0)}-\omega-\veps_n(1-i0))}\Bigr\}\,.
\ee
The term $\Delta H_{ik}^{\rm(2,irred)}$ approaches zero if
$E_i^{(0)}\rightarrow E_k^{(0)}\,$.
The expressions (\ref{irr}) and (\ref{irr2}) differ by the
term $\Delta H_{ik}^{\rm(2,irred)}$ which can be represented as
\be
\Delta H_{ik}^{\rm(2,irred)}=(E_i^{(0)}-E_k^{(0)}) O_{ik}^{(2)}.
\ee  
Here $O^{(2)}$ is an operator of  second order
in the perturbation parameter which we denote by $\lambda_0$ 
(for simplicity, we assume here that $\delta V$ and the
interelectronic-interaction operator $I(\omega)$ are characterized
by the same perturbation parameter).
This fact can be understood by observing that the integrand in
(\ref{delh}) is the sum of terms which contain singularities
from the external electron propagators only  above or below
of the real axis and, therefore, integrating over  $\omega$
cannot result in the appearance of contributions
 $\sim 1/(E_i^{(0)}-E_k^{(0)})$ that could compensate the
factor  $(E_i^{(0)}-E_k^{(0)})$.
In particular, it means that $O_{ik}^{(2)}$ remains finite
when $E_i^{(0)}\rightarrow E_k^{(0)}$.
 It can be shown that the
term  $\Delta H_{ik}^{\rm(2,irred)}$ 
 contributes only to  third and higher orders 
in $\lambda_0$ and, therefore, can be omitted if we
restrict our calculations to second order in $\lambda_0$.
Let us prove this fact for the case of two 
quasidegenerate levels. In this case the energy levels are 
determined from the equation
\be
(E-H_{11})(E-H_{22})-H_{12}H_{21}=0
\ee
which yields
\be
E_{1,2}=\frac{H_{11}+H_{22}}{2} \pm \frac{1}{2}
\sqrt{(H_{11}-H_{22})^2+4H_{12}H_{21}}\,.
\ee
If $E_1-E_2 \sim \lambda_0$, the proof of the statement
is evident. If $E_1-E_2 \gg \lambda_0$, the contribution
of the second-order off-diagonal matrix elements
is given by
\be
\Delta E_{1,2} \approx 
\pm (H_{12}^{(2)}H_{21}^{(1)}+H_{12}^{(1)}H_{21}^{(2)})
/(E_1^{(0)}-E_2^{(0)})\,.
\ee
>From this equation we find that the terms 
$\sim (E_i^{(0)}-E_k^{(0)})O_{ik}^{(2)}$
 in $H_{ik}^{(2)}$ contribute only to 
third and higher orders in $\lambda_0$.

Equations (\ref{hik})-(\ref{red}) will give a part of
the VP screening contribution, if we replace $\delta V$ by
$U_{\rm VP}$. The total contribution of the VP screening
diagrams for quasidegenerate states is derived in Ref. \cite{art00}.
Corresponding formulas for the SE screening diagrams
are derived in Ref. \cite{leb}. 

\subsection{Nuclear recoil corrections}

So far we considered the nucleus as a source of the external
Coulomb field $V_{\rm C}$. This consideration corresponds to the approximation
of an infinite nuclear mass. However, high precision calculations
of the energy levels in high-$Z$ few-electron atoms
must include also the nuclear recoil corrections to first
order in $m/M$ ($M$ is the nuclear mass) and to zeroth order
in $\alpha$ (but to all orders in $\alpha Z$). As was shown
in \cite{sh98}, these corrections can be included
 in calculations of the
energy levels by adding  an additional term
to the standard Hamiltonian
of the electron-positron field interacting with the quantized
electromagnetic field and with the Coulomb field of the nucleus
$V_{\rm C}$. In the Coulomb gauge and the Schr\"odinger representation,
 this term is given by
\begin{eqnarray} \label{hamm}
H_{M}&=&
\frac{1}{2M}
\int d{\bf x} \;
\psi^{\dag}({\bf x})(-i
\nabla_{\bf x})
\psi({\bf x})
\int d{\bf y} \;
\psi^{\dag}({\bf y})(-i
\nabla_{\bf y})
\psi({\bf y})\nonumber\\
&&-\frac{eZ}{M}
\int d{\bf x} \;
\psi^{\dag}({\bf x})(-i
\nabla_{\bf x})
\psi({\bf x}){\bf A}(0)+\frac{e^{2}Z^{2}}{2M}{\bf A}^{2}(0)\,.
\end{eqnarray}
$H_M$ taken in the interaction
representation must be added to the interaction Hamiltonian.
It gives the following additional lines and vertices to 
the Feynman rules (we assume that the Coulomb gauge and the
Furry picture are used).

\begin{enumerate}
\item {\it Coulomb contribution}.
An additional line ("Coulomb-recoil" line) appears to be
\newline\\
\setlength{\unitlength}{0.7mm}
\begin{picture}(60,5)(0,0)
  \multiput(15,2)(2,0){15}{\circle*{1} }
  \put(15,2){\circle*{2}}
  \put(45,2){\circle*{2}}
  \put(30,6){$\omega$}
  \put(12,-6){${\bf x}$}
  \put(45,-6){${\bf y}$}
  \label{intphotlinerec}
\end{picture}
          $ \frac{i}{2\pi} \frac{\delta_{kl}}{M}
\int_{-\infty}^{\infty}d\omega\,. $ \\
\newline
This line joins two vertices each of which corresponds to
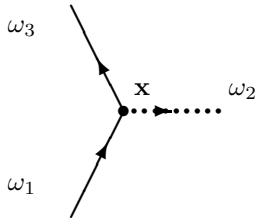
\newline\\
\begin{picture}(60,60)(0,0)
  \put(50,30){\line(-1,2){10}}
  \put(50,30){\line(-1,-2){10}}
  \multiput(50,30)(2,0){10}{\circle*{1}}
  \put(50,30){\circle*{2}}
  \put(52,33){{\bf x}}
  \put(70,33){$\omega_{2}$}
  \put(28,15){$\omega_{1}$}
  \put(28,45){$\omega_{3}$}
  \put(58,30){\vector(1,0){1}}
  \put(46,38){\vector(-1,2){1}}
  \put(46,22){\vector(1,2){1}}
\label{vertexrec}
\end{picture}  
       $ - 2\pi i\gamma^{0}\delta(\omega_{1}-\omega_{2}-\omega_{3})
            \int d{\bf x}\,p_{k} \;,$ \newline \\ 
\newline
where
 ${\bf p}=
-i\nabla_{\bf x}$ and $k=1,2,3$.

\item {\it One-transverse-photon contribution}.
An additional vertex on an electron line appears to be
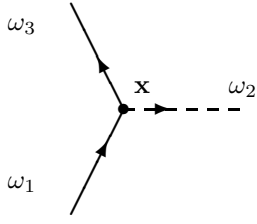
\newline
\begin{picture}(60,60)(0,0)
  \put(50,30){\line(-1,2){10}}
  \put(50,30){\line(-1,-2){10}}
  \multiput(50,30)(4,0){6}{\line(1,0){2}}
  \put(50,30){\circle*{2}}
  \put(52,33){{\bf x}}
  \put(70,33){$\omega_{2}$}
  \put(28,15){$\omega_{1}$}
  \put(28,45){$\omega_{3}$}
  \put(58,30){\vector(1,0){1}}
  \put(46,38){\vector(-1,2){1}}
  \put(46,22){\vector(1,2){1}}
\label{vertex1}
\end{picture}  
   $ - 2\pi i\gamma^{0}
\delta(\omega_{1}-\omega_{2}-\omega_{3})
       \frac{eZ}{M} \int d{\bf x}\,p_{k} \;,$ \newline \\ 
The transverse photon line attached to this vertex (at
the point ${\bf x}$) is
\newline\\
\begin{picture}(60,5)(0,0)
  \multiput(15,2)(4,0){9}{\line(1,0){2} }
  \put(15,2){\circle*{2}}
  \put(30,6){$\omega$}
  \put(12,-6){${\bf x}$}
  \put(47,-6){${\bf y}$}
  \label{intphotlinerec1}
\end{picture}
          $ \frac{i}{2\pi}
\int_{-\infty}^{\infty}d\omega \; D_{kl}(\omega,{\bf y})\,. $ \\
\newline\\
At the point ${\bf y}$ this line is to be attached to a usual
vertex, where we have $-2\pi i e\gamma^{0}\alpha_{l}2\pi
\delta(\omega_{1}-\omega_{2}-\omega_{3})\int d{\bf y}$.
$\alpha_{l}$ ($l=1,2,3$)
are the usual Dirac matrices.

\item {\it Two-transverse-photon contribution}.
An additional line ("two-transverse-photon-recoil" line)
appears to be
\newline\\
\begin{picture}(60,5)(0,0)
  \multiput(15,2)(4,0){9}{\line(1,0){2} }
  \put(32,2){\circle*{2}}
  \put(12,-6){${\bf x}$}
  \put(47,-6){${\bf y}$}
  \put(30,6){$\omega$}
  \label{intphotlinerec2}
\end{picture}
          $ \frac{i}{2\pi}\frac{e^{2}Z^{2}}{M}
\int_{-\infty}^{\infty}d\omega \;
 D_{il}(\omega,{\bf x})
 D_{lk}(\omega,{\bf y})\,. $ 
\newline\\
This line joins  usual vertices (see the previous item).
\end{enumerate}

Let us apply this formalism to the case of a single level $a$
in a one-electron atom.
To find the Coulomb nuclear recoil correction, we have to calculate 
the contribution of the diagram shown in Fig. 25. According to the
Feynman rules given above 
we obtain
\begin{eqnarray}
\Delta g_{aa}^{(1)}(E)
=\frac{1}{(E-E_{a}^{(0)})^{2}}\frac{1}{M}\frac{i}{2\pi}
\int_{-\infty}^{\infty}d\omega \; \sum_{n}\frac
{\langle a| p_{i}|n\rangle\langle n|p_{i}|a\rangle}
{\omega-\varepsilon_{n}(1-i0)}\,.
\end{eqnarray}
The formula (\ref{e3e11}) gives
\begin{eqnarray}
\Delta E_{\rm C}=\frac{1}{M}\frac{i}{2\pi}
\int_{-\infty}^{\infty}d\omega \; \sum_{n}\frac
{\langle a| p_{i}|n\rangle\langle n|p_{i}|a\rangle}
{\omega-\varepsilon_{n}(1-i0)}\,.
\end{eqnarray}
The one-transverse-photon nuclear recoil correction corresponds
to the diagrams shown in Fig. 26. A similar calculation yields
\begin{eqnarray}
\Delta E_{\rm tr(1)}&=&\frac{4\pi\alpha Z}{M}\frac{i}{2\pi}
\int_{-\infty}^{\infty}d\omega\,\sum_{n}\Biggl\{
\frac{
\langle a|p_{i}|n\rangle\langle n|
\alpha_{k} D_{ik}(\varepsilon_{a}-\omega)|a\rangle}
{\omega-\varepsilon_{n}(1-i0)}
\nonumber\\
&&
+\frac{
\langle a|
\alpha_{k} D_{ik}(\varepsilon_{a}-\omega)|n\rangle
\langle n|p_{i}|a\rangle}
{\omega-\varepsilon_{n}(1-i0)}\Biggr\}\,.
\end{eqnarray}
The two-transverse-photon nuclear recoil correction is defined
by the diagram shown in Fig. 27. We find
\begin{eqnarray}
\Delta E_{\rm tr(2)}&=&\frac{(4\pi\alpha Z)^{2}}{M}\frac{i}{2\pi}
\int_{-\infty}^{\infty}d\omega\,\sum_{n}\nonumber\\
&&\times\frac{
\langle a|\alpha_{i}D_{il}(\varepsilon_{a}-\omega)|n\rangle\langle n|
\alpha_{k} D_{lk}(\varepsilon_{a}-\omega)|a\rangle}
{\omega-\varepsilon_{n}(1-i0)}\,.
\end{eqnarray}
The sum of all the contributions is 
\begin{eqnarray} \label{reccom}
\Delta E&=&\frac{1}{M}\frac{i}{2\pi}\int_{-\infty}^{\infty}d\omega\,
\langle a|[p_{i} +4\pi\alpha Z\alpha_{l} D_{li}(\omega)]\nonumber\\
&&\times G_{\rm C}(\omega +\varepsilon_{a})
[ p_{i} +4\pi\alpha Z\alpha_{m} D_{mi}(\omega)]|a\rangle\,,
\end{eqnarray}
where $G_{\rm C}(\omega)$
is the Dirac-Coulomb Green function defined above.
For  practical calculations it is convenient to represent the
expression (\ref{reccom}) by the sum of a lower-order term
$\Delta E_{\rm L}$ and a higher-order term $\Delta E_{\rm H}$:
\be \label{recmy0}
\Delta E &=&\Delta E_{\rm L}+\Delta E_{\rm H}\,,\\
 \label{recmy1}
\Delta E_{\rm L}&=&\frac{1}{2M}\la a|[p_i^2-(D_i(0)p_i+p_i D_i(0))]|a\ra\,,
\\
\Delta E_{\rm H}&=&\frac{i}{2\pi M}\int_{-\infty}^{\infty} d\omega \;
\la a|\Bigl(D_i(\omega)-\frac{[p_i,V_{\rm C}]}{\omega+i0}\Bigr)
G_{\rm C}(\omega+\veps_a)\Bigl(D_i(\omega)+\frac{[p_i,V_{\rm C}]}{\omega+i0}\Bigr)
|a\ra\,, \label{recmy2}
\ee
where $D_i(\omega)=-4\pi \alpha Z \alpha_l D_{li}(\omega)$.
The term $\Delta E_{\rm L}$ contains all the recoil corrections
within the $(\alpha Z)^4 m^2/M$ approximation, while the term
$\Delta E_{\rm H}$ contains  the contribution of order
$(\alpha Z)^5 m^2/M$  and all contributions of higher orders
in $\alpha Z$ which are not included in  $\Delta E_{\rm L}$.
The formulas (\ref{recmy0})-(\ref{recmy2})
were first derived by a quasipotential method in Ref. \cite{sh85}
and subsequently rederived by other methods in Refs.
\cite{yelkh94,pach95}. 
The representation (\ref{reccom}) was  found in Ref. \cite{yelkh94}.

Consider now a two-electron atom. For simplicity, as usual,
 we  assume  that the unperturbed
wave function is a one-determinant function (\ref{e3e1}).
The nuclear recoil correction is the sum of the one-electron
and two-electron contributions. Using the Feynman rules
and the formula (\ref{e3e11}), one easily
finds that the one-electron contribution is equal to the sum
of the expressions ({\ref{reccom}) for the $a$ and $b$ states.
 The two-electron
contributions correspond to the diagrams shown in Figs. 28-30.
A simple calculation of these diagrams yields 
\begin{eqnarray} \label{recint}
\Delta E^{(\rm int)}&=&
\frac{1}{M} \sum_{P}(-1)^{P}
\langle Pa|[p_{i}+4\pi\alpha Z\alpha_{l}D_{li}
(\varepsilon_{Pa}-\varepsilon_{a})]
|a\rangle\nonumber\\
&&\times\langle Pb|[p_{i}+4\pi\alpha Z\alpha_{m}D_{mi}
(\varepsilon_{Pb}-\varepsilon_{b})]
|b\rangle\,.
\end{eqnarray}
The formula (\ref{recint}) was
first derived by a quasipotential method in Ref. \cite{sh88}.

\section{Transition probabilities and cross sections of
scattering processes}

According to the basic principles of quantum field theory 
\cite{bogolyubov}, the
number of the particles scattered into the interval 
$d{\bf p}_{1}^{\prime}\cdots
d{\bf p}_{r}^{\prime}$ for a  unit time  and in a unit volume is
\begin{eqnarray} \label{vergen}
dW_{p \rightarrow p'} = (2 \pi)^{3s+1}n_{1}\cdots n_{s}|{\tau}_
{p'p}|^{2}\delta(E_{p}-E_{p'})d{\bf p}_{1}^{\prime}\cdots d{\bf p}_{r}^
{\prime}\,,
\end{eqnarray}
where $p,p'$ are the initial and final states of the system,
respectively; ${\tau}_{p'p}$ is the amplitude of the process
defined by
\begin{eqnarray} \label{deftau}
\la p'|(S-I)|p\ra=2\pi i\, \delta(E_{p}-E_{p'}){\tau}_{p'p}\, ,
\end{eqnarray}
$S$ is the scattering operator, $I$ is the identity operator,
 $s$ is the number of initial particles,
$r$ is the number of final particles; $n_{1},...n_{s}$ are the
average numbers of particles per unit volume.

 We will consider
the scattering of photons and electrons by an atom that is located
at the origin of the coordinate frame. The differential
cross section is defined by
\begin{eqnarray}
d\sigma=\frac{dW_{p\rightarrow p'}}{j}\, ,
\end{eqnarray}
where $j$ is the current of initial particles (for photons
$j=nc$; for electrons $j=nv$, where $v$ is the velocity of the electrons in
the nucleus frame). The total
cross section can be found by integrating the differential
cross section over all  final states. The cross section
for the elastic scattering is 
\be
\sigma_{\rm tot}^{({\rm elast})}(p)=\frac{(2\pi)^4}{v}\int d{\bfp}'\;
\delta(E_p-E_{p'})|\tau_{p'p}|^2\,. 
\ee
The total (elastic plus inelastic)
 cross section can be found
by employing the optical theorem
\begin{eqnarray}
\sigma_{\rm tot}(p)=\frac{2(2 \pi)^{3}}{v} {\rm Im}{\tau}_{pp}\, .
\end{eqnarray}
In terms of the amplitude $f_{p'p}$ which is defined so that
$d\sigma=|f_{p'p}|^2 d\Omega$, the optical theorem
has a well-known form (see, e.g., \cite{berestetsky})
\begin{eqnarray}
\sigma_{\rm tot}(p)=\frac{4 \pi}{|\bfp|}\, {\rm Im}{f}_{pp}\, .
\end{eqnarray}
The aim of this section  is to derive formulas for
the transition and scattering amplitudes for various processes
in the framework of QED.

\subsection{Photon emission by an atom}

Consider the process of the photon emission by an atom.
According to the standard reduction technique (see,
e.g., \cite{itzykson,bjorken}), the atomic transition amplitude
from state $a$ to state $b$ accompanied by photon emission
with momentum $\bfk_f$ and polarization $\beps_f$ is
\begin{eqnarray} \label{red1}
S_{\gamma_{f},b;a}=\la b|a_{\rm out}(k_{f},{\eps}_{f})|a\ra
=-iZ_{3}^{-\frac{1}{2}}
\int d^{4}y \; \frac{{\eps}_{f}^{\nu *} \exp{(ik_{f}\cdot y)}}
{\sqrt{2k_{f}^{0}(2\pi )^3}} 
\la b|j_{\nu}(y)|a\ra\,.
\end{eqnarray}
Here $j_{\nu}(y)$ is the electron-positron
 current operator in the Heisenberg representation,
 $|a\ra$ and  $|b\ra$ are the
vectors of the initial and final states in the Heisenberg
representation, $Z_3$ is a renormalization constant,
$a\cdot b\equiv a_{\nu}b^{\nu}$,
 $\eps_f=(0,\beps_f)$,
$k_f=(k_f^0,\bfk_f)$, and $k_{f}^{0} \equiv |{\bf k}_{f}|$.
Employing the equation (see Appendix A)
\be
j^{\nu}(y)=\exp{(iHy^0)}j^{\nu}(0,\bfy)\exp{(-iHy^0)}\,,
\ee
we obtain
\begin{eqnarray} \label{red2}
S_{\gamma_{f},b;a}&=&-iZ_{3}^{-\frac{1}{2}}
\int d^{4}y \; \exp{[i(E_b+k_f^0-E_a)y^0]} A_f^{\nu *}(\bfy) 
\la b|j_{\nu}(0,\bfy)|a\ra\nonumber\\
&=&-2\pi iZ_{3}^{-\frac{1}{2}}\delta(E_b+k_f^0-E_a)
\int d\bfy \; A_f^{\nu *}(\bfy)
\la b|j_{\nu}(0,\bfy)|a\ra\,,
\end{eqnarray}
where
\begin{eqnarray}\label{defA}
A_{f}^{\nu}(\bfx)=
 \frac{{\eps}_{f}^{\nu} \exp{(i{\bf k}_{f}\cdot{\bf x})}}
{\sqrt{2k_{f}^{0}(2\pi )^3}}\,
\end{eqnarray}
is the wave function of the emitted photon.
Since $|a\ra$ and $|b\ra$ are bound states, 
equation (\ref{red2}) as well as the standard reduction technique
\cite{itzykson,bjorken} cannot be  used for a
direct evaluation of the amplitude. The desired
 calculation formula can be derived within the two-time Green
function formalism \cite{shab88b,shab90a,shab90b}.

To formulate the method for a general case,
we assume that in  zeroth approximation the state $a$
belongs to an $s_a$-dimensional subspace
 of unperturbed degenerate states
$\Omega _{a}$ and the state $b$ belongs to an 
$s_b$-dimensional subspace of 
unperturbed degenerate states $\Omega_{b}$. We denote
the projectors onto these subspaces  
 by $P^{(0)}_{a}$ and $P^{(0)}_{b}$, respectively.
We  denote the exact states originating from $\Omega _{a}$ 
by $|n_{a}\ra$ and
the exact states originating from $\Omega _{b}$
by $|n_{b}\ra$. We also assume that on an intermediate stage
of the  calculations
a non-zero photon mass $\mu$ is introduced. It is considered
to be larger than the energy splitting  of the
initial and final states under consideration and much smaller than the
distance to  other levels. 

We introduce
\begin{eqnarray} \label{gtranscal}
\lefteqn{
{\cal G}_{\gamma_f}(E',E;{\bf x}_1^{\prime},...{\bf x}_N^{\prime};
 {\bf x}_1,...{\bf x}_N) \delta (E^{\prime}+k^0-E)\;\;\;\;\;\;\;\;\;\;
\;\;\;\;\;\;\;\;\;\;\;\;\;\;\;\;\;\;
\;\;\;\;\;\;\;\;\;\;\;}\nonumber\\
& = &
\frac{1}{2\pi i}\frac{1}{2\pi}\frac{1}{N!}\int_{-\infty}^
{\infty}dx^0dx'^0 \; \int d^4y \; \exp{(iE^{\prime}x'^0-iEx^0) }
\exp{(ik^0y^0)}\nonumber\\
&&\times 
A_f^{\nu *}(\bfy)
\langle 0|T\psi (x'^0,{\bf x}_1
^{\prime})\cdots\psi (x'^0,{\bf x}_N^{\prime})
\nonumber\\
&&\times
j_{\nu}(y)
\overline{\psi}
(x^0,{\bf x}_N)\cdots\overline{\psi}
(x^0,{\bf x}_1)|0\rangle\,,
\end{eqnarray}
where, as in the previous section, $\psi(x)$ is the electron-positron
field operator in the Heisenberg representation. 
Let us investigate the singularities
 of ${\cal G}_{\gamma_f}$
in the vicinity of the points $E'\approx E_b^{(0)}$ and $E\approx E_a^{(0)}$.
Using the transformation rules
\be
\psi(x^0,\bfx)&=&
\exp{(iHy^0)}\psi(x^0-y^0,\bfx)\exp{(-iHy^0)}\,,\nonumber\\
j(y^0,\bfy)&=&
\exp{(iHy^0)}j(0,\bfy)\exp{(-iHy^0)}\,,
\ee
we obtain
\begin{eqnarray} \label{gtrans12}
\lefteqn{
{\cal G}_{\gamma_f}(E',E;{\bf x}_1^{\prime},...{\bf x}_N^{\prime};
 {\bf x}_1,...{\bf x}_N) \delta (E^{\prime}+k^0-E)\;\;\;\;\;\;\;\;\;\;
\;\;\;\;\;\;\;\;\;\;\;\;\;\;\;\;\;\;\;\;
\;\;\;\;\;\;\;\;\;\;\;\;\;\;\;\;\;\;\;\;\;\;\;\;\;\;\;\;\;}\nonumber\\
& = &
\frac{1}{2\pi i}\frac{1}{2\pi}\frac{1}{N!}\int_{-\infty}^
{\infty}dtdt' \;\int d^4y \; \exp{(iE^{\prime}t'-iEt) }
\exp{[i(E'+k^0-E)y^0]}\nonumber\\
&&\times A_f^{\nu *}(\bfy)
\langle 0|T\psi (t',{\bf x}_1
^{\prime})\cdots\psi (t',{\bf x}_N^{\prime})
\nonumber\\
&&\times
j_{\nu}(0,\bfy)
\overline{\psi}
(t,{\bf x}_N)\cdots\overline{\psi}
(t,{\bf x}_1)|0\rangle \nonumber\\
& = &
\frac{1}{2\pi i}\delta(E'+k^0-E)\frac{1}{N!}\int_{-\infty}^
{\infty}dtdt' \;\int d\bfy \; \exp{(iE^{\prime}t'-iEt) }\nonumber\\
&&\times A_f^{\nu *}(\bfy)
\langle 0|T\psi (t',{\bf x}_1
^{\prime})\cdots\psi (t',{\bf x}_N^{\prime})
\nonumber\\
&&\times
j_{\nu}(0,\bfy)
\overline{\psi}
(t,{\bf x}_N)\cdots\overline{\psi}
(t,{\bf x}_1)|0\rangle\,.
\end{eqnarray}
Using again the time-shift transformation rules,
we obtain
\begin{eqnarray} \label{gtrans13}
\lefteqn{
{\cal G}_{\gamma_f}(E',E;{\bf x}_1^{\prime},...{\bf x}_N^{\prime};
 {\bf x}_1,...{\bf x}_N) \;\;\;\;\;\;\;\;\;\;
\;\;\;\;\;\;\;\;\;\;\;\;\;\;}\nonumber\\
& = &
\frac{1}{2\pi i}\frac{1}{N!}\int_{-\infty}^
{\infty}dtdt' \;\int d\bfy \;\exp{(iE^{\prime}t'-iEt) }
\sum_{n_1,n_2}  A_f^{\nu *}(\bfy)
\nonumber\\
&&\times \exp{(-iE_{n_1}t')}
\exp{(iE_{n_2}t)} \theta(t') \theta(-t)
\langle 0|T\psi (0,{\bf x}_1
^{\prime})\cdots\psi (0,{\bf x}_N^{\prime})|n_1\ra
\nonumber\\
&&\times
\la n_1|j_{\nu}(0,\bfy)|n_2\ra \la n_2|
\overline{\psi}
(0,{\bf x}_N)\cdots\overline{\psi}
(0,{\bf x}_1)|0\rangle + {\cdots}\,.
\end{eqnarray}
Here we have assumed $E_0=0$, as in the previous section.
Taking into account the identities
\be
\int_{0}^{\infty}dt \;\exp{[i(E'-E_{n_1})t]}&=&\frac{i}{E'-E_{n_1}+i0}\,,
\nonumber\\
\int_{-\infty}^{0}dt\;\exp{[i(-E+E_{n_2})t]}&=&\frac{i}{E-E_{n_2}+i0}\,,
\ee
we find
\begin{eqnarray} \label{gtrans14}
\lefteqn{
{\cal G}_{\gamma_f}(E',E;{\bf x}_1^{\prime},...{\bf x}_N^{\prime};
 {\bf x}_1,...{\bf x}_N) \;\;\;\;\;\;\;\;\;\;
\;\;\;\;\;\;\;\;\;\;\;\;\;\;\;\;\;\;\;\;\;\;\;\;\;\;\;}\nonumber\\
& = &
\frac{i}{2\pi }\frac{1}{N!}\sum_{n_1,n_2}\int d\bfy \;  A_f^{\nu *}(\bfy)
\frac{1}{E'-E_{n_1}+i0}\frac{1}{E-E_{n_2}+i0}
\nonumber\\
&&\times 
\langle 0|T\psi (0,{\bf x}_1
^{\prime})\cdots\psi (0,{\bf x}_N^{\prime})|n_1\ra \la n_1|
j_{\nu}(0,\bfy)|n_2\ra\nonumber\\
&&\times \la n_2| \overline{\psi}
(0,{\bf x}_N)\cdots\overline{\psi}
(0,{\bf x}_1)|0\rangle + {\cdots}\,.
\end{eqnarray}
We are interested in the analytical properties of
 ${\cal G}_{\gamma_f}$ as a function of the two complex
variables $E'$ and $E$ in the region
 $E'\approx E_b^{(0)}$, $E\approx E_a^{(0)}$.
These properties can be studied using the double spectral
representation of this type of Green function which is given
in Appendix E (a similar representation was derived in 
Ref. \cite{faustov}).
As it follows from the spectral representation,
the terms which are omitted
in equation (\ref{gtrans14}) are regular functions
of $E'$ or $E$
if  $E'\approx E_b^{(0)}$ and $E\approx E_a^{(0)}$.
The equation (\ref{gtrans14})
and the spectral representation given in Appendix E 
show that, for a non-zero photon
mass $\mu$, the Green function ${\cal G}_{\gamma_f}(E',E)$
has isolated poles in the variables $E'$ and $E$ at the points
$E'=E_{n_b}$ and $E=E_{n_a}$, respectively.
Let us now introduce a Green function 
$g_{\gamma_f,b;a}(E',E)$ by
\be \label{gtrans}
 g_{\gamma_f,b;a}(E',E)=P_b^{(0)}{\cal G}_{\gamma_f}(E',E)\gamma_1^0\cdots
\gamma_N^0 P_a^{(0)}\,,
\ee
where, as in (\ref{gsmall}),
the integration over the electron coordinates is implicit.
According to equation (\ref{gtrans14}) (see also Appendix E)
the Green function $g_{\gamma_f,b;a}(E',E)$
can be written as
\begin{eqnarray} \label{gtrans16}
g_{\gamma_f,b;a}(E',E)& = &
\frac{i}{2\pi }\sum_{n_a=1}^{s_a}
\sum_{n_b=1}^{s_b}
\frac{1}{E'-E_{n_b}}\frac{1}{E-E_{n_a}} \varphi_{n_b}
\int d\bfy \; A_f^{\nu *}(\bfy)
\la n_b|j_{\nu}(0,\bfy)|n_a\ra \varphi_{n_a}^{\dag}\nonumber\\
&& + \mbox{ terms that are regular functions of $E'$ or $E$ if
 $E'\approx E_b^{(0)}$}\nonumber\\
&&\,\,\;\;\mbox{ and $E\approx E_a^{(0)}$}\,,
\end{eqnarray}
where the vectors
 $\varphi_{k}$ are defined by equation
(\ref{varphi}).
Let the contours $\Gamma_{a}$ and $\Gamma_{b}$ surround the poles
corresponding to the 
initial and final levels, respectively, and keep
outside other singularities of $g_{\gamma_{f},b;a}(E',E)$
including the cuts starting from the lower-lying bound states.
Comparing  equation (\ref{gtrans16}) with equation (\ref{red2})
and taking into account the biorthogonality condition (\ref{biort}),
we obtain the desired formula \cite{shab88b}
\be \label{transvv}
S_{\gamma_f,b;a}=Z_3^{-1/2}\delta(E_b+k_f^0-E_a)
\oint_{\Gamma_b}dE' \;\oint_{\Gamma_a}dE \; v_b^{\dag}g_{\gamma_f,b;a}
(E',E)v_a\,,
\ee
where  by $a$ we imply one of the initial states
and by $b$ one of the final states under consideration.
The vectors $v_k$ are determined from equations (\ref{kkpp})-
(\ref{normgen}).

In the case of a single initial state ($a$) and a single final
state ($b$), the vectors $v_a$ and $v_b$ simply become normalization
factors. So, for the initial state,
\be
v_a^*P_a v_a=v_a^*\frac{1}{2\pi i}\oint_{\Gamma_a} dE \; g_{aa}(E) v_a=1
\ee
and, therefore,
\be
|v_a|^2=\Bigl[\frac{1}{2\pi i}\oint_{\Gamma_a}dE \; g_{aa}(E)\Bigr]^{-1}\,.
\ee
Choosing
\be
v_a=\Bigl[\frac{1}{2\pi i}\oint_{\Gamma_a}dE \; g_{aa}(E)\Bigr]^{-1/2}\,,\;
\;\;\;\;\;\;\;\;\;\;
v_b=\Bigl[\frac{1}{2\pi i}\oint_{\Gamma_b}dE \; g_{bb}(E)\Bigr]^{-1/2}\,,
\ee
we obtain
\be \label{transfin}
S_{\gamma_f,b;a}&=&Z_3^{-1/2}\delta(E_b+k_f^0-E_a)
\oint_{\Gamma_b}dE' \;\oint_{\Gamma_a}dE \; g_{\gamma_f,b;a}
(E',E)\nonumber\\
&&\times \Bigl[\frac{1}{2\pi i}\oint_{\Gamma_b}dE \; g_{bb}(E)\Bigr]^{-1/2}
\Bigl[\frac{1}{2\pi i}\oint_{\Gamma_a}dE \; g_{aa}(E)\Bigr]^{-1/2}
\,.
\ee

For practical calculations of the transition amplitude, it is
convenient to express the Green function
$g_{\gamma_f,b;a}(E',E)$ in terms of the Fourier transform
of the 2$N$-time Green function,
\begin{eqnarray}       
 g_{\gamma_f,b;a}(E',E) \delta(E'+k^0-E)  &=&
 \frac{1}{N!}
     \int_{-\infty}^{\infty}dp_{1}^{0}\cdots dp_{N}^{0}
    dp_{1}^{\prime 0}\cdots dp_{N}^{\prime 0}\nonumber\\
&& \times \delta(E-p_{1}^{0} -\cdots - p_{N}^{0})
            \delta(E'-p_{1}^{\prime 0} -\cdots - p_{N}^{\prime 0}) 
\nonumber\\  
 && \times  P_b^{(0)}
  G_{\gamma_f}
(p_{1}^{\prime 0} ,\ldots ,p_{N}^{\prime 0};k^0;p_{1}^{0},
\ldots ,p_{N}^{0})
	     \gamma_{1}^{0} \ldots \gamma_{N}^{0}
	     P_a^{(0)}\,,
       \label{gtrans22}
\end{eqnarray}
where
\be \label{gtrans23}
\lefteqn{  G_{\gamma_f}
((p_{1}^{\prime 0},\bfx_1') ,\ldots ,(p_{N}^{\prime 0},\bfx_N');k^0;
(p_{1}^{0},\bfx_1),\ldots ,(p_{N}^{0},\bfx_N))} \nonumber\\
&=&\frac{2\pi}{i}\frac{1}{(2\pi)^{2N+1}}
\int_{-\infty}^{\infty}dx_{1}^{0}...
dx_{N}^{0}
dx_{1}^{\prime 0}...dx_{N}^{\prime 0} \, \int d^4y\nonumber\\
&&\times\exp{(ip_{1}^{\prime 0}x_{1}^{\prime 0}+...+ip_{N}^{\prime 0}x_{N}^
{\prime 0}-
ip_{1}^{ 0}x_{1}^{0}-...-ip_{N}^{ 0}
x_{N}^{0}+ik^0y^0)}\nonumber\\
&&\times A_f^{\nu *}(\bfy)
\la 0|T\psi(x_{1}^{\prime})\cdots \psi(x_{N}^{\prime})j_{\nu}(y) 
\overline {\psi}(x_{N})\cdots \overline {\psi} (x_{1})|0\ra\, .
\ee
The  Green function $G_{\gamma_f}$ is constructed by perturbation
theory after the transition in (\ref{gtrans23}) to the interaction
representation and using the Wick theorem. The Feynman rules
for $G_{\gamma_f}$ differ from those for $G$ considered in the
previous section only by the presence of an outgoing  photon line
that corresponds to the wave function of the emitted photon
$A_f^{\nu *}(\bfx)$.
The Feynman rule for a vertex in which
a real photon is  emitted remains the same as
for a virtual photon vertex. The energy variable
of the emitted photon ($k^0$) in the expression corresponding
to a real photon vertex is considered
as a free variable ($k^0\ne k_f^0=|\bfk_f|$)
which, due to the energy conservation,
can be expressed in terms of the initial ($E$) and final ($E'$)
atomic energy variables.

\subsection{Transition probability in a one-electron atom}

To demonstrate the practical ability of the formalism,
in this section  we derive formulas for
the transition probability in a one-electron
atom to zeroth and first orders in $\alpha$.
An application of the method for two-electron atoms
is considered in \cite{ind00}.

\subsubsection{ Zeroth order approximation}

To zeroth order the transition amplitude
is described by the diagram shown in Fig. 31.
The formula (\ref{transfin}) gives 
\begin{eqnarray}\label{zer01}
S^{(0)}_{\gamma_f,b;a}=\delta(E_{b}
+k_{f}^0-E_{a})\oint_{\Gamma_b}dE' \;\oint_{\Gamma_a}dE 
\; g^{(0)}_{\gamma_{f},b;a}(E',E)\,,
\end{eqnarray}
where the superscript indicates the order in $\alpha$.
Here we have taken into account that
\begin{eqnarray}\label{zer2}
\frac{1}{2\pi i}\oint_{\Gamma_a} dE\;g_{aa}^{(0)}(E)=
\frac{1}{2\pi i}\oint_{\Gamma_a} dE \; \frac{1}{E-\veps_{a}}=1
\end{eqnarray}
and a similar equation exists for the $b$ state.
According to the Feynman rules, we have
\begin{eqnarray}\label{zer3}
G_{\gamma_{f}}^{(0)}((E,'\bfx');k^0;(E,\bfx))&=&
\int d\bfy \; \frac{i}{2\pi} S(E',\bfx',\bfy)
 (-2\pi ie\gamma^{\nu})
\delta(E'+k^{0}-E)\nonumber\\
&&\times
A_{f,\nu}^{*}(\bfy)
\frac{i}{2\pi} S(E,\bfy,\bfx)\,.
\end{eqnarray}
Substituting the expression (\ref{zer3}) into 
the definition of $g_{\gamma_f,b;a}(E',E)$ (see equation
(\ref{gtrans22})), we obtain
\begin{eqnarray}\label{zer5}
 g^{(0)}_{\gamma_f,b;a}(E',E)=\frac{i}{2\pi}\frac{1}{E'-\veps_b}
\la b|e\alpha^{\nu}A_{f,\nu}^{*}
|a\ra \frac{1}{E-\veps_{a}}\,.
\end{eqnarray}
The equations (\ref{zer01}), (\ref{zer5}) yield
\begin{eqnarray}\label{zer11}
S^{(0)}_{\gamma_f,b;a}=-2\pi i\,
\delta(E_{b}
+k_{f}^0-E_a)
\la b|e\alpha^{\nu}A_{f,\nu}^{*}
|a\ra\,
\end{eqnarray}
or, in accordance with the definition (\ref{deftau}),
\begin{eqnarray}\label{tau0}
\tau^{(0)}_{\gamma_f,b;a}=-\la b|e\alpha^{\nu}A_{f,\nu}^{*}
|a\ra=\la b|e\balpha\cdot {\bf A}_{f}^{*}
|a\ra\,.
\end{eqnarray}
The transition probability to zeroth order is
\begin{eqnarray}\label{inf10a}
dW^{(0)}_{\gamma_f,b;a}&=&
2\pi |\tau^{(0)}_{\gamma_f,b;a}|^2 \delta(E_b+k_f^0-E_a)
d\bfk_f\nonumber\\
&=&2\pi \frac{e^2}{2k_f^0(2\pi)^3}|\la b|\eps_f^{\nu *}\alpha_{\nu}
\exp{(-i\bfk_f\cdot\bfx)}|a\ra|^2
\delta(E_b+k_f^0-E_a)
d\bfk_f\,.
\end{eqnarray}
Integrating over the photon energy, we obtain
\be \label{ver00}
dW^{(0)}_{\gamma_f,b;a}=
\alpha\,\frac{k_f^0}{2\pi}\,|\beps_f^* \cdot {\bf j}_{ba}(\bfk_f)|^2
d\Omega_f\,,
\ee
where
\be
{\bf j}_{ba}(\bfk_f)=\la b|\balpha \exp{(-i\bfk_f\cdot \bfx)}|a\ra\,.
\ee

\subsubsection{ QED corrections of first order in $\alpha$}

The QED corrections of first order in $\alpha$ are 
defined by the diagrams shown  in Fig. 32. 
Let us consider  the derivation of the formulas for
the self-energy (SE) corrections (the diagrams (a)-(c)) in detail. 
The formula (\ref{transfin}) gives in the order under consideration
\begin{eqnarray}\label{first1}
S^{(1)}_{\gamma_f,b;a}&=&\delta(E_{b}
+k_{f}^0-E_a)\;\Bigl\{
\oint_{\Gamma_b}dE' \; \oint_{\Gamma_a}dE
 \; g_{\gamma_{f},b;a}^{(1)}(E',E)\nonumber\\
&&-\frac{1}{2}
\oint_{\Gamma_b}dE' \; \oint_{\Gamma_a}dE
 \; g_{\gamma_{f},b;a}^{(0)}(E',E)\Bigl[
\frac{1}{2\pi i}\oint_{\Gamma_b} dE\; g_{bb}^{(1)}(E)
+\frac{1}{2\pi i}\oint_{\Gamma_a} dE\; g_{aa}^{(1)}(E)
\Bigr] \Bigr\}\,,
\end{eqnarray}
where $g_{aa}^{(1)}(E)$ and $g_{bb}^{(1)}(E)$ are defined by the first-order
self-energy diagram.
Here we have omitted a term of first order in $\alpha$ which comes
 from the factor $Z_{3}^{-1/2}$ since
it has to be combined with
 the vacuum-polarization (VP) correction.  
Consider first the diagram (a).
According to the Feynman rules,
we have
\begin{eqnarray}\label{first2}
G_{\gamma_{f}}^{(1,a)}((E',\bfx');k^0;(E,\bfx))&=&
\delta(E'+k^{0}-E)
\int d\bfy d\bfy' d\bfz \,
\frac{i}{2\pi} S(E',\bfx',\bfy)
 \frac{2\pi}{i}\gamma^{0}\Sigma (E',\bfy',\bfy)\nonumber\\
&&\times
\frac{i}{2\pi } S(E',\bfy,\bfz)
A_{f,\nu}^{*}(\bfz)(-2\pi ie\gamma^{\nu})
\frac{i}{2\pi} S(E,\bfz,\bfx)
\,,
\end{eqnarray}
where
\begin{eqnarray}\label{first3}
\Sigma(E',\bfy',\bfy)=e^2\frac{i}{2\pi}\int d\omega \,
\gamma^{0}\gamma^{\rho}
S(E'-\omega,\bfy', \bfy)\gamma^{\sigma}D_{\rho \sigma}(\omega,\bfy'-\bfy)
\end{eqnarray}
is the kernel of the  self-energy operator  and $  D_{\rho \sigma} 
(\omega,{\bf y}'-{\bf y})$ 
is the photon propagator for a non-zero photon mass.
According to the definition of $g_{\gamma_f,b;a}(E',E)$ (see equation
(\ref{gtrans22})), we find
\begin{eqnarray}\label{first5}
g_{\gamma_{f},b;a}^{(1,a)}(E',E)=\frac{i}{2\pi}\sum_{n}
\frac{\la b|\Sigma(E')|n\ra 
\la n|e\alpha^{\nu}A_{f,\nu}^{*}
|a\ra
}{(E'-\veps_{b})(E'-\veps_{n})(E-\veps_a)}\,
\end{eqnarray}
and
\begin{eqnarray}\label{first6}
\oint_{\Gamma_b} dE' \; \oint_{\Gamma_a} dE
\; g_{\gamma_{f},b;a}^{(1,a)}(E',E)
&=&-2\pi i\Bigl[
\sum_n^{n\not = b}
\frac{\la b|\Sigma(\veps_{b})|n\ra 
\la n|e\alpha^{\nu}A_{f,\nu}^{*}
|a\ra}
{\veps_{b}-\veps_{n}}\nonumber\\
&&+\la b|\Sigma'(\veps_{b})|b\ra 
\la b|e\alpha^{\nu}A_{f,\nu}^{*}
|a\ra \Bigr]\,,
\end{eqnarray}
where $\Sigma'(\veps_{b})\equiv \frac{d\Sigma(\veps)}{d\veps}
\Bigr|_{\veps=\veps_{b}}\,$. 
A similar calculation of the diagram (b)  gives
\begin{eqnarray}\label{first7}
\oint_{\Gamma_b} dE' \; \oint_{\Gamma_a} dE
\;g_{\gamma_{f},b;a}^{(1,b)}(E',E)
&=&-2\pi i\Bigl[
\sum_n^{n\not = a}
\frac{\la b|e\alpha^{\nu}A_{f,\nu}^{*}
|n\ra
\la n|\Sigma(\veps_{a})|a\ra}
{\veps_{a}-\veps_{n}}\nonumber\\
&&+\la b|e\alpha^{\nu}A_{f,\nu}^{*}
|a\ra \la a|\Sigma'(\veps_{a})|a\ra  \Bigr]\,.
\end{eqnarray}
The second ("reducible") terms in equations
(\ref{first6}) and (\ref{first7}) have to be combined with
the second term in equation (\ref{first1}).
Taking into account that
\begin{eqnarray}
\frac{1}{2\pi i}\oint_{\Gamma_a}
 dE\; g_{aa}^{(1)}(E)=\la a|\Sigma'(\veps_{a})|a\ra\,,\\
\frac{1}{2\pi i}\oint_{\Gamma_b}
 dE\; g_{bb}^{(1)}(E)=\la b|\Sigma'(\veps_{b})|b\ra\,,
\end{eqnarray}
we obtain
\begin{eqnarray}\label{first11}
-\frac{1}{2}
\oint_{\Gamma_b}dE' \;\oint_{\Gamma_a}dE
 \; g_{\gamma_{f},b;a}^{(0)}(E',E)
\Bigl[\frac{1}{2\pi i}\oint_{\Gamma_b}dE\;g_{bb}^{(1)}(E)
+\frac{1}{2\pi i}\oint_{\Gamma_a}dE \;
g_{aa}^{(1)}(E)\Bigr]\nonumber\\
=2\pi i \Bigl[\frac{1}{2}\la b|e\alpha^{\nu}A_{f,\nu}^{*}|a\ra 
(\la b|\Sigma'(\veps_{b})|b\ra +\la a|\Sigma'(\veps_{a})|a\ra)\Bigr]\,.
\end{eqnarray}
For the diagram (c) we find
\be \label{first12}
g_{\gamma_f,b;a}^{(1,c)}(E',E)&=& \frac{i}{2\pi}\frac{1}{E'-\veps_b}
\int d\bfx d\bfy d\bfz \; \overline{\psi}_b(\bfx)\frac{i}{2\pi}
\int_{-\infty}^{\infty} d\omega \; \gamma^{\rho} 
S(E'-\omega,\bfx,\bfz)\nonumber\\
&&\times e A_{\nu}^{*}(\bfz)\gamma^{\nu}
S(E-\omega,\bfz,\bfy)\gamma^{\sigma}e^2
D_{\rho \sigma}(\omega,\bfx-\bfy)\psi_a(\bfy)
\frac{1}{E-\veps_a}\,.
\ee
Substituting (\ref{first12}) into (\ref{first1}),
we obtain
\begin{eqnarray}\label{first8}
\oint_{\Gamma_b} dE' \; \oint_{\Gamma_a} dE
\;g_{\gamma_{f},b;a}^{(1,c)}(E',E)
&=&
=-2\pi i \int d \bfz\,
e A_{f,\nu}^{*}(\bfz)
\Lambda^{\nu}(\veps_{b},\veps_a,\bfz)\,,
\end{eqnarray}
where
\begin{eqnarray}\label{first9}
\Lambda^{\nu}(\veps_b,\veps_a,\bfz)&=&
e^{2}\frac{i}{2\pi}\int_{-\infty}^{\infty}d\omega\,
\int d\bfx d\bfy \, \overline{\psi}_{b}(\bfx)\gamma^{\rho}
 S(\veps_b-\omega,\bfx,\bfz)\gamma^{\nu}
 S(\veps_a-\omega,\bfz,\bfy)\nonumber\\
&&\times \gamma^{\sigma}
D_{\rho\sigma}(\omega,\bfx-\bfy)
\psi_a(\bfy)\,.
\end{eqnarray}
Summing all the first order SE contributions derived above
and adding the contribution of the mass-counterterm diagrams
(Fig. 33), we find
\begin{eqnarray}\label{first15}
S^{(1,{\rm SE})}_{\gamma_f,b;a}&=&-2\pi i\,
\delta(E_{b}+k_{f}^0-E_a)\;\Bigl[
\sum_n^{n\not = b}
\frac{\la b|\Sigma(\veps_{b})-\beta\delta m|n\ra 
\la n|e\alpha^{\nu}A_{f,\nu}^{*}
|a\ra}
{\veps_{b}-\veps_{n}}\nonumber\\
&&+\sum_n^{n\not = a}
\frac{\la b|e\alpha^{\nu}A_{f,\nu}^{*}
|n\ra \la n|\Sigma(\veps_{a})-\beta\delta m |a\ra}
{\veps_{a}-\veps_{n}}\nonumber\\
&&+\int d \bfz\,
e A_{f,\nu}^{*}(\bfz)
\Lambda^{\nu}(\veps_{b},\veps_a,\bfz)\nonumber\\
&&+\frac{1}{2}\la b|e\alpha^{\nu}A_{f,\nu}^{*}|a\ra
(\la b|\Sigma'(\veps_{b})|b\ra + \la a|\Sigma'(\veps_{a})|a\ra)
\Bigr]\,.
\end{eqnarray}

A similar calculation of the vacuum-polarization diagrams
(Fig. 32,  diagrams (d)-(f)) gives
\begin{eqnarray}\label{first16}
S^{(1,{\rm VP})}_{\gamma_f,b;a}&=&-2\pi i\,
\delta(E_{b}+k_{f}^0-E_a)\;\Bigl[
\sum_n^{n\not = b}
\frac{\la b|U_{\rm VP}|n\ra 
\la n|e\alpha^{\nu}A_{f,\nu}^{*}
|a\ra}
{\veps_{b}-\veps_{n}}\nonumber\\
&&+\sum_n^{n\not = a}
\frac{\la b|e\alpha^{\nu}A_{f,\nu}^{*}
|n\ra \la n|U_{\rm VP} |a\ra}
{\veps_{a}-\veps_{n}}\nonumber\\
&&+\int d \bfz\,
e A_{f,\nu}^{*}(\bfz)
Q^{\nu}(k_f^0,\bfz)
+(Z_{3}^{-1/2}-1)
\la b|e\alpha^{\nu}A_{f,\nu}^{*}|a\ra
\Bigr]\,,
\end{eqnarray}
where
\begin{eqnarray}
U_{\rm VP}(\bfx)=\frac{\alpha}{2\pi i}\int d\bfy
\,\frac{1}{|\bfx-\bfy|}\int_{-\infty}^{\infty}
d\omega\, {\rm Tr}[S(\omega,\bfy,\bfy)\gamma^0]
\end{eqnarray}
is the VP potential and
\begin{eqnarray}
Q^{\nu}(k^0,\bfz)&=&-e^2\int d{\bfx} d{\bfy}\,\overline{\psi}_{b}
(\bfx)\gamma^{\rho}\psi_{a}(\bfx)
D_{\rho \sigma}(k^0,\bfx-\bfy) \nonumber\\
&&\times\frac{i}{2\pi}\int_{-\infty}
^{\infty} d\omega \, {\rm Tr}[\gamma^{\sigma}S(\omega, \bfy,\bfz)
\gamma^{\nu} S(\omega+k^0,\bfz,\bfy)]\,.
\end{eqnarray}

Some individual terms in equations (\ref{first15}) and
(\ref{first16}) contain ultraviolet divergences. 
These divergences  arise solely from the zero- and one-potential
terms in the expansion of the electron propagators
in powers of the Coulomb potential. Using  the standard
 expressions for the divergent parts of the zero- and
one-potential SE terms (see, e.g., \cite{itzykson})
 and the Ward  identity ($Z_{1}=Z_{2}$)
one easily finds that the ultraviolet divergences cancel each
other in  equation (\ref{first15}).
As to  equation (\ref{first16}), the divergent parts incorporate
into the charge renormalization factor ($e=Z_{3}^{1/2}e_{0}$).

An alternative approach to the renormalization problem consists
in using the renormalized field operators
$\psi_{R}=Z_{2}^{-1/2}\psi$, $A_{R}=Z_{3}^{-1/2}A$, the renormalized
electron charge $e=e_{0}+\delta e=Z_{1}^{-1}Z_{2}Z_{3}^{1/2}e_{0}$,
and, respectively, the renormalized Green functions from the very
beginning. In that approach,
additional counterterms arise for the Feynman rules.

The vertex and reducible contributions to the SE corrections
(third and fourth terms in equation (\ref{first15}))
contain infrared divergences which cancel each other 
when being considered together.

In addition to the QED corrections derived in this subsection,
we must take into account the contribution originating from
changing the photon energy in  the zeroth-order transition
probability (\ref{ver00}) due to the QED correction 
to the energies of the bound states $a$ and $b$.
 It follows that the total QED
correction to the transition probability of
  first order in $\alpha$ 
 is given by
\begin{eqnarray}\label{form34}
dW_{\gamma_f,b;a}^{(1)}
=2\pi(k_f^0)^2
2{\rm Re}\Bigl\{\tau_{\gamma_f,b;a}^{(0)*}
\tau_{\gamma_f,b;a}^{(1)}\Bigr\}d\Omega_f+
\Bigl[dW_{\gamma_f,b;a}^{(0)}\Bigr|_{k_{f}^0=E_a-E_b}
-dW_{\gamma_f,b;a}^{(0)}\Bigr|_{k_{f}^0=\veps_a-\veps_{b}}
\Bigr]\,.
\end{eqnarray} 
Here $\tau_{\gamma_f,b;a}^{(1)}=
\tau_{\gamma_f,b;a}^{\rm (1,SE)}+\tau_{\gamma_f,b;a}^{\rm (1,VP)}$
is the QED correction given by equations (\ref{first15}) and 
(\ref{first16})
in  accordance with the definition (\ref{deftau}). $E_a$,
 $E_b$ and $\veps_a$, $\veps_b$
 are the energies of the bound states $a$ and $b$
with and without the
QED correction, respectively.

\subsection{Radiative recombination of an electron with an atom}

In calculations of processes which contain a free electron 
in the initial or final or in both states we consider 
that the interaction with the Coulomb field of the nucleus
$V_{\rm C}({\bfx})$  is included in the source  $\tilde{j}(x)$
which leads to a scattering \cite{bjorken},
\be
(i\not\stackrel{}{\partial}-m)\psi(x)=\tilde{j}(x) \,.
\ee
It means that, after the transition to the "in" operators,
 the unperturbed Hamiltonian does not contain the interaction 
with the Coulomb potential. As a result, the Feynman rules
contain free-electron propagators, instead of the bound-electron
ones, and the vertices corresponding to the interaction of
electrons with $V_{\rm C}(\bfx)$ appear.  Since we consider the case 
of a strong Coulomb field, we will sum up infinite sequences of
Feynman diagrams describing the interaction of electrons
 with the Coulomb  potential. As a result of this summation,
the free-electron propagators are replaced by the bound-electron
propagators,
\be \label{propbound}
[p^0-H(1-i0)]^{-1}&=&
[p^0-H_0(1-i0)]^{-1}\nonumber\\
&&+ [p^0-H_0(1-i0)]^{-1}V_{\rm C} [p^0-H_0(1-i0)]^{-1}
+\cdots\,,
\ee
where $H=H_0+V_{\rm C}$ and $ H_0=-i\balpha \cdot \bnabla+\beta m$,
and the free-electron wave functions are replaced by 
the wave functions in the Coulomb field. For instance, the
wave function of an incident electron  with momentum $\bfp_i$
and polarization $\mu_i$ is
\be \label{wavebound}
\psi_{p_i\mu_i(+)}&=&U_{p_i \mu_i}+[p_i^0-H_0(1-i0)]^{-1} V_{\rm C}
 U_{p_i \mu_i}
\nonumber\\
&&+[p_i^0-H_0(1-i0)]^{-1} V_{\rm C} [p_i^0-H_0(1-i0)]^{-1}
U_{p_i \mu_i} +\cdots
\nonumber\\
&=&U_{p_i\mu_i}+ [p_i^0-H(1-i0)]^{-1} V_{\rm C} U_{p_i\mu_i}\,,
\ee
where $p_i=(p_i^0,\bfp_i)$, $p_i^0=\sqrt{\bfp_i^2+m^2}$,
\be \label{defU}
 U_{p_i\mu_i}= \frac{u(p_i,\mu_i) 
\exp{(i\bfp_i\cdot \bfx)}}
{\sqrt{(p_i^0/m)(2\pi)^3}}
\ee
is the free wave function of  the
incident electron, and $u(p_i,\mu_i)$ is
 normalized by the condition $\overline{u}u=1$.

Consider the process of the radiative recombination of an electron
with  momentum ${\bf p}_{i}$ and polarization ${\mu}_{i}$
with an $(N-1)$-electron atom $X^{(Z-N+1)+}$ in a
state $a$. As a result of the process, the N-electron
 atom $X^{(Z-N)+}$ in
a state $b$ and a photon with momentum ${\bf k}_{f}$ and 
polarization $\eps_f=(0,{\beps}_{f})$ arise,
\widetext
\begin{eqnarray}
e^{-}({p}_{i},{\mu}_{i}) +X^{(Z-N+1)+}(a) \rightarrow
\gamma({k}_{f},\eps_{f})+X^{(Z-N)+}(b) \,, 
\end{eqnarray}
where $k_f=(k_f^0,\bfk_f)$ and $k_f^0=|\bfk_f|$.
In this section we will assume that we consider a non-resonant
process. It means that the total initial energy ($p_i^0+E_a$)
is not close to any excited-state energy of the $N$-electron atom.
As to the resonance recombination, a detailed description of this 
process is given in \cite{sh94b} (see also the next sections 
of the present paper).

According to the  standard reduction technique \cite{itzykson,bjorken},
the amplitude of the process is 
\begin{eqnarray}
S_{\gamma_{f},b;e_{i},a}&=&\la b|a_{\rm out}(k_{f},{\eps}_{f})
b_{\rm in}^{\dag}
(p_{i},{\mu}_{i})|a\ra\nonumber\\
&=&(-iZ_{3}^{-\frac{1}{2}})(-iZ_{2}^
{-\frac{1}{2}})
\int d^{4}y d^{4}z \; \frac{{\eps}_{f}^{\nu *} \exp{(ik_{f}\cdot y)}}
{\sqrt{2k_{f}^{0}(2\pi )^3}}   \nonumber\\
&&\times 
\la b|Tj_{\nu}(y)\overline \psi(z)|a\ra
(-i\not\stackrel{\leftarrow}
{\partial}_{z}-m)\frac{u(p_{i},{\mu}_{i})
\exp{(-ip_{i}\cdot z)}} {\sqrt{\frac{p_{i}^
{0}}{m}(2\pi )^3}}\, ,
\end{eqnarray}
where $|a\ra,|b\ra$ are the vectors of the initial and
 final states in the Heisenberg
representation.
 Taking into account that
\begin{eqnarray}
\overline{\psi}(z)(-i\not\stackrel{\leftarrow}{\partial}_{z}-m)=
e\overline{\psi}(z)A(z)\equiv \overline {\eta}(z) 
\end{eqnarray}
we obtain
\begin{eqnarray} \label{srec2}
S_{\gamma_{f},b;e_{i},a}&=&2\pi \delta (E_{b}+k_{f}^{0}-E_{a}-p_{i}^
{0})(-iZ_{3}^{-\frac{1}{2}})(-iZ_{2}^{-\frac{1}{2}}) \int d{\bf y}
d{\bf z} \; A_f^{\nu *} (\bfy)  
\nonumber\\
&&\times\{ \int_{0}^{\infty}d\tau \;\la b|j_{\nu}(\tau,{\bf y})\overline
{\eta}(0,{\bf z})|a\ra \exp{(ik_{f}^{0}\tau)} \nonumber\\
&&+\int_{0}^{\infty}d \tau \;\la b|\overline{\eta}(\tau,{\bf z})
j_{\nu}(0,{\bf y})|a\ra \exp{(-ip_{i}^{0}\tau)} \nonumber\\
&&-i\la b|[\psi^{\dag}(0,{\bf z}),\, j_{\nu}(0,{\bf y})]|a\ra \}
U_{p_i\mu_i}(\bfz)
\, ,
\end{eqnarray}
where $A_f^{\nu} (\bfy)$ is the wave function of the emitted
photon defined by equation (\ref{defA}) and $U_{p_i\mu_i}(\bfz)$
is the free wave function of the incident electron defined
by equation (\ref{defU}).
To construct the perturbation theory for  the amplitude
of the process, we introduce the Fourier transform of the 
corresponding two-time Green function,
\begin{eqnarray}
\lefteqn{{\cal G}_{\gamma_f; e_{i}}(E',E,p^{0};{\bf x}_{1}^
{\prime},...{\bf x}_{N}^{\prime};{\bf x}_{1},...{\bf x}_{N-1})
\delta(E'+k^{0}-E-p^{0})}\nonumber\\
&=&\Bigl(\frac{i}{2\pi}\Bigr)^{2}\frac{1}{\sqrt{N!}}\frac{1}{\sqrt{(N-1)!}}
\int_{-\infty}^{\infty}dx^0 dx^{\prime 0}\;\exp{(iE'x^{\prime 0}-
iEx^{0})}\nonumber\\
&&\times\int d^{4}yd^{4}z \;\exp{(ik^{0}y^{0}-ip^{0}z^{0})}
A_f^{\nu *}(\bfy)
\nonumber\\
&&\times \la 0|T\psi(x^{\prime 0},{\bf x}_{1}^{\prime 0})\cdots
\psi(x^{\prime 0},{\bf x}_{N}^{\prime})j_{\nu}(y)\overline{\psi}
(z)\overline{\psi}(x^{0},{\bf x}_{N-1})\cdots \overline{\psi}
(x^{0},{\bf x}_{1})|0\ra\nonumber\\
&&\times (-i\not\!{\stackrel{\leftarrow}{\partial}}_{z}-m)
U_{p_i\mu_i}(\bfz)
\, .
\end{eqnarray}
As in the derivation of the formulas for the transition
amplitudes, we will assume
that in zeroth approximation the initial and final states
are degenerate in energy with some other states
 and we will
use the same notations for these states as in
section III(A).
As usual, we also assume that
a non-zero photon mass $\mu$ is introduced. 
The spectral representation of
${\cal G}_{\gamma_{f};e_{i}}(E',E,p^{0})$ can be derived
in the same way as for the Green function describing the
transition amplitude (see section III(A) and Appendix E). It gives
\widetext
\begin{eqnarray} \label{spectrec1}
\lefteqn{{\cal G}_{\gamma_{f};e_{i}}(E',E,p^{0};{\bf x}_{1}^{\prime},
...{\bf x}_{N}^{\prime};{\bf x}_{1},...{\bf x}_{N-1})
\;\;\;\;\;\;\;\;\;\;\;\;\;\;\;\;\;\;\;\;\;\;\;\;\;}\nonumber\\
&&=\frac{1}{2\pi}
\frac{1}{\sqrt{N!}}\frac{1}{\sqrt{(N-1)!}}\
\sum_{n_a=1}^{s_{a}}\sum_{n_b=1}^{s_{b}}
\frac{1}{E'-E_{n_{b}}}\frac{1}
{E-E_{n_{a}}}\nonumber\\
&&\times \int d{\bf y}d{\bf z}\;
A_f^{\nu *}(\bfy)
\la 0|T\psi(0,{\bf x}_{1}^{\prime 0})\cdots
\psi(0,{\bf x}_{N}^{\prime})|n_{b}\ra \nonumber\\
&&\times\{ \int_{0}^{\infty}d\tau \;
\la n_{b}|j_{\nu}(\tau,{\bf y})\overline
{\eta}(0,{\bf z})|n_{a}\ra \exp{(i(E-E_{n_{b}}+p^{0})\tau)} 
\nonumber\\
&&+\int_{0}^{\infty}d \tau \;\la n_{b}|\overline{\eta}(\tau,{\bf z})
j_{\nu}(0,{\bf y})|n_{a}\ra\exp{(i(E'-E_{n_{b}}-p^{0})\tau)} \nonumber\\
&&-i\la n_{b}|[\psi^{\dag}(0,{\bf z}),\, j_{\nu}(0,{\bf y})]|n_{a}\ra \}
\la n_{a}|\overline{\psi}(0,{\bf x}_{N-1})\cdots \overline{\psi}
(0,{\bf x}_{1})|0\ra 
U_{p_i\mu_i}(\bfz) \nonumber\\
&&+\mbox{  terms that are regular functions of $E'$ or $E$ if
 $E'\approx E_{b}^{(0)}$ }\nonumber\\
&&\;\;\; \mbox{ and } E \approx E_{a}^{(0)}\, .
\end{eqnarray}
We introduce the Green function $g(E',E,p^{0})$ by
\begin{eqnarray}
g_{\gamma_{f},b;e_{i},a}
(E',E,p^{0})=P^{(0)}_{b}{\cal G}_{\gamma_{f};e_{i}}(E',E,p^{0})\gamma_{1}^
{0}\cdots \gamma_{N-1}^{0}P^{(0)}_{a}\, .
\end{eqnarray}
>From equation (\ref{spectrec1}) we have
\begin{eqnarray} \label{spectrec2}
g_{\gamma_{f},b; e_{i},a}(E',E,p^{0})&=&\frac{1}{2\pi}
\sum_{n_a=1}^{s_{a}}\sum_{n_b=1}^{s_{b}}\frac{\varphi_{n_{b}}}
{E'-E_{n_{b}}}\int d{\bf y}d{\bf z}\;
A_f^{\nu *} (\bfy)
\nonumber\\
&&\times\{ \int_{0}^{\infty}d\tau \;\la n_{b}|j_{\nu}(\tau,{\bf y})\overline
{\eta}(0,{\bf z})|n_{a}\ra \exp{(i(E-E_{n_{b}}+p^{0})\tau)} 
\nonumber\\
&&+\int_{0}^{\infty}d \tau \;\la n_{b}|\overline{\eta}(\tau,{\bf z})
j_{\nu}(0,{\bf y})|n_{a}\ra\exp{(i(E'-E_{n_{a}}-p^{0})\tau)}
\nonumber\\ 
&&-i\la n_{b}|[\psi^{\dag}(0,{\bf z}),\, j_{\nu}(0,{\bf y})]|n_{a}\ra \}
U_{p_i\mu_i}(\bfz)
\frac{\varphi_{n_{a}}^{\dag}}{E-E_{n_{a}}}\nonumber\\
&&+\mbox{  terms that are regular functions of $E'$ or $E$ 
if  $E'\approx E_b^{(0)}$}\nonumber\\
&&\;\;\;\mbox{ and $E\approx E_a^{(0)}$}\,  ,
\end{eqnarray}
where $\varphi_{n_{a}}$ and $\varphi_{n_{b}}$ are the wave functions
of the initial and final states as defined above.
Let the contours $\Gamma_{a}$ and $\Gamma_{b}$ surround the poles
corresponding to the initial and final levels, respectively, and keep
outside other singularities of $g_{\gamma_{f},b; e_{i},a}(E',E,p^{0})$.
This can be performed if the photon mass $\mu$ is chosen as indicated 
in section III(A).
Taking into account the biorthogonality condition (\ref{biort})
and comparing (\ref{spectrec2}) with (\ref{srec2}) we obtain 
the desired formula \cite{sh94b}
\begin{eqnarray} \label{recfor}
S_{\gamma_{f},b;e_{i},a}= (Z_{2}Z_{3})^{-\frac{1}{2}}
\delta (E_{b}+k_{f}^{0}-E_{a}-p_{i}^{0})
\oint_{\Gamma_{b}}dE' \;\oint_{\Gamma_
{a}}dE\; v_{b}^{\dag}g_{\gamma_{f},b;e_{i},a}(E',E,p_{i}^{0})v_{a}\,,
\end{eqnarray}
where by $a$ we imply  one of the initial states and  by $b$ 
one of the final states under consideration.
In the case of a single initial state ($a$) and a single final
state ($b$), it yields
\be \label{recfor1}
S_{\gamma_f,b;e_{i}, a}&=&
(Z_{2}Z_{3})^{-\frac{1}{2}}
\delta (E_{b}+k_{f}^{0}-E_{a}-p_{i}^{0})
\oint_{\Gamma_b}dE' \;\oint_{\Gamma_a}dE \; g_{\gamma_f,b;e_i,a}
(E',E,p_i^0)\nonumber\\
&&\times \Bigl[\frac{1}{2\pi i}\oint_{\Gamma_b}dE\; g_{bb}(E)\Bigr]^{-1/2}
\Bigl[\frac{1}{2\pi i}\oint_{\Gamma_a}dE \; g_{aa}(E)\Bigr]^{-1/2}
\,.
\ee

For practical calculations by perturbation theory, it is convenient
to express the Green function $g_{\gamma_{f},b;e_{i},a}$
in terms of the Fourier transform of the $(2N-1)$-time Green function,
\begin{eqnarray} \label{recgsmall}
\lefteqn{g_{\gamma_{f},b;e_{i},a}(E',E,p^{0})\delta (E'+k^{0}-E-p^{0})
\;\;\;\;\;\;\;\;\;\;\;\;\;\;\;\;\;\;\;\;\;\;} 
\nonumber\\
&&=\frac{1}{\sqrt{N!(N-1)!}}\int_{-\infty}^
{\infty} dp_{1}^{0}\cdots dp_{N-1}^{0}dp_{1}^{\prime 0}\cdots dp_{N}^
{\prime 0}\;\delta(E-p_{1}^{0}\cdots -p_{N-1}^{0})\delta(E'-p_{1}^
{\prime 0}\cdots -p_{N}^{\prime 0})\nonumber\\
&&\times P^{(0)}_{b}G_{\gamma_{f};e_{i}}
(p_{1}^{\prime 0},...,p_{N}^{\prime 0};k^{0},p^{0};p_{1}^{0},...,
p_{N-1}^{0})\gamma_{1}^{0}\cdots \gamma_{N-1}^{0}P^{(0)}_{a}\, ,
\end{eqnarray}
where
\begin{eqnarray} \label{greenrec3}
\lefteqn{G_{\gamma_{f};e_{i}}((p_{1}^{\prime 0},
{\bf x}_{1}^{\prime}),...,(p_{N}^{\prime 0},
{\bf x}_{N}^{\prime});k^{0},p^{0};(p_{1}^{ 0},
{\bf x}_{1}),...,(p_{N-1}^{0},{\bf x}_{N-1}))
\;\;\;\;\;\;\;\;\;\;\;\;\;\;\;\;\;\;\;\;\;\;\;\;\;\;}\nonumber\\
&&=\Bigl (\frac{2 \pi}{i}\Bigr )^{2}\frac{1}{(2\pi)^{2N+1}}
\int_{-\infty}^{\infty}dx_{1}^{0}\cdots
dx_{N-1}^{0}
dx_{1}^{\prime 0}\cdots dx_{N}^{\prime 0}\nonumber\\
&&\times\exp{(ip_{1}^{\prime 0}x_{1}^{\prime 0}+
\cdots +ip_{N}^{\prime 0}x_{N}^
{\prime 0}-
ip_{1}^{ 0}x_{1}^{0}-\cdots -ip_{N-1}^{ 0}
x_{N-1}^{0})}\nonumber\\
&&\times \int d^{4}y d^{4}z \; \exp{(ik^{0}y^{0}-ip^{0}z^{0})}\;
A_f^{\nu *}(\bfy)
\nonumber\\ 
&&\times\la 0|T\psi(x_{1}^{\prime})\cdots \psi(x_{N}^{\prime})j_{\nu}(y) 
\overline{\psi}(z)
\overline {\psi}(x_{N-1})\cdots \overline {\psi} (x_{1})|0\ra \nonumber\\
&&\times (-i\not\stackrel{\leftarrow}
{\partial}_{z}-m) 
U_{p_i\mu_i}(\bfz)
\, .
\end{eqnarray}
The Green function $G_{\gamma_{f},e_{i}}$ is constructed using the Wick
theorem after the transition in (\ref{greenrec3}) 
to the interaction representation. Since we have not included
the interaction with $V_{\rm C}$ in the unperturbed Hamiltonian,
the Feynman rules contain the free electron propagators
and the vertices corresponding to the interaction with $V_{\rm C}$.
Summing over all the insertions of the vertices with $V_{\rm C}$
in the electron lines we replace the free electron propagators
and wave functions by the electron propagators and wave functions
in the Coulomb field, according to equations (\ref{propbound}) and 
(\ref{wavebound}). The free  wave function of the incident electron  
$U_{p_i \mu_i}(\bfx)$
is replaced by the wave function
$\psi_{p_i\mu_i(+)}({\bf x})$ that can be determined
 from the equation
\begin{eqnarray} \label{psip+}
\psi_{p_i,\mu_i(+)}=U_{p_i\mu_i}
+(p_{i}^{0}-H_{0}(1-i0))
^{-1}V_{\rm C}\psi_{p_i\mu_i(+)}\, .
\end{eqnarray}
 The wave function $\psi_{p_i\mu_i(+)}
({\bf x})$ is a solution of the Dirac equation  with the Coulomb
potential $V_{\rm C}(\bfx)$
 which satisfies the boundary condition \cite{akhiezer}
\begin{eqnarray}
\psi_{p\mu(+)}({\bfx}) \;\sim \; U_{p\mu}(\bfx)+
G^{+}({\bf n})\frac{\exp{(i|\bfp||\bfx|)}}{|\bfx|}\,,\;\;
\;\;\;|\bfx|\rightarrow\infty\, ,
\end{eqnarray}
where  $G^{+}({\bf n})$ is a bispinor depending
on ${\bf n}\equiv \bfx/|\bfx|$.
The apparent expressions for 
$\psi_{p\mu(+)}(\bfx)$ are given, e.g., in \cite{eichler}. Thus,
we again revert to the Furry picture. The Feynman rules
for $G_{\gamma_{f};e_{i}}$  differ from those for $G_{\gamma_f}$
 only by
a replacement of one of the incoming electron propagators
$\frac{i}{2\pi}S(\omega,{\bf x},{\bf y})$ by the wave function
$\psi_{p_i\mu_i(+)}({\bf x})$.

In calculations by perturbation theory a problem appears which is
caused by the fact that in zeroth approximation the energy
of the $(N-1)$-electron atom may coincide with the difference of the
 $N$-electron energy and the one-electron energy.
As a result, some of the terms which are omitted in  equation
(\ref{spectrec1}) are singular functions of $E'$ and $E$
if $ E'\approx E_{b}^{(0)}$ and $E \approx E_{a}^{(0)}$.
A special analysis of the complete spectral representation
of the Green function ${\cal G}_{\gamma_f,e_i}$ shows 
that to eliminate these  terms in  calculations
by formula (\ref{recfor}) the following prescription
should be used:
integrate first over $E$
and keep the point $E=E_a^{(0)}+(E'-E_b^{(0)})$ outside the
contour $\Gamma_a$.

In Refs. \cite{sh00,yer00} this method was used to derive formulas for 
the QED corrections to the radiative recombination of an electron 
with a bare nucleus and for the interelectronic-interaction
corrections to the radiative recombination of an electron
with a high-$Z$  heliumlike atom.
These formulas are presented in sections IV(D1), IV(D2).
 Here we consider another application
of the method.

\subsection{Radiative recombination of an electron with
a high-$Z$ hydrogenlike atom}

As a practical application of the method, in this section we derive
formulas for the radiative recombination of an electron
with a high-$Z$ hydrogenlike ion to zeroth and first orders in
$1/Z$. 
 We will assume that
 the final state of the heliumlike
ion is described by a one-determinant wave function
\be \label{recwf}
u_b (\bfx_1,\bfx_2)=\frac{1}{\sqrt{2}}
\sum_P(-1)^P \psi_{Pb_1}(\bfx_1)\psi_{Pb_2}(\bfx_2)
\ee
and the one-electron state 
 $|b_1\ra$ coincides with the initial state
$|a\ra$ of the hydrogenlike ion.

\subsubsection{Zeroth order approximation}

To zeroth order, the radiative recombination amplitude
is described by the diagram shown in Fig. 34.
The formula (\ref{recfor1}) gives 
\begin{eqnarray}\label{zer1}
S^{(0)}_{\gamma_f,b;e_i,a}=\delta(E_{b}
+k_{f}^0-E_{a}-p_i^0)\oint_{\Gamma_b}dE' \; \oint_{\Gamma_a}dE 
\; g^{(0)}_{\gamma_{f},b;e_i,a}(E',E,p_i^0)\,,
\end{eqnarray}
where the superscript indicates the order in $\alpha$.
According to the definition (\ref{recgsmall})
and the Feynman rules, we have 
\be
\lefteqn{g^{(0)}_{\gamma_f,b;e_i,a}(E',E,p_i^0)
\delta(E'+k^0-E-p_i^0)}\nonumber\\
&&=\frac{i}{2\pi}\sum_{P}(-1)^P\int_{-\infty}^{\infty} dp'^0_1
dp'^0_2 \;
\delta(E'-p'^0_1-p'^0_2)\frac{1}{p'^0_2-\veps_{Pb_2}+i0}
\nonumber\\
&&\times \la Pb_2|e\alpha_{\nu}A_f^{\nu *}|p\ra
\delta(p'^0_2+k^0-p_i^0)\frac{1}{p'^0_1-\veps_{a}+i0}
\delta(p'^0_1-E)\la Pb_1|a\ra\,,
\ee
where $|p\ra\equiv |p_i,\mu_i\ra $ denotes the state vector of
the incident electron. We obtain
\be
g^{(0)}_{\gamma_f,b;e_i,a}(E',E,p_i^0)=
\frac{i}{2\pi}\frac{1}{E'-E-\veps_{b_2}}
\la b_2|e\alpha_{\nu}A_f^{\nu *}|p\ra
\frac{1}{E-\veps_{a}}\,.
\ee
Following to the rule for the integration over $E$ and $E'$ 
formulated at the end of section III(C),
 we find
\be
\oint_{\Gamma_b}dE'\; \oint_{\Gamma_a} dE \;
g^{(0)}_{\gamma_f,b;e_i,a}(E',E,p_i^0)=-2\pi i \la b_2|
e\alpha_{\nu}A_f^{\nu *}|p\ra\,.
\ee
It yields
\be
S_{\gamma_f,b;e_i,a}^{(0)}=-2\pi i \delta(E_b+k_f^0-E_a-p_i^0)
\la b_2|e\alpha_{\nu}A_f^{\nu *}|p\ra
\ee
or, according to the definition (\ref{deftau}),
\be
\tau_{\gamma_f,b;e_i,a}^{(0)}=- \la b_2|
e\alpha_{\nu}A_f^{\nu *}|p\ra\,.
\ee
The corresponding cross section is
\begin{eqnarray}\label{inf10ab}
\frac{d\sigma^{(0)}}{d\Omega_{f}}=
\frac{(2\pi)^4}{v_{i}}\,
 \bfk_{f}^2\,|\tau_{\gamma_f,b;e_i,a}^{(0)}|^2\,,
\end{eqnarray}
where $v_i$ is the velocity of the
incident electron in the frame of the nucleus.

\subsubsection{Interelectronic-interaction corrections
of first order in 1/Z}

The interelectronic-interaction corrections of first order 
in $1/Z$ are defined by the diagrams shown in Fig. 35.
In the order under consideration,
the formula (\ref{recfor1}) gives 
\begin{eqnarray}\label{recfirst1}
S^{(1)}_{\gamma_f,b;e_i,a}&=&\delta(E_{b}
+k_{f}^0-E_a-p_i^0)\;\Bigl[
\oint_{\Gamma_b}dE' \; \oint_{\Gamma_a}dE
 \;g_{\gamma_{f},b;e_i,a}^{(1)}(E',E,p_i^0)\nonumber\\
&&-\frac{1}{2}
\oint_{\Gamma_b}dE'\; \oint_{\Gamma_a}dE
 \; g_{\gamma_{f},b;e_i,a}^{(0)}(E',E,p_i^0)\;
\frac{1}{2\pi i}\oint_{\Gamma_b} dE\; g_{bb}^{(1)}(E)
 \Bigr]\,,
\end{eqnarray}
where $g_{bb}^{(1)}(E)$ is defined by the first-order
interelectronic-interaction diagram (see Fig. 20).
According to the definition (\ref{recgsmall})
and the Feynman rules, we have 
\be
\lefteqn{g^{(1)}_{\gamma_f,b;e_i,a}(E',E,p_i^0)\delta(E'+k^0-E-p_i^0)
\;\;\;\;\;\;\;\;\;\;\;\;\;\;\;\;\;\;\;\;\;\;\;\;\;\;\;}
\nonumber\\
&&=\Bigl(\frac{i}{2\pi}\Bigr)^2
\sum_{P}(-1)^P\int_{-\infty}^{\infty} dp'^0_1
dp'^0_2\;
\delta(E'-p'^0_1-p'^0_2)\nonumber\\
&&\times \frac{1}{p'^0_1-\veps_{Pb_1}+i0}
\frac{1}{p'^0_2-\veps_{Pb_2}+i0}
\frac{1}{E-\veps_{a}+i0}
\int_{-\infty}^{\infty}dq^0 d\omega\nonumber\\
&&\times \sum_n \Bigl[ \la Pb_1 Pb_2|I(\omega)|an\ra
\delta(p'^0_2+\omega-q^0)
\delta(p'^0_1-\omega-E)\nonumber\\
&&\times\frac{1}{q^0-\veps_n(1-i0)}\la n |e \alpha_{\nu} A_f^{\nu *}|p\ra
\delta(q^0+k^0-p_i^0)\nonumber\\
&&+ \la Pb_1 Pb_2|I(\omega)|np\ra
\delta(p'^0_2+\omega-p_i^0)
\delta(p'^0_1-\omega-q^0)\nonumber\\
&&\times \frac{1}{q^0-\veps_n(1-i0)}\la n |e \alpha_{\nu} A_f^{\nu *}|a\ra
\delta(q^0+k^0-E)\nonumber\\
&&+\la Pb_2 |e \alpha_{\nu} A_f^{\nu *}|n\ra
\frac{1}{q^0-\veps_n(1-i0)}
\delta(p'^0_2+k^0-q^0)\nonumber\\
&&\times \la Pb_1 n|I(\omega)|ap\ra
\delta(q^0+\omega-p_i^0)
\delta(p'^0_1-\omega-E)\nonumber\\
&&+\la Pb_1 |e \alpha_{\nu} A_f^{\nu *}|n\ra
\frac{1}{q^0-\veps_n(1-i0)}
\delta(p'^0_1+k^0-q^0)\nonumber\\
&&\times \la n Pb_2 |I(\omega)|ap\ra
\delta(p'^0_2+\omega-p_i^0)
\delta(q^0-\omega-E)\Bigr]\,.
\ee
We obtain
\be \label{recint2}
g^{(1)}_{\gamma_f,b;e_i,a}(E',E,p_i^0)&=&
\Bigl(\frac{i}{2\pi}\Bigr)^2
\frac{1}{E'-E_b^{(0)}}\frac{1}{E-\veps_a}
\sum_{P}(-1)^P \sum_n
\int_{-\infty}^{\infty} dp'^0_1
\nonumber\\
&&\times \Bigl(\frac{1}{p'^0_1-\veps_{Pb_1}+i0}+
\frac{1}{E'-p'^0_1-\veps_{Pb_2}+i0}\Bigr)\nonumber\\
&&\times\Bigr[\la Pb_1 Pb_2|I(p'^0_1-E)|an\ra
\frac{1}{E'-E-\veps_n(1-i0)}\la n |e \alpha_{\nu} A_f^{\nu *}|p\ra
\nonumber\\
&&+\la Pb_2 |e \alpha_{\nu} A_f^{\nu *}|n\ra
\frac{1}{E+p_i^0-p'^0_1-\veps_n(1-i0)}\nonumber\\
&&\times \la Pb_1 n|I(p'^0_1-E)|ap\ra \Bigr]\nonumber\\
&&+\Bigl(\frac{i}{2\pi}\Bigr)^2
\frac{1}{E'-E_b^{(0)}}\frac{1}{E-\veps_a}
\sum_{P}(-1)^P \sum_n\int_{-\infty}^{\infty} dp'^0_2
\nonumber\\
&&\times \Bigl(\frac{1}{p'^0_2-\veps_{Pb_2}+i0}+
\frac{1}{E'-p'^0_2-\veps_{Pb_1}+i0}\Bigr)\nonumber\\
&&\times\Bigr[\la Pb_1 Pb_2|I(p_i^0-p'^0_2)|np\ra
\frac{1}{E'-p_i^0-\veps_n(1-i0)}\la n |e \alpha_{\nu} A_f^{\nu *}|a\ra
\nonumber\\
&&+\la Pb_1 |e \alpha_{\nu} A_f^{\nu *}|n\ra
\frac{1}{E+p_i^0-p'^0_2-\veps_n(1-i0)}\nonumber\\
&&\times \la n Pb_2|I(p_i^0-p'^0_2)|ap\ra \Bigr]\,.
\ee
The first term in equation (\ref{recint2}) is conveniently
divided into an irreducible ($\veps_n\not=E_b^{(0)}-\veps_a=\veps_{b_2}$)
and a reducible ($\veps_n=\veps_{b_2}$) part.
Since we consider the case of a single final state described
by the one-determinant wave function (\ref{recwf}) with
$|b_1\ra=|a\ra$, the condition
$\veps_n=\veps_{b_2}$ means $|n\ra=|b_2\ra$
(otherwise $\la Pb_1 Pb_2|I(\omega)|an\ra =0$).
Substituting (\ref{recint2}) into (\ref{recfirst1}),
integrating over $E$ and $E'$
according to the rule formulated at the end of section III(C), 
 and using the identity
(\ref{e3e17}), we find for the irreducible contribution 
\be \label{recirred}
S_{\gamma_{f},b;e_i,a}^{(1,{\rm irred})}&=&
\delta(E_b+k_f^0-E_a-p_i^0)
\oint_{\Gamma_b}dE' \; \oint_{\Gamma_a}dE
 \;g_{\gamma_{f},b;e_i,a}^{(1,{\rm irred})}(E',E,p_i^0)\nonumber\\
&=&-2\pi i \delta(E_b+k_f^0-E_a-p_i^0)
\sum_{P}(-1)^P \nonumber\\
&& \times \Bigr[
\sum_n^{n\not=b_2}
\la Pb_1 Pb_2|I(\veps_{Pb_1}-\veps_a)|an\ra
\frac{1}{\veps_{b_2}-\veps_n}\la n |e \alpha_{\nu} A_f^{\nu *}|p\ra
\nonumber\\
&&+\sum_n \la Pb_2 |e \alpha_{\nu} A_f^{\nu *}|n\ra
\frac{1}{\veps_a+p_i^0-\veps_{Pb_1}-\veps_n}
\la Pb_1 n|I(\veps_{Pb_1}-\veps_a)|ap\ra \nonumber\\
&&+\sum_n \la Pb_1 Pb_2|I(p_i^0-\veps_{Pb_2})|np\ra
\frac{1}{E_b^{(0)}-p_i^0-\veps_n}\la n |e \alpha_{\nu} A_f^{\nu *}|a\ra
\nonumber\\
&&+\sum_n \la Pb_1 |e \alpha_{\nu} A_f^{\nu *}|n\ra
\frac{1}{\veps_a+p_i^0-\veps_{Pb_2}-\veps_n}
\la n Pb_2|I(p_i^0-\veps_{Pb_2})|ap\ra \Bigr]\,.
\ee
For the reducible part we have
\be
\lefteqn{\oint_{\Gamma_b}dE' \;\oint_{\Gamma_a}dE
 \;g_{\gamma_{f},b;e_i,a}^{(1,{\rm red})}(E',E,p_i^0)}\nonumber\\
&&=2\pi i \Bigl(\frac{i}{2\pi}\Bigr)^2\oint_{\Gamma_b} dE' \;
\sum_{P}(-1)^P \frac{1}{(E'-E_b^{(0)})^2}\int_{-\infty}^{\infty}
dp'^0_1 \;
\Bigl( \frac{1}{p'^0_1-\veps_{Pb_1}+i0}\nonumber\\
&&+\frac{1}{E'-p'^0_1-\veps_{Pb_2}+i0}\Bigr)
\la Pb_1 Pb_2|I(p'^0_1-\veps_a)|ab_2\ra
\la b_2 | e\alpha_{\nu}A_f^{\nu *}|p\ra \nonumber \\
&&=-\sum_P(-1)^P\int_{-\infty}^{\infty} d\omega \;
\frac{1}{(\omega-\veps_{Pb_1}-i0)^2}
 \la Pb_1 Pb_2|I(\omega-\veps_a)|b_1 b_2\ra
\la b_2 | e\alpha_{\nu}A_f^{\nu *}|p\ra\,.
\label{recred00}
\ee
The reducible contribution must be considered together
with the second term in formula (\ref{recfirst1}).
Taking into account that
\be
\frac{1}{2\pi i}\oint_{\Gamma_b} dE \; g_{bb}^{(1)}(E)
&=&-\frac{i}{2\pi}\Bigl[ 2\int_{-\infty}^{\infty} dp'^0_1 \;
\frac{1}{(p'^0-\veps_{b_1}-i0)^2} \la b_1 b_2|I(p'^0_1-\veps_{b_1})
|b_1 b_2\ra\nonumber\\
&& - \int_{-\infty}^{\infty} dp'^0_1 \;
\frac{1}{(p'^0_1-\veps_{b_2}-i0)^2}
 \la b_2 b_1|I(p'^0_1-\veps_{b_1})
|b_1 b_2\ra\nonumber\\
&& - \int_{-\infty}^{\infty} dp^0_1 \;
\frac{1}{(p^0_1-\veps_{b_1}-i0)^2}
 \la b_2 b_1|I(p^0_1-\veps_{b_2})
|b_1 b_2\ra\,,
\ee
we find
\be \label{recred0}
\lefteqn{-\frac{1}{2}\oint_{\Gamma_b}dE' \; \oint_{\Gamma_a}dE
\;g_{\gamma_{f},b;e_i,a}^{(0)}(E',E,p_i^0)
\frac{1}{2\pi i}\oint_{\Gamma_b} dE \; g_{bb}^{(1)}(E)}
\nonumber\\
&&=\frac{1}{2}\la b_2|e\alpha_{\nu}A_f^{\nu *}|p\ra\Bigl[
2\int_{-\infty}^{\infty} d\omega \;\frac{\la b_1 b_2|I(\omega)|b_1 b_2\ra}
{(\omega-i0)^2}\nonumber\\
&&-\int_{-\infty}^{\infty} d\omega \; \la b_2 b_1|I(\omega)|b_1 b_2\ra
\Bigl(\frac{1}{(\omega-\Delta_b-i0)^2}+\frac{1}{(\omega+\Delta_b-i0)^2}
\Bigr)\Bigr]\,,
\ee
where $\Delta_b\equiv \veps_{b_2}-\veps_{b_1}$.
Summing (\ref{recred00}) and (\ref{recred0}), we obtain
for the total reducible contribution 
\be
S_{\gamma_f,b;e_i,a}^{(1,{\rm red})}&=&-2\pi i 
\delta(E_b+k_f^0-E_a-p_i^0)
\frac{1}{2}
\la b_2 |e\alpha_{\nu} A_f^{\nu *}|p\ra \frac{i}{2\pi}
\int_{-\infty}^{\infty} d\omega \; \la b_2 b_1|I(\omega) |b_1 b_2\ra
\nonumber\\
&&\times \Bigr[\frac{1}{(\omega +\Delta_b+i0)^2}
-\frac{1}{(\omega+\Delta_b-i0)^2}\Bigr]\,.
\ee
Using the identity
\be
\frac{1}{(\omega+i0)^2}-
\frac{1}{(\omega-i0)^2} = -\frac{2\pi}{i}\frac{d}{d\omega}
\delta(\omega)
\ee
and integrating by parts, we obtain
\be \label{recred}
S_{\gamma_f,b;e_i,a}^{(1,{\rm red})}=
2\pi i \delta(E_b+k_f^0-E_a-p_i^0)
\frac{1}{2}
\la b_2 |e\alpha_{\nu} A_f^{\nu *}|p\ra 
 \la b_2 b_1|I'(\Delta_b) |b_1 b_2\ra\,.
\ee

In addition to the interelectronic-interaction
 correction derived in this subsection,
we must take into account the contribution originating from
changing the photon energy in  the zeroth-order cross
section (\ref{inf10ab}) due to the interelectronic-interaction
 correction to the energy
of the bound state $b$. It follows that the total 
interelectronic-interaction
correction to the cross section of  first order in $1/Z$ 
is given by
\begin{eqnarray}\label{form34a}
\frac{d\sigma^{(1)}}{d\Omega_{f}}
=\frac{(2\pi)^4}{v_{i}}
\bfk_{f}^2\,2{\rm Re}{\Bigl\{\tau_{\gamma_f,b;e_i,a}^{(0)*}
\tau_{\gamma_f,b;e_i,a}^{(1)}}\Bigr\}+
\Bigl[\frac{d\sigma^{(0)}}{d\Omega_{f}}
\Bigr|_{k_{f}^0=p_{i}^0+\veps_{a}-E_b}
-\frac{d\sigma^{(0)}}{d\Omega_{f}}\Bigr|_{k_{f}^0=p_{i}^0+
\veps_{a}-E_b^{(0)}}
\Bigr]\,.
\end{eqnarray} 
Here $\tau_{\gamma_f,b;e_i,a}^{(1)}=
\tau_{\gamma_f,b;e_i,a}^{(1,{\rm irred})}+
\tau_{\gamma_f,b;e_i,a}^{(1,{\rm red})}$
is the interelectronic-interaction
 correction given by equations (\ref{recirred}) and (\ref{recred})
in accordance with the definition (\ref{deftau}).
 $E_b$ and
 $E_{b}^{(0)}$ are the energies of the bound state $b$
with and without the interelectronic-interaction
 correction, respectively.

\subsection {Photon scattering by an atom}

Let us consider the scattering of a photon  with momentum
$\bfk_i$ and polarization $\eps_i=(0,\beps_i)$  by
an atom which is initially  in a state $a$. As a result of
the scattering, a photon with momentum $\bfk_f$ and polarization
$\eps_f=(0,\beps_f)$ is emitted and the atom is left in a state $b$.
In this section we  consider the non-resonant photon scattering.
This means that the total initial energy of the system
($E_a+k_i^0$) is not close to any excited-state energy of the atom.
The resonant photon scattering will be considered in detail in the
next section.

According to the standard reduction technique (see, e.g.,
\cite{itzykson}), the amplitude of the process is
\begin{eqnarray} \label{sphot1}
S_{\gamma_{f},b;\gamma_{i},a}&=&\la b|a_{\rm out}(k_{f},{\eps}_{f})
a_{\rm in}^{\dag}
(k_{i},{\eps}_{i})|a\ra\nonumber\\
&=& {\rm Disconnected \;\;\, term\,}
-Z_{3}^{-1}
\int d^{4}y d^{4}z \; \frac{{\eps}_{f}^{\nu} \exp{(ik_{f}\cdot y)}}
{\sqrt{2k_{f}^{0}(2\pi )^3}}   \nonumber\\
&&\times 
\la b|Tj_{\nu}(y)j_{\rho}(z)|a\ra
\frac{{\eps}_{i}^{\rho} \exp{(-ik_{i}\cdot z)}}
{\sqrt{2k_{i}^{0}(2\pi )^3}}
\, ,
\end{eqnarray}
where $|a\ra$ and $|b\ra$ are the vectors of the initial and
 final states in the Heisenberg
representation; $k_f=(k_f^0,\bfk_f)$ and
$k_i=(k_i^0,\bfk_i)$. The first term
 on the right-hand side of equation (\ref{sphot1})
 corresponds to a non-scattering process, i.e., the photon does not interact
with the atom. We are interested in the second term
which describes the photon scattering by the atom.
Let us designate this term by $S_{\gamma_f,b;\gamma_i,a}^{\rm scat}$.

To derive the calculation
 formula for  $S_{\gamma_f,b;\gamma_i,a}^{\rm scat}$,
 in the scattering amplitude we isolate  a term which describes
the scattering of the photon by the Coulomb potential $V_{\rm C}$.
With this in mind, we write
\be \label{sphot2}
\la b|Tj(y)j(z)|a\ra=\la b|[Tj(y)j(z)-\la 0|Tj(y)j(z)|0\ra]|a\ra
+\delta_{ab}\la 0|Tj(y)j(z)|0\ra\,.
\ee
The second term on the right-hand side of this equation
corresponds to the photon scattering by the Coulomb field
and the first term describes the photon scattering  
by the electrons of the atom. To the first term, 
 only diagrams contribute where
the incident photon is connected to at least one 
external electron line. We denote this term
by $S_{\gamma_f,b;\gamma_i,a}^{\rm con}$.
We have
\begin{eqnarray} \label{sphot3}
S_{\gamma_{f},b;\gamma_{i},a}^{\rm con}
&=&-Z_{3}^{-1}
\int d^{4}y d^{4}z \; \frac{{\eps}_{f}^{\nu} \exp{(ik_{f}\cdot y)}}
{\sqrt{2k_{f}^{0}(2\pi )^3}}  
\la b|[Tj_{\nu}(y)j_{\rho}(z)\nonumber\\
&&-\la 0|Tj_{\nu}(y)j_{\rho}(z)|0\ra]|a \ra
\frac{{\eps}_{i}^{\rho} \exp{(-ik_{i}\cdot z)}}
{\sqrt{2k_{i}^{0}(2\pi )^3}}\nonumber\\
&=&-Z_{3}^{-1} 2\pi \delta(E_b+k_f^0-E_a-k_i^0)
\int_{-\infty}^{\infty} dt \; \exp{(ik_f^0 t)}
\int d\bfy d\bfz \; A_f^{\nu *}(\bfy)\nonumber\\  
&&\times\la b|[Tj_{\nu}(t,\bfy)j_{\rho}(0,\bfz)
-\la 0|Tj_{\nu}(t,\bfy)j_{\rho}(0,\bfz)|0\ra]|a\ra
A_i^{\rho}(\bfz)
\, .
\end{eqnarray}
Here the second term is zero if $a\not=b$. 

To construct the perturbation theory for  the amplitude
of the process, we introduce the Fourier transform of the 
corresponding two-time Green function
\begin{eqnarray}
\lefteqn{{\cal G}^{\rm con}
_{\gamma_f; \gamma_{i}}(E',E,k^{\prime 0};{\bf x}_{1}^
{\prime},...{\bf x}_{N}^{\prime};{\bf x}_{1},...{\bf x}_{N})
\delta(E'+k^{\prime 0}-E-k^{0})
\;\;\;\;\;\;\;\;\;\;\;\;\;\;\;\;\;\;\;\;\;\;\;\;\;\;}\nonumber\\
&=&\Bigl(\frac{i}{2\pi}\Bigr)^{2}\frac{1}{N!}
\int_{-\infty}^{\infty}dx^0dx^{\prime 0} \;
\int d^{4}yd^{4}z
\;\exp{(iE'x^{\prime 0}+ik^{\prime 0}y^0-
iEx^{0}-ik^0z^0)}\nonumber\\
&&\times
A_f^{\nu *}(\bfy) A_i^{\rho}(\bfz)
\la 0|T\psi(x^{\prime 0},{\bf x}_{1}^{\prime})\cdots
\psi(x^{\prime 0},{\bf x}_{N}^{\prime})
[j_{\nu}(y)j_{\rho}(z)\nonumber\\
&&-\la 0|T j_{\nu}(y)j_{\rho}(z)|0\ra]
\overline{\psi}(x^{0},{\bf x}_{N})\cdots \overline{\psi}
(x^{0},{\bf x}_{1})|0\ra
\, .
\end{eqnarray}
Introducing $t'=x'^{0}-z^0,\; t=x^0-z^0$, and $\tau=y^0-z^0$
and integrating over $z^0$, we obtain
\begin{eqnarray}
\lefteqn{{\cal G}^{\rm con}
_{\gamma_f; \gamma_{i}}(E',E,k^{\prime 0};{\bf x}_{1}^
{\prime},...{\bf x}_{N}^{\prime};{\bf x}_{1},...{\bf x}_{N})
\delta(E'+k^{\prime 0}-E-k^{0})
\;\;\;\;\;\;\;\;\;\;\;\;\;\;\;\;\;\;\;\;\;\;}\nonumber\\
&=&-\delta(E'+k^{\prime 0}-E-k^{0})
\frac{1}{2\pi}\frac{1}{N!}
\int_{-\infty}^{\infty}dtdt'd\tau \;
\exp{(iE't^{\prime}+ik^{\prime 0}\tau-
iE\tau)}\nonumber\\
&&\times\int d\bfy d\bfz \;
A_f^{\nu *}(\bfy) A_i^{\rho}(\bfz)
\la 0|T\psi(t^{\prime},{\bf x}_{1}^{\prime})\cdots
\psi(t^{\prime },{\bf x}_{N}^{\prime})
[j_{\nu}(\tau,\bfy)j_{\rho}(0,\bfz)\nonumber\\
&&-\la 0|T j_{\nu}(\tau,\bfy)j_{\rho}(0,\bfz)|0\ra]
\overline{\psi}(t,{\bf x}_{N})\cdots \overline{\psi}
(t,{\bf x}_{1})|0\ra
\, .
\end{eqnarray}
Integrating over $t$ and $t'$ we find
\begin{eqnarray} \label{greenph4}
\lefteqn{{\cal G}^{\rm con}
_{\gamma_f; \gamma_{i}}(E',E,k^{\prime 0};{\bf x}_{1}^
{\prime},...{\bf x}_{N}^{\prime};{\bf x}_{1},...{\bf x}_{N})}
\nonumber\\
&=&-\frac{1}{2\pi}\frac{1}{N!}\sum_{n,m}
\frac{i}{E'-E_n+i0}\frac{i}{E-E_m+i0} 
\int_{-\infty}^{\infty}d\tau \;
\exp{(ik^{\prime 0}\tau)}\nonumber\\
&&\times\int d\bfy d\bfz \;
A_f^{\nu *}(\bfy) A_i^{\rho}(\bfz)\{
\theta(\tau)\exp{[i\tau(E'-E_n)]}
\la 0|T\psi(0,{\bf x}_{1}^{\prime})\cdots
\psi(0,{\bf x}_{N}^{\prime})|n\ra\nonumber\\
&&\times \la n|[j_{\nu}(\tau,\bfy)j_{\rho}(0,\bfz)
-\la 0|T j_{\nu}(\tau,\bfy)j_{\rho}(0,\bfz)|0\ra]|m\ra\nonumber\\
&&\times
\la m|\overline{\psi}(0,{\bf x}_{N})\cdots \overline{\psi}
(0,{\bf x}_{1})|0\ra 
+\theta(-\tau)\exp{[i\tau(E_m-E)]}\nonumber\\
&&\times \la 0|T\psi(0,{\bf x}_{1}^{\prime})\cdots
\psi(0,{\bf x}_{N}^{\prime})|n\ra
\la n|[j_{\rho}(0,\bfz)j_{\nu}(\tau,\bfy)\nonumber\\
&&-\la 0|T j_{\nu}(\tau,\bfy)j_{\rho}(0,\bfz)|0\ra]|m\ra
\la m|\overline{\psi}(0,{\bf x}_{N})\cdots \overline{\psi}
(0,{\bf x}_{1})|0\ra +\cdots\}
\, .
\end{eqnarray}
It can be shown that the terms which are omitted in equation
(\ref{greenph4}) are regular functions of $E'$ or $E$ if
 $E'\approx E_b$ and $E\approx E_a$.
As in the previous sections of this paper, we assume that
 in the zeroth approximation the initial
and final states of the atom are degenerate in energy
with some other states and use the same notations
for these states as above.
As usual, we also assume that
a non-zero photon mass $\mu$ is introduced.  
We define the Green function $g^{\rm con}_{\gamma_f,b;\gamma_i,a}
(E',E,k'^{0})$ by
\begin{eqnarray}
g_{\gamma_{f},b;\gamma_{i},a}
(E',E,k'^{0})=P^{(0)}_{b}{\cal G}^{\rm con}
_{\gamma_{f};\gamma_{i}}(E',E,k'^{0})\gamma_{1}^
{0}\cdots \gamma_{N}^{0}P^{(0)}_{a}\, .
\end{eqnarray}
From  equation (\ref{greenph4}) we have
\begin{eqnarray} \label{greenph5}
g^{\rm con}_{\gamma_f,b; \gamma_{i},a}(E',E,k^{\prime 0})
&=&\frac{1}{2\pi}\sum_{n_a,n_b}
\frac{\varphi_{n_b}}{E'-E_{n_b}} 
\int d\bfy d\bfz \;
A_f^{\nu *}(\bfy) A_i^{\rho}(\bfz)
\int_{-\infty}^{\infty}d\tau \;
\exp{(ik^{\prime 0}\tau)}\nonumber\\
&&\times\{\theta(\tau)\exp{[i\tau(E'-E_{n_b})]}
\la n_b|[j_{\nu}(\tau,\bfy)j_{\rho}(0,\bfz)\nonumber\\
&&-\la 0|T j_{\nu}(\tau,\bfy)j_{\rho}(0,\bfz)|0\ra]|n_a\ra
+\theta(-\tau)\exp{[i\tau(E_{n_a}-E)]}\nonumber\\
&&\times\la n_b|[j_{\rho}(0,\bfz)j_{\nu}(\tau,\bfy)
-\la 0|T j_{\nu}(\tau,\bfy)j_{\rho}(0,\bfz)|0\ra]|n_a\ra\}
\frac{\varphi_{n_a}^{\dag}}{E-E_{n_a}}\nonumber\\
&&+\mbox{ terms that are regular functions of $E'$ or $E$ 
if  $E'\approx E_b^{(0)}$}\nonumber\\
&&\;\;\;\mbox{ and $E\approx E_a^{(0)}$}\, .
\end{eqnarray}
Taking into account the biorthogonality condition (\ref{biort})
and comparing (\ref{sphot3}) with (\ref{greenph5}), we obtain 
the desired formula \cite{shab88b}
\begin{eqnarray} \label{sphotfin}
S^{\rm con}
_{\gamma_{f},b;\gamma_{i},a}=Z_{3}^{-1}
 \delta (E_{b}+k_{f}^{0}-E_{a}-k_{i}^
{0})\oint_{\Gamma_{b}}dE' \;\oint_{\Gamma_
{a}}dE\; v_{b}^{\dag}g_{\gamma_{f},b;\gamma_{i},a}^{\rm con}
(E',E,k_{f}^{0})v_{a}\,,
\end{eqnarray}
where we imply by $a$  one of the initial states and  by $b$ 
one of the final states under consideration.
In the case of a single initial state ($a$) and a single final
state ($b$) it yields
\be \label{photscat2}
S_{\gamma_f,b;\gamma_{i}, a}^{\rm con}&=&
Z_{3}^{-1}
\delta (E_{b}+k_{f}^{0}-E_{a}-k_{i}^{0})
\oint_{\Gamma_b}dE'\;
\oint_{\Gamma_a}dE \; g_{\gamma_f,b;\gamma_i,a}^{\rm con}
(E',E,k_f^0)\nonumber\\
&&\times \Bigl[\frac{1}{2\pi i}\oint_{\Gamma_b}dE \;
 g_{bb}(E)\Bigr]^{-1/2}
\Bigl[\frac{1}{2\pi i}\oint_{\Gamma_a}dE \; g_{aa}(E)\Bigr]^{-1/2}
\,.
\ee

The disconnected term describing the scattering of the photon
by the Coulomb field is calculated by the formula
\begin{eqnarray} \label{sphotcoul}
S^{\rm discon}_{\gamma_{f},b;\gamma_{i},a}=
-Z_{3}^{-1}\delta_{ab}
\int d^{4}y d^{4}z \;
\frac{{\eps}_{f}^{\nu} \exp{(ik_{f}\cdot y)}}
{\sqrt{2k_{f}^{0}(2\pi )^3}} 
\la 0|Tj_{\nu}(y)j_{\rho}(z)|0\ra
\frac{{\eps}_{i}^{\rho} \exp{(-ik_{i}\cdot z)}}
{\sqrt{2k_{i}^{0}(2\pi )^3}}
\, .
\end{eqnarray}

For practical calculations by perturbation theory it is convenient
to express the Green function $g^{\rm con}_{\gamma_{f},b;\gamma_{i},a}$
in terms of the Fourier transform of the $2N$-time Green function,
\begin{eqnarray}
\lefteqn{g^{\rm con}_{\gamma_{f},b;
\gamma_{i},a}(E',E,k'^{0})\delta (E'+k'^{0}-E-k^{0})
\;\;\;\;\;\;\;\;\;\;\;\;\;\;\;\;\;\;\;\;\;\;\;\;\;\;} 
\nonumber\\
&&=\frac{1}{N!}\int_{-\infty}^
{\infty} dp_{1}^{0}\cdots dp_{N}^{0}dp_{1}^{\prime 0}\cdots dp_{N}^
{\prime 0}\;\delta(E-p_{1}^{0}\cdots -p_{N}^{0})\delta(E'-p_{1}^
{\prime 0}\cdots -p_{N}^{\prime 0})\nonumber\\
&&\times P^{(0)}_{b}G^{\rm con}_{\gamma_{f};\gamma_{i}}
(p_{1}^{\prime 0},...,p_{N}^{\prime 0};k'^{0},k^{0};p_{1}^{0},...,
p_{N}^{0})\gamma_{1}^{0}\cdots \gamma_{N}^{0}P^{(0)}_{a}\, ,
\end{eqnarray}
where
\begin{eqnarray} \label{greenph7}
\lefteqn{G^{\rm con}_{\gamma_{f};\gamma_{i}}((p_{1}^{\prime 0},
{\bf x}_{1}^{\prime}),...,(p_{N}^{\prime 0},
{\bf x}_{N}^{\prime});k'^{0},k^{0};(p_{1}^{ 0},
{\bf x}_{1}),...,(p_{N}^{0},{\bf x}_{N}))
\;\;\;\;\;\;\;\;\;\;\;\;\;\;\;\;\;\;\;\;\;\;\;\;}\nonumber\\
&&=-\frac{1}{(2\pi)^{2N}}
\int_{-\infty}^{\infty}dx_{1}^{0}\cdots
dx_{N}^{0}
dx_{1}^{\prime 0}\cdots dx_{N}^{\prime 0}\nonumber\\
&&\times\exp{(ip_{1}^{\prime 0}x_{1}^{\prime 0}+\cdots
+ip_{N}^{\prime 0}x_{N}^
{\prime 0}-
ip_{1}^{ 0}x_{1}^{0}-\cdots-ip_{N}^{ 0}
x_{N}^{0})}\nonumber\\
&&\times \int d^{4}y d^{4}z \; \exp{(ik'^{0}y^{0}-ik^{0}z^{0})}\;
A_f^{\nu *}(\bfy)
\nonumber\\ 
&&\times\la 0|T\psi(x_{1}^{\prime})\cdots \psi(x_{N}^{\prime})
[j_{\nu}(y) j_{\rho}(z)-\la 0|Tj_{\nu}(y)j_{\rho}(z)|0\ra]
\nonumber\\
&&\times
\overline {\psi}(x_{N})\cdots \overline {\psi} (x_{1})|0\ra 
A_i^{\rho}(\bfz)
\, .
\end{eqnarray}
The Green function $G^{\rm con}_{\gamma_{f},\gamma_{i}}$
 is constructed using the Wick
theorem after the transition in (\ref{greenph7}) 
to the interaction representation. The Feynman rules for
$G^{\rm con}_{\gamma_{f},\gamma_{i}}$ differ from those for
$G_{\gamma_{f}}$ only by the presence of the incoming photon
line which corresponds to the incident photon wave function
$A_i^{\rho}(\bfx)$.

\subsection {Resonance scattering: Spectral line shape}

The formulas  for the scattering amplitudes derived above
allow one to perform calculations by perturbation
theory in the case where the total initial energy of the system
is not close to the energy of an intermediate quasistationary
state. This is the so-called non-resonant scattering. In the case
of resonance scattering, when the initial energy of the system 
is close to an intermediate-state energy, the direct calculation
by perturbation theory according to the formulas derived above
leads to some singularities in the scattering amplitude.
It means that those formulas cannot be directly applied to
the resonance processes. In this section we formulate a method
which allows one to calculate the 
resonance-scattering amplitudes. This method was worked out in Ref.
 \cite{shab91}. Another approach, which is 
limited to the one-electron atom case,  was previously developed 
in Ref. \cite{low52}. An attempt to describe a decay process
within QED was undertaken in Ref. \cite{braun88}.

The photon scattering by an atom that is initially
in its ground state $a$ is now considered for
 the case of resonance 
$E_a+k_i^0\sim E_{d}(d=1,...,s)$, where $E_a$ is the
ground state energy of the atom, $k_i^0$ is the incident photon energy, 
and $E_{d}(d=1,...,s)$ are  the energies of intermediate 
atomic states which in zeroth order approximation
are equal to the unperturbed energy $E_d^{(0)}$ of a degenerate
level. We consider that, as a result of the scattering, a photon
 of energy $k_f^0=k_i^0$
is emitted and the atom returns to its ground state $a$.
 The calculation of the photon scattering amplitude by the
formula derived above leads to a singularity which is caused by 
the fact  that in any finite order of  perturbation theory one
of the energy denominators of an intermediate Green function
is close to zero. Therefore, in the calculation of the
Green function  $g^{\rm con}_{\gamma_f,a;\gamma_i,a}(E',E,k'^0)$
we have to  go beyond the finite-order approximation.
With this in mind, let us represent
 $g^{\rm con}_{\gamma_f,a;\gamma_i,a}(E',E,k'^0)$
as
\be \label{grres}
g^{\rm con}_{\gamma_f,a;\gamma_i,a}(E',E,k'^0)&=&
\frac{i}{2\pi} g_a(E') R_{\gamma_f}^{(-)}(E',k'^0,E+k^0)
\frac{i}{2\pi} g_d(E+k^0) R_{\gamma_i}^{(+)}(E+k^0,k^0,E)
\frac{i}{2\pi} g_a(E)\nonumber\\ 
&&+\Delta g^{\rm con}_{\gamma_f,a;\gamma_i,a}(E',E,k'^0)\,,
\ee
where $g_a$ are $g_d$ are the Green functions defined by equation
(\ref{gsmall2})
 with the projectors $P_a^{(0)}$ and $P_d^{(0)}$, respectively,
$k^0=k'^0+E'-E$, and
$\Delta g^{\rm con}_{\gamma_f,a;\gamma_i,a}(E',E,k'^0)$
is a part of the Green function 
$ g^{\rm con}_{\gamma_f,a;\gamma_i,a}(E',E,k'^0)$ which is
regular at $E+k^0\sim E_{d}(d=1,...,s)$. 
The operators $R_{\gamma_f}^{(-)}$ and $R_{\gamma_i}^{(+)}$
are constructed by perturbation theory from equation
(\ref{grres}) which must be considered as their definition.
Taking into account  equation (\ref{spect5}) und using
the formula (\ref{sphotfin}), we obtain
\be \label{sres1}
S^{\rm con}_{\gamma_f,a;\gamma_i,a}&=&Z_3^{-1} \delta(k_f^0-k_i^0)
\varphi_a^{\dag} R_{\gamma_f}^{(-)}(E_a,k_f^0,E_a+k_i^0)
\frac{i}{2\pi}g_d(E_a+k_i^0)R_{\gamma_i}^{(+)}(E_a+k_i^0,k_i^0,E_a)
\varphi_a\nonumber\\
&&+Z_3^{-1}\delta(k_f^0-k_i^0)
\oint_{\Gamma_a}dE' \;\oint_{\Gamma_a} dE \;
v_a^{\dag}
\Delta g^{\rm con}_{\gamma_f,a;\gamma_i,a}(E',E,k_f^0) v_a\,.
\ee
Consider now  how the intermediate Green function $g_{d}
(E_a+k_i^0)$ can be calculated.
Let us introduce a quasipotential $V_{d}(E)$ by
\begin{eqnarray} \label{quasipot}
g_{d}(E)=g_{d}^{(0)}(E)+g_{d}^{(0)}(E)V_{d}(E)g_{d}(E)\, ,
\end{eqnarray}
where $g_{d}^{(0)}=P_{d}^{(0)}/(E-E_{d}^{(0)})$. The quasipotential
$V_{d}(E)$ is constructed by perturbation theory
according to equation (\ref{quasipot}) which must be considered as its 
definition. This equation  yields
\be \label{quasipot1}
V_d(E)=[g_d^{(0)}(E)]^{-1}-[g_d(E)]^{-1}&=&[g_d^{(0)}(E)]^{-1}
-[g_d^{(0)}(E)+g_d^{(1)}(E)+\cdots ]^{-1}\nonumber\\
&=&[g_d^{(0)}(E)]^{-1}g_d^{(1)}(E)[g_d^{(0)}(E)]^{-1}+\cdots\,.
\ee
If the quasipotential $V_d(E)$ is constructed from equation
(\ref{quasipot1}) to a finite order of perturbation theory,
  the Green function ${g}_{d}$ is determined by
\begin{eqnarray} \label{quasipot2}
g_{d}(E)=[E-E_{d}^{(0)}-V_{d}(E)]^{-1}\, .
\end{eqnarray}
The Green function $g_d(E)$ has poles on the second sheet
of the Riemann surface, slightly below the right-hand real
semiaxis (see Fig. 5), and has no singularities for real $E$
when $E\sim E_d^{(0)}$.
It means, in particular, that if we take the quasipotential
at least to the lowest order approximation $(\;V(E)\approx V(E_{d}
^{(0)})\;)$, the Green function ${g}_{d}(E)$  calculated by
equation (\ref{quasipot2}) has no singularities 
at $E\sim E_d (d=1,...,s)$, and, therefore, the calculation of
the resonance-scattering amplitude by equation (\ref{sres1})
will be correct.
The calculation of ${g}_{d}(E)$ by equation (\ref{quasipot2})
 effectively  corresponds to  summing an infinite
subsequence of Feynman diagrams.

For the calculation of $g_d(E)$ to
the lowest order approximation it is convenient to introduce
an operator $\cal{H}$ by
\begin{eqnarray}
{\cal H}\equiv E_{d}^{(0)}+V_{d}(E_{d}^{(0)})\, .
\end{eqnarray}
The operator $\cal{H}$ is not Hermitian and  has complex
eigenvalues. We assume that $\cal{H}$ is a simple matrix, i.e.,
its eigenvectors form a complete basis in the space 
 of the unperturbed $d$-states.
We denote its eigenvalues by ${\cal E}_{d}=E_{d}-i\Gamma_{d}/2$,
the right eigenvectors by $|d_{R}\ra$, and
the left eigenvectors by $|d_{L}\ra$. It means
\begin{eqnarray}
{\cal H}|d_{R}\ra={\cal E}_{d}|d_{R}\ra \,,
\qquad\;\;\;\la d_{L}|{\cal H}=\la d_{L}|{\cal E}
_{d}\, .
\end{eqnarray}
It is convenient to normalize the vectors $|d_{R}\ra$, $|d_{L}\ra$ by
the condition 
\begin{eqnarray}
\la d_{L}|d_{R}^{\prime}\ra=\delta_{dd'}\, .
\end{eqnarray}
They satisfy the completeness condition
\begin{eqnarray}
\sum_{d=1}^{s}|d_{R}\ra\la d_{L}|=I\, .
\end{eqnarray}
For $g_{d}(E)$ we obtain
\begin{eqnarray}  \label{gresap}
g_{d}(E)\approx (E-{\cal H})^{-1}=
\sum_{d=1}^{s}\frac{|d_{R}\ra\la d_{L}|}{E-{\cal E}_{d}}\, .
\end{eqnarray}
In fact, due to $T$-invariance, ${\cal H}_{ik}={\cal H}_{ki}$. 
For this reason the components of the vector $\la d_{L}|$ can be chosen to
be equal to the corresponding components of the vector $|d_{R}\ra$.
In other words, the components of the vector $|d_{R}\ra$ are equal to
the complex conjugated components of the vector $|d_{L}\ra$.
If  the $d$ states  have different quantum numbers, such as
 the total angular momentum or
the parity, one finds
$|d_{R}\ra=|d_{L}\ra\equiv |d\ra$.

Substituting the lowest order approximation for
 $g_d(E)$  given by equation (\ref{gresap}) into  
(\ref{sres1}), we obtain in the resonance approximation
\be \label{sres3}
S^{\rm con}_{\gamma_f,a;\gamma_i,a}\approx \frac{i}{2\pi}
 \delta(k_f^0-k_i^0)\sum_{d=1}^{s}
\frac{\la a|R_{\gamma_f}^{(-)}|d_R\ra \la d_L|R_{\gamma_i}^{(+)}|a\ra}
{E_a+k_i^0-E_d+i\Gamma_d/2} \,.
\ee
In the resonance approximation, it is sufficient to evaluate
 the operators $R_{\gamma_f}^{(-)}$,
$R_{\gamma_i}^{(+)}$ to the lowest order of 
perturbation theory. They are determined directly from
equation (\ref{grres}).

To demonstrate how the method works we consider below 
 the resonance photon scattering on a one-electron atom.
A more general case of a few-electron atom is considered in 
\cite{shab91}.

\subsection{Resonance photon scattering by a one-electron atom}

To study the resonance photon scattering by a one-electron atom, in
the lowest order we must consider the diagram
shown in Fig. 36. The contribution of this diagram
 to the Green function $G_{\gamma_f,\gamma_i}$ is
\be
G_{\gamma_f,\gamma_i}^{\rm con}((E',\bfx');k'^0,k^0;(E,\bfx))
&=& \int d\bfy d\bfz \;
\int_{-\infty}^{\infty} dp^0 \;
\frac{i}{2\pi} S(E',\bfx',\bfy)
\frac{2\pi}{i} e\gamma_{\nu} \delta(E'+k'^0-p^0)\nonumber\\
&&\times A_f^{\nu *}(\bfy) \frac{i}{2\pi} S(p^0,\bfy,\bfz)
\frac{2\pi}{i} e\gamma_{\rho} \delta(p^0-k^0-E)\nonumber\\
&&\times A_i^{\rho}(\bfz)\frac{i}{2\pi} S(E,\bfz,\bfx)\,.
\ee
For  $g^{\rm con}_{\gamma_f,a;\gamma_i,a}$ we obtain
\be
g^{\rm con}_{\gamma_f,a;\gamma_i,a}(E',E,k'^0)
&=&\frac{i}{2\pi}\frac{|a\ra \la a|}
{E'-\veps_a}\frac{2\pi}{i} e\alpha_{\nu} A_f^{\nu *}\nonumber\\
&&\times\frac{i}{2\pi} \sum_n
\frac{|n\ra \la n|}{E+k^0-\veps_n(1-i0)}
\frac{2\pi}{i} e\alpha_{\rho}A_i^{\rho}\frac{i}{2\pi}
\frac{|a\ra \la a|}{E-\veps_a}\,.
\ee
We consider that $E_a+k_i^0\sim E_d$ and, therefore, represent
$g^{\rm con}_{\gamma_f,a;\gamma_i,a}(E',E,k'^0)$ as the sum of
two terms
\be
g^{\rm con}_{\gamma_f,a;\gamma_i,a}(E',E,k'^0)&=&
\frac{i}{2\pi}g_a^{(0)}(E')\frac{2\pi}{i} e \alpha_{\nu}
A_f^{\nu *}\frac{i}{2\pi}
g_d^{(0)}(E+k^0)
\nonumber\\
&&\times \frac{2\pi}{i} e\alpha _{\rho} A_i^{\rho}
\frac{i}{2\pi} g_a^{(0)}(E)
+
\frac{i}{2\pi}g_a^{(0)}(E')\frac{2\pi}{i} e \alpha_{\nu}
A_f^{\nu *}\nonumber\\
&&\times\sum_n^{\veps_n\not =\veps_d}\frac{|n\ra \la n|}
{E+k^0-\veps_n(1-i0)}
 e\alpha _{\rho} A_i^{\rho} g_a^{(0)}(E)\,.
\ee
Comparing this equation with equation (\ref{grres}),
we derive
\be \label{rrres}
R_{\gamma_f}^{(-)}(\bfy)=\frac{2\pi}{i}e\alpha_{\nu}A_f^{\nu *}(\bfy)\,,
\;\;\;\;\;\;\;\;\;\;\;\;\;
R_{\gamma_i}^{(+)}(\bfz)=\frac{2\pi}{i}e\alpha_{\rho}A_i^{\rho}(\bfz)\,.
\ee
Let us derive now the quasipotential $V_d(E)$. To the lowest order
of perturbation theory it is defined by the SE and VP diagrams
(see Figs. 12,13).  As was derived above (see section II(E)),
 the contribution of these diagrams to $g_d(E)$ is
\be
g_d^{(1)}(E)=g_d^{(0)}(\Sigma_{\rm SE}(E)+U_{\rm VP})g_d^{(0)}.
\ee
Therefore,
\be
V_d^{(1)}(E)=[g_d^{(0)}(E)]^{-1}g_d^{(1)}(E)[g_d^{(0)}(E)]^{-1}
=P_d^{(0)}(\Sigma_{\rm SE}(E)+U_{\rm VP})P_d^{(0)}\,.
\ee
The operator $P_d^{(0)}\Sigma_{\rm SE}(E)P_d^{(0)}$
contains a non-Hermitian part which is responsible for
the imaginary part of the energy. The operator
\be
{\cal H}=\veps_d+V_d^{(1)}(\veps_a+k_i^0)
\ee
acts in the $s$-dimensional space of the unperturbed states.
In reality, due to the fact that the operators 
$\Sigma_{\rm SE}$ and $U_{\rm VP}$ do not mix states with
different quantum numbers and among the degenerate one-electron 
states there are no states with the same quantum numbers,
in the case under consideration the right eigenvectors of
${\cal H}$ coincide with the left eigenvectors,
$|d_R\ra=|d_L\ra\equiv |d\ra$. However, to keep a general form
of the equations, below we will use  the right and left eigenvectors.
In the resonance approximation, the amplitude of the process
is defined by equation (\ref{sres3}), where the operators
$R_{\gamma_f}^{(-)}$ and $R_{\gamma_i}^{(+)}$ are given
by equations (\ref{rrres}), $E_a=\veps_a+\la a|\Sigma_{\rm SE}(\veps_a)
+U_{\rm VP}|a\ra$ is the ground state energy including
the QED corrections of first order in $\alpha$, $E_d$ and 
$-\Gamma_d/2$ are the real and imaginary parts of an
eigenvalue of ${\cal H}$.

For the differential cross section in the resonance approximation,
we obtain
\be \label{sigdif}
d\sigma&=&(2\pi)^4\delta(k_f^0-k_i^0)\Bigl[
\sum_{d=1}^{s}
\frac{|\la a|e\alpha_{\nu}A_f^{\nu *}|d_R\ra
 \la d_L|e\alpha_{\rho}A_i^{\rho}|a\ra|^2}
{(E_a+k_i^0-E_d)^2+\Gamma_d^2/4}\nonumber\\
&&+2{\rm Re}\sum_{d<d'}
\frac{\la a|e\alpha_{\nu}A_f^{\nu *}|d_R\ra
 \la d_L|e\alpha_{\rho}A_i^{\rho}|a\ra|^2}
{E_a+k_i^0-E_d+i\Gamma_d/2}\nonumber\\
&&\times \frac{\la a|e\alpha_{\nu}A_f^{\nu *}|d'_R\ra^*
 \la d'_L|e\alpha_{\rho}A_i^{\rho}|a\ra^*}
{E_a+k_i^0-E_{d'}-i\Gamma_{d'}/2}\Bigr] d\bfk_f\,.
\ee
For the total cross section, using the optical theorem,
we find
\be \label{sigtot}
\sigma_{\rm tot}&=&2(2\pi)^3
\sum_{d=1}^{s}\Bigl[
\frac{{\rm Re}(\la a|e\alpha_{\nu}A_i^{\nu *}|d_R\ra
 \la d_L|e\alpha_{\rho}A_i^{\rho}|a\ra)(\Gamma_d/2)}
{(E_a+k_i^0-E_d)^2+\Gamma_d^2/4}\nonumber\\
&&+\frac{{\rm Im}(\la a|e\alpha_{\nu}A_i^{\nu *}|d_R\ra
 \la d_L|e\alpha_{\rho}A_i^{\rho}|a\ra)(E_d-E_a-k_i^0)}
{(E_a+k_i^0-E_d)^2+\Gamma_d^2/4}\Bigr]\,.
\ee
Let us discuss, for simplicity, the case $s=2$.
Only if the $d$
levels  have the same quantum numbers,
  the second term on the right-hand side of equation
(\ref{sigtot}) is not equal to zero. 
In the opposite case,
which takes place in the process under consideration,
$|d_R\ra=|d_L\ra$ and, therefore,
\be
{\rm Im}(\la a|e\alpha_{\nu}A_i^{\nu *}|d\ra
 \la d|e\alpha_{\rho}A_i^{\rho}|a\ra)=0\,.
\ee
It follows that the the total cross section given by
equation (\ref{sigtot}) is the sum of Lorentz-type terms.
As to the differential cross section, the second
term in equation (\ref{sigdif}) is not equal to zero
even if the states $d=1,2$ have different quantum numbers.

Levels close to each other with identical quantum numbers
can appear among doubly excited states of  high-$Z$
few-electron atoms \cite{gorshkov}. 
As an example, we can consider
 the $(2s,2s)_0$ and $(2p_{1/2},2p_{1/2})_0$
states of a heliumlike ion which can arise in the process
of recombination of an electron with a hydrogenlike ion.
A detailed theory of this process was given in Ref.
\cite{sh94b}. The related numerical calculations were presented 
in Refs. \cite{kar92,nef94}. The results obtained
in these papers are discussed in section IV(D3).


\section{Numerical evaluations of  QED and 
interelectronic-interaction corrections in heavy ions}

\subsection{ Methods of numerical evaluations
and renormalization procedure}

In the previous sections we have demonstrated
how formulas for the energy shifts and
the transition and scattering amplitudes can be derived
from the first principles of QED.
These formulas usually contain infinite summations over
intermediate electron states (summations over the bound states
and integrations over the continuum). 
These sums are generally evaluated by using analytical expressions
for the Dirac-Coulomb Green function 
\cite{mohr98a,zapryagaev85,wich56,man73,mohr74,gyul75,soff88}
or by using  relativistic finite basis set methods 
\cite{drake81,grant82,johnson88,salomon89,sapir96,grant00}.
In some cases the summation can be performed analytically
by employing the generalized virial relations for
the Dirac equation \cite{shab91vir}.

Calculations of most QED corrections require
 the application of a renormalization procedure.
To first order in $\alpha$, one has to renormalize
the self-energy and vacuum-polarization diagrams
(Figs. 12, 13). 

The renormalized expression for
the SE correction is given by equation (\ref{eq102})
which implies using the same covariant regularization
for both terms on the right-hand side.
For the numerical evaluation of this correction,
 it is convenient
to analytically isolate the ultraviolet divergence 
in the $\la a|\Sigma(\veps_a)|a\ra$ term and cancel
it by the counterterm. This can be performed by expanding
the Dirac-Coulomb Green function in terms of the free
Dirac Green function,
\be \label{expgr}
[\omega - H(1-i0)]^{-1}&=&[\omega - H_0(1-i0)]^{-1}+
[\omega - H_0(1-i0)]^{-1}V_{\rm C}
[\omega - H_0(1-i0)]^{-1}\nonumber\\
&&+[\omega - H_0(1-i0)]^{-1}V_{\rm C}
[\omega - H(1-i0)]^{-1}V_{\rm C}
[\omega - H_0(1-i0)]^{-1}\,,
\ee
where $H_0=\balpha\cdot \bfp+\beta m$ is the free Dirac
Hamiltonian. The three terms in equation (\ref{expgr}) 
inserted into  $\la a|\Sigma(\veps_a)|a\ra$ divide 
the SE correction into zero-potential, one-potential,
and many-potential terms, respectively.   The ultraviolet
divergences in the zero- and one-potential terms and in the
counterterm can be cancelled analytically (see Refs.
\cite{baranger53,snyder91,yerokh99} for details). As to the many-potential
term, it can easily be shown
 that it does not contain any ultraviolet
divergences. For an overview of other methods
of carrying out mass renormalization in numerical calculations, 
we refer to \cite{mohr98a}.

The vacuum-polarization correction is determined by equation
(\ref{vp1}) with the VP potential defined by (\ref{vppot}). 
The expression (\ref{vppot}) is
ultraviolet divergent. The simplest way to renormalize
this expression is to expand the vacuum loop in powers of 
the Coulomb potential by employing equation (\ref{expgr}).
According to the Furry theorem, contributions of 
diagrams with odd numbers of vertices in a vacuum loop
(with free Dirac propagators) are equal to zero.
Therefore, the first non-zero contribution results from 
second term in the expansion (\ref{expgr}). Only
this contribution, which is called the Uehling term,
is ultraviolet divergent.  This term becomes finite by
charge renormalization. The renormalized expression for the Uehling
potential is given by
 \begin {eqnarray} \label{uehlexpr}
U_{\rm Uehl}(r)&=&-\alpha Z
\frac{2\alpha}{3\pi}\int\limits_0^\infty dr'\; 4\pi r'\rho
(r')
\int\limits_1^\infty dt \;
(1 +\frac{1}{2t^2})
\frac{\sqrt{t^2-1}}{t^{2}}\nonumber \\
&&\times \frac{[\exp{(-2m|r-r'|t)}-\exp{(-2m(r+r')t)}]}
{4mrt} \,,
\end{eqnarray}
where $|e|Z\rho(r)$ is the density of the nuclear charge distribution
($\int \rho(r) d{\bf r}=1$).
The higher order (in $V_{\rm C}$) terms  are
finite and their sum is called the Wichmann-Kroll
correction \cite{wich56}.
 However, the regularization is still required in the second
non-zero term due to a spurious gauge dependent piece of
the light-by-light scattering contribution.
As was shown in \cite{gyul75,soff88,rink75}, 
in the calculation of
the vacuum polarization charge density based on
the partial wave
 expansion of the Dirac-Coulomb Green function,
the spurious term does not contribute if
the sum over the angular momentum quantum number $\kappa$
 is restricted to a finite number of terms ($|\kappa|\leq K$).
  Thus the Wichmann-Kroll
contribution can be calculated by summing up the partial 
differences between the full contribution and the Uehling term.

The renormalization procedures described above 
can be adopted for calculations of the 
self-energy and vacuum-polarization screening diagrams
\cite{yer97,art97,art99,yer99,pers96a,sun98a}
as well as for the QED corrections to the hyperfine
splitting and the bound-electron $g$ factor
\cite{beier00,pers96b,blund97,shab97hf,pers97,shab98hf,sun98b,bei00}.

\subsection {Energy levels in heavy ions}

\subsubsection{Hydrogenlike ions}

The relativistic  energies  of a hydrogenlike ion 
are determined
by the Dirac equation (\ref{dirac}).
For the point-nucleus case, the Dirac equation can 
be solved analytically 
and  the binding energy is given by
\begin{eqnarray}
E_{nj}-mc^2=-\frac{(\alpha Z)^2}{2\nu^2}\frac
{2}{1+(\alpha Z/\nu)^2 +\sqrt{1+(\alpha Z/\nu)^2}}\,mc^2\,,
\end{eqnarray}
where $\nu=n+\sqrt{(j+1/2)^2-(\alpha Z)^2}-(j+1/2)$,
 $n$ is the principal quantum number, and $j$ is 
the total angular momentum.
To obtain the binding energy to higher accuracy, 
 QED and nuclear effects must be taken into account.

The finite nuclear size correction is calculated
by  solving the Dirac equation with the potential
of an extended nucleus and by taking the difference 
between the energies for the 
extended and point nucleus models. This can be
performed numerically (see, e.g.,
\cite{franosch91,and00}) or, with a good accuracy, analytically
\cite{shab93fs}. With a relative accuracy of
$\sim 0.2\%$ for $Z=1-100$,
  this correction is given by the following
approximate formulas \cite{shab93fs}
\begin{eqnarray} \label{fns}
\Delta E_{ns}&=&\frac{(\alpha Z)^2}{10 n}[1+(\alpha Z)^2 f_{ns}(\alpha Z)]
\Bigl (2\frac{\alpha Z}{n}\frac{R}{(\hbar/mc)}\Bigr )^{2\gamma}mc^2\,,\\
\Delta E_{np_{1/2}}&=&\frac{(\alpha Z)^4}{40 }\frac{n^2-1}{n^3}
[1+(\alpha Z)^2 f_{np_{1/2}}(\alpha Z)]
\Bigl (2\frac{\alpha Z}{n}\frac{R}{(\hbar/mc)}\Bigr )^{2\gamma}mc^2\,,
\end{eqnarray}
where $\gamma=\sqrt{1-(\alpha Z)^2}$,
\be
f_{1s}(\alpha Z)&=&1.380-0.162\alpha Z+1.612(\alpha Z)^2\,,\nonumber\\
f_{2s}(\alpha Z)&=&1.508+0.215\alpha Z+1.332(\alpha Z)^2\,,\nonumber\\
f_{2p_{1/2}}(\alpha Z)&=&1.615+4.319\alpha Z-9.152(\alpha Z)^2
+11.87(\alpha Z)^3\,,\nonumber
\ee
and $R$ is an effective radius of the nuclear charge distribution
 defined by
\begin{eqnarray}
R=\Bigl\{\frac{5}{3}\langle r^2\rangle \Bigl [1-\frac{3}{4}
(\alpha Z)^2\Bigl (\frac{3}{25}\frac{\langle r^4\rangle}
{\langle r^2\rangle ^2}-\frac{1}{7}\Bigl)\Bigl]\Bigl\}^{1/2}\,.
\end{eqnarray}
For the Fermi model of the nuclear charge distribution,
\begin{eqnarray}
\rho (r)=\frac{N}{1+\exp{[(r-c)/a]}}\,,
\end{eqnarray}
it is possible to obtain with a very high precision
\begin{eqnarray}
N&=&\frac{3}{4\pi c^3}\Bigl (1+\frac{\pi ^2 a^2}{c^2}\Bigr)^{-1}\,,\\
\langle r^2 \rangle&=&\frac{3}{5}c^2+\frac{7}{5}\pi^2a^2\,,\\
\langle r^4 \rangle&=&\frac{3}{7}c^4
+\frac{18}{7}\pi^2a^2c^2+\frac{31}{7}\pi^4a^4
\,.
\end{eqnarray}
The expectation values for powers of $r$ for a great variety
of nuclear charge distribution models are given
in Ref. \cite{and00}.

Next the QED corrections of first order
in $\alpha$ should be taken into account.
 To this order, the QED correction is defined
by the self-energy and vacuum-polarization diagrams 
(Figs. 12,13). 
The energy shift from the self-energy diagram (Fig. 12)
 combined with the
related mass counterterm (Fig. 14) is
determined by equation (\ref{eq102}).
The self-energy correction for
heavy ions was first evaluated by Desederio and Johnson 
\cite{desid71} who employed a method suggested by Brown, Langer,
and Schaefer \cite{brown59}. Later, a more efficient method
was developed by Mohr \cite{mohr74} who calculated
this correction to a very high accuracy in a wide interval
of $Z$. The method of the potential expansion of the 
bound-electron propagator for the calculation of the SE correction
to all orders in $\alpha Z$ was developed by Snyderman
\cite{snyder91} and numerically realized first by Blundell and 
Snyderman \cite{blund91}.  A very efficient procedure
for the self-energy calculations which is closely
related to the methods of Snyderman and Mohr 
 was developed
by Yerokhin and Shabaev \cite{yerokh99}.
An approach in which the ultraviolet divergences
are removed by subtractions in coordinate space was
worked out by Indelicato and Mohr \cite{ind92}.
 The method of the partial-wave
renormalization for the calculation of the first-order
SE correction was developed
 by Persson, Lindgren, and
Salomonson \cite{pers93} and by Quiney and Grant \cite{quiney}.
 To date, the most accurate calculations of the
 SE correction  to all orders in $\alpha Z$
were performed by Mohr \cite{mohr74,mohr92} and by Indelicato and Mohr
\cite{indelicato98} for the point nucleus case and by
 Mohr and Soff \cite{mohr93} for the extended nucleus case.
This correction was comprehensively tabulated
by Beier and co-workers \cite{beier98}
for finite nuclear radii.
The highest accuracy for low-$Z$ atoms was gained
by Jentschura et al. \cite{jent99,jent00}.

The VP correction (Fig. 13) is the sum of the Uehling
and  Wichmann-Kroll contributions. The first
contribution can easily be calculated using
the expression (\ref{uehlexpr}) for the Uehling
potential.
Calculations of the Wichmann-Kroll contribution
to all orders
in $\alpha Z$ were performed first by Soff and Mohr
\cite{soff88}
for the extended nucleus case and by Manakov et al.
\cite{manakov89} for the point nucleus case.
A comprehensive tabulation of this correction
for extended nuclei was presented in Ref. \cite{beier97a}.
The most accurate calculations for some specific ions
were accomplished by Persson et al. \cite{persson93}.

The QED corrections of second order in $\alpha$ 
have not yet been calculated completely.
Most VP-VP and SE-VP diagrams
can be evaluated by the methods developed for the
first-order SE and VP corrections (see 
\cite{mohr98a,beier97b} and references therein).
The most difficult task consists in the evaluation
of the SE-SE contribution. The simplest part of
this contribution, the loop-after-loop diagram,
was calculated 
by Mitrushenkov and co-workers \cite{mitrushenkov95} and
by Mallampalli and Sapirstein \cite{mallampalli98a}
for high-$Z$ ions.
These numerical calculations were extended to low-$Z$ atoms
by Mallampalli and Sapirstein \cite{mallampalli98b}
and by Yerokhin \cite{yerokh00a}.
Recently they were confirmed by analytical
calculations of Yerokhin \cite{yerokh00b}.
   As to the  residual SE-SE contribution,
a specific part of it was evaluated by Mallampalli and
Sapirstein \cite{mallampalli98a} and an estimate
of the complete SE-SE contribution is in progress \cite{goi00}.
For the current status of the corresponding calculations
for low-$Z$ atoms, we refer to \cite{pach98,pach00}.

The calculations of the corrections discussed above
are based on quantum electrodynamics within the
external field approximation. It means that 
in these calculations the nucleus is considered
only as a source of the external Coulomb field
$V_{\rm C}$.
The first step beyond this  approximation consists
in evaluating the nuclear recoil correction.
This correction is given by the sum
of the lower-order term (\ref{recmy1}) and the higher-order
term (\ref{recmy2}).
For the point nucleus case, an analytical calculation of the
lower-order term 
employing the virial relations for the Dirac equation
 yields  \cite{sh85}
\begin{eqnarray}
\Delta E_{\rm L}=
\frac{m^{2}-\veps_{a}^{2}}{2M}\,.
\end{eqnarray}
The higher-order term  was numerically evaluated
to all orders in $\alpha Z$ for point nuclei in Refs. 
\cite{artemyev95a,artemyev95b}.
The corresponding calculations for extended nuclei were carried out
in Refs. \cite{sh98recpra,sh99recpst}.
In the case of hydrogen, the highest accuracy was gained
in Ref. \cite{sh98recjpb}.

Finally, the nuclear polarization correction should be taken
into account. This
correction results from diagrams describing the
interaction of the electron 
with the nucleus where
the intermediate states of the nucleus are excited. 
It was evaluated by Plunien and Soff 
\cite{plunien95} and by Nefiodov et al. \cite{nefiodov96}.

In Table 1 we present
the individual contributions to the ground-state Lamb shift
in $^{238}{\rm U}^{91+}$. 
The uncertainty of the Dirac binding energy results
from the uncertainty of the Rydberg constant \cite{mohr00a}.
As can be deduced from the table, the present status of 
the experimental precision 
on the ground-state Lamb shift in hydrogenlike
uranium  \cite{beyer95a,beyer95b,stoehlker00}
provides a test
of QED in first order in $\alpha$ on the level of about 5\%.

\subsubsection{Heliumlike ions}

In heavy heliumlike ions, in addition to the one-electron
contributions considered in the previous subsection, 
the two-electron corrections have to be taken into account.
To lowest order in $\alpha$, this correction is
defined by the one-photon exchange diagram (Fig. 20).
The calculation of this diagram causes no problem.
To second order in $\alpha$, we should account for
the two-photon exchange diagrams (Fig. 21),
the self-energy screening  diagrams (Fig. 22),
 and the vacuum-polarization screening diagrams (Fig. 23).
For the ground state of a heliumlike ion,
the contribution of the two-photon exchange diagrams
is defined by equations (\ref{e3e24}), (\ref{eq139}),
and (\ref{eq140}). The corresponding expressions for the case
of a single excited state  were obtained  in Ref. \cite{sh94a}.
The derivation of the related formulas 
for the case of degenerate and quasidegenerate states
by the TTGF method also causes no difficulties.
The self-energy and vacuum-polarization screening contributions
 are given by expressions (\ref{sescr})-(\ref{vpscr3})
(the renormalization of these expressions is considered in detail 
in Refs. \cite{yer97,art97,art99,yer99}).
For the case of quasidegenerate states,
the corresponding formulas are derived by the TTGF method
in Refs. \cite{art00,leb}. 

The two-photon exchange contribution is conveniently
divided into two parts. The first part is the 
one which can be derived
from the Breit equation. For the ground state, it is well
determined by the lowest-order terms of the $\alpha Z$-expansion
 series \cite{drake88},
\be 
\Delta E^{(\rm Breit)}=\alpha^2[-0.15766638-0.6356(\alpha Z)^2]m\,.
\ee
The second part is the remaining one. Evaluated  in Refs. 
\cite{blund93,lind95}, it was found to be much smaller
than the first part.
 The self-energy
and vacuum-polarization screening diagrams were
evaluated in
Refs. \cite{yer97,art97,pers96a,sun98a}.
The related calculations for excited states of heliumlike
ions were performed for the vacuum-polarization 
screening diagrams
\cite{art00} and, in the case of non-mixed states,
for the two-photon exchange diagrams \cite{mohr00b}.

Today, the theoretical uncertainty of the ground-state energy
in heavy heliumlike ions is completely defined by the uncertainty
of the one-electron contribution. In this connection,
a direct measurement of the 
two-electron contribution to the ground-state energy
in heliumlike ions performed in \cite{marrs95}
turns out to be rather important.
In Table 2 we present the individual two-electron
 contributions to the ground state energy in heliumlike 
bismuth. The two-photon exchange contribution 
 is divided into two parts as described above.
The three- and more photon contribution is evaluated within
the Breit approximation by summing the $Z^{-1}$ expansion terms 
for the ground state energy beginning from $Z^{-3}$ 
\cite{yer97}. For the zeroth order in $\alpha Z$,
the coefficients of this expansion are taken
 from Ref. \cite{sanders69}
and for the second order in $\alpha Z$ from Ref. \cite{drake88}.
The uncertainty of this contribution results from
 QED corrections yet uncalculated.
As one can see from the table, to test the  second-order
QED effects that result from the theory beyond the
Breit approximation,
 the experimental precision has to be improved
 by an order of magnitude.

\subsubsection{Lithiumlike ions}

To date, the highest experimental accuracy was reached in 
Lamb-shift experiments of lithiumlike ions 
\cite{schweppe91,beiersdorfer1,staude98,bosselmann99}.
In these ions, in addition to the one-
and two-electron contributions,
the three-electron corrections have to be calculated.
To second order in $\alpha$, the three-electron
contribution is determined by six diagrams that describe
two-photon exchange involving all three electrons.
One of these diagrams is shown in Fig. 37.
In the case of one electron over the closed $(1s)^2$
shell, the unperturbed wave function is a one-determinant
function,
\be
u(\bfx_1,\bfx_2,\bfx_3)=\frac{1}{\sqrt{3!}}
\sum_{P}(-1)^P\psi_{Pa}(\bfx_1)
\psi_{Pb}(\bfx_2)\psi_{Pv}(\bfx_3)\,,
\ee
where
$a$ and $b$ denote the core states with opposite signs 
of the angular momentum projection,  $v$ indicates the
valence state. It can be shown that the derivation of the
expressions for the one- and two-electron corrections 
in three-electron atoms is easily reduced to the 
derivation in one- and two-electron atoms, respectively.
As to the three-electron correction of second order
in $\alpha$, it can be derived by the TTGF method
using the identity
(\ref{iden3}) and the general rules formulated in section
II(E5). Such a derivation for the irreducible part yields
\cite{y0a}
\be
\Delta E^{\rm irred}=\sum_{PQ}(-1)^{P+Q}\sum_{n}{}^{'}\;
\frac{I_{PbPvnQv}(\veps_{Qv}-\veps_{Pv})-
I_{PanQaQb}(\veps_{Pa}-\veps_{Qa})}
{\veps_{Qa}+\veps_{Qb}-\veps_{Pa}-\veps_n}\,,
\ee
where $P$ and $Q$ denote the permutations
over the outgoing and incoming electron states, respectively;
$I_{abcd}(\omega)\equiv \la ab|I(\omega)|cd\ra$.
The prime at the sum indicates that terms with vanishing
denominator have to be omitted in the summation.
The reducible part of the three-electron contribution
is \cite{y0a}
\be \label{3red}
\Delta E^{\rm red}&=&\sum_{\mu_a}
[I'_{vaav}(\Delta)(I_{ab;ab}-I_{bv;bv})\nonumber\\
&&+\frac{1}{2}I'_{bv\overline{v}a}(\Delta)
I_{a\overline{v};bv}
\frac{1}{2}I'_{a\overline{v}vb}(\Delta)
I_{vb;\overline{v}a}]\,,
\ee
where $I_{ab;cd}=I_{abcd}(\veps_b-\veps_d)-
I_{bacd}(\veps_a-\veps_d)$, $\Delta=\veps_v-\veps_a$,
$\mu_a$ is the angular momentum projection of the
$a$ electron, $\mu_b=-\mu_a$, $\overline{v}$ 
is the valence electron with the opposite sign of
 the angular momentum projection. In derivation
of equation (\ref{3red}) by the TTGF  method, some
terms containing the $\overline{v}$ electron have been
cancelled with the corresponding terms from the reducible
two-electron contribution of the two-photon exchange
diagrams (see Ref. \cite{y0a} for details).
 The accurate QED calculations of all the two- and 
three-electron  corrections to the $2p_{1/2}-2s$ transition
energy up to second order in $\alpha$
 were performed in Refs.
\cite{art99,yer99,y0a}.
Approximate evaluations of these corrections were 
 previously considered in Refs. 
\cite{ind91,cheng91,mohr93t,blund93a,lind93,chen95}.

In Table 3 the individual contributions to
the $2p_{1/2}-2s$ transition energy in lithiumlike uranium
are presented.
The total theoretical value of the transition energy,
280.46(9) $\pm$ 0.20 eV, is in  agreement with 
the related experimental value, 280.59(10) eV
 \cite{schweppe91}.
As can be seen from the table, the first-order QED contribution
is -42.93 eV while
 the total second-order QED contribution
beyond the Breit approximation amounts to 1.33 $\pm$ 0.20 eV.
Comparing these values with the total theoretical and experimental
uncertainties indicates that
the present status of the  theory for lithiumlike uranium
 provides a test of the QED effects of first order in $\alpha$
on the level of about 0.5\% and of the  QED effects
of  second order in $\alpha$, which result from the
theory beyond the Breit approximation, 
on the level of about 15\%.

\subsection{Hyperfine splitting and bound-electron $g$ factor}

\subsubsection{Hyperfine splitting in hydrogenlike ions}

The ground-state hyperfine splitting  of a hydrogenlike ion
is conveniently written as \cite{shab94hf}
\begin{eqnarray}
\Delta E_{\mu}&=&\frac{4}{3}\alpha(\alpha Z)^{3}\frac{\mu}{\mu_{N}}
\frac{m}{m_{p}}\frac{2I+1}{2I}
mc^{2}\nonumber\\
& &\times\{A(\alpha Z)(1-\delta)(1-\varepsilon)+x_{\rm rad}\}\;.
\end{eqnarray}
Here $m_{p}$ is the proton mass,
$\mu$ is the nuclear magnetic moment, $\mu_{N}$ is the nuclear
magneton, and $I$ is the nuclear spin.
 $A(\alpha Z)$ denotes
the relativistic factor
\begin{eqnarray}
A(\alpha Z)=\frac{1}{\gamma(2\gamma-1)}=1+\frac{3}{2}(\alpha Z)^{2}
+\frac{17}{8}(\alpha Z)^{4}+\cdots\,\,,
\end{eqnarray}
where $\gamma=\sqrt{1-(\alpha Z)^{2}}$.
 $\delta$ is the nuclear charge
 distribution
correction,
 $\varepsilon$ is the
nuclear  magnetization distribution correction (the Bohr-Weisskopf
correction),
and $x_{\rm rad}$ is the QED correction.
The formulas for the first-order QED corrections to the
hyperfine splitting are derived in the same way as formulas
(\ref{abxx1}), (\ref{abxx2}) for the $\delta V$-SE corrections.
For instance, the SE correction is simply determined by
equations (\ref{abxx1}), (\ref{abxx2}), if $\delta V$
is replaced by the hyperfine interaction operator
\be
H_{\rm hfs}=\frac{|e|}{4\pi}\frac{(\balpha \cdot [\bmu\times{\bf r}])}
{r^3}
\ee
and the electron wave functions are replaced by the wave 
functions of the whole (electron plus nucleus) atomic system.
The most complete calculations of the QED and nuclear corrections
to the hyperfine splitting
were presented in \cite{shab97hf,sun98b} (see also
 a recent review  in \cite{shab99hf}).
Table 4 shows the individual contributions to the hyperfine
splitting in hydrogenlike ions. The total theoretical values are
compared with the experimental results obtained in 
\cite{klaft1,crespo1,crespo2,seelig1}.
 The uncertainty of the
theoretical predictions is mainly determined by
the uncertainty of the Bohr-Weisskopf correction which
was evaluated within the single-particle nuclear model
\cite{shab97hf}.
 This uncertainty should be considered only as 
an estimate of the order of magnitude of the real error.
  Except for Ho, the nuclear 
magnetic moments are taken from Ref. \cite{raghavan1}.
 For Ho the value recommended in \cite{gustavsson1} is used.
In case of $^{207}{\rm Pb}$, 
in Ref. \cite{raghavan1} two values of
 the nuclear magnetic moment are given. One
 ($\mu= 0.592583(9)\mu_{N}$)
 was measured by the nuclear magnetic resonance (NMR) method 
 \cite{lutz1} while another 
($\mu= 0.58219(2)\mu_{N}$) results from an optical pumping (OP) 
experiment \cite{gibbs1}. As was shown in \cite{sushkov1},
the OP value turns out to be very close to that obtained by
 NMR if it is corrected for an atomic effect (see the related
discussion in \cite{shabaev00a}). Therefore, in Table 4
the NMR value for the nuclear magnetic moment of lead is used. 

Taking into account that the theoretical uncertainties indicated 
in Table 4 should be considered only as an order of magnitude of
the real errors, one can deduce that the total theoretical values 
are in reasonable agreement with the experimental ones.
However,  remeasurements of the nuclear  magnetic moments
 by employing modern experimental technique and calculations of the 
Bohr-Weisskopf effect within many-particle nuclear models are
required to promote investigations of the hyperfine splitting in 
 hydrogenlike ions. 

\subsubsection{Hyperfine splitting in lithiumlike ions}

The energy difference between the ground-state hyperfine
 splitting components of a lithiumlike ion
is conveniently written as \cite{sh95}
\begin{eqnarray}
\Delta E_{(1s)^{2}2s}&=&\frac{1}{6}\alpha (\alpha Z)^{3}\frac{m}{m_{p}}
\frac{\mu}{\mu_{N}}\frac{2I+1}{2I} mc^{2}\Bigl\{
[A^{(2s)}(\alpha Z)(1-\delta^{(2s)})(1-\varepsilon^{(2s)})\nonumber\\
&&+
x_{\rm rad}^{(2s)}]+
\frac{1}{Z}B(\alpha Z)+\frac{1}{Z^{2}}C(\alpha Z)+\cdots\Bigr\}\;.
\end{eqnarray}
Here
$A^{(2s)}(\alpha Z)$ denotes the one-electron relativistic factor
for the 2s state,
\begin{eqnarray}
A^{(2s)}(\alpha Z)=\frac{2[2(1+\gamma)+\sqrt{2(1+\gamma)}]}{(1+\gamma)^{2}
\gamma (4\gamma^{2}-1)}=1+\frac{17}{8}(\alpha Z)^{2}+\frac{449}{128}
(\alpha Z)^{4}+\cdots\;,
\end{eqnarray}
  $\delta^{(2s)}$
 is the one-electron
nuclear charge distribution correction,
$\varepsilon^{(2s)}$ is the one-electron
nuclear magnetization distribution correction, and
 $x_{\rm rad}^{(2s)}$ is the one-electron
QED correction. The terms $B(\alpha Z)/Z$
and $C(\alpha Z)/Z^{2}$
 describe
the interelectronic-interaction corrections of  first
and second orders
in $1/Z$, respectively.
The first-order interelectronic-interaction correction
is conveniently derived using the TTGF method with
the closed $(1s)^2$ shell regarded as belonging to the vacuum
(see section II(E3)). Such a derivation yields 
\cite{sh95}
\be
\Delta E_{FM_FIj}&=&\sum_{M_I,m}\sum_{M'_I,m'}
C_{IM'_Ijm'}^{FM_F} C_{IM_Ijm}^{FM_F}
\chi_{IM'_I}\sum_{\mu_c}\nonumber\\
&&\times\Bigl\{\sum_{P}(-1)^P\sum_n^{\veps_n\not =\veps_v}
\frac{\la Pv' Pc|I(\Delta_{Pc c})|nc\ra \la n|H_{\rm hfs}|v\ra}
{\veps_v-\veps_n}\nonumber\\
&&+\sum_{P}(-1)^P\sum_n^{\veps_n\not =\veps_v}
\frac{\la v'|H_{\rm hfs}|n\ra
\la n c|I(\Delta_{Pc c})|Pv Pc\ra} 
{\veps_v-\veps_n}\nonumber\\
&&+\sum_{P}(-1)^P\sum_n^{\veps_n\not =\veps_c}
\frac{\la Pv' Pc|I(\Delta_{Pv' v})|vn\ra \la n|H_{\rm hfs}|c\ra}
{\veps_c-\veps_n}\nonumber\\
&&+\sum_{P}(-1)^P\sum_n^{\veps_n\not =\veps_c}
\frac{\la c|H_{\rm hfs}|n\ra
\la v' n|I(\Delta_{Pv v'})|Pv Pc\ra} 
{\veps_c-\veps_n}\nonumber\\
&&-\la v'|H_{\rm hfs}|v\ra \la  c v|I'(\Delta_{v c})|v c\ra 
\nonumber\\
&&+\sum_{\mu_{c'}}
\la c|H_{\rm hfs}|c'\ra \la  v' c'|I'(\Delta_{v c})|c v\ra
\Bigr\}\chi_{IM_I}\,,
\ee
where $F$ and $M_F$ are the total angular momentum of the atom
and its projection, $v$ and $v'$ are the valence states 
of electron with quantum numbers $(jm)$ and $(j'm')$,
respectively; $C_{IM_Ijm}^{FM_F}$ is the Clebsch-Gordan
coefficient, $\chi_{IM_I}$ is the nuclear wave function,
$c$ and $c'$  denote the core states, $\mu_c$
indicates the angular momentum projection of the core
electron, and $\Delta_{ab}=\veps_a-\veps_b$.
Calculations of the nuclear, QED, and interelectronic-interaction
corrections to the hyperfine splitting in
heavy lithiumlike ions were performed in Refs. 
\cite{sh95,shab98hf,shab99hf,shab98hfb,shabaeva99,shab00hf,bou00,zher00hf,art00hf,sapir00hf}.
As for hydrogenlike ions, the uncertainty of the theoretical
values is mainly determined by the uncertainty of the Bohr-Weisskopf
effect evaluated
 within the single particle nuclear model.
However, in Refs. \cite{shab98hf,shab98hfb} it was
found that this uncertainty can be considerably reduced
if the experimental value of the 1s hyperfine splitting in
 the corresponding
hydrogenlike ion is known.
The basic idea of this method is the following. 
It can be shown that, with a good precision,
the ratio of the 2s$-$Bohr-Weisskopf correction to 
the 1s$-$Bohr-Weisskopf correction is 
a function of the atomic  structure only and does not depend
on the nuclear structure,
\be
\frac{\varepsilon^{(2s)}}{\varepsilon^{(1s)}}=f(\alpha Z)\,.
\ee
For Z=83, $f(\alpha Z)$=1.078 and, therefore, $\varepsilon^{(2s)}=1.078\,
\varepsilon^{(1s)}$.
The 1s$-$Bohr-Weisskopf correction
 can be found by the equation
\be
\varepsilon^{(1s)}=\frac{\Delta E_{\rm Dirac}^{(1s)}
+\Delta E_{\rm QED}^{(1s)}
-\Delta E_{\rm exp}^{(1s)}}{\Delta E_{\rm Dirac}^{(1s)}}\,,
\ee
where $\Delta E_{\rm Dirac}^{(1s)}$ is the relativistic value
of the 1s hyperfine splitting including the nuclear charge
distribution correction, $\Delta E_{\rm QED}^{(1s)}$ is the
1s$-$QED correction, and
$\Delta E_{\rm exp}^{(1s)}$ is the experimental value of the
1s hyperfine splitting. For Z=83, this method predicts 
the ground-state hyperfine
splitting in lithiumlike bismuth to be 0.7971(2) eV \cite{shab00hf}
(for comparison, the direct evaluation based on 
the single-particle nuclear model gives 0.800(4) eV).
Recently, this value was confirmed  
by Sapirstein and Cheng \cite{sapir00hf} who obtained
0.79715(13) eV.
Both  theoretical values agree with the experimental one
 of 0.820(26) eV \cite{beiersdorfer1}.

\subsubsection{Bound-electron $g$ factor}

The bound-electron $g$ factor in a hydrogenlike ion
 is defined by
\be
g^{(e)}=-\frac{\langle JM_{J}|\mu_{z}^{(e)}|J M_{J}\rangle}
{\mu_{B} M_{J}}\,,
\ee
where $\bmu^{(e)}$ is the operator of the magnetic
moment of electron, $\mu_{B}$ is the Bohr magneton,
$J$ is the total angular momentum of the electron,
and $M_J$ is its projection.
For the ground state, 
a simple relativistic calculation based on the Dirac 
equation yields \cite{breit1} 
\be
g_0=2-(4/3)(1-\sqrt{1-(\alpha Z)^2})
\ee
The QED and nuclear effects give some corrections to
this value:
\be
g^{(e)}=g_0+\Delta g_{\rm QED} +\Delta g_{\rm NS}\,.
\ee
Calculation of the nuclear size correction causes no
difficulties. For low-Z atoms, this correction can be
found by a simple analytical formula \cite{kar00a}.
 The QED correction of first order in $\alpha$
 was evaluated without expansion in $\alpha Z$
 in Refs. \cite{blund97,pers97,bei00} (see also \cite{beier00}).

Direct measurements of the bound-electron $g$ factor
in hydrogenlike ions are presently being performed
by a GSI - Universit\"at Mainz collaboration 
\cite{quint1,hermanspahn1,hermanspahn2,haffner}.  
To date, the experimental result
obtained for hydrogenlike carbon (C$^{5+}$) \cite{haffner}
amounts to  $g_{\rm exp}=2.001041596(5)$ and agrees with the
theoretical predictions of Refs. \cite{beier00,bei00},
$g_{\rm theo}=2.001041591(7)$, and of Ref. \cite{kar00b},
$g_{\rm theo}=2.001041590(2)$.
The collaboration plans to extend these measurements 
to heavy ions.

Another possibility for investigations of the bound-electron
$g$ factor was recently proposed in \cite{shabaev98can}.
In this work it was shown that the transition
probability between the ground-state hyperfine splitting
components of a hydrogenlike ion, including the first-order
QED and nuclear corrections, is given by
\begin{eqnarray} \label{trans}
w=\frac{\alpha}{3}\,\frac{\omega^3}{m^2}\,\frac{I}{2I+1}
\Bigl[g^{(e)}-g_{I}^{(n)}\frac{m}{m_{p}}\Bigr]^2\,,
\end{eqnarray}
where $\omega$ is the transition frequency, $g^{(e)}$
is the bound-electron $g$ factor defined above, and
$g_{I}^{(n)}$ is the nuclear $g$ factor (both $g$ factors are
defined to be positive).
Formula (\ref{trans}) allows
a simple calculation of the QED and nuclear
corrections to the transition probability using
 the corresponding
corrections to the bound-electron $g$ factor.
In \cite{shabaev98can}
 it was found that in the experimentally interesting 
cases of Pb and Bi,
the  QED and nuclear corrections increase the transition probability by
about 0.3\%. In \cite{winter1} the lifetime of the upper
hyperfine splitting component in $^{209}{\rm Bi}^{82+}$
was measured to be $\tau_{\rm exp}$=397.5(1.5) 
$\mu$s. This result is in good agreement 
with the theoretical predictions of Ref. \cite{shabaev98can},
$\tau_{\rm theo}=399.01(19) \mu$s,
and of Ref. \cite{pal00},
$\tau_{\rm theo}=398.89 \mu$s.
Using formula (\ref{trans}) and the experimental values of the
hyperfine splitting and the transition probability in
$^{209}{\rm Bi}^{82+}$ \cite{klaft1,winter1},
the experimental value of the bound-electron $g$ factor
in $^{209}{\rm Bi}^{82+}$
 is found to be 1.7343(33). The corresponding
theoretical value  is  1.7310.
 The individual contributions
 to the bound-electron $g$ factor in $^{209}{\rm Bi}^{82+}$
 are given in Table 5.
>From this table, it is clear
 that the QED correction has to be included in order
to obtain agreement between theory and experiment.

\subsection{Radiative recombination of an electron with a heavy
ion}

In  an energetic collision between a high-$Z$ 
ion and a low-$Z$
target atom,  an electron may
be captured by the projectile, while a simultaneously emitted 
photon carries
away the excess energy and momentum. This process is denoted 
as  radiative
electron capture (REC). Since a loosely bound target electron can
be considered as quasi-free, this process
 is essentially equivalent 
to radiative
recombination (RR) or its time-reversed analogon, 
the photoelectric effect.
 The relativistic theory of REC was considered
 in detail  in \cite{eichler,ichihara94,eichler95a},
 and the results of
this theory are in excellent agreement with experiments
 \cite{stoehlker95,stoehlker99}.

A systematic QED theory of the RR process 
is described in detail in sections III(D,E) of the present paper.
In Refs. \cite{kar92,sh94b,nef94},
this theory was employed to study the
 process of  resonance recombination
of an electron with a heavy hydrogenlike ion
in the case of
resonance with doubly excited $(2s,2s)_{0}$, $(2p_{1/2},2p_{1/2})_0$,
$(2s,2p_{1/2})_{0,1}$ states of the corresponding 
heliumlike ion.
Later, this theory was used
to  evaluate QED corrections  to  radiative recombination of
 an electron with a bare nucleus \cite{sh00}
and interelectronic-interaction
corrections to radiative recombination of an electron 
with a heavy heliumlike ion \cite{yer00}.
The results  of these investigations are briefly 
discussed below. More details can be found in 
the original papers \cite{kar92,sh94b,sh00,yer00,nef94}.

\subsubsection{QED corrections to the radiative recombination
of an electron with a bare nucleus}

We consider the radiative recombination of an electron with
momentum $\bfp_{i}$ and polarization  $\mu_{i}$ with a bare nucleus
that is placed at the origin of the coordinate frame.
This corresponds to the projectile system if we study the radiative
recombination of a free target electron with  a bare heavy
projectile.
To zeroth order in $\alpha$, the cross section is
\begin{eqnarray}\label{inf10arec}
\frac{d\sigma^{(0)}}{d\Omega_{f}}=
\frac{(2\pi)^4}{v_{i}}\,
 \bfk_{f}^2\,|\tau^{(0)}|^2\,,
\end{eqnarray}
where
\begin{eqnarray}\label{tau0rec}
\tau^{(0)}=-\la a|e\alpha^{\nu}A_{f,\nu}^{*}
|p\ra=\la a|e\balpha\cdot {\bf A}_{f}^{*}
|p\ra\,,
\end{eqnarray}
$|p\ra\equiv |p_i,\mu_i\ra$ is the wave function of the
incident electron defined by equation (\ref{psip+}),
$p_i=(p_i^0,\bfp_i)$,
 $p_{i}^0=\sqrt{\bfp_{i}^2+m^2}\,$ is the energy of the incident 
electron, $a$ is the final state 
of the one-electron atom,
$k_f=(k_f^0,{\bf k}_f)$ with
 $k_{f}^0$ and $\bfk_f$ being the photon energy
and momentum, respectively, $v_{i}$ is the velocity of the incident
 electron in the nucleus frame.

The QED corrections of first order in $\alpha$ are 
defined by diagrams similar to those 
shown  in Fig. 32. 
The direct calculation by the TTGF method yields
for the self-energy correction  to the amplitude of the process
\begin{eqnarray}\label{first15rec}
\tau_{\rm SE}^{(1)}&=&
-\Bigl[\sum_n^{n\not = a}
\frac{
\la a|\Sigma(\veps_{a})-\beta\delta m|n\ra 
\la n|e\alpha^{\nu}A_{f,\nu}^{*}
|p\ra}
{\veps_{a}-\veps_{n}}\nonumber\\
&&+
\frac{1}{2}\la a|\Sigma'(\veps_{a})|a\ra 
\la a|e\alpha^{\nu}A_{f,\nu}^{*}
|p\ra\nonumber\\
&&+\sum_{n}\frac{
\la a|e\alpha^{\nu}A_{f,\nu}^{*}
|n\ra
\la n|\Sigma(p_{i}^{0})-\beta \delta m
|p\ra}
{p_{i}^0-\veps_{n}(1-i0)}\nonumber\\
&&+\int d \bfz\,
e A_{f,\nu}^{*}(\bfz)
\Lambda^{\nu}(\veps_{a},p_{i}^{0},\bfz)
+(Z_{2}^{-1/2}-1)
\la a|e\alpha^{\nu}A_{f,\nu}^{*}
|p\ra
 \Bigr]\,,
\end{eqnarray}
where the mass counterterm has been added and
\begin{eqnarray}\label{first999}
\Lambda^{\nu}(\vare,p^{0},\bfz)&=&
e^{2}\frac{i}{2\pi}\int_{-\infty}^{\infty}d\omega\,
\int d\bfx d\bfy \, \overline{\psi}_{a}(\bfx)\gamma^{\rho}
 S(\vare-\omega,\bfx,\bfz)\gamma^{\nu}
 S(p^{0}-\omega,\bfz,\bfy)\nonumber\\
&&\times \gamma^{\sigma}
D_{\rho\sigma}(\omega,\bfx-\bfy)
\psi_{p_{i}\mu_{i}(+)}(\bfy)\,.
\end{eqnarray}
A similar calculation of the VP correction gives
\begin{eqnarray}\label{first16rec}
\tau_{\rm VP}^{(1)}&=&
-\Bigl[\sum_n^{n\not = a}
\frac{
\la a|U_{\rm VP}|n\ra 
\la n|e\alpha^{\nu}A_{f,\nu}^{*}
|p\ra}
{\veps_{a}-\veps_{n}}\nonumber\\
&&+\sum_{n}\frac{
\la a|e\alpha^{\nu}A_{f,\nu}^{*}
|n\ra
\la n|U_{\rm VP}
|p\ra}
{p_{i}^0-\veps_{n}(1-i0)}\nonumber\\
&&+\int d \bfz \,
e A_{f,\nu}^{*}(\bfz)
Q^{\nu}(k_{f}^0,\bfz)
+(Z_{3}^{-1/2}-1)
\la a|e\alpha^{\nu}A_{f,\nu}^{*}
|p\ra
 \Bigr]\,.
\end{eqnarray}
Here $U_{\rm VP} (\bfx)$ is the vacuum-polarization
potential defined by equation (\ref{vppot}) and
\begin{eqnarray}
Q^{\nu}(k^0,\bfz)&=&-e^2\int d{\bfx} d{\bfy}\,\overline{\psi}_{a}
(\bfx)\gamma^{\rho}\psi_{p_{i}\mu_{i}(+)}(\bfx)
D_{\rho \sigma}(k^0,\bfx-\bfy) \nonumber\\
&&\times\frac{i}{2\pi}\int_{-\infty}
^{\infty} d\omega \, {\rm Tr}[\gamma^{\sigma}S(\omega, \bfy,\bfz)
\gamma^{\nu} S(\omega+k^0,\bfz,\bfy)]\,.
\end{eqnarray}
In addition to these corrections,
we have to take into account a contribution originating from
changing the photon energy in  the zeroth-order cross
section (\ref{inf10arec}) due to the QED correction to the energy
of the bound state $a$. It follows that the total QED
correction of  first order in $\alpha$ to the cross section
is given by
\begin{eqnarray}\label{form34rec}
\frac{d\sigma_{\rm QED}^{(1)}}{d\Omega_{f}}
=\frac{(2\pi)^4}{v_{i}}
\bfk_{f}^2\,2{\rm Re}{\Bigl\{\tau^{(0)*}\tau_{\rm QED}^{(1)}}\Bigr\}+
\Bigl[\frac{d\sigma^{(0)}}{d\Omega_{f}}\Bigr|_{k_{f}^0=p_{i}^0-E_{a}}
-\frac{d\sigma^{(0)}}{d\Omega_{f}}\Bigr|_{k_{f}^0=p_{i}^0-\veps_{a}}
\Bigr]\,.
\end{eqnarray} 
Here $\tau_{\rm QED}^{(1)}=\tau_{\rm SE}^{(1)}+\tau_{\rm VP}^{(1)}$
is the QED correction given by equations 
(\ref{first15rec}) and (\ref{first16rec}). $E_a$ and
 $\veps_{a}$ are the energies of the bound state $a$
with and without the QED correction, respectively.

The expressions (\ref{first15rec}) and (\ref{first16rec})
contain ultraviolet and infrared divergences. While
the ultraviolet divergences can be eliminated 
by the standard renormalization procedure (see the related
discussion in section III(B2)), the infrared divergences are 
more difficult to remove.

The infrared-divergent part of $\tau$
results from the region of small momenta
of the virtual photon and is regularized by a non-zero
photon mass $\mu$. An evaluation of this part yields
\begin{eqnarray}
\tau_{\rm infr}=\tau^{(0)}\frac{\alpha}{\pi}\Bigl[
-\log{(\mu/m)}-\frac{1}{2}
\,\log{(\mu/m)}
\sqrt{1+\frac{m^2}{\bfp_{i}^2}}\log{\Bigl(\frac
{\sqrt{\bfp_{i}^2+m^2}-|\bfp_{i}|}
{\sqrt{\bfp_{i}^2+m^2}+|\bfp_{i}|}\Bigr)}\Bigr]\,.
\end{eqnarray}
The related contribution to the cross section is
\begin{eqnarray}\label{inf10rec}
\frac{d\sigma_{\rm infr}}{d\Omega_{f}}
=\frac{d\sigma^{(0)}}{d\Omega_{f}}\,
\frac{\alpha}{\pi}\Bigl[
-2\log{(\mu/m)}-
\log{(\mu/m)}
\sqrt{1+\frac{m^2}{\bfp_{i}^2}}\,\log{\Bigl(\frac
{\sqrt{\bfp_{i}^2+m^2}-|\bfp_{i}|}
{\sqrt{\bfp_{i}^2+m^2}+|\bfp_{i}|}\Bigr)}\Bigr]\,,
\end{eqnarray}
where $\frac{d\sigma^{(0)}}{d\Omega_{f}}$
is the cross section in the zeroth-order approximation
defined by equation (\ref{inf10arec}).
To cancel the infrared divergent contribution
(\ref{inf10rec}), we have to take into account that any experiment
 has a finite  energy resolution $\Delta E$. It means that any
numbers of photons of the total energy less than $\Delta E$ 
can be emitted in the process. It follows that to find the total
cross section in the order under consideration, we must include
diagrams in which one photon of energy $k^0=\sqrt{\bfk^2+\mu^2}
\le \Delta E$ is emitted along with the emission of the photon with the
energy $k_{f}^0\approx p_{i}^0-\veps_{a}$
(we assume $\Delta E \ll k_f^0$). 
These diagrams are shown in Fig. 38.
Assuming that the energy resolution is sufficiently high  
($\Delta E\ll k_f^0, m$), we retain only those contributions
 from the diagrams shown in Fig. 38 which dominate at 
$\Delta E\rightarrow 0$. 
Using the standard technique and omitting terms which approch
zero  at $\mu\rightarrow 0$, we
obtain for this contribution
\begin{eqnarray}\label{cross1rec}
\frac{d\sigma_{\gamma}(\Delta E)}{d\Omega_{f}}&=&
\frac{d\sigma^{(0)}}{d\Omega_{f}}\,
\frac{\alpha}{\pi}\Bigl[1-2\log{2}
-2\log{(\Delta E/\mu)}
-\sqrt{1+\frac{m^2}{\bfp_{i}^2}}\,F(|\bfp_{i}|/p_{i}^0)
\nonumber\\
&&-\Bigl(\frac{1}{2}+\log{(\Delta E/\mu)} \Bigr)
\sqrt{1+\frac{m^2}{\bfp_{i}^2}}\,
\log{\Bigl(\frac
{\sqrt{\bfp_{i}^2+m^2}-|\bfp_{i}|}
{\sqrt{\bfp_{i}^2+m^2}+|\bfp_{i}|}\Bigr)}\Bigr]\,,
\end{eqnarray}
where
\begin{eqnarray}
F(a)=\int_{0}^{\infty}dx\, \frac{x}{x^2+1}\log{\Bigr[
\frac{(1+a)(\sqrt{x^2+1}-ax)}
{(1-a)(\sqrt{x^2+1}+ax)}\Bigl]}\,.
\end{eqnarray}
We can see that the infrared divergent parts
in the equations (\ref{inf10rec}) and (\ref{cross1rec})
cancel each other,
\begin{eqnarray}\label{cross2rec}
\frac{d\sigma_{\rm infr}}{d\Omega_{f}}+
\frac{d\sigma_{\gamma}(\Delta E)}{d\Omega_{f}}&=&
\frac{d\sigma^{(0)}}{d\Omega_{f}}\,
\frac{\alpha}{\pi}\Bigl[1-2\log{2}
-2\log{(\Delta E/m)}
-\sqrt{1+\frac{m^2}{\bfp_{i}^2}}\,F(|\bfp_{i}|/p_{i}^0)
\nonumber\\
&&-\Bigl(\frac{1}{2}+\log{(\Delta E/m)}\Bigr)
\sqrt{1+\frac{m^2}{\bfp_{i}^2}}\log{\Bigl(\frac
{\sqrt{\bfp_{i}^2+m^2}-|\bfp_{i}|}
{\sqrt{\bfp_{i}^2+m^2}+|\bfp_{i}|}\Bigr)}\Bigr]\,.
\end{eqnarray}
According to this equation,
at a fixed energy of the incident electron,
 the QED correction
depends on the photon-energy resolution $\Delta E$
and becomes infinite when 
$\Delta E\rightarrow 0$.
It means that the validity of this equation is restricted
by the condition $\frac{\alpha}{\pi}|\log{(\Delta E/m)}|\ll 1$ .
For extension of the theory beyond this limit it is
necessary to include the radiative corrections
of higher orders in $\alpha$ (see, e.g., \cite{Jauch,Yennie}).
It results in an "exponentiation" of the radiative corrections
and removes the singularity for $\Delta E\rightarrow 0$.

In the derivation of the formulas (\ref{cross1rec}),  (\ref{cross2rec})
it has been assumed that the incident electrons have a fixed energy.
These formulas remain also valid in the case when 
 the energy spread of the incident electrons
 is much smaller than the energy interval $\Delta E$
in which the photons are detected.
 However, this is not the case for the present
REC experiments \cite{stoehlker95,stoehlker99}, where the energy
spread of a quasi-free target electron is much larger than
the finite photon-energy resolution. In that case, the QED correction
to the total RR cross section depends on the form of the energy
distribution of the target electron. Since the form of this distribution
is not well determined, the only way to study the QED effects in REC 
processes is to investigate the cross section into a photon-energy
interval which is chosen to be much larger than the effective
energy spread of the quasi-free target electrons and much
smaller than the energy of the emitted photon.

The Uehling part of the
VP correction to the RR cross section
 and a part of the SE correction were numerically
evaluated in Ref. \cite{sh00}.
 Expressed in terms of the unperturbed
cross section, the individual QED corrections to the total cross
section for the radiative recombination into the K-shell of
bare uranium are  presented in the Table 6. 
The correction $\sigma^{(1)}_{\rm en}$ results from changing the bound-state 
energy. It is determined by the term in the square brackets
of equation (\ref{form34rec}).
The correction $\sigma^{(1)}_{\rm bw}$ corresponds to the irreducible part
of the diagrams describing the first-order QED effect on
the bound-state electron wave function.
The correction $\sigma^{(1)}_{\rm cw}$ results from the diagrams
describing
the QED effect on the continuum-state wave function.

As in bound-state QED, 
the Uehling approximation may be expected to
account for a dominant part of the 
 VP correction. As to the self-energy correction, we  
expect that the terms calculated in Ref. \cite{sh00} give  
a reasonable estimate of the order of magnitude of the self-energy
correction which is beyond the correction depending on the
photon-energy interval $\Delta E$ 
(see equation (\ref{cross2rec})). The relative value of the last correction,
which we denote by 
$\delta(\Delta E)$,
 is defined by
\begin{eqnarray}\label{dele}
\delta(\Delta E)=\frac{\alpha}{\pi}\Bigl[-2+\frac{1}{\beta}
\log{\frac{1+\beta}{1-\beta}}\Bigr]\log{\frac{\Delta E}{m}}\,,
\end{eqnarray}
where $\beta=v_i/c$. 
The photon-energy interval has to be  chosen
in the range $\Gamma \ll \Delta E \ll k_f^0, m$,
where $\Gamma$ characterizes the energy spread of the
incident electrons.

In the REC experiments which are performed at GSI,
 the effective electron-energy
 spread  is determined by the momentum distribution of the
quasi-free target electrons. The width  of this spread 
in the projectile system
 increases with increasing impact energy.
 In the case of a ${\rm N}_2$
gas target, which is presently being employed in the experiments,
 the effective energy spread in the projectile (heavy ion) frame
amounts to about 10-40 keV for the impact energy in the range
100-1000 MeV/u. This value  will be considerably
reduced in the experiments with a ${\rm H}_2$ gas target which
are  under preparation. In the case of a ${\rm H}_2$ gas target,
to  satisfy the conditions
on  $\Delta E$ given above we can choose $\Delta E$ to be 50 keV 
in the projectile frame for
 the impact energy 1 GeV/u. 
The corresponding photon-energy interval in the laboratory
(gas-target) frame is determined according to the Lorentz
transformation,
\begin{eqnarray}\label{Lorentz}
\Delta E_{\rm proj}=\Delta E_{\rm lab}(1-\beta 
\cos{\theta_{\rm lab}})/\sqrt{1-\beta^2}\,.
\end{eqnarray}
At a fixed $\Delta E_{\rm proj}$, 
$\Delta E_{\rm lab}$ as a function of the polar angle
can be found from this equation.
For the photon-energy interval chosen above,
 $\delta(\Delta E)$ amounts to
 $-0.59$\% for an impact energy 
of 1 GeV/u.
Adding the Uehling correction and the part of the SE correction 
presented in Table 6
to the correction  $\delta(\Delta E)$,
we find the QED correction to the total
 cross section amounts to  $-0.92$\% for an impact energy 
of 1 GeV/u. For a more accurate evaluation of this effect,
complete calculations of all the SE corrections are required.

The results of the numerical evaluation of the differential
cross section can be found in Ref. \cite{sh00}.
Here, we note only that the differential cross section
at the backward direction vanishes at an impact energy close
to 130 MeV/u. In particular, it results 
 in a relatively large contribution of the QED
correction to the backward cross section at an energy of 130 MeV/u. 
At this energy, 
the ${\rm QED}_{{\rm en+bw}+\Delta E}$ correction is about 0.022 mbarn/sr
while the zeroth-order cross section amounts only to 0.009 mbarn/sr.

\subsubsection{Interelectronic-interaction effect on the
radiative recombination of an electron with a heavy
heliumlike ion}

We consider the non-resonant radiative recombination of an electron
 with momentum $\bfp_{i}$ and
polarization  $\mu_{i}$ with a heavy heliumlike ion 
in the ground state which is placed at
the origin of the coordinate frame. 
The final state of the system is a lithiumlike ion
in the state $(1s)^2 v$, where $v$ denotes the valence state. 
This picture
corresponds to the projectile system if we study 
the radiative recombination of a free
target electron with a heavy heliumlike projectile.

 To zeroth order, the amplitude of the process is given by 
\begin{equation}\label{bas9int}
  \tau^{(0)}=\la v|e\balpha\cdot {\bf A}_{f}^{*} |p\ra \, ,
\end{equation}
where $|v\rbr$ denotes the wave function of the valence electron.
The interelectronic-interaction correction of first order in
$1/Z$ can easily be derived using the TTGF method with
the closed $(1s)^2$ shell regarded as belonging to the vacuum.
 Such a derivation
yields \cite{yer00}
\begin{equation}\label{bas3int}
  \tau^{(1)}_{\rm int} = \sum_{l=1}^4 \tau^{\rm int}_l \ ,
\end{equation}
where
\begin{eqnarray}
\tau^{\rm int}_{1} &=& \sum_{\mu_c} \sum_P (-1)^P 
\sum_n^{\vare_n\ne \vare_v}
        \frac{\lbr PvPc|
I(\Delta_{Pc\,c})| nc\rbr \lbr n|e\balpha\cdot {\bf A}_{f}^{*}|p\rbr}
    {\vare_v-\vare_n}  \nonumber \\
&&  + \sum_{\mu_c} (-\frac12) \lbr cv|I'(\Delta_{vc})|vc\rbr   
\label{bas4int}
    \lbr v| e\balpha\cdot {\bf A}_{f}^{*} |p\rbr \ ,   \\
\tau^{\rm int}_{2} &=& \sum_{\mu_c} \sum_P (-1)^P \sum_{n}   
\label{bas4aint}
        \frac{\lbr v|e\balpha\cdot 
{\bf A}_{f}^{*}|n\rbr \lbr nc|I(\Delta_{Pc\,c} )| PpPc\rbr}
    {p^0_i-\vare_n(1-i0)} \ , \\
    \label{bas4bint}
\tau^{\rm int}_{3} &=& \sum_{\mu_c} \sum_P (-1)^P \sum_{n}
        \frac{\lbr PvPc|I(\Delta_{p\,Pv})|
 pn\rbr \lbr n|e\balpha\cdot {\bf A}_{f}^{*}|c\rbr}
    {\vare_c-k^0_f-\vare_n(1-i0)} \ , \\   
        \label{bas5int}
\tau^{\rm int}_{4} &=& \sum_{\mu_c} \sum_P (-1)^P \sum_{n}
        \frac{\lbr c|e\balpha\cdot
 {\bf A}_{f}^{*}|n\rbr \lbr vn|I(\Delta_{Pp\,v})| PpPc\rbr}
    {\vare_c+k^0_f-\vare_n(1-i0)} \ .
\end{eqnarray}
Here $|v\ra$ and $|c\ra$ are the valence and core states,
respectively, $p_i^0=\sqrt{{\bf p}_i^2+m^2}$ is the energy
of the incident electron, $k_f^{0}=p_i^{0}-\veps_v$ is
the energy of the emitted photon, and $\mu_c$ indicates the
angular momentum projection of the core electron.
The expressions (\ref{bas4int}) -- (\ref{bas5int})
 represent the interelectronic-interaction
corrections to the amplitude of the process. 
The corresponding corrections to the
differential cross section are
\begin{equation}\label{bas8}
  \frac{d\sigma^{\rm int}_l}{d\Omega_f} = \frac{(2\pi)^4}{v_i} 
\bfk^2_f 2 {\rm Re}
    \left[ \tau^{(0)*}\tau^{\rm int}_l \right] \ .
\end{equation}
In addition to this, 
a contribution originating from a
modification of the energy of the emitted 
photon in the zeroth-order cross section due
to the interelectronic interaction should be taken into account
which is given by
\begin{equation}\label{bas10int}
  \frac{d\sigma^{\rm int}_{\rm en}}{d\Omega_f} =
    \frac{d\sigma^{(0)}}{d\Omega_{f}}\Bigr|_{k^0_f=p^0_i-E_{v}}
     -\frac{d\sigma^{(0)}}{d\Omega_{f}}\Bigr|_{k^0_f=p^0_i-\vare_{v}}
      \ .
 \end{equation}
Here $E_v = \vare_v+ 
\Delta E^{(1)}_{\rm int}$ is the energy of
the valence electron including the first-order 
interelectronic-interaction correction,
\begin{equation}\label{bas11int}
\Delta E^{(1)}_{\rm int} = \sum_{\mu_c}\sum_P(-1)^P
    \lbr PvPc|I(\Delta_{Pvv})|vc\rbr \ .
\end{equation}
The total interelectronic-interaction correction 
to the cross section in first order
in $1/Z$ is
\begin{equation}\label{bas12}
\frac{d\sigma_{\rm int}^{(1)}}{d\Omega_{f}} =
 \frac{d\sigma^{\rm int}_{\rm en}}{d\Omega_f}
 +  \sum_{l =1}^4
    \frac{d\sigma_l^{\rm int}}{d\Omega_{f}} \ .
\end{equation}

The direct part of the corrections $\tau^{\rm int}_1$
 and $\tau^{\rm
int}_2$ can be accounted for
by a modification of the incoming
and outcoming electron wave functions by the screening potential
\be
V_{\rm C}(x)\rightarrow V_{\rm C}(x)+ V_{\rm scr}(x)\,,
\ee
where
\begin{equation}\label{bas13}
  V_{\rm scr}(x) = 2\alpha \left\{
    \frac1x \int_0^x dy \, y^2 \left(g_{1s}^2(y) +f_{1s}^2(y) \right)
   +  \int_x^{\infty} dy \, y \left(g_{1s}^2(y) +f_{1s}^2(y) \right)
   \right\} \,.
\end{equation}
Here $g_{1s}$ and $f_{1s}$ are the upper and 
the lower components of the radial wave
function of the ground state, respectively. 
The screening-potential approximation allows one to account for
the dominant part of the interelectronic-interaction
effect and is widely used in practical calculations.

Numerical results
for the interelectronic-interaction 
correction to the total cross
section of radiative recombination 
of an electron with heliumlike uranium are presented
in Table 7 (for more extensive data we refer to \cite{yer00}).
The calculations are carried out  for a capture
into the $2s$, $2p_{1/2}$, and $2p_{3/2}$ states 
of lithiumlike uranium and for various projectile
energies. 
The results of the rigorous relativistic 
treatment are compared
 with the calculations based on the screening-potential
approximation. A deviation of the complete
relativistic results from the screening-potential
approximation is mainly determined by the term
$\sigma^{\rm int}_4$ which strongly increases when 
$p_i^0$  comes close to the  resonance condition 
($p_i^0-(\veps_v-\veps_c)\approx \veps_n$).
Numerical results for the differential cross section
are given in Ref. \cite{yer00}.

\subsubsection{Resonance recombination of an electron
with a heavy hydrogenlike ion}

We consider the process of recombination of an electron with
a very heavy $(Z\sim 70-110)$ hydrogenlike ion in
its ground state for the case of resonance with doubly excited
 $(2s\;2s)_{0}$, $(2p_{1/2}\;2p_{1/2})_{0}$,
 $(2s\;2p_{1/2})_{0,1}$ states of the corresponding
 heliumlike ion. We assume
that  the experimentally
 measured quantity is a part of the total cross section
 which corresponds to the emission of photons with an
energy $\omega \approx E_{d}-E_{r}$, where $E_{d}\;(d=1,2,3,4)$
are the energies of the doubly excited 
$(2s\;2s)_{0}$, $(2p_{1/2}\;2p_{1/2})_{0}$,
 $(2s\;2p_{1/2})_{0,1}$ states, respectively, and
$E_{r}(r=1,2,3,4)$ are the energies of singly excited
$(1s\;2p_{1/2})_{0,1}$, $(1s\;2s)_{0,1}$ states. The
amplitude of this process is given by the sum of the
amplitudes of the dielectronic recombination (DR)
 and radiative recombination (RR) processes
\[e^{-}(p_i^0) + X^{(Z-1)+}(1s) \rightarrow
\left\{\begin{array}{ll}
 X^{(Z-2)+}(d)^{**}\rightarrow X^{(Z-2)+}(r)^{*}+\gamma(\omega)
\rightarrow \cdots\\
 X^{(Z-2)+}(r)^{*}+\gamma(\omega)\rightarrow \cdots
\end{array}
\right. \]
Here $ p_i^0 $ is the energy of the incident electron.
Among the doubly excited states $\{d\}$, there are
 states with identical quantum numbers $\bigl(\;(2s\,2s)_{0},
\;(2p_{1/2}\;2p_{1/2})_{0}\;\bigr)$, while all
the singly excited states $\{r\}$ have different quantum
numbers. Main channels of the decay of $(1s\;2s)_{1},\;
(1s \;2p_{1/2})_{1}$ and $(1s\;2s)_{0},\;(1s\;2p_
{1/2})_{0}$ states are one- and
two-photon transitions to the ground state, respectively. 
Therefore, two- and
three-photon processes give the dominant contribution
to the cross section.
The cross section of the process can be derived using the
method described in section III(F). In the resonance approximation,
such a derivation yields (see Ref. \cite{sh94b} for details)
\begin{eqnarray}\label{resrec1}
\sigma(p_i^0)&=&\frac{2\pi^{2}}{2(j'+1)}\frac{1}{|{\bf p}_i|^{2}}\sum_
{jlJM}\Biggl \{\sum_{d}\frac{W_{dd}|\la d_{L}|\hat{I}|i_{J}\ra|^{2}}
{(E_{1s}+p_i^0-E_{d})^{2}+\Gamma_{d}^{2}/{4}}\nonumber\\
&&+2\,{\rm Re}\sum_{d<d'}\frac{W_{dd'}\la d_{L}|\hat{I}|i_{J}\ra
\la d_{L}^{\prime}|\hat{I}|i_{J}\ra
^{*}}{(E_{1s}+p_i^0-E_{d}+i\Gamma_{d}/{2})(E_{1s}+
p_i^0-E_{d'}-i\Gamma_{d'}/{2})}\nonumber\\
&&+W_{i_{J}i_{J}}+2\,{\rm Re}\sum_{d}\frac{W_{di_{J}}
\la d_{L}|\hat{I}|i_{J}\ra}
{E_{1s}+p_i^0 -E_{d}+i\Gamma_{d}/{2}}\Biggr \}\, ,
\end{eqnarray}
where $|i_{J}\ra \equiv |JMnj'l'p_i^0 jl\ra$,
$(n\;j'\;l')=(1\;1/2\; 0)$ are the quantum numbers of the
$1s$ state of the hydrogenlike ion, 
$\hat{I}\equiv I(\veps_{2s}-\veps_{1s})$,
\begin{eqnarray}
W_{dd'}=\sum_{r}W_{dd'}^{(r)}\, ,
\qquad\qquad W_{di_J}=\sum_{r}W_{di_J}^{(r)}\, ,
\qquad\qquad W_{i_Ji_J}=\sum_{r}W_{i_Ji_J}^{(r)}\, ,
\end{eqnarray}
\begin{eqnarray}
W_{dd'}^{(r)}&=&2\pi\omega^{2}\sum_{\beps}\int d\Omega_f \;
\la r|\hat{R}_{\gamma}|d_{R}\ra \la r
|\hat{R}_{\gamma}|d_{R}^{\prime}\ra^{*}\,, \\
W_{di_J}^{(r)}&=&2\pi\omega^{2}\sum_{\beps}\int d\Omega_f \;
\la r|\hat{R}_{\gamma}|d_{R}\ra \la r
|\hat{R}_{\gamma}|i_J\ra^{*}\,,\\
W_{i_Ji_J}^{(r)}&=&2\pi\omega^{2}\sum_{\beps}\int d\Omega_f\;
|\la r|\hat{R}_{\gamma}|i_J\ra|^{2}\,, \\
\hat{R}_{\gamma}&=&-\sum_{n=1}^{2}e\balpha \cdot {\bf A}_f^{*}({\bf x}_n)\,.
\ee
The cross section (\ref{resrec1}) 
consists of four terms, $\sigma=
\sum_{l=1}^{4}\sigma_{l}$. The first two terms ($\sigma_{1}$,
$\sigma_{2}$) correspond to the DR process, the third term ($\sigma_{3}$)
corresponds to the RR process, and $\sigma_{4}$ describes the
interference between the DR and RR processes. The term $\sigma_{2}$ is
caused by the interference of the DR amplitudes on the levels 
with identical quantum numbers $(d,d'=1,2)$. The magnitude of
this term is determined by the overlap of the levels $d,d'$ and can be
characterized by the nonortogonality integral $\la d_{R}^{\prime}|d_{R}\ra$
which is connected with $W_{dd'}$ by the  identity
\begin{eqnarray} \label{resrec2}
W_{dd'}=i({\cal E}_{d}-{\cal E}_{d'}^{*})\la d_{R}^{\prime}|d_{R}\ra\, .
\end{eqnarray}

Formula (\ref{resrec1})
 was used in \cite{kar92} for the  numerical calculation
of the cross section of the resonance
 recombination of an electron with hydrogenlike uranium. Later, a
more accurate calculation was performed by
Yerokhin \cite{yer93}. According to these calculations, the ratio
$\sigma_{2}/\sigma$
 which characterizes the overlap effect amounts
up to $30\%$ in the region between the maxima of the curve $\sigma
(p_i^0)$. For the parameters of the overlapping levels it
was found \cite{kar92}
\begin{eqnarray}
|\la d_{R}|d_{R}^{\prime}\ra|=0.180\, ,\qquad\qquad |W_{dd'}|/|E_{d}
-E_{d'}|=0.183 \,.
\end{eqnarray}
Using (\ref{resrec2}) and the identity
\begin{eqnarray} \label{resrec3}
W_{di}=i\la i|(\hat{I}-\hat{I}^{\dagger})|d_{R}\ra\,, 
\end{eqnarray}
the expression (\ref{resrec1}) can be transformed to the following
(see Ref. \cite{sh94b} for details)
\begin{eqnarray} \label{resrec4}
\sigma(p_i^0)&=&\frac{2\pi^{2}}{2(2j'+1)}\frac{1}{|{\bf p}_i|^{2}}
\sum_{jlJM}\Biggl \{
\sum_{d}\frac{\Gamma_{d}{\rm Re}\,(\la i_{J}|\hat{I}|d_{R}\ra
\la d_{L}|\hat{I}|i_{J}\ra)}{(E_{1s}+p_i^0-E_{d})^{2}+
\frac{\Gamma_{d}^{2}}{4}}\nonumber\\
&&-2\sum_{d}\frac{{\rm Im}\,(\la i_{J}|\hat{I}^{\dagger}|d_{R}\ra
\la d_{L}|\hat{I}|i_{J}\ra)(E_{1s}+p_i^0 -E_{d})}
{(E_{1s}+p_i^0-E_{d})^{2}+
\frac{\Gamma_{d}^{2}}{4}}+W_{i_{J}i_{J}}\nonumber\\
&&-2\sum_{d}\frac{{\rm Im}\,(\la i_{J}|
(\hat{I}-\hat{I}^{\dagger})|d_{R}\ra
\la d_{L}|\hat{I}|i_{J}\ra)(E_{1s}+p_i^0 -E_{d})}
{(E_{1s}+p_i^0-E_{d})^{2}+
\frac{\Gamma_{d}^{2}}{4}}\Biggr \}\, .
\end{eqnarray}
This expression also consists of four terms $\sigma=\sum_{l=1}^{4}
\overline{\sigma}_{l}$. But, in contrast to (\ref{resrec1}), here the
Breit-Wigner part of the cross section is completely contained
in the first term $(\overline{\sigma}_{1})$ which is a sum
of Lorentz-type terms.  The terms $\overline{\sigma}_{2}$, $\overline
{\sigma}_{4}$  again correspond to the interference of the DR amplitudes
on the levels $d,d'=1,2$ and the interference of the DR and RR
processes, respectively. But, unlike $\sigma_{2}$, $\sigma_{4}$,
the terms $\overline{\sigma}_{2}$, $\overline{\sigma}_{4}$ do not
contain any admixture of Lorentz-type terms. They are given by
sums of terms which are  odd functions of
$(E_{1s}+p_i^0-E_{d})$. (We should note
 that the admixture of  
Lorentz-type terms in $\sigma_{2}$, $\sigma_{4}$ is very small
and hardly affects the values  $\sigma_{2}/\sigma$ 
and $\sigma_{4}/\sigma$.) The terms $\sigma_{2}$,
$\sigma_{4}$ $\;\;\;(\overline{\sigma}_{2},\;\;\overline{\sigma}_{4})$
lead to a deviation of the shape of the individual resonances from
 the Lorentz shape. This deviation can be characterized by the Low
parameter \cite{low52}
\begin{eqnarray}
\delta=\frac{\sigma(E_{d}-E_{1s}+\frac{\Gamma_{d}}{2})-
\sigma(E_{d}-E_{1s}-\frac{\Gamma_{d}}{2})}{\sigma(E_{d}-E_{1s})}\, .
\end{eqnarray}
Calculation of this parameter for $d=2$ in the case of
recombination with ${\rm U}^{91+}$ using the results of Refs.
 \cite{kar92,yer93} yields
$\delta\approx -0.17$. The contributions to $\delta$
 from the terms $\overline
{\sigma}_{2}$ and $\overline{\sigma}_{4}$ are
equal $-0.24$ and $0.07$, respectively.

Summing the interference terms in (\ref{resrec4}),
we obtain a compact formula
for $\sigma$,
\begin{eqnarray} \label{resrec5}
\sigma(p_i^0)&=&\frac{2\pi^{2}}{2(2j'+1)}\frac{1}{|{\bf p}_i|^{2}}
\sum_{jlJM}\Biggl \{
\sum_{d}\frac{\Gamma_{d}{\rm Re}\,(\la i_{J}|\hat{I}|d_{R}\ra
\la d_{L}|\hat{I}|i_{J}\ra)}{(E_{1s}+p_i^0-E_{d})^{2}+
\frac{\Gamma_{d}^{2}}{4}}\nonumber\\
&&-2\sum_{d}\frac{{\rm Im}\,(\la i_{J}|\hat{I}|d_{R}\ra
\la d_{L}|\hat{I}|i_{J}\ra)(E_{1s}+p_i^0 -E_{d})}
{(E_{1s}+p_i^0-E_{d})^{2}+
\frac{\Gamma_{d}^{2}}{4}}+W_{i_{J}i_{J}}\Biggr \}\, .
\end{eqnarray}
The formulas (\ref{resrec4}) and (\ref{resrec5})
 are more convenient for the numerical
calculations than formula (\ref{resrec1}),
since they do not require any calculation of
 the radiative amplitudes. These formulas
were employed in Ref. \cite{nef94} to calculate
the resonance recombination of an electron
with hydrogenlike lead.

\section{Conclusion}

In the present paper we have considered in detail the two-time Green
function method for high-Z few-electron systems. 
This method allows one to formulate a perturbation theory
for calculations of various physical quantities in a rigorous and
systematic way in the framework of quantum electrodynamics.
To demonstrate 
the efficiency of the method, we have
derived formulas for QED and interelectronic-interaction
 corrections to the energy levels, transition
and scattering amplitudes in one-, two-, and three-electron atoms.

For the last decade, the TTGF method was intesively employed
 in calculations of QED effects in heavy ions. An overview
of these calculations was also given in the present paper.
Details of the calculations and other applications of the 
method can be found in Refs.
\cite{sh93,sh94a,sh94b,sh95,yer97,art97,sh98,art99,yer99,art00,sh00,yer00,y0a}.
In particular, in  Refs. \cite{art99,yer99} 
the vacuum-polarization and
 self-energy screening corrections
to the energies of lithiumlike ions were calculated.
The two-photon exchange diagrams for
lithiumlike ions were evaluated in Ref. \cite{y0a}.
The second-order two-electron diagrams for quasidegenerate
states of heliumlike ions are studied in 
\cite{art00,leb}.
In Ref. \cite{sh93}, the TTGF method was employed to construct
an effective-energy operator for a high-Z few-electron atom.
In Ref. \cite{sh00}, this method was used to evaluate
the QED corrections to the radiative recombination of an electron
with a bare nucleus. The interelectronic-interaction corrections
to the radiative electron capture for a heliumlike
ion were considered in Ref. \cite{yer00}. In Ref. \cite{ind00}, the 
interelectronic-interaction corrections to the transition probabilities
 in heliumlike ions are derived.

Concluding, the two-time Green function method provides 
a uniform and very efficient
 approach for deriving  QED and interelectronic-interaction
 corrections
to  energy levels, transition probabilities, and  cross sections
of scattering processes in high-Z few-electron atoms.
Using an effective potential  instead of the Coulomb potential
of the nucleus allows one to extend this approach to
many-electron atoms.

\section*{ Acknowledgements}
Many practical calculations by the TTGF method were
carried out in collaboration with
 Anton  Artemyev and Vladimir Yerokhin. Stimulating
discussions with  T. Beier, E.-O. Le Bigot, J. Eichler,
P. Indelicato, U. Jentschura, S. Karshenboim, I. Lindgren,
 P. Mohr, G. Plunien, J. Sapirstein, G. Soff, and S. Zschocke 
are gratefully acknowledeged.
Valuable conversations with O. Andreev, D. Arbatsky,  
I. Bednyakov, M. Sysak, and O. Zherebtsov 
are also acknowledged.
This work was supported in part by RFBR (Grant No. 
98-02-18350, Grant No. 98-02-0411, and Grant No. 01-02-17248),
 by the program
"Russian Universities - Basic Research" (project No. 3930),
by DFG (Grant No. 436 RUS 113/616/0-1), and by GSI.
 
\newpage 
\appendix
\section{QED in the Heisenberg representation}
In the Heisenberg representation, the basic equations of  quantum
electrodynamics in the presence of a classical time-independent field 
$A_{\rm cl}^{\nu}(\bfx)$ are
\be
(i\not\stackrel{}{\partial}-m-e\not\stackrel{}{A}_{\rm cl}(\bfx))
\psi(x)&=&e\not\stackrel{}{A}(x)\psi(x)-\delta m\psi(x)\,,
\nonumber\\
\Box A_{\nu}(x)&=&j_{\nu}(x)\,,
\ee
where $j_{\nu}(x)=(e/2)[\overline{\psi}(x)\gamma_{\nu},\psi(x)]$
is the electron-positron current operator. The state vectors
in the Heisenberg representation are time-independent
\be
\partial_t|\Phi\ra=0\,.
\ee
The physical state vectors have to obey a subsidiary condition
\be
(\partial_{\nu}A^{\nu}(x))^{(+)}|\Phi\ra=0\,,
\ee
where $(\partial_{\nu}A^{\nu}(x))^{(+)}$ is the positive-frequency part
of $\partial_{\nu}A^{\nu}(x)\,$. The Heisenberg operators
$\psi(x)$, $\overline{\psi}(x)$, and $A^{\nu}(x)$ obey the same equal-time
permutation relations as the corresponding free-field operators.
However, in contrast to the free fields, the permutation relations
for arbitrary times remain unknown. Due to the time-translation invariance,
Heisenberg operators obey the following transformation equation
\be
\exp{(iHt)}F(0,\bfx)\exp{(-iHt)}=F(t,\bfx)\,,
\ee
where $H$ is the Hamiltonian of the system in the Heisenberg
representation. For more details, see \cite{schweber61,bjorken,akhiezer}.

\section{Singularities of the two-time Green function in a finite
order of perturbation theory}

Let us investigate the singularities of the Green function ${\cal G}(E)$
in a finite order of perturbation theory. To $m$-th order 
in $e$, which corresponds to order $m/2$ in $\alpha$,
 it is given by
\begin{eqnarray} \label{greenB1}
\lefteqn{{\cal G}^{(m/2)}(E;\bfx_{1}^{\prime},
\dots \bfx_{N}^{\prime};\bfx_{1},\dots \bfx_{N})\delta(E-E')
\;\;\;\;\;\;\;\;\;\;\;\;\;\;\;\;\;\;\;}\nonumber\\
&=&\frac{1}{2\pi i}\frac{1}{N!}
\int_{-\infty}^{\infty} dt dt'\;\exp{(iE't'-iEt)}
\frac{(-i)^m}{m!}e^m
\int d^4y_1\cdots d^4y_m\nonumber\\
&&\times
\langle 0|T\psi_{\rm in}(t',\bfx_{1}^{\prime})\cdots
\psi_{\rm in}(t',\bfx_{N}^{\prime})
\overline{\psi
}_{\rm in}
(t,\bfx_{N})\cdots\overline{\psi}_{\rm in}(t,\bfx_{1})\nonumber\\
&&\times
\overline{\psi}_{\rm in}(y_1)\gamma_{\rho}\psi_{\rm in}(y_1)
A_{\rm in}^{\rho}(y_1)\cdots
\overline{\psi}_{\rm in}(y_m)\gamma_{\sigma}\psi_{\rm in}(y_m)
A_{\rm in}^{\sigma}(y_m)|0\ra_{\rm con}\,,
\end{eqnarray}
where the label "con" means that disconnected vacuum-vacuum
subdiagrams must be omitted.
For simplicity, we omit here the mass renormalization counterterm.
The presence of this term would not change the consideration
given below.

Let us consider the contribution of a diagram of $m$-th order
in $e$. This diagram is defined by a certain order of contractions
in equation (\ref{greenB1}). The contractions between the
 electron-positron fields and between the photon fields
give the propagators (\ref{propel}) and (\ref{propph}),
respectively. In this Appendix we will use the following
representation for these propagators
\be \label{propelB}
\la 0|T\psi_{\rm in}(x)\overline{\psi}_{\rm in}(y)|0\ra&=&
\theta(x^0-y^0)\sum_{\veps_n>0}\psi_n(\bfx)\overline{\psi}_n(\bfy)
\exp{[-i\veps_n(x^0-y^0)]}\nonumber\\
&&-\theta(y^0-x^0)\sum_{\veps_n<0}\psi_n(\bfx)\overline{\psi}_n(\bfy)
\exp{[-i\veps_n(x^0-y^0)]}\,, \\
\la 0|T A_{\rm in}^{\rho}(x)A_{\rm in}^{\sigma}(y)|0\ra &=&
-g^{\rho \sigma}\int \frac{d\bfk}{(2\pi)^3}\;
\frac{\exp{[-i\sqrt{\bfk^2+\mu^2}|x^0-y^0|]}\exp{[i\bfk\cdot(\bfx-\bfy)]}}
{2\sqrt{\bfk^2+\mu^2}}\,.
\label{propphB}
\ee
Here, by definition, $\theta(t)=(t+|t|)/(2t)$ for $t\ne 0$
and $\theta(0)=1/2$, and a non-zero photon mass is introduced.
Following  \cite{braun84,shir81}, 
 we will use the formalism of time-ordered
diagrams \cite{vanhove,hugenholtz}
to investigate the singularities
of ${\cal G}^{(m/2)}(E)$.
Let us consider a certain order of the time variables,
$$
y_{i_m}^0>y_{i_{m-1}}^0>\cdots y_{i_l}^0>t'>y_{i_{l-1}}^0>
\cdots y_{i_s}^0>t>y_{i_{s-1}}^0>\cdots y_{i_1}^0 
$$
which defines a time-ordered version of the Feynman diagram.
The contribution of the Feynman diagram is the sum
of all time-ordered versions. Each time-ordered version is
conveniently represented by a diagram in which the vertices
are ordered  upwards according to  
increasing time (see, for example, Fig. 39). According to
equations (\ref{propelB}) and (\ref{propphB}),
each electron propagator contains a sum over the whole 
electron-energy spectrum and each photon propagator
contains an integral over the photon momentum.
Let us place these sums and integrals in front of the expression
for the time-ordered version of the Feynman diagram under
consideration. Then, an electron line is characterized
by an electron energy $\veps_n$ and, according to (\ref{propelB}),
gives a factor  (here we are only interested  
in time-dependent terms) 
$$
\exp{[-i\veps_n(y_i^0-y_k^0)]}=
\exp{[-i|\veps_n|(y_i^0-y_k^0)]}
\,\;\;\;\;{\rm for}\;\;\;\veps_n>0
$$
and
$$
-\exp{[i\veps_n(y_i^0-y_k^0)]}=
-\exp{[-i|\veps_n|(y_i^0-y_k^0)]}
\,\;\;\;\;{\rm for}\;\;\;\veps_n<0\,,
$$
where in both cases we consider $y_i^0>y_k^0$.
A photon line gives a factor
$$
\exp{[-i\sqrt{\bfk^2+\mu^2}\,(y_i^0-y_k^0)]}\,,
$$
where we consider again $y_i^0>y_k^0$.
Each time $y_i^0$ in the diagram is marked by a horizontal dashed
line. These lines may intersect other electron and photon
lines (see Fig. 39). For the point of intersection with an 
electron line, we introduce a factor 
$\exp{(i|\veps_n|y_i^0)}\exp{(-i|\veps_n|y_i^0)}=1$,
where $\veps_n$ is the energy of the intersected electron.
For the point of intersection with a photon line we introduce
a factor $\exp{(i\sqrt{\bfk^2+\mu^2}\,y_i^0)}
\exp{(-i\sqrt{\bfk^2+\mu^2}\,y_i^0)}=1$, where $\bfk$ is the momentum
of the intersected photon.
In addition, we represent the factor $\exp{(iE't')}\exp{(-iEt)}$
as
\be
\exp{(iE't')}\exp{(-iEt)}&=&\exp{[i(E'-E)y_{i_m}^0]}\exp{[-i(E'-E)
y_{i_m}^0]} \cdots \nonumber\\
&&\times \exp{[i(E'-E)y_{i_l}^0]}\exp{[-i(E'-E)
y_{i_l}^0]}\nonumber\\
&&\times \exp{(iE't')}\exp{(-iEt')}\exp{(iEt')}
\exp{(-iEy_{i_{l-1}}^0)}\nonumber\\
&&\times \exp{(iEy_{i_{l-1}}^0)}
\exp{(-iEy_{i_{l-2}}^0)}\cdots \exp{(iEy_{i_s}^0)}
\exp{(-iEt)}\,.
\ee
As a result of all these representations, the integral over times
at fixed intermediate electron states ($n$) and  photon
momenta ($\bfk$) is
\be
I_m & \equiv& \int_{-\infty}^{\infty}dy_{i_m}^0 \;
\int_{-\infty}^{y_{i_m}^0}dy_{i_{m-1}}^0\cdots
\int_{-\infty}^{y_{i_l}^0}dt'\;
\int_{-\infty}^{t'}dy_{i_{l-1}}^0 \cdots
\int_{-\infty}^{y_{i_s}^0}dt \;
\int_{-\infty}^{t}dy_{i_{s-1}}^0\cdots
\int_{-\infty}^{y_{i_2}^0}dy_{i_{1}}^0\nonumber\\
&&\times \exp{[i(E'-E)y_{i_m}^0]}
\exp{[i(E-E'-\sum_{(m)}|\veps_n|-\sum_{(m)}\sqrt{\bfk^2+\mu^2})
(y_{i_m}^0-y_{i_{m-1}}^0)]}\cdots\nonumber\\
&&\times
\exp{[i(E-E'-\sum_{(l)}|\veps_n|-\sum_{(l)}\sqrt{\bfk^2+\mu^2})
(y_{i_l}^0-t')]}\nonumber\\
&&\times
\exp{[i(E-\sum_{(t')}|\veps_n|-\sum_{(t')}\sqrt{\bfk^2+\mu^2})
(t'-y_{i_{l-1}}^0)]}\cdots\nonumber\\
&&\times
\exp{[i(E-\sum_{(s)}|\veps_n|-\sum_{(s)}\sqrt{\bfk^2+\mu^2})
(y_{i_s}^0-t)]}\nonumber\\
&&\times
\exp{[i(-\sum_{(t)}|\veps_n|-\sum_{(t)}\sqrt{\bfk^2+\mu^2})
(t-y_{i_{s-1}}^0)]}\cdots\nonumber\\
&&\times
\exp{[i(-\sum_{(2)}|\veps_n|-\sum_{(2)}\sqrt{\bfk^2+\mu^2})
(y_{i_2}^0-y_{i_{1}}^0)]}\,.
\ee
Here $\sum_{(m)}|\veps_n|$ denotes the sum of the electron 
energies from the electron lines which are sandwiched
between the horizontal lines corresponding to the times $y_{i_m}^0$
and $y_{i_{m-1}}^0$. $\sum_{(m)}\sqrt{\bfk^2+\mu^2}$
denotes  the sum of the photon
energies from the photon lines which are sandwiched
between the horizontal lines corresponding to the times $y_{i_m}^0$
and $y_{i_{m-1}}^0$. 
Using the identity
\be
\int_{-\infty}^{0}dx\;\exp{(-i\alpha x)}=\frac{i}{\alpha+i0}\,,
\ee
we easily find
\be \label{b7}
I_m&=&2\pi \delta(E-E') \frac{i}
{-\sum_{(m)}|\veps_n|-\sum_{(m)}\sqrt{\bfk^2+\mu^2}}
\cdots \frac{i}{-\sum_{(l)}|\veps_n|-\sum_{(l)}\sqrt{\bfk^2+\mu^2}}
\nonumber\\
&&\times \frac{i}{E-\sum_{(t')}|\veps_n|-\sum_{(t')}\sqrt{\bfk^2+\mu^2}
+i0}\cdots
\frac{i}{E-\sum_{(s)}|\veps_n|-\sum_{(s)}\sqrt{\bfk^2+\mu^2}
+i0}\nonumber\\
&&\times
\frac{i}{-\sum_{(t)}|\veps_n|-\sum_{(t)}\sqrt{\bfk^2+\mu^2}}
\cdots
\frac{i}{-\sum_{(2)}|\veps_n|-\sum_{(2)}\sqrt{\bfk^2+\mu^2}}\,.
\ee
A similar calculation for $t'<t$ yields an expression which
is obtained from (\ref{b7}) by a replacement $E\rightarrow -E$
in all the denominators.

Because each photon line contracts two
vertices, at least $m/2$ denominators
in (\ref{b7}) have to contain the photon-energy terms and
therefore do not contribute to the singularities under
consideration. It follows that ${\cal G}^{(m/2)}(E)$
has isolated poles of all orders till $m/2+1$ at the unperturbed
positions of the bound state energies. The separation of these
poles from the related cuts is provided by keeping a non-zero
photon mass $\mu$. As to the cuts starting from the lower
energy levels, they can be turned down.

\section{Two-time Green function in terms of the Fourier transform
of the $2N$-time Green function}
To prove the equation (\ref{e2e3}) we have to show that
\begin{eqnarray}       
 {\cal G}(E) \delta(E-E')  &=&
  \frac{2\pi}{i}\frac{1}{N!}
           \int_{-\infty}^{\infty}dp_{1}^{0}\cdots dp_{N}^{0}
    dp_{1}^{\prime 0}\cdots dp_{N}^{\prime 0}\nonumber\\
&& \times \delta(E-p_{1}^{0} -\cdots - p_{N}^{0})
            \delta(E'-p_{1}^{\prime 0} -\cdots - p_{N}^{\prime 0}) 
\nonumber\\  
 && \times  
  G(p_{1}^{\prime 0} ,\ldots ,p_{N}^{\prime 0};p_{1}^{0},
\ldots ,p_{N}^{0}) \,,
       \label{C1}
\end{eqnarray}
where the coordinate variables are omitted, for brevity.
According to the definition of $G$ (see equation (\ref{green2np})),
 equation (\ref{C1}) is equivalent to
\begin{eqnarray}       
 {\cal G}(E) \delta(E-E')  &=&
  \frac{2\pi}{i}\frac{1}{N!}
           \int_{-\infty}^{\infty}dp_{1}^{0}\cdots dp_{N}^{0}
    dp_{1}^{\prime 0}\cdots dp_{N}^{\prime 0}\nonumber\\
&& \times \delta(E-p_{1}^{0} -\cdots - p_{N}^{0})
            \delta(E'-p_{1}^{\prime 0} -\cdots - p_{N}^{\prime 0}) 
\nonumber\\  
&&\times (2\pi)^{-2N} \int_{-\infty}^{\infty}
  dx_{1}^{0}\cdots dx_{N}^{0} dx_{1}^{\prime 0}\cdots dx_{N}^{\prime 0} 
      \nonumber  \\ & &
   \times \exp{(ip_{1}^{\prime 0}x_{1}^{\prime 0}+
\cdots+ip_{N}^{\prime 0}x_{N}^{\prime 0}-
  ip_{1}^{0}x_{1}^{0}-\cdots-ip_{N}^{0}x_{N}^{0} )} 
   \nonumber \\
 &&  \times 
\langle 0|  T\psi(x_{1}^{\prime}) \cdots \psi(x_{N}^{\prime})
   \overline{\psi}(x_{N}) \cdots 
   \overline{\psi}(x_{1})|0\rangle \nonumber\\
&=&
(2\pi)^{-2N} \frac{2\pi}{i}\frac{1}{N!}
           \int_{-\infty}^{\infty}dp_2^0\cdots dp_{N}^{0}
    dp_{2}^{\prime 0}\cdots dp_{N}^{\prime 0}\nonumber\\
&&\times \int_{-\infty}^{\infty}
  dx_{1}^{0}\cdots dx_{N}^{0} dx_{1}^{\prime 0}\cdots dx_{N}^{\prime 0} 
      \nonumber  \\ & &
   \times \exp{[i(E'-p_{2}^{\prime 0}-\cdots-p_N^{\prime 0})
 x_{1}^{\prime 0}+ip_2^{\prime 0}x_2^{\prime 0}
\cdots+ip_{N}^{\prime 0}x_{N}^{\prime 0}]}\nonumber\\
&&\times \exp{[-i(E-p_{2}^{0}-\cdots-p_N^{0})
 x_{1}^{0}
 -ip_{2}^{0}x_{2}^{0}-\cdots-ip_{N}^{0}x_{N}^{0} ]} 
   \nonumber \\
 &&  \times 
\langle 0|  T\psi(x_{1}^{\prime}) \cdots \psi(x_{N}^{\prime})
   \overline{\psi}(x_{N}) \cdots 
   \overline{\psi}(x_{1})|0\rangle \,.
  \label{C2}
\end{eqnarray}
Using the identity
\be
\frac{1}{2\pi}\int_{-\infty}^{\infty}d\omega \; \exp{(i\omega x)}
=\delta(x)\,,
\ee
we obtain
\begin{eqnarray}       
 {\cal G}(E) \delta(E-E')  &=&
(2\pi)^{-2} \frac{2\pi}{i}\frac{1}{N!}
          \int_{-\infty}^{\infty}
  dx_{1}^{0}\cdots dx_{N}^{0} dx_{1}^{\prime 0}\cdots dx_{N}^{\prime 0}\; 
 \delta(x_1'^0-x_2'^0)\cdots\delta(x_1'^0-x_N'^0)\nonumber\\
&&\times \delta(x_1^0-x_2^0)\cdots\delta(x_1^0-x_N^0)
\exp{(iE'x_1'^0-iEx_1^0)}  \nonumber \\
 &&  \times 
\langle 0|  T \psi(x_{1}^{\prime}) \cdots \psi(x_{N}^{\prime})
   \overline{\psi}(x_{N}) \cdots 
   \overline{\psi}(x_{1})|0\rangle \nonumber\\
&=&
\frac{1}{2\pi i}\frac{1}{N!}
          \int_{-\infty}^{\infty}
  dx^{0}dx^{\prime 0} \;
\exp{(iE'x'^0-iEx^0)}  \nonumber \\
 &&  \times 
\langle 0|  T\psi(x'^0,\bfx_{1}^{\prime})
 \cdots \psi(x'^0,\bfx_{N}^{\prime})
   \overline{\psi}(x^0,\bfx_{N}) \cdots 
   \overline{\psi}(x^0,\bfx_{1})|0\rangle \,.
  \label{C3}
\end{eqnarray}
The last equation exactly coincides with the definition
of ${\cal G}(E)$ given by (\ref{gfirst}).

\section{Matrix elements of the two-time Green function
between one-determinant wave functions}
 
To derive the equation (\ref{e2e13}) we use the following two
identities. First, if $A$ is a symmetric operator in the coordinates
of all electrons, we obtain (see, e.g., \cite{bethe})
\begin{eqnarray}
 A_{ik}\equiv \langle u_{i}|A|u_{k} \rangle &=& 
         \sum_{P}(-1)^{P}
         \psi_{P i_{1}}^{*}(\xi_{1}^{\prime})\cdots
         \psi_{P i_{N}}^{*}(\xi_{N}^{\prime})
    A(\xi_{1}^{\prime},\ldots,\xi_{N}^{\prime};\xi_{1},\ldots,\xi_{N})
\nonumber\\       
&&\times  \psi_{k_{1}}(\xi_{1})\cdots\psi_{k_{N}}(\xi_{N})\,,
  \label{a1}
\end{eqnarray}
where repeated variables $\{\xi\}$ imply integration (the integration
over ${\bf x}$ and the summation over $\alpha$ ) and 
$A(\xi_{1}^{\prime},\ldots,\xi_{N}^{\prime};\xi_{1},\ldots ,\xi_{N})$ is
the kernel of the operator $A$. Second, if the kernel of the operator
$A$ is represented in the form 
\begin{equation}
  A(\xi_{1}^{\prime},\ldots,\xi_{N}^{\prime};\xi_{1},\ldots ,\xi_{N})=
         \sum_{Q}(-1)^{Q}
    a(\xi_{Q1}^{\prime},\ldots,\xi_{QN}^{\prime};
\xi_{1},\ldots ,\xi_{N})\,,
 \label{a2}
\end{equation}
 we can find
\begin{eqnarray}
 A_{ik} &=& N! \sum_{P}(-1)^{P}
         \psi_{P i_{1}}^{*}(\xi_{1}^{\prime})\cdots
         \psi_{P i_{N}}^{*}(\xi_{N}^{\prime})
    a(\xi_{1}^{\prime},\ldots,\xi_{N}^{\prime};\xi_{1},\ldots ,\xi_{N})
\nonumber\\ 
&&\times   \psi_{k_{1}}(\xi_{1})\cdots\psi_{k_{N}}(\xi_{N})\,.
  \label{a3}
\end{eqnarray}
According to (\ref{e2e12}), we have
\begin{eqnarray}
{\cal G}(E)\gamma_1^0\cdots \gamma_N^0
\delta(E-E^{\prime}) &=&\frac{2\pi}{i}\frac{1}{N!}
          \int_{-\infty}^{\infty}dp_{1}^{0}\cdots dp_{N}^{0}
	     dp_{1}^{\prime 0}\cdots dp_{N}^{\prime 0}
\nonumber \\ 
&&  \times 
             \delta(E-p_{1}^{0} -\cdots - p_{N}^{0})
             \delta(E^{\prime}-p_{1}^{\prime 0} -\cdots - p_{N}^{\prime 0}) 
\nonumber \\ 
&& \times    
        \sum_{P}(-1)^{P}
     \widehat{G}((p_{P1}^{\prime 0},\xi_{P1}^{\prime})  ,
          \ldots ,(p_{PN}^{\prime 0},\xi_{PN}^{\prime});
          (p_{1}^{0},\xi_{1}),
           \ldots ,(p_{N}^{0},\xi_{N}))
\nonumber \\ 
& =& \frac{2\pi}{i}\frac{1}{N!}
       \sum_{P}(-1)^{P}
          \int_{-\infty}^{\infty}dp_{1}^{\prime 0}\cdots dp_{N}^{\prime 0}
	     dp_{1}^{0}\cdots dp_{N}^{0}
\nonumber \\
 && \times 
             \delta(E-p_{1}^{0} -\cdots - p_{N}^{0})
             \delta(E^{\prime}-p_{1}^{\prime 0} -\cdots - p_{N}^{\prime 0}) 
\nonumber \\ 
&& \times    
     \widehat{G}((p_{1}^{\prime 0},\xi_{P1}^{\prime})  ,
          \ldots ,(p_{N}^{\prime 0},\xi_{PN}^{\prime});
          (p_{1}^{0},\xi_{1}),
           \ldots ,(p_{N}^{0},\xi_{N}))
\nonumber \\ 
& \equiv & \sum_{P}(-1)^{P}\tilde{G}(\xi_{P1}^{\prime},
          \ldots ,\xi_{PN}^{\prime};
          \xi_{1},\ldots ,\xi_{N})\,.
    \label{a4}
\end{eqnarray}
Using (\ref{a1})-(\ref{a4}) we easily obtain (\ref{e2e13}).

\section{Double spectral representation for the two-time
Green function describing a transition process}

Let us consider the function $G(E',E)$ defined as
\be
G(E',E)=\int_{-\infty}^{\infty} dt dt'\; \exp{(iE't'-iEt)}
\la 0|T A(t') B(0) C(t)|0\ra\,.
\ee
Using the transformation rules for the Heisenberg operators
and integrating over the time variables, we can derive
the following double spectral representation for $G(E',E)$
\be
G(E',E)&=& -\int_{-\infty}^{\infty} dW'dW \;
\Bigl[\frac{K(W',W)}{(E'-W')(E-W)}+\frac{L(W',W)}{(E'+W')(E+W)}
\Bigr]\nonumber\\
&&+\int_{-\infty}^{\infty} dW'd\omega \;
\Bigl[\frac{M(W',\omega)}{(E'-W')(k^0+\omega)}
+\frac{N(W',\omega)}{(E'+W')(k^0-\omega)}
\Bigr]\nonumber\\
&& -\int_{-\infty}^{\infty} d\omega dW \;
\Bigl[\frac{P(\omega,W)}{(k^0-\omega)(E-W)}+\frac{Q(\omega,W)}
{(k^0+\omega)(E+W)}
\Bigr]\,,\nonumber\\
\ee
where $k^0=E-E'$,
\be
K(W',W)&=&\sum_{n,m}\delta(W'-E_n)\delta(W-E_m)\la 0|A(0)|n\ra
\la n|B(0)|m\ra \la m|C(0)|0\ra\,, \\
L(W',W)&=&\sum_{n,m}\delta(W'-E_n)\delta(W-E_m)\la 0|C(0)|n\ra
\la n|B(0)|m\ra \la m|A(0)|0\ra\,, \\
M(W',\omega)&=&\sum_{n,m}\delta(W'-E_n)\delta(\omega-E_m)
\la 0|A(0)|n\ra
\la n|C(0)|m\ra \la m|B(0)|0\ra\,, \\
N(W',\omega)&=&\sum_{n,m}\delta(W'-E_m)\delta(\omega-E_n)\la 0|B(0)|n\ra
\la n|C(0)|m\ra \la m|A(0)|0\ra\,, \\
P(\omega,W)&=&\sum_{n,m}\delta(\omega-E_n)\delta(W-E_m)\la 0|B(0)|n\ra
\la n|A(0)|m\ra \la m|C(0)|0\ra\,, \\
Q(\omega,W)&=&\sum_{n,m}\delta(\omega-E_m)\delta(W-E_n)\la 0|C(0)|n\ra
\la n|A(0)|m\ra \la m|B(0)|0\ra\,.
\ee

\newpage

\begin{table}
\centering
\caption{The ground-state Lamb shift 
in $^{238}{\rm U}^{91+}$, in eV.}
\begin{tabular}{l c } \hline
Point nucleus binding energy & $-$132279.92(1)  \\ 
Finite nuclear size \cite{yer97,franosch91} 
 &       198.81(38) \\ 
First-order SE  \cite{mohr93}& 355.05   \\                   
First-order VP \cite{pers93}  &    $-$88.60\\ 
Second-order QED   &      $\pm$ 1.5\\ 
Nuclear recoil \cite{sh98recpra} &  0.46   \\ 
Nuclear polarization \cite{plunien95,nefiodov96}     &    $-$0.20(10)\\
Lamb shift theory  &  465.52(39) $\pm$ 1.5 \\
Lamb shift experiment \cite{stoehlker00}   &468(13)\\
 \hline
\end{tabular}
\end{table}

\begin{table}
\centering
\caption{The two-electron contribution  to the ground-state 
energy in  $^{209}{\rm Bi}^{81+}$, in eV.}
\begin{tabular}{l d } \hline
One-photon exchange contribution & 1897.56(1)  \\ 
Two-photon exchange  &       \\ 
within the Breit approximation  &     $-$10.64\\ 
Two-photon exchange &  \\ 
beyond the Breit approximation & $-$0.30(1)   \\
SE screening & $-$6.73 \\
VP screening & 1.55 \\
Three- and more photon contribution & 0.06(7)\\ 
Total theory \cite{yer97}  &  1881.50(7)   \\
Experiment \cite{marrs95}  &1876(14)\\
 \hline
\end{tabular}
\end{table}

\begin{table}
\centering
\caption{The $2p_{1/2}-2s$ transition energy  
 in $^{238}{\rm U}^{89+}$, in eV.}
\begin{tabular}{l c} \hline
One-photon exchange \cite{yer99} &     368.83 \\
One-electron nuclear size \cite{yer99}  & $-$33.35(6)  \\ 
First-order QED \cite{pers93,mohr93}  & $-$42.93\\
Two-photon exchange within & \\
the Breit appoximation \cite{y0a} & $-$13.54\\
Two-photon exchange beyond & \\
the Breit approximation\cite{y0a} &   0.17\\ 
Self energy screening \cite{yer99} &  1.52   \\ 
Vacuum polarization screening \cite{art99} & $-$0.36 \\
Three- and more photon exchange \cite{sh00hi} & 0.16(7)\\
Nuclear recoil \cite{artemyev95a} &$-$0.07\\
Nuclear polarization \cite{plunien95,nefiodov96}     &  
  0.03(1)\\
One-electron second-order QED & $\pm$0.20\\
Total theory  &  280.46(9) $\pm$ 0.20    \\
Experiment \cite{schweppe91}   &280.59(10)\\
 \hline
\end{tabular}
\end{table}

\footnotesize
\begin{table}
\caption
{Theoretical contributions to the ground state
 hyperfine splitting in hydrogenlike ions (in eV).}
\begin{tabular}{ l c c c c c}         \hline
Ion &$^{165}$Ho$^{66+}$&$^{185}$Re$^{74+}$ &$^{187}$Re$^{74+}$&
$^{207}$Pb$^{81+}$&$^{209}$Bi$^{82+}$\\ \hline   
$\mu/\mu_{N}$ &4.177(5)&3.1871(3) &3.2197(5) &0.592583(9)&4.1106(2)\\ 
Relativistic value &&&&&\\
for point nucleus &2.326(3)&3.010&3.041 &1.425 & 5.839\\
Finite nuclear charge & $-$0.106(1)& $-$0.213(2) &$-$0.215(2)&
$-$0.149&$-$0.649(2)\\
Bohr-Weisskopf effect& $-$0.020(6)&$-$0.034(10)&
$-$0.035(10)&$-$0.053(5)&$-$0.061(27)\\ 
QED&$-$0.011&$-$0.015&$-$0.015&$-$0.007&$-$0.030\\  
Total theory \cite{shab00hf}
&2.189(7)&2.748(10)&2.776(10)&1.215(5)&5.100(27)\\ 
Experiment \cite{klaft1,crespo1,crespo2,seelig1}
&2.1645(5) &2.719(2)&2.745(2)&1.2159(2)&5.0840(8)\\ \hline
\end{tabular}
\end{table}
\normalsize

 \begin{table}
\caption{The bound-electron $g$ factor in  $^{209}{\rm Bi}^{82+}$.}
\begin{tabular}{ l l } \hline
Relativistic value& 1.7276 \\
QED&0.0029\\
Nuclear size correction &0.0005\\
Total theory \cite{shab00hf} & 1.7310\\
Experiment \cite{winter1}&1.7343(33)\\ \hline
\end{tabular}
\end{table}

\begin{table}
\caption{The relative values of the
 QED corrections to the total cross section for the
radiative recombination into the K-shell of bare uranium
[45],
 expressed in \%.}
\begin{tabular}{c|c d d}
Impact energy [MeV/u]&
Correction & Vacuum polarization, in \% & Self energy, in \%\\
\hline 
100 &$\sigma^{(1)}_{\rm en}+\sigma^{(1)}_{\rm bw}$
&0.126&$-$0.390\\ 
&$\sigma^{(1)}_{\rm cw}$&$-$0.006&?\\ 
&Total&0.120&$-$0.390\\ \hline 
300 &$\sigma^{(1)}_{\rm en}+
\sigma^{(1)}_{\rm bw}$&0.175&$-$0.513\\ 
&$\sigma^{(1)}_{\rm cw}$&$-$0.003&?\\ 
&Total&0.173&$-$0.513\\ \hline
1000 &$\sigma^{(1)}_{\rm en}+
\sigma^{(1)}_{\rm bw}$&0.220&$-$0.591\\ 
&$\sigma^{(1)}_{\rm cw}$&0.043&?\\ 
&Total&0.263&$-$0.591\\ \hline
\end{tabular}

\end{table}

\begin{table}
\caption{The zeroth-order cross section $\sigma^{(0)}$ and the first-order
interelectronic-interaction correction
calculated in Ref. [46], in barns.
 $\sigma^{(1)}_{\rm scr}$ denotes the interelectronic-interaction
correction calculated in the screening potential approximation
and  $\sigma^{(1)}_{\rm int}$
indicates the results of the rigorous relativistic calculation. }
\begin{tabular}{dddd}
 Impact energy [Mev/u]
    &$\sigma^{(0)}$
             &$\sigma^{(1)}_{\rm scr}$
                                   & $\sigma^{(1)}_{\rm int}$
                                                      \\  \hline
\multicolumn{4}{l}{$2s$-state:} \\
100 & 41.203   &  $-$1.393   &  $-$2.055    \\
300 &  9.105  &  $-$0.3345  &  $-$0.3755    \\
700 &  2.457  &  $-$0.0979  &  $-$0.1051    \\
        \hline
\multicolumn{4}{l}{$2p_{1/2}$-state:} \\
100 & 33.041  &  $-$2.535   &  $-$3.088   \\
300 &  5.042  &  $-$0.4538  &  $-$0.3864  \\
700 &  1.065  &  $-$0.1022  &  $-$0.0861  \\
        \hline
\multicolumn{4}{l}{$2p_{3/2}$-state:} \\
100 & 31.489   & $-$2.275   &  $-$2.896   \\
300 &  3.646  & $-$0.3132  &  $-$0.2804  \\
700 &  0.622  & $-$0.0568  &  $-$0.0489  \\
\end{tabular}
\end{table}

\newpage

 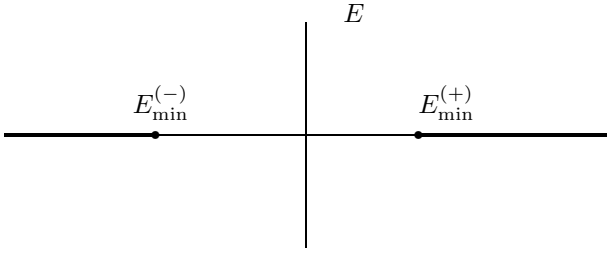
\begin{figure}
\setlength{\unitlength}{1mm}\thinlines
\begin{picture}(100,50)
\put(50,10){\line(0,1){30}}
\put(10,25){\line(1,0){80}}
\put(30,25){\circle*{1}}
\put(65,25){\circle*{1}}
\thicklines
\put(10,25){\line(1,0){20}}
\put(65,25){\line(1,0){25}}
\put(55,40){$E$}
\put(27,28){$E_{\rm min}^{(-)}$}
\put(65,28){$E_{\rm min}^{(+)}$}
\end{picture} 
\caption{Singularities of the two-time Green function in the complex
$E$ plane.}
 \end{figure}

 \begin{figure}
\setlength{\unitlength}{1mm}\thinlines
\begin{picture}(100,50)
\put(10,10){\line(0,1){30}}
\put(0,25){\line(1,0){70}}
\put(30,25){\circle*{1}}
\put(50,25){\circle*{1}}
\put(65,25){\circle*{1}}
\thicklines
\put(15,40){$E$}
\end{picture} 
\caption{Singularities of the two-time Green function 
in the bound-state region,
if the interaction between the electron-positron
 field and the electromagnetic field is switched off.}
 \end{figure}
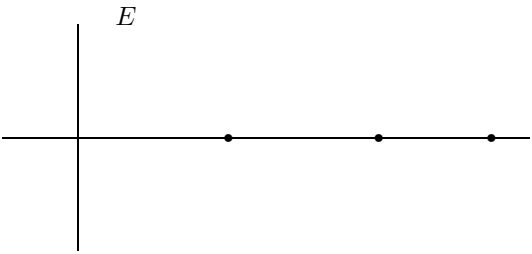

 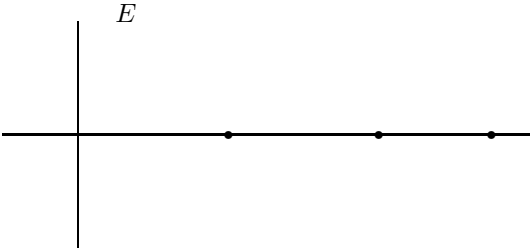
\begin{figure}
\setlength{\unitlength}{1mm}\thinlines
\begin{picture}(100,50)
\put(10,10){\line(0,1){30}}
\put(0,25){\line(1,0){70}}
\put(30,25){\circle*{1}}
\put(50,25){\circle*{1}}
\put(65,25){\circle*{1}}
\thicklines
\put(30,25){\line(1,0){40}}
\put(15,40){$E$}
\end{picture} 
\caption{Singularities of the two-time Green function
in the bound-state region,
disregarding the instability of excited states.}
 \end{figure}

 \begin{figure}
\setlength{\unitlength}{1mm}\thinlines
\begin{picture}(100,50)
\put(10,10){\line(0,1){30}}
\put(0,25){\line(1,0){70}}
\put(30,25){\circle*{1}}
\put(50,25){\circle*{1}}
\put(65,25){\circle*{1}}
\thicklines
\put(30,25){\line(0,-1){15}}
\put(50,25){\line(0,-1){15}}
\put(65,25){\line(0,-1){15}}
\put(15,40){$E$}
\end{picture} 
\caption{Singularities of the two-time Green function 
in the bound-state region if the cuts are turned down, 
to the second sheet of the Riemann
surface. The instability of excited states is disregarded.}
 \end{figure}
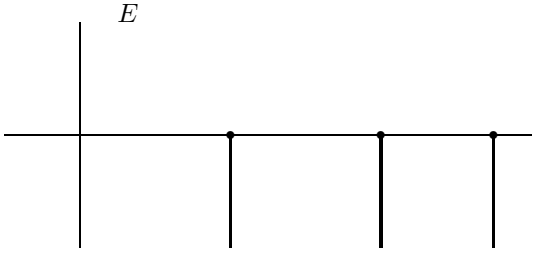

 \begin{figure}
\setlength{\unitlength}{1mm}\thinlines
\begin{picture}(100,50)
\put(10,10){\line(0,1){30}}
\put(0,25){\line(1,0){70}}
\put(30,25){\circle*{1}}
\put(50,23){\circle*{1}}
\put(65,23){\circle*{1}}
\thicklines
\put(30,25){\line(0,-1){15}}
\put(50,23){\line(0,-1){13}}
\put(65,23){\line(0,-1){13}}
\put(15,40){$E$}
\end{picture} 
\caption{Singularities of the two-time Green function 
in the bound-state region if the cuts are turned down, 
to the second sheet of the Riemann
 surface. The instability of excited states is taken 
into account.}
 \end{figure}
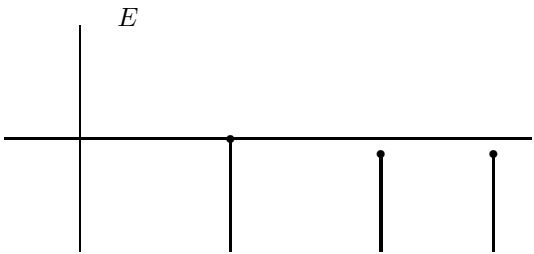

 \begin{figure}
\setlength{\unitlength}{1mm}\thinlines
\begin{picture}(100,50)
\put(10,10){\line(0,1){30}}
\put(0,25){\line(1,0){70}}
\put(30,25){\circle*{1}}
\put(32,25){\circle*{1}}
\put(34,25){\circle*{1}}
\put(50,25){\circle*{1}}
\put(52,25){\circle*{1}}
\put(54,25){\circle*{1}}
\put(65,25){\circle*{1}}
\put(67,25){\circle*{1}}
\put(69,25){\circle*{1}}
\thicklines
\put(32,25){\line(0,-1){15}}
\put(34,25){\line(0,-1){15}}
\put(52,25){\line(0,-1){15}}
\put(54,25){\line(0,-1){15}}
\put(67,25){\line(0,-1){15}}
\put(69,25){\line(0,-1){15}}
\put(15,40){$E$}
\end{picture} 
\caption{Singularities of the two-time Green function 
in the bound state  region for a non-zero photon mass, 
including one- and two-photon spectra, if the cuts are turned 
down, to the second sheet of the Riemann surface.
The instability of excited states is disregarded.}
 \end{figure}
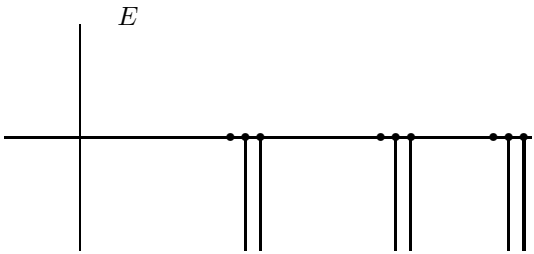

 \begin{figure}
\setlength{\unitlength}{1mm}\thinlines
\begin{picture}(100,50)
\put(30,5){
\put(0,25){\line(1,0){80}}
\put(5,25){\circle*{1}}
\put(10,25){\circle*{1}}
\put(15,25){\circle*{1}}
\put(60,25){\circle*{1}}
\put(65,25){\circle*{1}}
\put(70,25){\circle*{1}}
\thicklines
\put(10,25){\line(0,-1){15}}
\put(15,25){\line(0,-1){15}}
\put(65,25){\line(0,-1){15}}
\put(70,25){\line(0,-1){15}}
\put(60,25){\circle{6}}
\put(60,28){\vector(-1,0){1}}
\put(60,30){$\Gamma$}
}
\end{picture} 
\caption{The contour $\Gamma$ surrounds the pole corresponding
to the level under consideration and keeps outside all other 
singularities. For simplicity, only one- and two-photon spectra
are displayed.}
 \end{figure}
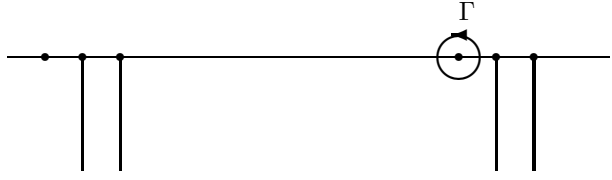

 \begin{figure}
\setlength{\unitlength}{1mm}\thinlines
\begin{picture}(100,50)
\put(50,10){\line(0,1){30}}
\put(10,25){\line(1,0){80}}
\put(47,26){\circle*{1}}
\put(53,24){\circle*{1}}
\thicklines
\put(47,26){\line(-1,0){37}}
\put(53,24){\line(1,0){37}}
\put(55,40){$\omega$}
\put(43,28){$-\mu+i0$}
\put(51,21){$\mu-i0$}
\end{picture} 
\caption{Singularities of the photon propagator in the complex
$\omega$ plane for a non-zero photon mass $\mu$.}
 \end{figure}
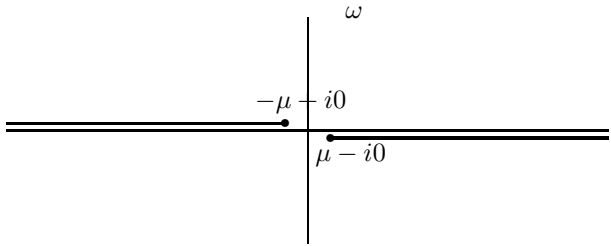

 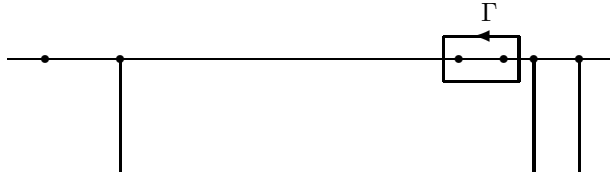
\begin{figure}
\setlength{\unitlength}{1mm}\thinlines
\begin{picture}(100,50)
\put(30,5){
\put(0,25){\line(1,0){80}}
\put(5,25){\circle*{1}}
\put(15,25){\circle*{1}}
\put(60,25){\circle*{1}}
\put(66,25){\circle*{1}}
\put(70,25){\circle*{1}}
\put(76,25){\circle*{1}}
\thicklines
\put(15,25){\line(0,-1){15}}
\put(70,25){\line(0,-1){15}}
\put(76,25){\line(0,-1){15}}
\put(58,28){\line(1,0){10}}
\put(58,22){\line(1,0){10}}
\put(58,22){\line(0,1){6}}
\put(68,22){\line(0,1){6}}
\put(63,28){\vector(-1,0){1}}
\put(63,30){$\Gamma$}
}
\end{picture} 
\caption{The contour $\Gamma$ surrounds the poles corresponding
to the quasidegenerate
levels under consideration and keeps outside all other 
singularities. For simplicity, only one-photon spectra
are displayed.}
 \end{figure}

 \begin{figure}
\setlength{\unitlength}{1mm}\thinlines
\begin{picture}(100,50)
\put(30,5){
\put(0,25){\line(1,0){80}}
\put(5,25){\circle*{1}}
\put(8,25){\circle*{1}}
\put(60,25){\circle*{1}}
\put(66,25){\circle*{1}}
\put(63,25){\circle*{1}}
\put(69,25){\circle*{1}}
\thicklines
\put(8,25){\line(0,-1){15}}
\put(63,25){\line(0,-1){15}}
\put(69,25){\line(0,-1){15}}
\put(58,28){\line(1,0){10}}
\put(58,22){\line(1,0){3.5}}
\put(64.5,22){\line(1,0){3.5}}
\put(58,22){\line(0,1){6}}
\put(61.5,22){\line(0,1){4.5}}
\put(68,22){\line(0,1){6}}
\put(64.5,22){\line(0,1){4.5}}
\put(61.5,26.5){\line(1,0){3}}
\put(63,28){\vector(-1,0){1}}
\put(63,30){$\Gamma$}
}
\end{picture} 
\caption{ A deformation of the contour $\Gamma$ 
that allows drawing the cuts to the related poles
in the case of quasidegenerate states
when $\mu \rightarrow 0$. For simplicity, only one-photon spectra
are displayed.}
 \end{figure}
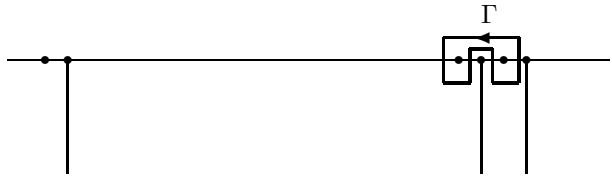

 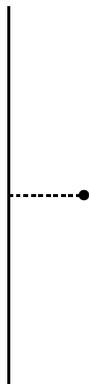
\begin{figure}
\setlength{\unitlength}{0.5mm}
\begin{picture}(120,120)(0,0)
  \put(100,10){\electronline}
  \put(100,60){\hyperfine}
\end{picture}  
\caption{The interaction with an external potential $\delta V({\bf x})$.}
 \end{figure}

 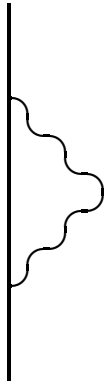
\begin{figure}
\setlength{\unitlength}{0.5mm}
\begin{picture}(120,120)(0,0)
  \put(100,0){\electronline}    
 \put(100,30){\pphotarctop}    
\end{picture}  
\caption{The first-order self-energy diagram.}
 \end{figure}

 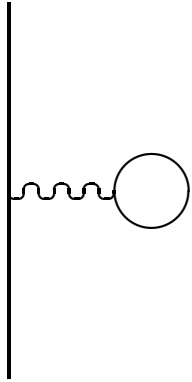
\begin{figure}
\setlength{\unitlength}{0.5mm}
\begin{picture}(120,120)(0,0)
  \put(100,0){\electronline}    
      \put(102,50){\photonvac}       
	  \put(138,50){\circle {19}}    
\end{picture}  
\caption{The first-order vacuum-polarization diagram.}
 \end{figure}

 \begin{figure}
\setlength{\unitlength}{0.5mm}
\begin{picture}(120,120)(0,0)
  \put(100,0){\electronline}    
  \put(96,46){\line(1,1){8}}       
  \put(96,54){\line(1,-1){8}}       
\end{picture}  
\caption{The mass-counterterm diagram.}
 \end{figure}
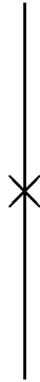

\begin{figure}    
\setlength{\unitlength}{0.5mm}
  \begin{picture}(300,120)     
    \put(180,20){    
      \put(40,0){\electronline}    
      \put(40,30){\pphotarcbottom}    
      \put(40,50){\hyperfine}}    
    \put(0,20){    
      \put(40,0){\electronline}    
      \put(40,20){\pphotarcbottom}    
     \put(40,85){\hyperfine}    
 }    
    \put(90,20){    
      \put(40,0){\electronline}    
      \put(40,40){\pphotarcbottom}    
     \put(40,15){\hyperfine}    
}    
\put(40,5) {a}
\put(130,5) {b}
\put(220,5) {c}
\end{picture}    
\caption{$\delta V$ - self-energy diagrams.}    
\end{figure}
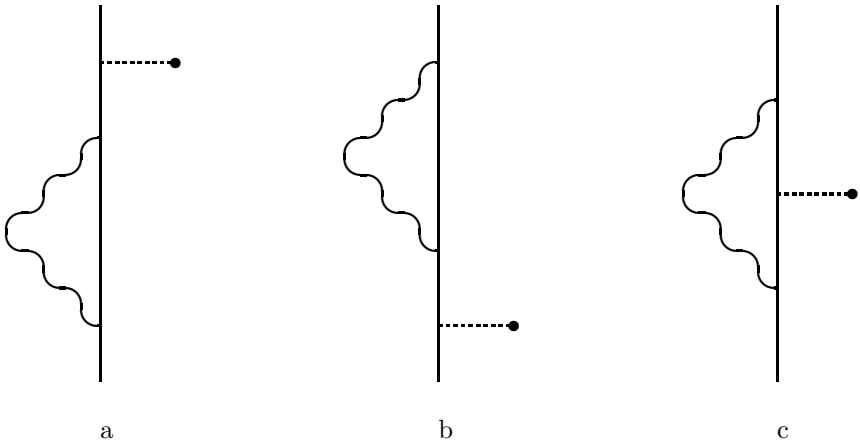

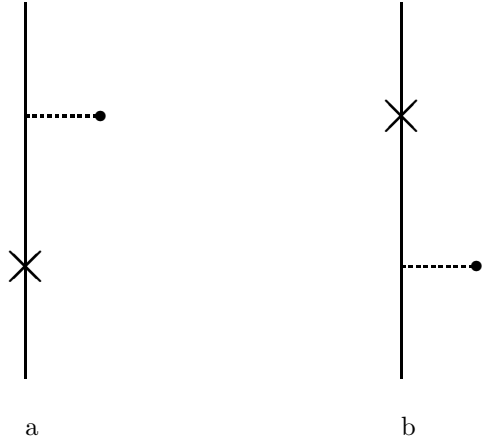
\begin{figure}    
\setlength{\unitlength}{0.5mm}
  \begin{picture}(300,150)        
    \put(50,30){    
      \put(40,0){\electronline}    
  \put(36,26){\line(1,1){8}}       
  \put(36,34){\line(1,-1){8}} 
     \put(40,70){\hyperfine}    
 }    
    \put(150,30){    
      \put(40,0){\electronline}        
  \put(36,66){\line(1,1){8}}       
  \put(36,74){\line(1,-1){8}} 
     \put(40,30){\hyperfine}    
}    
\put(90,15) {a}
\put(190,15) {b}
\end{picture}    
\caption{$\delta V$ - mass-counterterm diagrams.}    
\end{figure}

 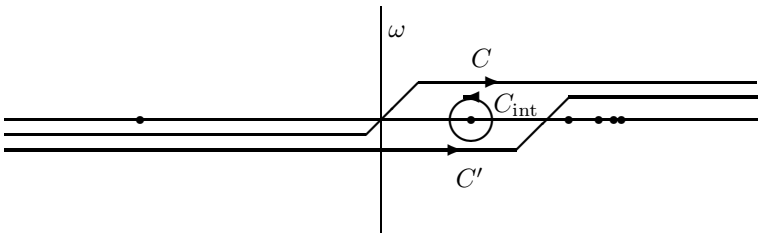
\begin{figure}
\setlength{\unitlength}{1mm}\thinlines
\begin{picture}(100,50)
\put(50,10){\line(0,1){30}}
\put(0,25){\line(1,0){100}}
\put(62,25){\circle*{1}}
\put(75,25){\circle*{1}}
\put(79,25){\circle*{1}}
\put(81,25){\circle*{1}}
\put(82,25){\circle*{1}}
\put(18,25){\circle*{1}}
\thicklines
\put(82,25){\line(1,0){18}}
\put(18,25){\line(-1,0){18}}
\put(48,23){\line(1,1){7}}
\put(0,23){\line(1,0){48}}
\put(55,30){\line(1,0){45}}
\put(68,21){\line(1,1){7}}
\put(0,21){\line(1,0){68}}
\put(75,28){\line(1,0){25}}
\put(65,30){\vector(1,0){1}}
\put(62,32){$C$}
\put(60,21){\vector(1,0){1}}
\put(60,16){$C'$}
\put(62,28){\vector(-1,0){1}}
\put(65,26){$C_{\rm int}$}
\put(51,36){$\omega$}
\put(62,25){\circle{6}}
\end{picture} 
\caption{$C$ is the contour of the integration
over the electron energy $\omega$ in the formalism with 
the standard vacuum.  $C'$ is the integration contour
for the vacuum with the $(1s)^2$ shell included.
 The integral
along the contour $C_{\rm int}=C'-C$ describes the interaction 
of the valence electron with the $(1s)^2-$shell electrons.}
 \end{figure}

 \begin{figure}
\setlength{\unitlength}{0.5mm}
\begin{picture}(120,115)(0,0)
  \put(80,10){\electronline}    
 \put(80,40){\pphotarcbottom}    
  \put(110,10){\electronline}    
  \put(170,10){\electronline}    
  \put(200,10){\electronline}    
 \put(200,40){\pphotarctop}    
\end{picture}  
\caption{The first-order self-energy diagrams for a two-electron atom.}
 \end{figure}
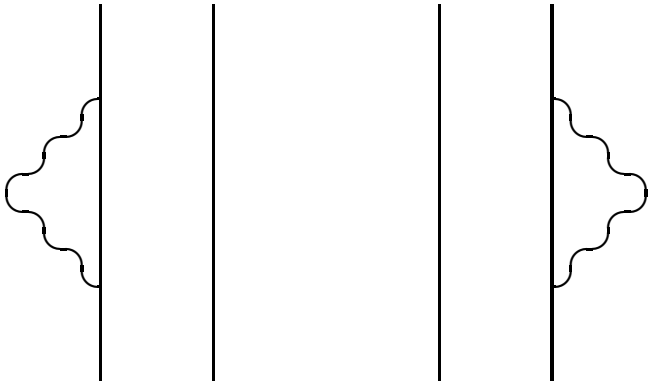

 \begin{figure}
\setlength{\unitlength}{0.5mm}
\begin{picture}(120,115)(0,0)
  \put(80,10){\electronline}        
  \put(54,60){\photonvac}       
  \put(42,60){\circle {19}}    
  \put(110,10){\electronline}    
  \put(170,10){\electronline}    
  \put(200,10){\electronline}    
   \put(202,60){\photonvac}       
   \put(238,60){\circle {19}}    
\end{picture}  
\caption{The first-order vacuum-polarization
 diagrams for a two-electron atom.}
 \end{figure}
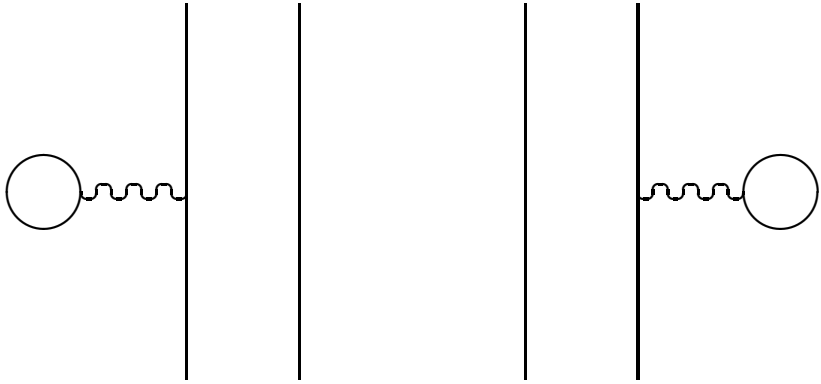

 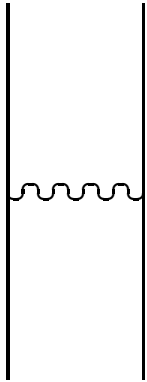
\begin{figure}
\setlength{\unitlength}{0.5mm}
\begin{picture}(120,120)(0,0)
  \put(124,10){\electronline}    
   \put(126,60){\photonlineright}       
  \put(160,10){\electronline}    
\end{picture}  
\caption{One-photon exchange diagram.}
 \end{figure}

 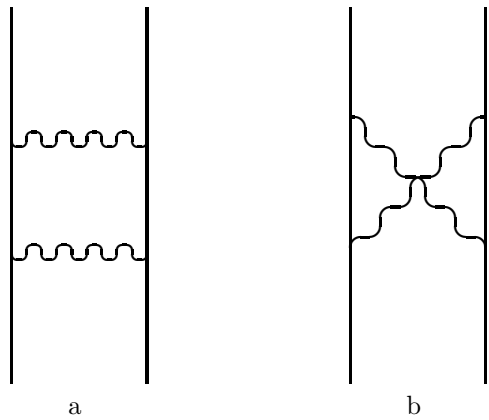
\begin{figure}
\setlength{\unitlength}{0.5mm}
\begin{picture}(120,120)(0,0)
  \put(80,10){\electronline}        
  \put(82,45){\photonlineright}       
  \put(82,75){\photonlineright}       
  \put(116,10){\electronline}    
  \put(170,10){\electronline}    
  \put(174,45){\photonrightup}       
  \put(202,45){\photonleftup}       
  \put(206,10){\electronline}    
  \put(95,2){a}
  \put(185,2){b}
\end{picture}  
\caption{Two-photon exchange diagrams.}
 \end{figure}

 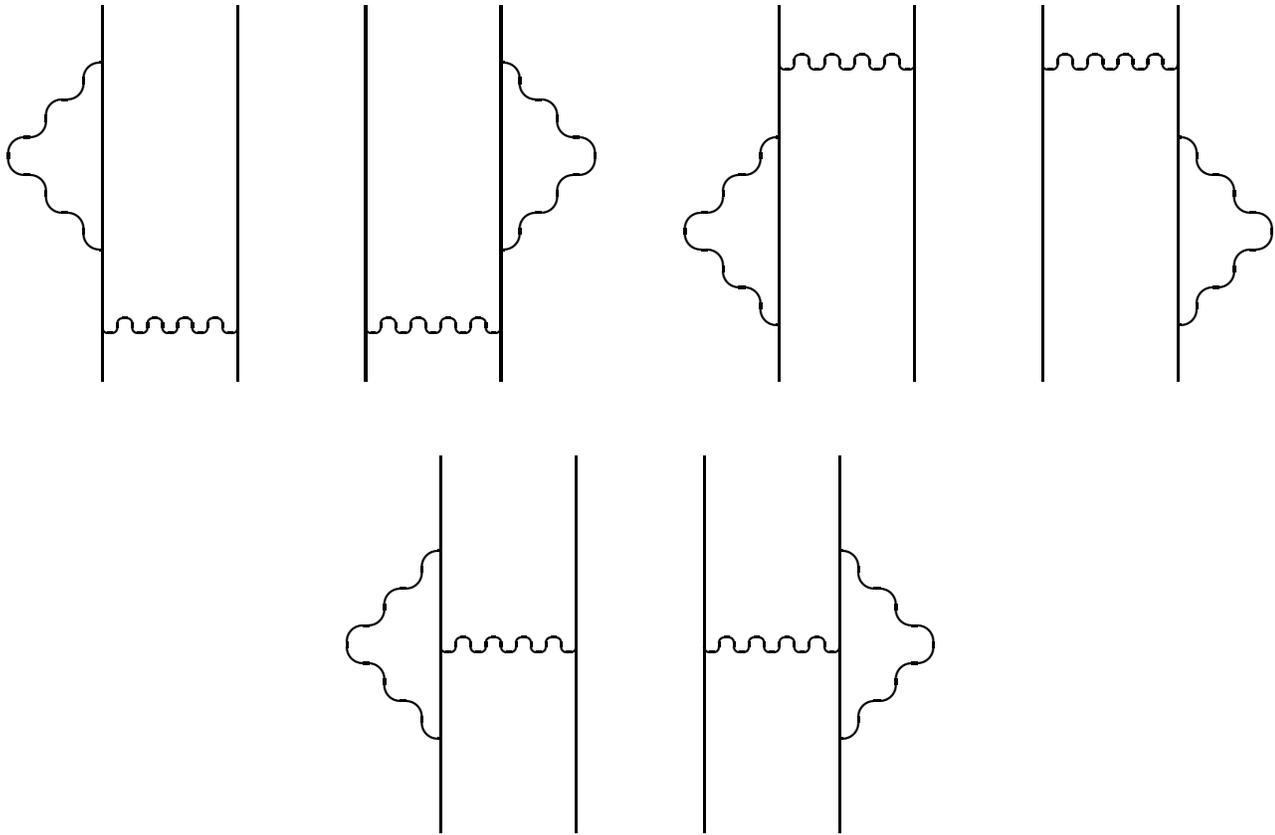
\begin{figure}
\setlength{\unitlength}{0.5mm}
\begin{picture}(360,240)(0,0)
\put(10,120)
{  \put(0,10){\electronline}    
 \put(0,50){\pphotarcbottom}    
  \put(2,25){\photonlineright} 
  \put(36,10){\electronline}    
  \put(70,10){\electronline}    
  \put(106,10){\electronline}    
 \put(106,50){\pphotarctop}    
  \put(72,25){\photonlineright} 
}
\put(190,120)
{  \put(0,10){\electronline}    
 \put(0,30){\pphotarcbottom}    
  \put(2,95){\photonlineright} 
  \put(36,10){\electronline}    
  \put(70,10){\electronline}    
  \put(106,10){\electronline}    
 \put(106,30){\pphotarctop}    
  \put(72,95){\photonlineright} 
}
\put(100,0)
{  \put(0,10){\electronline}    
 \put(0,40){\pphotarcbottom}    
  \put(2,60){\photonlineright} 
  \put(36,10){\electronline}    
  \put(70,10){\electronline}    
  \put(106,10){\electronline}    
 \put(106,40){\pphotarctop}    
  \put(72,60){\photonlineright} 
}

\end{picture}  
\caption{The self-energy screening diagrams.}
 \end{figure}

 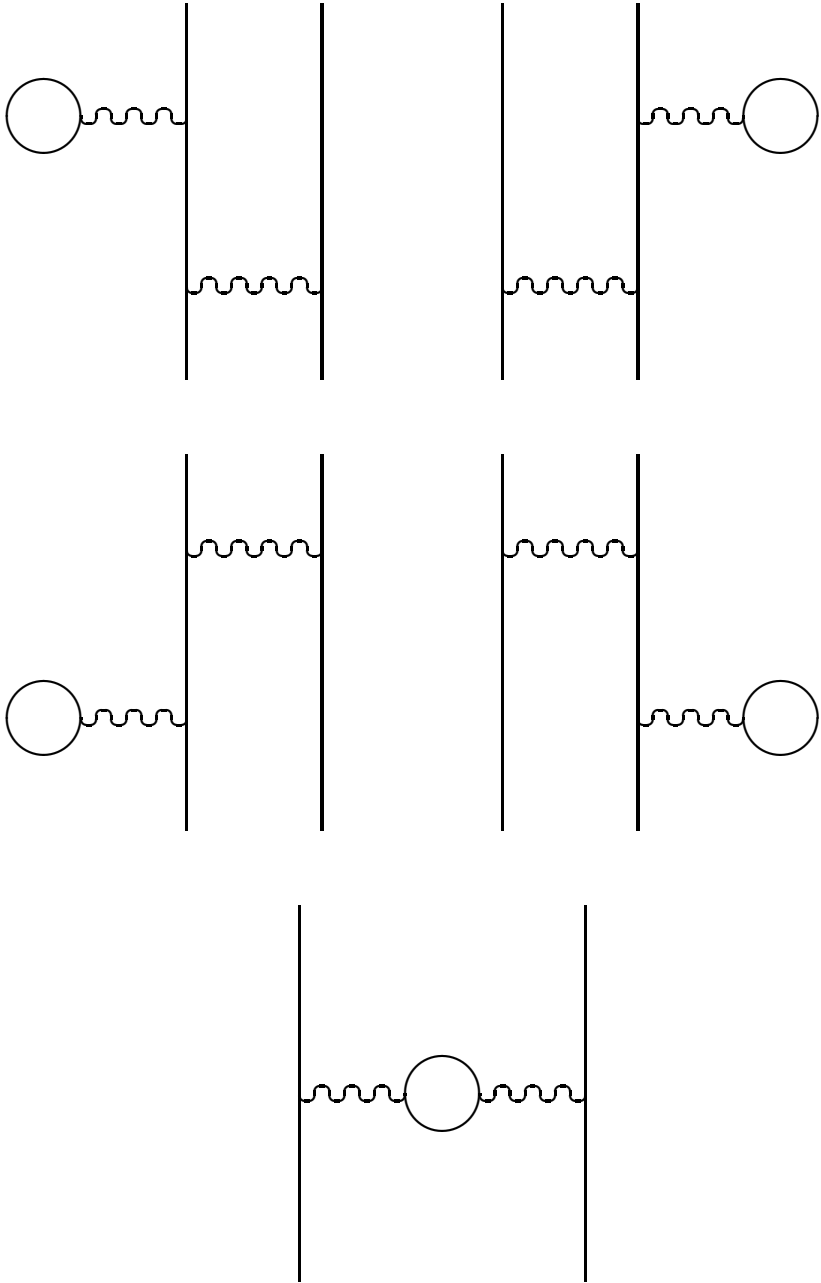
\begin{figure}
\setlength{\unitlength}{0.5mm}
\begin{picture}(360,360)(0,0)
\put(0,240) 
{ \put(80,10){\electronline}    
  \put(82,35){\photonlineright}     
  \put(54,80){\photonvac}       
  \put(42,80){\circle {19}}    
  \put(116,10){\electronline}    
  \put(164,10){\electronline}    
  \put(166,35){\photonlineright} 
  \put(200,10){\electronline}    
   \put(202,80){\photonvac}       
   \put(238,80){\circle {19}}
}
\put(0,120) 
{ \put(80,10){\electronline}    
  \put(82,85){\photonlineright}     
  \put(54,40){\photonvac}       
  \put(42,40){\circle {19}}    
  \put(116,10){\electronline}    
  \put(164,10){\electronline}    
  \put(166,85){\photonlineright} 
  \put(200,10){\electronline}    
   \put(202,40){\photonvac}       
   \put(238,40){\circle {19}}
}
\put(20,0) 
{ 
  \put(140,60){\photonvac}       
  \put(128,60){\circle {19}}    
  \put(90,10){\electronline}    
  \put(166,10){\electronline}    
   \put(92,60){\photonvac}       
}
    
\end{picture}  
\caption{The vacuum-polarization screening
 diagrams.}
 \end{figure}

 \begin{figure}
\setlength{\unitlength}{0.5mm}
\begin{picture}(120,120)(0,0)
  \put(20,10){\electronline}        
  \put(22,45){\photonlineright}       
  \put(20,75){\hyperfineleft}       
  \put(56,10){\electronline}
  \put(100,10){\electronline}        
  \put(102,45){\photonlineright}       
  \put(136,75){\hyperfine}       
  \put(136,10){\electronline}
  \put(180,10){\electronline}    
  \put(182,75){\photonlineright}
  \put(180,45){\hyperfineleft}                   
  \put(216,10){\electronline}    
  \put(260,10){\electronline}    
  \put(296,45){\hyperfine}       
  \put(262,75){\photonlineright}            
  \put(296,10){\electronline}    
\end{picture}  
\caption{$\delta V$ - interelectronic-interaction diagrams.}
 \end{figure}
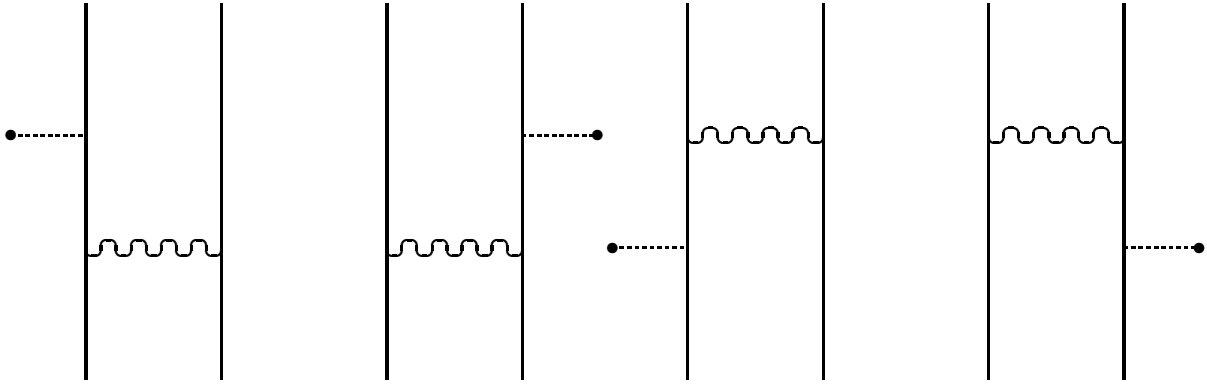

\begin{figure}
\setlength{\unitlength}{1mm}
\begin{center}
\begin{picture}(100,60)(-10,0)
   \put(40,10){\line(0,1){30}}
   \multiput(40,17)(-2,2){5}{\circle*{0.8}}
   \multiput(40,33)(-2,-2){5}{\circle*{0.8}}
   \put(40,33){\circle*{1.6}}
   \put(40,17){\circle*{1.6}}
   \end{picture}
\caption{ Coulomb nuclear recoil diagram.}
\end{center}
\end{figure}
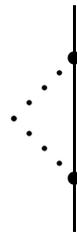

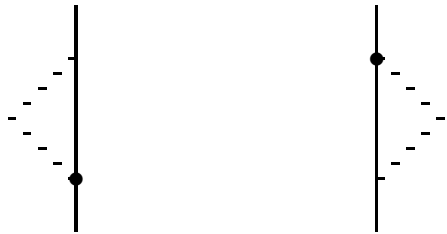
\begin{figure}
\setlength{\unitlength}{1mm}
\begin{center}
\begin{picture}(100,60)(-10,0)
   \put(20,10){\line(0,1){30}}
  \put(60,10){\line(0,1){30}}
   \multiput(20,17)(-2,2){5}{\line(-1,0){1}}
   \multiput(20,33)(-2,-2){5}{\line(-1,0){1}}
   \multiput(60,17)(2,2){5}{\line(1,0){1}}
   \multiput(60,33)(2,-2){5}{\line(1,0){1}}
   \put(60,33){\circle*{1.6}}
   \put(20,17){\circle*{1.6}}
   \end{picture}
\caption{ One-transverse-photon nuclear recoil diagrams.}
\end{center}
\end{figure}

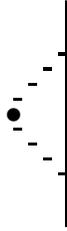
\begin{figure}
\setlength{\unitlength}{1mm}
\begin{center}
\begin{picture}(100,60)(-10,0)
   \put(40,10){\line(0,1){30}}
   \multiput(40,17)(-2,2){4}{\line(-1,0){1}}
   \multiput(40,33)(-2,-2){4}{\line(-1,0){1}}
   \put(33,25){\circle*{1.6}}
   \end{picture}
\caption{ Two-transverse-photon nuclear recoil diagram.}
\end{center}
\end{figure}

\begin{figure}
\setlength{\unitlength}{1mm}
\begin{center}
\begin{picture}(100,60)(-15,0)
  \put(20,10){\line(0,1){30}}
   \put(40,10){\line(0,1){30}}
   \multiput(20,25)(2,0){10}{\circle*{0.8}}
   \put(20,25){\circle*{1.6}}
   \put(40,25){\circle*{1.6}}
   \end{picture}
\caption{Two-electron Coulomb nuclear recoil diagram.}
\end{center}
\end{figure}
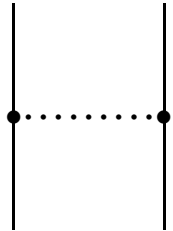

\begin{figure}
\setlength{\unitlength}{1mm}
\begin{center}
\begin{picture}(120,60)(-10,0)
  \put(20,10){\line(0,1){30}}
  \put(40,10){\line(0,1){30}}
  \multiput(20,25)(2,0){10}{\line(1,0){1}}
   \put(20,25){\circle*{1.6}}
  \put(60,10){\line(0,1){30}}
  \put(80,10){\line(0,1){30}}
  \multiput(60,25)(2,0){10}{\line(1,0){1}}
   \put(80,25){\circle*{1.6}}
\end{picture}
\caption{Two-electron one-transverse-photon nuclear recoil diagrams.}
\end{center}
\end{figure}
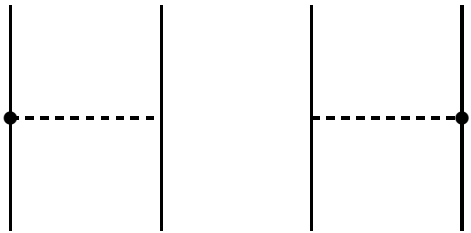

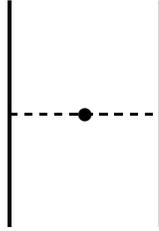
\begin{figure}
\setlength{\unitlength}{1mm}
\begin{center}
\begin{picture}(100,60)(-15,0)
  \put(20,10){\line(0,1){30}}
   \put(40,10){\line(0,1){30}}
   \multiput(20,25)(2,0){10}{\line(1,0){1}}
   \put(30,25){\circle*{1.6}}
   \end{picture}
\caption{Two-electron two-transverse-photon nuclear recoil diagram.}
\end{center}
\end{figure}

 \begin{figure}
\setlength{\unitlength}{0.5mm}
\begin{picture}(120,120)(0,0)
  \put(100,0){\electronline}    
 \put(102,50){\photonlineright}
  \put(121,52){\vector(1,0){3}}
\end{picture}  
\caption{The photon emission by a one-electron atom
 in zeroth-order approximation.}
 \end{figure}
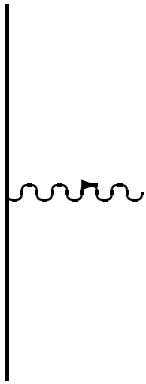

\begin{figure}    
\setlength{\unitlength}{0.5mm}
  \begin{picture}(300,280)    
    \put(180,160){    
      \put(40,0){\electronline}    
      \put(40,30){\pphotarcbottom}    
      \put(42,50){\photonlineright}
     \put(61,52){\vector(1,0){3}}
}    
    \put(90,160){    
      \put(40,0){\electronline}    
      \put(40,20){\pphotarcbottom}    
     \put(42,85){\photonlineright}    
     \put(61,87){\vector(1,0){3}} }    
    \put(0,160){    
      \put(40,0){\electronline}    
     \put(40,40){\pphotarcbottom}    
     \put(42,15){\photonlineright}    
     \put(61,17){\vector(1,0){3}}}    
    \put(0,20){    
      \put(40,0){\electronline}    
     \put(42,70){\photonvac}    
     \put(78,70){\circle {19}}
     \put(42,30){\photonlineright}    
     \put(61,32){\vector(1,0){3}}}    
    \put(90,20){    
      \put(40,0){\electronline}    
     \put(42,30){\photonvac}    
     \put(78,30){\circle {19}}
     \put(42,70){\photonlineright}    
     \put(61,72){\vector(1,0){3}}}    
    \put(180,20){    
      \put(40,0){\electronline}    
      \put(42,50){\photonvac} 
     \put(78,50){\circle {19}}
      \put(90,50){\photonlineright}
     \put(109,52){\vector(1,0){3}}
}    
\put(40,145) {a}
\put(130,145) {b}
\put(220,145) {c}
\put(40,5) {d}
\put(130,5) {e}
\put(220,5) {f}
\end{picture}    
\caption{The first-order QED corrections to the photon emission
by a one-electron atom.}    
\end{figure}
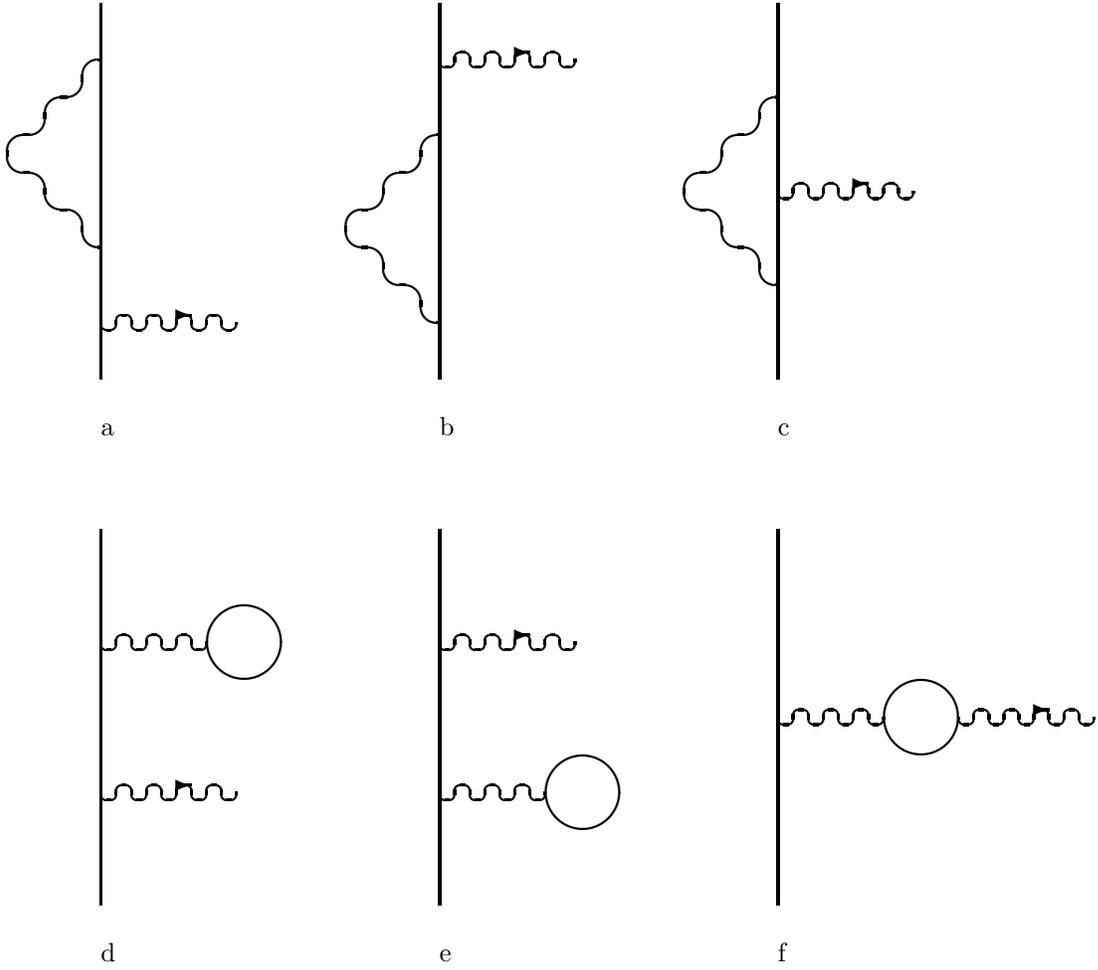

\begin{figure}    
\setlength{\unitlength}{0.5mm}
  \begin{picture}(300,150)    
    \put(150,30){    
      \put(40,0){\electronline}    
  \put(36,26){\line(1,1){8}}       
  \put(36,34){\line(1,-1){8}} 
     \put(42,70){\photonlineright}    
     \put(61,72){\vector(1,0){3}}   
 }    
    \put(50,30){    
      \put(40,0){\electronline}       
  \put(36,66){\line(1,1){8}}       
  \put(36,74){\line(1,-1){8}} 
    \put(42,30){\photonlineright}    
     \put(61,32){\vector(1,0){3}}   
}    

\put(90,15) {a}
\put(190,15) {b}
\end{picture}    
\caption{The mass-counterterm corrections to the photon emission
by a one-electron atom.}    
\end{figure}
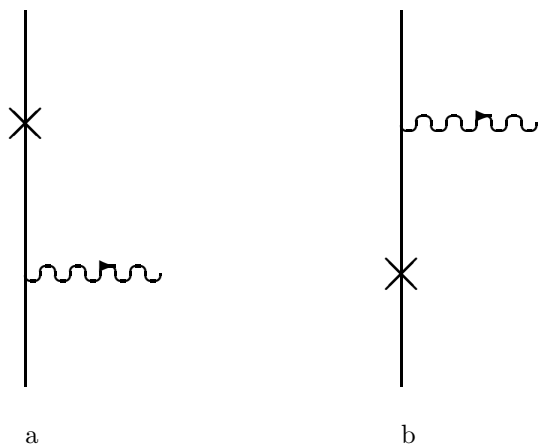

 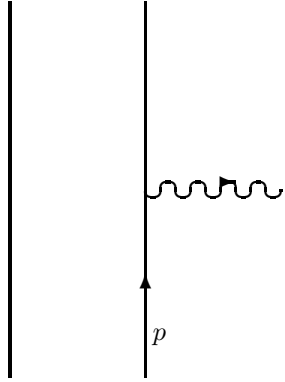
\begin{figure}
\setlength{\unitlength}{0.5mm}
\begin{picture}(120,120)(0,0)
  \put(124,10){\electronline}    
   \put(162,60){\photonlineright}       
  \put(160,10){\electronline}    
  \put(181,62){\vector(1,0){3}}
 \put(160,35){\vector(0,1){3}}
  \put(162,20){$p$}
\end{picture}  
\caption{The radiative recombination of an electron with a 
hydrogenlike atom in zeroth-order approximation.}
 \end{figure}

 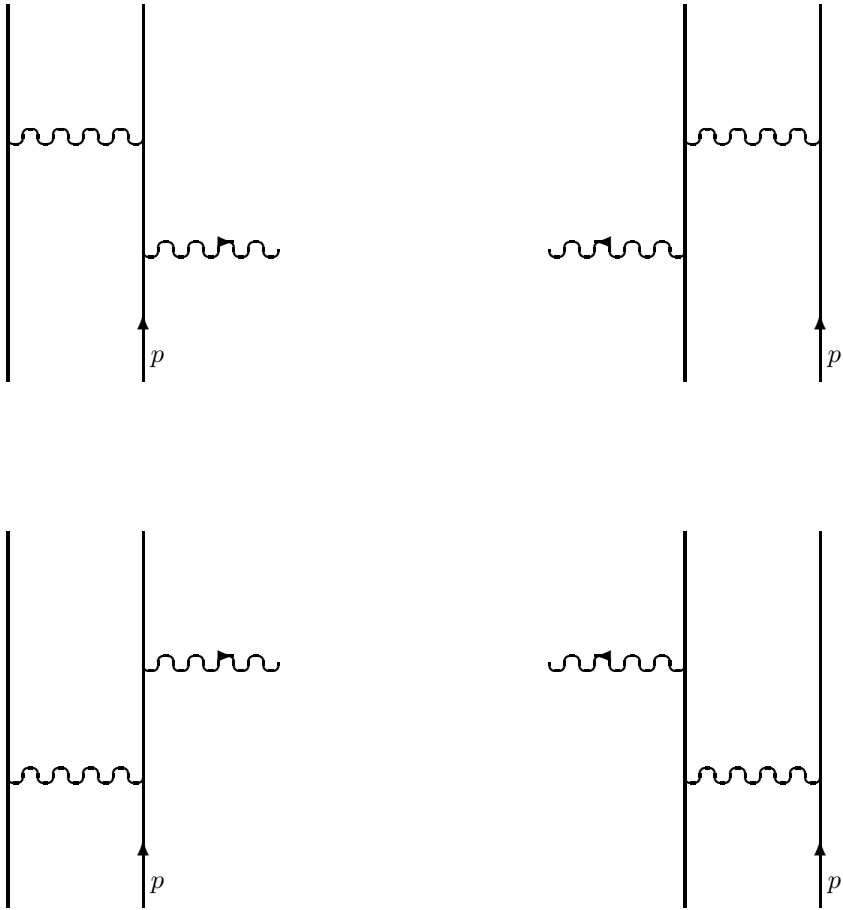
\begin{figure}
\setlength{\unitlength}{0.5mm}
\begin{picture}(300,280)(0,0)
\put(0,10)
{
  \put(200,10){\electronline}    
  \put(202,45){\photonlineright}       
  \put(166,75){\photonlineright}       
  \put(179,77){\vector(-1,0){3}}
   \put(236,10){\electronline}
   \put(236,25){\vector(0,1){3}}
    \put(238,15){$p$}
  \put(20,10){\electronline}        
  \put(22,45){\photonlineright}       
  \put(58,75){\photonlineright}
  \put(77,77){\vector(1,0){3}}
  \put(56,10){\electronline}
  \put(56,25){\vector(0,1){3}}
    \put(58,15){$p$}
}
\put(0,150)
{   
  \put(200,10){\electronline}    
  \put(202,75){\photonlineright}
  \put(166,45){\photonlineright}       
  \put(179,47){\vector(-1,0){3}}                 
  \put(236,10){\electronline}    
  \put(236,25){\vector(0,1){3}}
    \put(238,15){$p$}
  \put(20,10){\electronline}    
  \put(22,75){\photonlineright}       
  \put(77,47){\vector(1,0){3}}
  \put(58,45){\photonlineright}            
  \put(56,10){\electronline}    
  \put(56,25){\vector(0,1){3}}
    \put(58,15){$p$}
}
\end{picture}  
\caption{The interelectronic-interaction corrections
of first order in $1/Z$ to the radiative recombination of an electron
with a hydrogenlike atom.}
 \end{figure}

 \begin{figure} 
\setlength{\unitlength}{0.5mm}
\begin{picture}(120,120)(0,0)    
  \put(140,10){\electronline}    
   \put(142,80){\photonlineright}       
   \put(142,40){\photonlineright}       
  \put(161,82){\vector(1,0){3}}
  \put(163,42){\vector(-1,0){3}}
\end{picture}  
\caption{The photon scattering on a one-electron atom in 
zeroth-order approximation.}
 \end{figure}
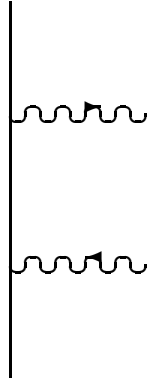

 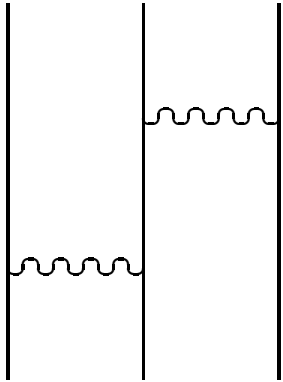
\begin{figure}
\setlength{\unitlength}{0.5mm}
\begin{picture}(120,120)(0,0)
  \put(124,10){\electronline}    
   \put(126,40){\photonlineright}       
  \put(160,10){\electronline}    
   \put(162,80){\photonlineright}       
  \put(196,10){\electronline}    
\end{picture}  
\caption{A typical diagram describing 
two-photon exchange between three electrons
in a lithiumlike atom.}
 \end{figure}

 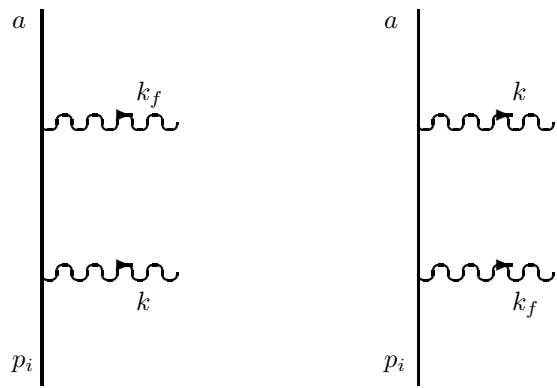
\begin{figure} 
\setlength{\unitlength}{0.5mm}
\begin{picture}(240,120)(-80,0)    
  \put(40,10){\electronline}    
   \put(42,80){\photonlineright}       
   \put(42,40){\photonlineright}       
  \put(61,82){\vector(1,0){3}}
  \put(61,42){\vector(1,0){3}}
   \put(32,15){$p_i$}
   \put(32,105){$a$}
   \put(65,30){$k$}
   \put(65,86){$k_f$}
  \put(140,10){\electronline}    
   \put(142,80){\photonlineright}       
   \put(142,40){\photonlineright}       
  \put(162,82){\vector(1,0){3}}
  \put(162,42){\vector(1,0){3}}
   \put(131,15){$p_i$}
   \put(131,105){$a$}
   \put(165,30){$k_f$}
   \put(165,86){$k$}
\end{picture}  
\caption{The radiative recombination accompanied by emission
of a soft photon.}
 \end{figure}

\newpage

 \begin{figure}
\setlength{\unitlength}{0.5mm}
\begin{picture}(120,140)(0,0)
  \put(120,10){\electronline}    
  \put(124,45){\photonrightup}       
  \put(152,55){\photonleftup}       
  \put(156,10){\electronline}    
  \put(120,20){\pphotarcbottom}    
\put(80,10){\timeline}
\put(80,110){\timeline}
\put(80,15){\timeline}
\put(80,47){\timeline}
\put(80,57){\timeline}
\put(80,65){\timeline}
\put(80,81){\timeline}
\put(80,91){\timeline}
\put(60,0) {\line(0,1){130}}
\put(60,130){\vector(0,1){3}}
\put(54,128){$t$}
\put(200,7){$t_1$}
\put(200,15){$t_2$}
\put(200,45){$t_3$}
\put(200,55){$t_4$}
\put(200,65){$t_5$}
\put(200,79){$t_6$}
\put(200,90){$t_7$}
\put(200,108){$t_8$}
\end{picture}  
\caption{A time-ordered version of a Feynman diagram.}
 \end{figure}


\begin{thebibliography}{99}                          

\bibitem{beyer97}
H.F. Beyer, H.-J. Kluge, V.P. Shevelko,
X-ray Radiation of Highly Charged Ions, Springer, Berlin, 1997.
\bibitem{beyer99} H.F. Beyer, V.P. Shevelko (eds.), 
 Atomic Physics with Heavy Ions, Springer, Berlin, 1999.
\bibitem{shabaev00a}
V.M. Shabaev, A.N. Artemyev, V.A. Yerokhin, Phys.
Scr. T 86 (2000) 7.
\bibitem{mohr98a}
P.J. Mohr, G. Plunien, G. Soff, Phys. Rep. 293 (1998) 227.
\bibitem{sapir99}
J. Sapirstein, in: D.H.E. Dubin, D. Schneider (eds.),
Trapped Charged Particles and Fundamental Physics, 
American Institute of Physics Conference Proceedings 457, 
1999, p. 3.
\bibitem{beier00}
T. Beier, Phys. Rep., 339 (2000) 79.
\bibitem{lindgren99}
I. Lindgren, Phys. Scr. T 80 (1999) 131.
 \bibitem{gellmann}
M. Gell-Mann, F. Low, 1951 Phys. Rev. 84 (1951) 350.
\bibitem{sucher} J. Sucher, Phys. Rev. 107 (1957) 1448.
\bibitem{labz70} L.N. Labzowsky,
Zh. Eksp. Teor. Fiz. 59 (1970) 168 [Sov. Phys. JETP 32 (1970) 94].
\bibitem{braun80} M.A. Braun, A.D. Gurchumelia, 
Teor. Mat. Fiz. 45 (1980) 199
[Theor. Math. Phys. 45 (1980) N 2].
\bibitem{dmit84} Yu.Yu. Dmitriev, G.L. Klimchitskaya,
L.N. Labzowsky,  Relativistic Effects
 in Spectra of Atomic Systems, Energoatomizdat, Moscow, 1984.
\bibitem{braun84} M.A. Braun, A.D. Gurchumelia, U.I. Safronova, 
 Relativistic Atom Theory, Nauka, Moscow, 1984.
\bibitem{mohr89} P.J. Mohr, in: W.R. Johnson, P.J. Mohr,
J. Sucher (eds.),
 Relativistic, Quantum Electrodynamics and Weak Interaction
Effects in Atoms, American Institute
of Physics Conference Series N 189, 1989, p. 47.
\bibitem{lindgren89} I. Lindgren, in: W.R. Johnson, P.J. Mohr,
J. Sucher (eds.), Relativistic, Quantum Electrodynamics and 
Weak Interaction Effects in Atoms, American Institute
of Physics Conference Series N 189, 1989, p. 371.
\bibitem{labz93} L. Labzowsky, G. Klimchitskaya,
Yu. Dmitriev, Relativistic Effects in Spectra of
Atomic Systems, IOP Publishing, Bristol, 1993.
\bibitem{sapir98}
J. Sapirstein, Rev. Mod. Phys. 70 (1998) 55.
\bibitem{dyson49} F.J. Dyson, Phys. Rev. 75 (1949) 486, 1736.
\bibitem{bethe51} E.E. Salpeter, H.A. Bethe, Phys. Rev.
84 (1951) 1232. 
\bibitem{hubbard57} J. Hubbard, Proc. Roy. Soc. A 240 (1957)  539. 
\bibitem{schweber61} S.S. Schweber,  An Introduction
to Relativistic Quantum Field Theory, Harper \& Row,
New York, 1961.
\bibitem{vasil'ev75} A.N. Vasil'ev, A.L. Kitanin, 
Teor. Mat. Fiz. 24 (1975) 219 
[Theor. Math. Phys. 24 (1975) N 2].
\bibitem{zapryagaev85} S.A. Zapryagaev, N.L. Manakov,
V.G. Pal'chikov, Theory of One- and Two-Electron Multicharged Ions,
Energoatomizdat, Moscow, 1985.
\bibitem{sapirstein87} J. Sapirstein, Phys. Scr.
36 (1987) 801.
\bibitem{dmi97}
Yu.Yu. Dmitriev, T.A. Fedorova, Phys. Lett. A 225 (1997) 296.
\bibitem{dmi98}
Yu.Yu. Dmitriev, T.A. Fedorova, Phys. Lett. A 245 (1998) 555.
\bibitem{lin00}
I. Lindgren, Molecular Physics 98 (2000) 1159.
\bibitem{itzykson} C. Itzykson, J.-B. Zuber,
Quantum Field Theory, McGraw-Hill, New York, 1980.  
\bibitem{shab88a} V.M. Shabaev, in : U.I. Safronova (ed.),
Many-particles Effects in Atoms,  AN SSSR,
 Nauchnyi Sovet po Spektroskopii, Moscow, 1988, p. 15.
\bibitem{shab88b}
 V.M. Shabaev, in : U.I. Safronova (ed.),
Many-particles Effects in Atoms,  AN SSSR,
 Nauchnyi Sovet po Spektroskopii, Moscow, 1988, p. 24.
\bibitem{shab90a}
 V.M. Shabaev, Izv. Vuz. Fiz. 
33 (1990) 43 [Sov. Phys. Journ. 33 (1990) 660].
\bibitem{shab90b}
 V.M. Shabaev, Teor. Mat. Fiz.
82 (1990) 83 [Theor. Math. Phys. 82 (1990) 57].
\bibitem{shab91} 
V.M. Shabaev, J. Phys. A 24 (1991) 5665.
\bibitem{kar92}
 V.V. Karasiov, L.N. Labzowsky, A.V. Nefiodov,
V.M. Shabaev, Phys. Lett. A 161 (1992) 453.
\bibitem{sh93} 
 V.M. Shabaev, J. Phys. B 26 (1993) 4703.
\bibitem{sh94a} V.M. Shabaev, I.G. Fokeeva,
Phys. Rev. A 49 (1994) 4489.
\bibitem{sh94b} 
V.M. Shabaev,  Phys. Rev. A 50 (1994) 4521.
\bibitem{sh95} M.B. Shabaeva, V.M. Shabaev, Phys. Rev. A 52
(1995) 2811. 
\bibitem{yer97}
 V.A. Yerokhin, A.N. Artemyev,
V.M. Shabaev, Phys. Lett. A 234 (1997) 361.
\bibitem{art97}  A.N. Artemyev, V.M. Shabaev, 
V.A. Yerokhin, Phys. Rev. A 56 (1997) 3529.
\bibitem{sh98} 
V.M. Shabaev, Phys. Rev. A 57 (1998) 59.
\bibitem{art99}
A.N. Artemyev, T. Beier, G. Plunien, 
V.M. Shabaev, G. Soff, V.A. Yerokhin,
Phys. Rev. A 60 (1999) 45.
\bibitem{yer99} 
V.A. Yerokhin, A.N. Artemyev, T. Beier, G. Plunien,
V.M. Shabaev, G. Soff, 
 Phys. Rev. A 60 (1999) 3522.
\bibitem{art00}
A.N. Artemyev, T. Beier, G. Plunien,
V.M. Shabaev, G. Soff, V.A. Yerokhin,
Phys. Rev. A 62 (2000) 022116.
\bibitem{sh00} 
V.M. Shabaev, V.A. Yerokhin, T. Beier,
J. Eichler,  Phys. Rev. A 61 (2000) 052112.
\bibitem{yer00} 
V.A. Yerokhin, V.M. Shabaev, T. Beier,
J. Eichler, Phys. Rev. A 62 (2000) 042712.
\bibitem{br77} 
M.A. Braun, A.S. Shirokov,  
 Izv. Akad. Nauk SSSR: Ser. Fiz. 41 (1977) 2585
[Bull. Acad. Sci. USSR: Phys. Ser. 41 (1977) 109].
\bibitem{br86} 
M.A. Braun, Kh. Parera,
Izv. Akad. Nauk SSSR: Ser. Fiz.
50 (1986) 1303 [Bull. Acad. Sci. USSR: Phys. Ser.
50 (1986)  59]
\bibitem{br87}
M.A. Braun, Teor. Mat. Fiz. 72 (1987) 394 
[Theor. Math. Phys. 72 (1987) 958]. 
\bibitem{sh85} 
V.M. Shabaev, Teor. Mat. Fiz.
63 (1985) 394 [Theor. Math. Phys. 63 (1985) 588].
\bibitem{sh88} 
V.M. Shabaev, Yad. Fiz.
47 (1988) 107 [Sov. J. Nucl. Phys. 47 (1988) 69].
\bibitem{zap87} 
S.A. Zapryagaev, D.I. Morgulis 
Yad. Fiz. 45 (1987) 716
[Sov. J. Nucl. Phys. 45 (1987) N 3].
\bibitem{fel87} 
G. Feldman, T. Fulton,
Ann. Phys. (N.Y.) 179 (1987) 20.
\bibitem{ful89}
T. Fulton, in: W.R. Johnson, P.J. Mohr,
J. Sucher (eds.), Relativistic, Quantum Electrodynamics and 
Weak Interaction Effects in Atoms, American Institute
of Physics Conference Series N 189, 1989, p. 429.
\bibitem{yen89}  
D.R. Yennie, in: W.R. Johnson, P.J. Mohr,
J. Sucher (eds.), Relativistic, Quantum Electrodynamics and 
Weak Interaction Effects in Atoms, American Institute
of Physics Conference Series N 189, 1989, p. 393.
\bibitem{bjorken}
J.D. Bjorken, D. Drell,
Relativistic Quantum Fields, McGraw-Hill, New York, 1965.
\bibitem{mohr85}
P.J. Mohr, Phys. Rev. A 32 (1985) 1949.
\bibitem{adkins1}
G.S. Adkins, Phys. Rev. D  36, 1929 (1987).
\bibitem{adkins2}
G.S. Adkins, Phys. Rev. D 27, 1814 (1983).
\bibitem{bonch}
V.L. Bonch-Bruevich, S.V. Tyablikov,
The Green Function Method in Statistical Mechanics,
 North-Holland Publishing Company, Amsterdam, 1962.
\bibitem{thouless}
D.J. Thouless, The Quantum Mechanics of Many-Body
Systems, Academic Press, New York, 1961.
\bibitem{migdal}
A.B. Migdal, Theory of Finite Fermi Systems
and Properties of Atomic Nuclei, Nauka, Moscow, 1983.
\bibitem{logunov} 
A.A. Logunov, A.N. Tavkhelidze,
Nuovo Cim. 29 (1963) 380. 
\bibitem{faustov}
R.N. Faustov, Teor. Mat. Fiz. 3 (1970) 240 
[Theor. Math. Phys. 3 (1970) N 2].
\bibitem{nagy} 
B. Sz-Nagy, Comm. Math. Helv. 19 (1946/47) 347.
\bibitem{kato1}
T. Kato, Progr. Theor. Phys. 4 (1949) 514.
\bibitem{kato2}
T. Kato, Perturbation Theory
for Linear Operators, Springer, Berlin, 1966.
\bibitem{messia} 
A. Messiah, Quantum Mechanics, Vol. 2,
Wiley, New York, 1962.
\bibitem{reed}  
M. Reed, B. Simon,
Methods of Modern Mathematical Physics, Vol. 4:
Analysis of Operators, Academic Press, New York, 1978.
\bibitem{lepage} 
G.P. Lepage, Phys. Rev. A 16 (1977) 863.
\bibitem{blund93} 
S.A. Blundell, P.J. Mohr, W.R. Johnson,
J. Sapirstein, Phys. Rev. A 48 (1993) 2615.
\bibitem{grad} 
I.S. Gradshtein, I.N. Ryzhyk, Tables of Integrals, 
Sums, Series, and Products, Nauka, Moscow, 1977.
\bibitem{lind95}
I.  Lindgren, H. Persson, S. Salomonson, 
 L.N.  Labzowsky,
 Phys. Rev. A  51 (1995) 1167.
\bibitem{mittleman}
M.H. Mittleman, Phys. Rev. A 5 (1972) 2395. 
\bibitem{leb}
E.-O. Le Bigot, P. Indelicato, V.M. Shabaev, Phys. Rev. A 63 (2001) 040501(R).
\bibitem{yelkh94}
A.S. Yelkhovsky, Budker Institute of Nuclear Physics,
Report No. BINP 94-27, Novosibirsk, 1994;
arXive: hep-th/9403095.
\bibitem{pach95}
K. Pachucki, H. Grotch, Phys. Rev. A 51 (1995) 1854.
\bibitem{bogolyubov}
N.N. Bogolyubov, D.V. Shirkov, Introduction to the Theory of
Quantized Fields, Nauka, Moscow, 1984.
\bibitem{berestetsky}
V.B. Berestetsky, E.M. Lifshitz, L.P. Pitaevsky,
Quantum Electrodynamics, Pergamon Press, Oxford, 1982.
\bibitem{ind00}
P. Indelicato, V.M. Shabaev, to be published.
\bibitem{akhiezer} 
A.I. Akhiezer, V.B. Berestetsky,
Quantum Electrodynamics, Nauka, Moscow, 1969.
\bibitem{eichler}
J. Eichler, W. Meyerhof, Relativistic Atomic Collisions,
Academic Press, San Diego, 1995.
\bibitem{low52}
F.E. Low, Phys. Rev. 88 (1952) 53.
\bibitem{braun88}
M.A. Braun, Zh. Eksp. Teor. Fiz. 94 (1988) 145.
\bibitem{gorshkov}
V.G. Gorshkov, L.N. Labzowsky, A.A. Sultanaev,
Zh. Eksp. Teor. Fiz. 96 (1989) 53
[Sov. Phys. JETP 69 (1989) 28].
\bibitem{nef94}
A.V. Nefiodov, V.V. Karasiov, V.A. Yerokhin,
 Phys. Rev. A 50 (1994) 4975.
\bibitem{wich56}
E.H. Wichmann, N.M. Kroll, Phys. Rev. 101 (1956) 843.
\bibitem{man73}
N.L. Manakov, L.P. Rapoport, S.A. Zapryagaev,
Phys. Lett. A 43 (1973) 139.
\bibitem{mohr74}
P.J. Mohr, Ann. Phys. 88 (1974) 26, 52.
\bibitem{gyul75}
M. Gyulassy, Nucl. Phys. A 244 (1975) 497.
\bibitem{soff88}
G. Soff, P. Mohr,  Phys. Rev. A 38 (1988) 5066.
\bibitem{drake81}
G.W.F. Drake, S.P. Goldman, Phys. Rev. A 23 (1981) 2093.
\bibitem{grant82}
I.P. Grant, Phys. Rev. A 25 (1982) 1230.
\bibitem{johnson88}
W.R. Johnson, S.A. Blundell, J. Sapirstein,
Phys. Rev. A 37 (1988) 307.
\bibitem{salomon89}
S. Salomonson,  P. Oster, Phys. Rev. A 40 (1989) 5548.
\bibitem{sapir96}
J. Sapirstein, W.R. Johnson, J. Phys. B 29 (1996) 5213.
\bibitem{grant00}
I.P. Grant, H.M. Quiney, Phys. Rev. A 62 (2000) 022508.
\bibitem{shab91vir}
V.M. Shabaev, J. Phys. B 24 (1991) 4479.
\bibitem{baranger53}
M. Baranger, H.A. Bethe, and R.P. Feynman,
Phys. Rev. 92 (1953) 482.
\bibitem{snyder91}
N.J. Snyderman, Ann. Phys. 211 (1991) 43.
\bibitem{yerokh99}
V.A. Yerokhin,  V.M. Shabaev, Phys. Rev. A 60 (1999) 800.
\bibitem{rink75}
G.A. Rinker and L. Wilets, Phys. Rev. A 12 (1975) 748.
\bibitem{pers96a}
H. Persson, S. Salomonson, P. Sunnergren, and
I. Lindgren,  Phys. Rev. Lett.  76 (1996) 204.
\bibitem{sun98a}
P. Sunnergren, Ph. D. thesis, G\"oteborg University and Chalmers
University of Technology, G\"oteborg, 1998.
\bibitem{pers96b}
H. Persson, S.M. Schneider, G. Soff, W. Greiner,
 I. Lindgren, Phys. Rev. Lett. 76 (1996) 1433.
\bibitem{blund97}
S.A. Blundell, K.T. Cheng, J. Sapirstein,
  Phys. Rev. A {55} (1997) 1857.
\bibitem{shab97hf}
V.M. Shabaev, M. Tomaselli, T. K\"uhl, A.N. Artemyev,
 V.A. Yerokhin, Phys. Rev. A {56} (1997) 252.
\bibitem{pers97}
H. Persson, S. Salomonson, P. Sunnergren, I. Lindgren,
{Phys. Rev. A} {56} (1997) R2499.
\bibitem{shab98hf}
V.M. Shabaev, M.B. Shabaeva, I.I. Tupitsyn, V.A. Yerokhin, 
A.N. Artemyev, T. K\"uhl, M. Tomaselli,   O.M.  Zherebtsov, 
{Phys. Rev. A} {57}, (1998) 149;
{58} (1998) 1610.
\bibitem{sun98b}
P. Sunnergren, H. Persson, S. Salomonson, S. M. Schneider,
I. Lindgren,  G. Soff, Phys. Rev. A { 58} (1998) 1055.
\bibitem{bei00}
T. Beier, I. Lindgren, H. Persson, S. Salomonson,  P.
Sunnergren, H. H\"affner, N. Hermanspahn,
 Phys. Rev. A 62 (2000) 032510.
\bibitem{franosch91}
T. Franosch, G. Soff, Z. Phys. D 18 (1991) 219.
\bibitem{and00}
D. Andrae, Phys. Rep. 336 (2000) 413.
\bibitem{shab93fs}
V.M. Shabaev, J. Phys. B 26 (1993) 1103.
\bibitem{desid71}
A.M. Desiderio,  W.R. Johnson, Phys. Rev. A 3 (1971) 1287.
\bibitem{brown59}
G.E. Brown, J.S. Langer,  G.W Schaefer, Proc. R. Soc.
London, Ser. A 251 (1959) 92.
\bibitem{blund91}
S.A. Blundell,  N.J. Snyderman, Phys. Rev. A 44 (1991)
R1427.
\bibitem{ind92}
P. Indelicato, P.J. Mohr, Phys. Rev. A 46 (1992) 172.
\bibitem{pers93}
H. Persson, I. Lindgren,  S. Salomonson, Phys. Scr. T 46
(1993) 125.
\bibitem{quiney}
H.M. Quiney, I.P. Grant, Phys. Scr. T 46 (1993) 132;
J. Phys. B 27 (1994) L299.
\bibitem{mohr92}
P.J. Mohr, Phys.~Rev.~A  46 (1992) 4421.
\bibitem{indelicato98}
 P. Indelicato, P.J. Mohr, Phys.~Rev.~A  58
 (1998) 165.
\bibitem{mohr93}
P.J. Mohr, G. Soff, Phys. Rev. Lett  70 (1993) 158.
\bibitem{beier98}
T. Beier, P.J. Mohr, H. Persson,  G. Soff,
Phys. Rev. A 58 (1998) 954.
\bibitem{jent99}
U.D. Jentschura, P.J. Mohr,  G. Soff,
Phys. Rev. Lett. 82 (1999) 53.
\bibitem{jent00}
U.D. Jentschura, P.J. Mohr, G. Soff,
Phys. Rev. A 63 (2001) 042512.
\bibitem{manakov89}
N.L. Manakov, A.A. Nekipelov, A.G. Fainstein,
Sov. Phys. JETP { 68} (1989) 673 [Zh. Eksp. Teor. Fiz. 
95 (1989) 1167].
\bibitem{beier97a}
T. Beier, G. Plunien, M. Greiner, G. Soff, J. Phys. B 30
(1997) 2761.
\bibitem{persson93}
H. Persson, I. Lindgren, S. Salomonson, P. Sunnergren,
Phys. Rev. A 48 (1993) 2772.
\bibitem{beier97b}
T. Beier, P.J. Mohr, H. Persson, G. Plunien,
M. Greiner, G. Soff, Phys. Lett. A 236
(1997) 329.
\bibitem{mitrushenkov95}
A. Mitrushenkov, L.N. Labzowsky, I. Lindgren,
H. Persson, S. Salomonson, Phys. Lett. A 200
(1995) 51.
\bibitem{mallampalli98a}
S. Mallampalli,  J. Sapirstein, Phys. Rev. A { 57}
(1998) 1548.
\bibitem{mallampalli98b}
S. Mallampalli, J. Sapirstein, Phys. Rev. Lett. {80}
(1998) 5297.
\bibitem{yerokh00a}
V.A. Yerokhin, Phys. Rev. A 62 (2000) 012508.
\bibitem{yerokh00b}
V.A. Yerokhin, Phys. Rev. Lett. 86 (2001) 1990.
\bibitem{goi00}
I. Goidenko, L. Labzowsky, A. Nefiodov, G. Plunien,
G. Soff, S. Zschocke, Hyperfine Interactions 127 (2000) 293.
\bibitem{pach98}
K. Pachucki, Hyperfine Interactions 114 (1998) 55.
\bibitem{pach00}
K. Pachucki, Phys. Rev. A 63 (2001) 042503.
\bibitem{artemyev95a}
V.M. Artemyev, V.M. Shabaev,  V.A Yerokhin, 
{Phys. Rev. A} { 52} (1995) 1884.
\bibitem{artemyev95b}
V.M. Artemyev, V.M. Shabaev, 
V.A Yerokhin, 
{J. Phys. B} { 28} (1995) 5201.
\bibitem{sh98recpra}
V.M. Shabaev, A.N. Artemyev, T. Beier, G. Plunien, 
V.A. Yerokhin,  G. Soff, Phys. Rev. A  57
(1998) 4235.
\bibitem{sh99recpst}
V.M. Shabaev, A.N. Artemyev, T. Beier, G. Plunien, 
V.A. Yerokhin,  G. Soff, Phys. Scr. T  80
(1999) 493.
\bibitem{sh98recjpb}
V.M. Shabaev, A.N. Artemyev, T. Beier,  G. Soff, 
J. Phys. B  31 (1998) L337.
\bibitem{plunien95}
G. Plunien, G. Soff,  Phys. Rev. A 51 (1995) 1119;
 53 (1996) 4614.
\bibitem{nefiodov96}
A.V.  Nefiodov, L.N.  Labzowsky, G. Plunien, G. Soff, 
Phys. Lett. A  222 (1996) 227.
\bibitem{mohr00a}
P.J. Mohr, B.N. Taylor, Rev. Mod. Phys. 72 (2000) 351.
\bibitem{beyer95a}
H.F. Beyer, IEEE Trans. Instrum. Meas. 44 (1995) 510.
\bibitem{beyer95b}
H.F. Beyer, G. Menzel, D. Liesen, A. Gallus, F. Bosch,
R. Deslattes, P. Indelicato, T. St\"ohlker,
O. Klepper, R. Moshammer, F. Nolden, H. Eickhoff, B. Franzke,
M. Steck, Z. Phys. D 35 (1995) 169.
\bibitem{stoehlker00}
T. St\"ohlker, P.H. Mokler, F. Bosch, R.W.  Dunford, 
O. Klepper,
C. Kozhuharov,  T. Ludziejewski,
F. Franzke, F. Nolden, H. Reich,
P. Rymuza, Z. Stachura, M. Steck,
P. Swiat, A. Warczak, Phys. Rev. Lett. 85 (2000) 3109.
\bibitem{drake88}
G.W. Drake, Can. J. Phys. 66 (1988) 586.
\bibitem{mohr00b}
P.J. Mohr, J. Sapirstein, Phys. Rev. A 62 (2000) 052501.
\bibitem{marrs95}
R.E. Marrs, S.R. Elliot, T. St\"ohlker,
Phys. Rev. A 52 (1995) 3577.
\bibitem{sanders69}
F.C. Sanders, C.W. Scherr, Phys. Rev. 181 (1969) 84.
\bibitem{schweppe91}
J. Schweppe, A. Belkacem, L. Blumenfeld, N. Claytor, B. Feinberg,
H. Gould, V. E. Kostroun, L. Levy, S. Misawa, J. R. Mowat,  M.
H. Prior, Phys. Rev. Lett.  66 (1991) 1434 .
\bibitem{beiersdorfer1}
P. Beiersdorfer, A.L. Osterheld, J.H. Scofield, 
J.R. Crespo-L\'opez-Urrutia, K. Widmann,
{ Phys. Rev. Lett.} {80} (1998) 3022.
\bibitem{staude98}
U. Staude, Ph. Bosselman, R. B\"uttner, D. Horn, K.-H. Schartner, F.
Folkmann, A.E. Livingston, T. Ludziejewski, P.H. Mokler,
Phys. Rev. A  58 (1998) 3516.
\bibitem{bosselmann99}
Ph. Bosselmann, U. Staude, D. Horn, K.-H. Schartner, F. Folkmann, A. E.
Livingston, P.H. Mokler, Phys. Rev. A 59 (1999) 1874.
\bibitem{y0a}
V.A. Yerokhin, A.N. Artemyev, V.M. Shabaev, M.M. Sysak,
O.M. Zherebtsov, G. Soff, Phys. Rev. Lett. 85 (2000) 4699.
\bibitem{ind91}
P. Indelicato, P.J. Mohr, Theor. Chem. Acta 80 (1991) 207.
\bibitem{cheng91}
K.T. Cheng, W.R. Johnson, J. Sapirstein,
Phys. Rev. Lett. 66 (1991) 2960.
\bibitem{mohr93t}
P.J. Mohr, Phys. Scr. T 46 (1993) 44.
\bibitem{blund93a}
S.A. Blundell, Phys. Rev. A 47 (1993) 1790.
\bibitem{lind93}
I. Lindgren, H. Persson, 
S. Salomonson, A. Ynnerman, Phys. Rev. A 47
(1993) R4555.
\bibitem{chen95}
M.H. Chen, K.T. Cheng, W.R. Johnson, J. Sapirstein,
Phys. Rev. A 52 (1995) 266.
\bibitem{sh00hi}
V.M. Shabaev, V.A. Yerokhin, O.M. Zherebtsov, A.N. Artemyev,
M.M. Sysak, G. Soff, Hyperfine Interactions 132 (2001) 339.
\bibitem{shab94hf}
V.M. Shabaev, {J. Phys. B} { 27} (1994) 5825.
\bibitem{shab99hf}
V.M. Shabaev, In H.F. Beyer, V.P. Shevelko (eds.), 
 Atomic Physics with Heavy Ions,
Springer, 1999, p. 139.
\bibitem{klaft1}
I. Klaft, S. Borneis, T. Engel, B. Fricke, R. Grieser, 
G. Huber, T. K\"uhl, D. Marx, R. Neumann, S. Schr\"oder,
P. Seelig, L. V\"olker, {Phys. Rev. Lett.} {73} (1994) 2425.
\bibitem{crespo1}
J.R. Crespo Lopez-Urrutia, P. Beiersdorfer, D. Savin, 
 K. Widmann, {Phys. Rev. Lett.} {77} (1996) 826.
\bibitem{crespo2}
J.R. Crespo Lopez-Urrutia, P. Beiersdorfer, K. Widmann, 
B. Birket, A.-M. M\aa rtensson-Pendrill,  M.G.H. Gustavsson, 
{Phys. Rev. A} {57} (1998) 879.
\bibitem{seelig1}
P. Seelig, S. Borneis, A. Dax, T. Engel, S. Faber, M. Gerlach, 
C. Holbrow, G. Huber, T. K\"uhl, D. Marx, K. Meier, P. Merz, 
W. Quint, F. Schmitt, M. Tomaselli, L. V\"olker, M. W\"urtz, 
K. Beckert, B. Franzke, F. Nolden, H. Reich, M. Steck, 
T. Winkler, {Phys. Rev. Lett.} {81} (1998) 4824.
\bibitem{raghavan1}
P. Raghavan, {At. Data Nucl. Data Tables} {\bf 42}, 189
(1989).
\bibitem{gustavsson1}
M.G.H. Gustavsson,  A.-M. M\aa rtensson-Pendrill,
{ Phys. Rev. A}, {58} (1998) 3611. 
\bibitem{lutz1}
O. Lutz, G. Stricker, Phys. Lett. A {35} (1971) 397.
\bibitem{gibbs1}
H.M. Gibbs, C.M. White, Phys. Rev. A {188} (1969) 180.
\bibitem{sushkov1}
O.P. Sushkov, V.V. Flambaum, I.B. Khriplovich,
Opt. Spectr. {44} (1978) 2.
\bibitem{shab98hfb}
V.M.  Shabaev, M.B. Shabaeva, I.I Tupitsyn,   V.A. Yerokhin,
Hyperfine Interactions {\bf 114}, 129 (1998).
\bibitem{shabaeva99}
M.B. Shabaeva, Opt. Spectr. 86 (1999) 368.
\bibitem{shab00hf}
V.M. Shabaev, A.N. Artemyev, O.M. Zherebtsov, V.A. Yerokhin, G. Plunien,
 G. Soff, Hyperfine Interactions 27 (2000) 279.
\bibitem{bou00}
S. Boucard, P. Indelicato, Eur. Phys. J. D 8 (2000) 59.
\bibitem{zher00hf}
O.M. Zherebtsov, V.M. Shabaev, Can. J. Phys. 78 (2000) 701.
\bibitem{art00hf}
A.N. Artemyev, V.M. Shabaev, G. Plunien, G. Soff, V.A. Yerokhin,
Phys. Rev. A, in press.
\bibitem{sapir00hf}
J. Sapirstein, K.T. Cheng,  Phys. Rev. A 63 (2001) 032506.
\bibitem{breit1}
G. Breit, Nature (London) {122} (1928) 649.
\bibitem{kar00a}
S.G. Karshenboim, Phys. Lett. A 266 (2000) 380.
\bibitem{quint1}
W. Quint, Phys. Scr. {59} (1995) 203.
\bibitem{hermanspahn1}
N. Hermanspahn, W. Quint, S. Stahl, 
M. T\"onges, G. Bollen,
H.-J. Kluge, R. Ley, R. Mann, G. Werth, Hyperfine
Interactions {99} (1996) 91.
\bibitem{hermanspahn2}
N. Hermanspahn, H. H\"affner,
H.-J. Kluge, W. Quint, S. Stahl, J. Verdu,
 and G. Werth, Phys. Rev. Lett. 84 (2000) 427.
\bibitem{haffner}
H. H\"affner, T. Beier, N. Hermanspahn, H.-J. Kluge,
W. Quint, S. Stahl, J. Verdu, G. Werth,
Phys. Rev. Lett. 85 (2000) 5308.
\bibitem{kar00b}
S.G. Karshenboim, in: S. Karshenboim et al. (eds.), The Hydrogen Atom,
Berlin, Springer, 2001, p.651; arXiv: hep-ph/0008227 (2000)
(http://xxx.lanl.gov).
\bibitem{shabaev98can}
V.M.  Shabaev, Can. J. Phys. {76} (1998) 907.
\bibitem{winter1}
H. Winter, S. Borneis, A. Dax, S. Faber, T. K\"uhl, 
D. Marx, F. Schmitt, P. Seelig, W. Seelig, V.M. Shabaev,
M. Tomaselli,  M. W\"urtz, in
GSI scientific report 1998, edited by U. Grundinger (GSI,
Darmstadt, Germany, 1999), p. 87.
\bibitem{pal00}
V.G. Pal'chikov, Hyperfine Interactions 127 (2000) 287.
\bibitem{ichihara94}
A. Ichihara, T. Shirai, J. Eichler, Phys. Rev. A 49
(1994) 1875.
\bibitem{eichler95a}
J. Eichler, A. Ichihara, T. Shirai, Phys. Rev. A
51 (1995) 3027.
\bibitem{stoehlker95} Th. St\"ohlker, C. Kozhuharov, P.H. Mokler, A.
    Warczak, F. Bosch, H. Geissel, R. Moshammer, C.Scheidenberger, J.
    Eichler, A. Ichihara, T. Shirai, Z. Stachura, P. Rymuza, Phys.
    Rev. A 51 (1995) 2098.
\bibitem{stoehlker99}
Th. St\"ohlker, T. Ludziejewski, F. Bosch, R.W. Dunford,
C. Kozhuharov, P.H. Mokler, H.F. Beyer, O. Brinzanescu, F. Franzke,
J. Eichler, A. Griegal, S. Hagmann, A. Ichihara, A. Kr\"amer,
D. Liesen, H. Reich, P. Rymuza, Z. Stachura, M. Steck, P. Swiat,
A. Warczak, Phys. Rev. Lett. 82 (1999) 3232.
\bibitem{Jauch}
J.M. Jauch, F.
Rohrlich, The Theory of Photons and Electrons,
Springer-Verlag, Berlin, 1976.
\bibitem{Yennie}
 D.R. Yennie, S.C. Frautschi, H. Suura,
Ann. Phys. {13} (1961) 379.
%
\bibitem{yer93}
V.A.Yerokhin, Diploma thesis, St.Petersburg State University, 
St.Petersburg, 1993.
\bibitem{shir81}
A.S. Shirokov, Candidate of Science thesis,
 Leningrad State University, Leningrad, 1981.
\bibitem{vanhove}
L. Van Hove, Physica 21 (1955) 901.
\bibitem{hugenholtz}
N.M. Hugenholtz, Physica 23 (1957) 481.
\bibitem{bethe}
H.A. Bethe,  Intermediate Quantum Mechanics,
 Benjamin, New York, 1964.

\end{thebibliography}
\end{document}